\documentclass[11pt, a4paper]{ociamthesis}
\pdfoutput=1

\usepackage[sumlimits]{amsmath}
\usepackage{amsfonts,amssymb, amsthm, mathrsfs, mathtools, amscd}
\usepackage{graphicx, longtable, caption, afterpage, float}
\usepackage{xspace}
\usepackage[latin1]{inputenc}
\usepackage[usenames,dvipsnames]{color}
\usepackage{rotating}
\usepackage{tikz}
\usetikzlibrary{arrows,shapes}
\usepackage{setspace}
\usepackage{epsfig,cancel,cite}
\usepackage[linktoc=page]{hyperref}

\usepackage{array}

\font\twentyonerm=cmr12 at 21pt
\font\eightrm=cmr8 at 8pt

\definecolor{dg}{rgb}{.1,0.5,.5}

\newcommand{\varstr}[2]{\vrule height #1 depth #2 width0pt}
\renewcommand{\a}{\alpha}

\newcommand{\D}{\Delta}

\renewcommand{\o}{\omega}

\newcommand{\calO}{\mathcal{O}}

\newcommand{\cA}{\mathcal{A}}

\newcommand{\cM}{\mathcal{M}}
\newcommand{\cN}{\mathcal{N}}
\newcommand{\cO}{\mathcal{O}}

\newcommand{\cV}{\mathcal{V}}

\newcommand{\IC}{\mathbb{C}}
\newcommand{\IH}{\mathbb{H}}
\newcommand{\IP}{\mathbb{P}}
\newcommand{\IR}{\mathbb{R}}

\newcommand{\IZ}{\mathbb{Z}}

\newcommand{\cicy}[2]{\begin{matrix} #1\end{matrix}\!\left[\begin{matrix}#2 \end{matrix}\right]}

\newcommand{\ba}{\begin{array}}
\newcommand{\ea}{\end{array}}

\newcommand{\beqn}{\begin{equation*}}
\newcommand{\eeqn}{\end{equation*}}

\theoremstyle{plain}

\newtheorem*{thmm}{Theorem}

\theoremstyle{definition}

\theoremstyle{remark}

\newcommand{\beq}{\begin{equation}}
\newcommand{\eeq}{\end{equation}}
\newcommand{\beqnn}{\begin{equation*}}
\newcommand{\eeqnn}{\end{equation*}}

\newcommand{\fref}[1]{Figure~\!\ref{#1}}
\newcommand{\tref}[1]{Table~\ref{#1}}

\newcommand{\hodgenos}{(h^{1,1},\,h^{2,1})}
\newcommand{\one}{{\hskip-0.75pt\bf 1}}                                     

\def\place#1#2#3{\vbox to0pt{\kern-\parskip\kern-7pt
                             \kern-#2truein\hbox{\kern#1truein #3}
                             \vss}\nointerlineskip}
                        
\newcommand{\capt}[3]{\parbox{#1}{\renewcommand{\baselinestretch}{1.0}
                                                           \caption{\label{#2}\small\it #3}}}

\newcommand{\cy}{Calabi-Yau\xspace}

\newcommand{\cys}{Calabi-Yau manifolds\xspace}

\newcommand{\SDelta}{\Delta^{^{\!\!*}}}

\newcommand{\Top}[1]{\langle #1 |}    
\newcommand{\Bot}[1]{| #1 \rangle}    
\newcommand{\topbot}[2]{\langle #1 | #2 \rangle}

\newcommand{\twoheadlongrightarrow}{\relbar\joinrel\twoheadrightarrow}

\newcommand{\Bigcheck}{\lower2pt\hbox{\smash{\hbox{{\twentyonerm \v{}}}}}}
\newcommand{\Bighat}{\lower3.8pt\hbox{\smash{\hbox{{\twentyonerm \^{}}}}}}

\newcommand{\Xsharp}{\mathscr{X}^{\raisebox{2pt}{$\scriptstyle\sharp$}}}

\newcommand{\Xcheck}{\kern2pt\hbox{\Bigcheck\kern-15pt{$\mathscr{X}$}}}
\newcommand{\Xhat}{\kern2pt\hbox{\Bighat\kern-12pt{$\mathscr{X}$}}}

\newcommand*{\myalign}[2]{\multicolumn{1}{#1}{#2}}

\renewcommand{\baselinestretch}{1.1}

\numberwithin{equation}{section}

\title{Heterotic String Models  on Smooth Calabi-Yau Threefolds}
\author{Andrei Constantin}
\college{University College}
\degree{Doctor of Philosophy}
\degreedate{Trinity Term 2013}

\begin{document}

\maketitle

\setstretch{1.12}
{\small
\begin{abstractseparate}

This thesis contributes with a number of topics to the subject of string compactifications, especially in the instance of the $E_8\times E_8$ heterotic string theory compactified on smooth Calabi-Yau threefolds. 

In the first half of the work, I discuss the Hodge plot associated with Calabi-Yau threefolds that are hypersurfaces in toric varieties. The intricate structure of this plot is explained by the existence of certain webs of elliptic-$K3$ fibrations, whose mirror images are also elliptic-K3 fibrations. Such manifolds arise from reflexive polytopes that can be cut into two parts along slices corresponding to the $K3$ fiber. Any two half-polytopes over a given slice can be combined into a reflexive polytope. This fact, together with a remarkable relation on the additivity of Hodge numbers, give to the Hodge plot the appearance of a fractal. 

Moving on, I discuss a different type of web of manifolds, by looking at smooth $\IZ_3$-quotients of Calabi-Yau three-folds realised as complete intersections in products of projective spaces.  Non-simply connected Calabi-Yau three-folds provide an essential ingredient in heterotic string compactifications. Such manifolds are rare in the classical constructions, but they can be obtained as quotients of homotopically trivial Calabi-Yau three-folds by free actions of finite groups. Many of these quotients are connected by conifold transitions. 

In the second half of the work, I explore an algorithmic approach to constructing $E_8\times E_8$ heterotic compactifications using holomorphic and poly-stable sums of line bundles over complete intersection Calabi-Yau three-folds that admit freely acting discrete symmetries. 
Such Abelian bundles lead to  $\cN=1$  supersymmetric GUT theories with gauge group $SU(5)\times U(4)$ and matter fields in the $\mathbf{10}$, $\mathbf{\overline{10}}$, $\mathbf{\overline{5}}$, $\mathbf{5}$ and $\mathbf{1}$ representations of $SU(5)$. The extra $U(1)$ symmetries are generically Green-Schwarz anomalous and, as such, they survive in the low energy theory only as global symmetries. These, in turn, constrain the low energy theory and in many cases forbid the existence of undesired operators, such as dimension four or five proton decay operators. The line bundle construction allows for a systematic computer search resulting in a plethora of models with the exact matter spectrum of the Minimally Supersymmetric Standard Model, one or more pairs of Higgs doublets and no exotic fields charged under the Standard Model group. In the last part of the thesis I focus on the case study of a Calabi-Yau hypersurface embedded in a product of four $\IC\IP^1$ spaces, referred to as the tetraquadric manifold.  I address the question of the finiteness of the class of consistent and physically viable line bundle models constructed on this manifold. 

Line bundle sums are part of a moduli space of non-Abelian bundles and they provide an accessible window into this moduli space. I explore the moduli space of heterotic compactifications on the tetraquadric hypersurface around a locus where the vector bundle splits as a direct sum of line bundles, using the monad construction. The monad construction provides a description of poly-stable $S\left(U(4)\times U(1) \right)$--bundles leading to GUT models with the correct field content in order to induce standard-like models. These deformations represent a class of consistent non-Abelian models that has co-dimension one in K\"ahler moduli space.

\end{abstractseparate}
}
\setstretch{1.5}
{\setstretch{1.54}
\begin{acknowledgements}
\vspace{-12pt}
This thesis has been shaped and grew along with its author inspired by the joint guidance of Prof.~Philip Candelas and Prof.~Andr\'e Lukas. I am very much indebted to them, considering myself blessed to be able to learn and work in their company. 

\bigskip
The ideas presented here wouldn't have acquired their fullness without the creative work of my co-laborators, Dr.~Lara Anderson, Dr.~Evgeny Buchbinder, Dr.~James Gray, Dr.~Eran Palti and Dr.~Harald Skarke. I am grateful for their collaboration and friendship.
Special thanks go to Dr.~Volker Braun, Dr.~Rhys Davis, Prof.~Xenia de la Ossa, Prof.~Yang Hui-He and Prof.~Graham Ross for their advice, inspiration and accurate knowledge shared with me in our conversations. 
I am grateful for the friendship and support of many of my colleagues (some of which have already graduated), especially of Maxime Gabella, Georgios Giasemidis, Michael Klaput, Cyril Matti, Challenger Mishra, Chuang Sun, Eirik Svanes and Lukas Witkowski.

\bigskip
My DPhil studies and the present thesis  wouldn't have been possible without the generous support received through the Bobby Berman scholarship offered by University~College, Oxford and complemented by the STFC. The final year of my studies was partly supported through the College Lectureship offered by Brasenose College and I would like to thank Prof.~Laura Herz and Prof.~Jonathan Jones for giving me the chance to partake in this facet of Oxford's academic life. 

\bigskip
Last but not least, I would like to express my sincere gratitude towards my wife Carmen Maria for her constant love, friendship and support and for being my constitutive Other; to our children Elisabeta, Clara Theodora and Cristian for brightening up my life in countless ways; to my parents Carmen and Marin, my sister Alina and our extended family for their unconditional love and support. There are of course many more people who contributed to my becoming during these years and close friends whose advice, help and presence meant a lot for me. To all these people I am deeply grateful and indebted. 


\end{acknowledgements}
\begin{dedication}

{\itshape To my family.}

\end{dedication}
\begin{romanpages}
\tableofcontents
\end{romanpages}
}

\chapter{Introduction}

The way in which string theory has come to light and developed is, undoubtedly, one of the most curious ones in the history of science \cite{Cappelli:2012cto}. Brought about by accident through Veneziano's scattering amplitude \cite{Veneziano:1968yb}, re-born by the discovery of anomaly cancellation via the Green-Schwarz mechanism~\cite{Green:1984sg}, revolutionised by the discovery of string dualities and the invention of D-branes \cite{Horava:1995qa, Polchinski:1995mt, Maldacena:1997re, Witten:1998qj}, a myth for many, scorned as the stumbling block of modern physics by others \cite{Woit:2006js, Smolin}, string theory marks a very particular phenomenon of the mathematical culture of our time. 

It is difficult (and perhaps pointless) to attempt to evaluate the relevance of string theory for physics or mathematics. And it is certainly beyond the intention of this thesis to do that. But it is hard to refrain from making a few remarks. The point of view adopted by the author is that the appearance of string theory is a phenomenon in modern science just as necessary as the departure from the attempt of explicitly solving polynomial equations which led Galois to lay the foundations of group theory. It has been argued\footnote{See e.g.~Smolin's book `The Problem with Physics' \cite{Smolin}.} that string theory has pushed to the extreme a paradigm initiated with the uprise of quantum (field) theory, in which one would evade understanding the deep physical meaning of mathematical equations in favour of deriving the ultimate consequences of an essentially hermetical theory. But this captures only one of the many facets of string theory.

`Theory' is, in fact, a bad word in this context, as string theory is more of a locus in which many, perhaps all ideas of modern physics and mathematics meet each other, collide, interact, reshape, react, fuse and produce new syntheses. For this matter, it is rather a `theoretical' laboratory reflecting the dynamics of our mathematical culture. It is a nexus of ideas where quantum gravity and gauge theories react with algebra and geometry in order to shed light upon the very nature of these theories. It is this perspective that motivated the present work. 

Mathematics has departed once and for all from attempting to solve problems explicitly and created in the last century a what looks to the uninitiated as a labyrinth of abstractions. These abstractions became part of our culture, and in this sense became realities, just as a polynomial equation has its own reality. Referring to this, C.~Yang (the co-autor of the famous Yang-Mills equations) confesses in a conversation with Chern:\footnote{quoted from \cite{Monastyrsky}, p.135} ``I was struck that gauge fields, in particular, connections on fibre bundles, were studied by mathematicians without any appeal to physical realities. I added that it is mysterious and incomprehensible how you mathematicians would think this up out of nothing. To this Chern immediately objected. `No, no, this concept is not invented - it is natural and real.'~"

It was impossible that these new mathematical realities which form the largest portion of our current mathematical culture, would not `contaminate' the realm of theoretical physics - mathematicians and physicists, after all, live in the same universe. The result of this interaction is most visible in string theory.

\section{The Heterotic String}
One of the earliest questions in string theory has been: what kind of geometry realises the Standard Model of Particle Physics as a four-dimensional, low-energy limit of string theory\footnote{For general material on String Theory, see the classical texts by Green, Schwarz and Witten \cite{GSW}, Polchinski \cite{Polchinski} and Zwiebach \cite{Zwiebach}. For more recent accounts, see \cite{Dine, Kiritsis, BBS, BLT}.}. And one of the earliest proposals for an answer to this question was the heterotic setup. In 1985, the creators of this framework wrote: ``Although much work remains to be done, there seem to be no insuperable obstacles to deriving all of known physics from the $E_8\times E_8$ heterotic string." \cite{Gross:1985fr} Almost 30 years later, we don't have much to boast about in this respect: we are unable to derive the mass of the electron or any other non-integer parameter in the Standard Model.

On the other hand, it is precisely because of this failure that string theory acquired its perplexing richness of today. The work collected in this thesis will reflect the pattern described above. A good portion of the present work is aimed at constructing heterotic string models exhibiting many features of the (supersymmetric) Standard Model. The other part has the flavour of solving a mathematical puzzle. But the two parts are interconnected. The puzzle emerged through the work of many researchers on the heterotic string. In turn, our work on heterotic model building generated a few other questions of mathematical interest; whenever this happens, I will stop to point them out.

\subsection{The Basics}
In 1985, David Gross, Jeffrey Harvey, Emil Martinec, and Ryan Rohm published a series of papers in which the 26-dimensional bosonic string and the 10-dimensional superstring were combined in a theory of first quantised, orientable closed strings, known ever since as the heterotic string \cite{Gross:1985fr, Gross:1985rr, Gross:1984dd}. The construction treats left and right moving modes separately. This separation is justified due to the fact that the quantum states of the closed string, as well as the one-particle operators and the vertex operators do not mix the left and the right moving modes. 

Several appealing features characterise this mixed setup. On one hand, it satisfies the requirements of the absence of ghost states and tachyons, as well as Lorenz invariance, just as in the supersymmetric string theory. On the other hand, it comes with a bonus. The bosonic modes require 26 dimensions, while the supersymmetric ones require only 10. In this setup, gauge and gravitational anomalies are absent only when the mismatched 16 dimensions are compactified  on a maximal torus\footnote{A maximal torus is a product of circles of equal radii.} on which points are identified on an integral, even, self-dual lattice\footnote{An integral lattice is a free abelian group of finite rank with an integral symmetric bilinear form $(\cdot,\cdot)$; it is even if $(a,a)$ is even.}. In 16 dimensions there are two possible such lattices: the lattice of weights of $Spin(32)/\mathbb{Z}_2$ and the direct product of two copies of the lattice of weights of $E_8$. Choosing one lattice or another, leads to two types of the heterotic string, whose low-energy limits are 10 dimensional $\cN=1$ supergravity coupled to super Yang-Mills theory with gauge groups $Spin(32)/\mathbb{Z}_2$ and $E_8\times E_8$, respectively. The result of this is that the heterotic string comes with a natural gauge group which is large enough to contain the Standard Model gauge group as a subgroup. For this purpose, the $E_8\times E_8$ version of the heterotic string is the most appealing. 

One of the great virtues of the heterotic string is that it leads to a chiral low-energy theory. As usual in quantum field theory, the presence of chiral fermions leads to potential anomalies. Intriguingly enough, such anomalies are absent in the 10 dimensional $\cN=1$ supergravity theory coupled to super Yang-Mills theory precisely when the gauge group is $Spin(32)/\mathbb{Z}_2$ or $E_8\times E_8$.\footnote{This result was obtained about a year before the formulation of the heterotic string. The supergravity anomaly cancellation could only be linked with the type I superstring with gauge group $SO(32)$ known at the time.} This cancellation happens in a non-trivial way due to the presence of a Chern-Simons term in the effective action of the heterotic string theory. The anomaly cancellation mechanism was uncovered by Michael Green and John Schwarz and goes under their name \cite{Green:1984sg}. 

{\setstretch{1.42}

\subsection{From 10 to 4 Dimensions}
The heterotic string employs the old idea of Kaluza and Klein \cite{Kaluza:1921tu, Klein:1926tv, Klein:1926fj, Duff:1994tn}, known as dimensional reduction, in order to bring the theory from its natural realm of existence to the four dimensional, low-energy context. In doing this, it requires the choice of a compact six dimensional manifold~$X$. Moreover, as the heterotic string comes with a natural set of gauge degrees of freedom, one can divide these into an `internal' part, associated with a vector bundle $V\rightarrow X$, and an `external' part. The latter represents the gauge degrees of freedom in the 4 dimensional theory, which usually, in the low-energy limit, corresponds to one of the known GUTs. The quantum numbers of the low-energy states are then determined by topological invariants of $X$ and $V$. 

There are many constraints on $X$ and $V$ that enter in this construction. I'll mention a number of these here. If one requires that the 4 dimensional space is maximally symmetric (as our current cosmological models suggest) and if one is keen to preserve $\cN=1$ supersymmetry in 4 dimensions (at least at the energy scale at which the compactification takes place), then $X$ must be a Calabi-Yau three-fold \cite{Candelas:1985en}.\footnote{However, there are conceivable ways to evade this theorem. One can give up for a moment the requirement of maximally symmetric 4d space-time and construct non-Calabi-Yau compactifications \cite{Lukas:2010mf, Klaput:2011mz, Klaput:2012vv}  (or even Calabi-Yau compactifications involving fluxes \cite{Klaput:2013nla}), hoping that maximal symmetry can be restored by non-perturbative effects, such as gaugino condensation or worldsheet instantons.} $\cN=1$ supersymmetry in 4 dimensions requires also that the vector bundle $V\rightarrow X$ is holomorphic and satisfies the hermitian Yang-Mills equations. Finally, the theory is anomaly-free if and only if $V$ and the tangent bundle $TX$ satisfy a certain topological constraint. 

If these requirements are satisfied, one can proceed to the second step in which the GUT group is reduced to the Standard Model gauge group by completing the vector bundle with a Wilson line. Should this be the case, the Calabi-Yau manifold must allow the existence of non-contractible loops. Calabi-Yau manifolds with non-trivial fundamental group are rare and the classical constructions provide a very limited number of examples. One way to obtain homotopically non-trivial Calabi-Yau manifolds, is the construction in which the points of a Calabi-Yau three-fold $X$ with trivial fundamental group are identified by the holomorphic action of a finite group~$\Gamma$. If this action is fixed point free, the quotient manifold $X/\Gamma$ is  Calabi-Yau and its fundamental group is isomorphic to $\Gamma$. This further increases the complexity of the set-up. In order to construct holomorphic vector bundles on the quotient manifold, one usually starts by constructing vector bundles $V\rightarrow X$ and then ensures that $V$ descends to a holomorphic bundle $\widetilde{V} \rightarrow X/\Gamma$. If this can be done the necessary descent data is given by a so-called equivariance structure of $V$. 

The work presented in this thesis relies on this construction. On one hand it consists of the construction of a large class of heterotic models; on the other hand it is a study of properties of Calabi-Yau manifolds, including manifolds realised as quotients by discrete group actions. 

\section{Calabi-Yau Manifolds}

The following story, heard from S.-T.~Yau\cite{Yau}, exposes the manner in which Calabi-Yau manifolds came into existence. Yau asked the question of whether there is such a thing as a (complex or supersymmetric) smooth manifold with no matter, but with gravity. In other words: is it possible to have a smooth manifold with vanishing Ricci tensor, but with non-trivial Riemann tensor? Based on intuitive arguments from general relativity, Yau attempted to prove that such mathematical objects cannot exist. On the contrary, Calabi had conjectured that such objects exist and moreover, that any K\"ahler manifold with vanishing first Chern class provides an example \cite{Calabi1954, Calabi1957}. Yau attempted to disprove Calabi's conjecture in concrete examples, but soon he realised that his arguments would always fail. This struggle determined him to revert his perspective and concentrate all efforts on proving Calabi's conjecture, rather than disproving it. The fruit of this work, published in \cite{Yau:1977}, is the following

\begin{thmm}[Calabi, Yau]
A compact, K\"ahler manifold with vanishing first Chern class admits a unique Ricci-flat K\"ahler metric in every K\"ahler class.
\end{thmm}

Calabi-Yau manifolds are ubiquitous in string theory. They come in two flavours: compact and non-compact, both of which have their place in the theory.\footnote{Non-compact Calabi-Yau manifolds appear in the study of gauge/gravity correspondence. Example: the gravity side given by type IIB string theory on $\text{AdS}_5$ times a 5 dimensional space with $S^2\times S^3$ topology is dual to the gauge theory corresponding to $N$ $\text{D}3$-branes at a singular conifold geometry. The conifold is a prototypical example of a non-compact Calabi-Yau geometry.} The heterotic string takes interest in the compact case. In (complex) dimension $1$, there exists only one type of compact Calabi-Yau manifolds: the complex elliptic curve. In two dimensions, we have the $K3$ manifold. On the other hand, the number of diffeomorphism classes of three dimensional compact Calabi-Yau manifolds is unknown.

\section{Overview and Summary}
I decided to organise the material of this thesis into two parts, `The Manifold' and `The Bundle'. The first part of `The Manifold' contains a study of Calabi-Yau three-folds realised as hypersurfaces in toric varieties which are $K3$ fibrations over $\IC\IP^1$. In the second part, I discuss quotients of Calabi-Yau three-folds realised as complete intersections of hypersurfaces in products of projective spaces by freely acting finite groups. `The Bundle' exposes the construction of heterotic string models on smooth Calabi-Yau manifolds involving vector bundles that are direct sums of holomorphic line bundles. In connection to this construction, I discuss the computation of line bundle cohomology on projective varieties. Let me briefly summarise below the results contained in this thesis.

\bigskip
To date, the largest class of compact Calabi-Yau three-folds consists of hypersurfaces in toric varieties. We know about 470 million examples of such manifolds\footnote{In fact, this is the number of isomorphism classes of reflexive polytopes in four dimensions. Calabi-Yau manifolds are obtained from reflexive polytopes in two steps: one can construct a toric variety from the fan over a triangulation of the surface of the polytope, and a Calabi-Yau hypersurface in this toric variety as the zero locus of a polynomial whose monomials are in one-to-one correspondence with the lattice points of the dual polytope. The triangulation process is, however, highly non-unique, except for the very simple polytopes. Thus the actual number of Calabi-Yau three-folds constructed in this class greatly exceeds 470 millions. To date, nobody took the pain of explicitly constructing all possible triangulations for each of the 470 million polytopes in the Kreuzer-Skarke list.}, due to Maximilian Kreuzer and Harald Skarke \cite{Kreuzer:2000xy, Kreuzer:2000qv, Kreuzer}. These manifolds were constructed in accordance with Batyrev's prescription involving four dimensional reflexive polytopes. 

\begin{figure}[h]
\begin{center}
\includegraphics[width=6.5in]{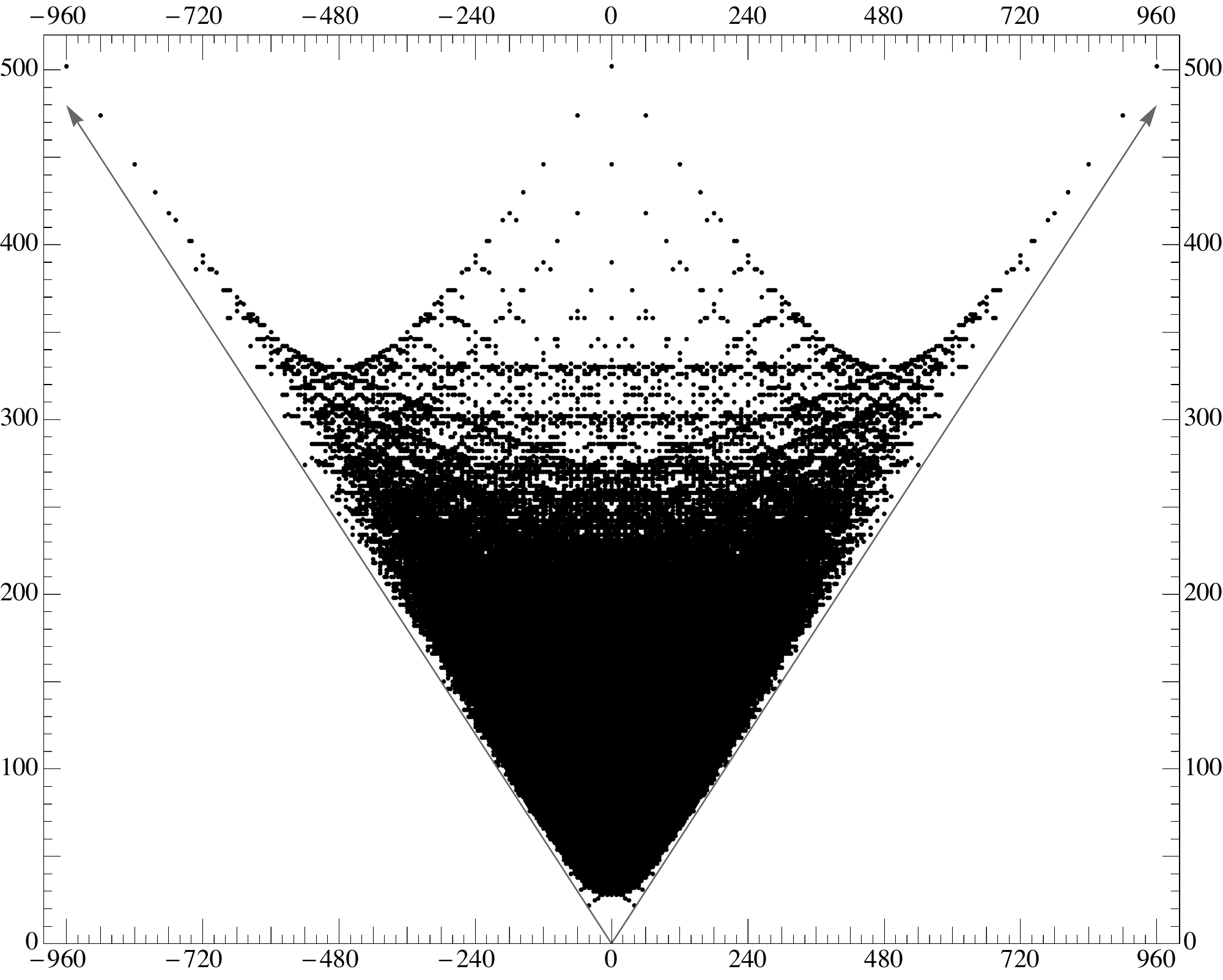}
\capt{6.5in}{BasicKSPlot1}{The Hodge plot for the list or reflexive 4-polytopes. The Euler number
$\chi = 2\left( h^{1,1}-h^{1,2} \right)$ is plotted against the height $y=h^{1,1}+h^{1,2}$. The oblique axes correspond to $h^{1,1}=0$ and $h^{1,2}=0$.}
\end{center}
\end{figure}

Certain topological invariants of these manifolds, such as the Hodge numbers $h^{1,1}$ and $h^{1,2}$ play an important role in the classification of Calabi-Yau manifolds as well as in the construction of string compactifications. A plot of the the Euler number $\chi = 2(h^{1,1} - h^{1,2})$ against $h^{1,1} + h^{1,2}$ reveals an intriguing structure, containing intricately self-similar nested patterns within patterns which give the appearance of a fractal. This is the plot shown in \fref{BasicKSPlot1}. Each point of the plot corresponds to one or several reflexive polytopes from the Kreuzer-Skarke list, as illustrated in \fref{occupationplot}, in which the colour code indicates the occupation number of each site. 

\begin{figure}[h]
\begin{center}
\includegraphics[width=6.5in]{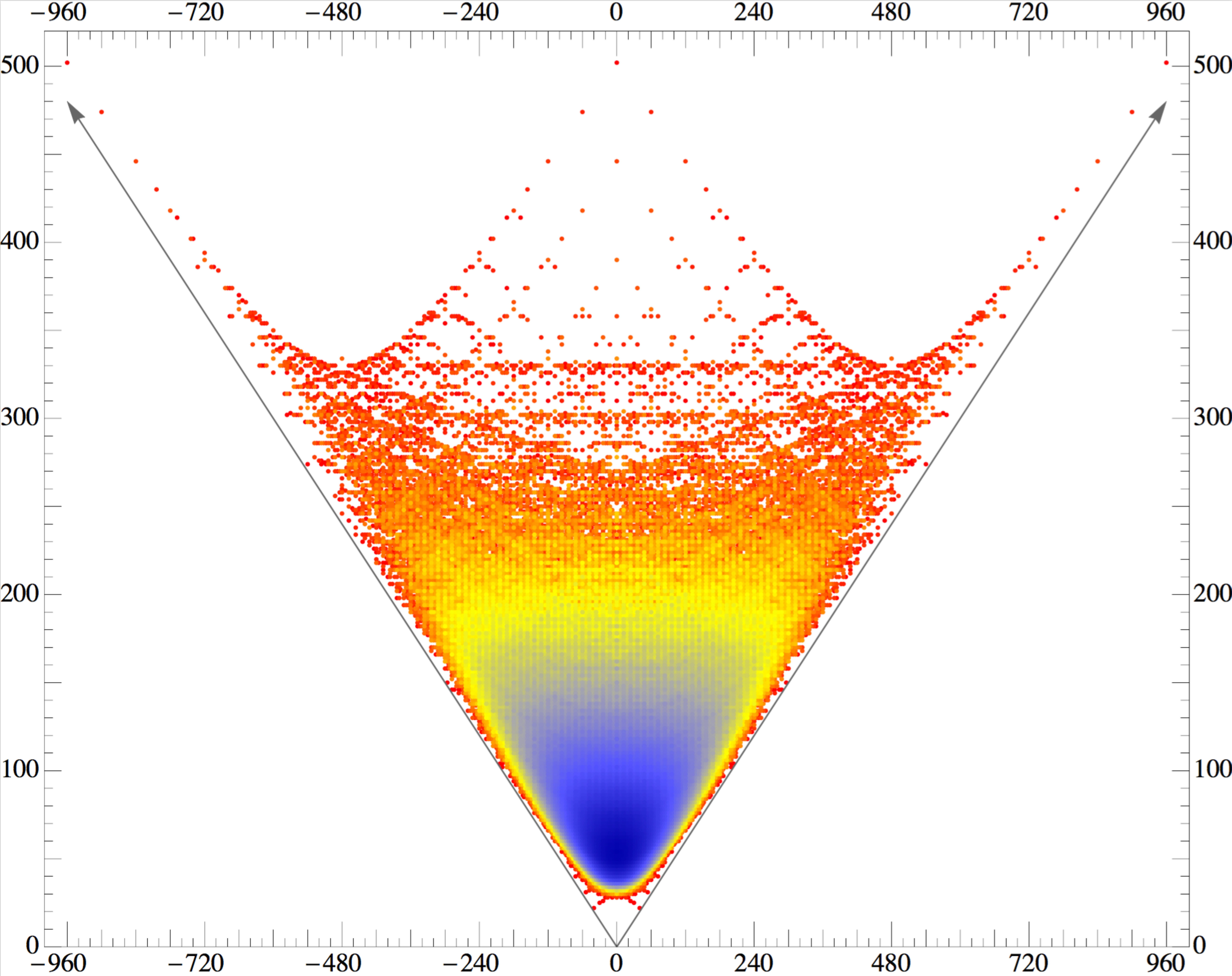}
\capt{6.5in}{occupationplot}{The Hodge plot for the list or reflexive 4-polytopes. Different colours correspond to different occupation numbers: the points in the top region of the plot correspond to one or very few reflexive polytopes, while the blue region in the vicinity of the tip contains points with occupation numbers of order of one million.}
\end{center}
\end{figure}

The structure of this plot has been mysterious for more than two decades. The distribution of points in the top region of the plot is symmetric with respect to the axes $\chi=\mp 480$. Inspired by the fact that the exact symmetry around the axis $\chi=0$ corresponds to mirror symmetry, we name this partial symmetry `half-mirror symmetry'. Another striking feature of the plot is that in the top middle region, the points are arranged into a grid-like structure. 

In Chapter~\ref{ToricCY}, we find that the generic features of the plot, as well as the structures mentioned above are explained as an overlapping of several webs formed by repeating a fundamental structure along many translation vectors. These webs correspond to Calabi-Yau manifolds fibered over $\IC\IP^1$ for which the fiber is a $K3$ manifold. Different types of $K3$ fibers give rise to distinct webs. Along with this interpretation, we discover and prove an additivity property for the Hodge numbers, valid for many types of $K3$ fibered Calabi-Yau three-folds. This additivity formula explains the existence of the translation vectors and is related to a new kind of geometrical transitions, which in the language of toric geometry can be understood as follows. 

The fibration structure of a Calabi-Yau $n$-fold is manifest in the structure of the reflexive $(n+1)$-polytope that defines it: a fibration structure exists if the $(n+1)$-polytope contains a $n$-dimensional reflexive polytope as a `slice' and as a `projection'. In this situation, the $n$-dimensional reflexive polytope defining the fiber divides the $(n+1)$-dimensional polytope into two parts, a top and a bottom. The fan corresponding to the toric variety of the base space is obtained by projecting the higher dimensional fan along the fiber. 

Given a slice, any two half-polytopes $\Top{A}$ and $\Top{B}$ that project onto it can be combined into a reflexive polytope $\Delta$.   We express this as $\Delta=\Top{A}\cup\Bot{B} = \topbot{A}{B}$. If $\Top{A}$ and $\Top{C}$ are tops and $\Bot{B}$ and $\Bot{D}$ are bottoms over a given $K3$ slice, then, under the assumption that the $K3$ slice obeys a specific condition, the Hodge numbers $h^{1,1}$ and $h^{1,2}$ satisfy the following additive formula:
$$
\ \  h^{\bullet\bullet}\big(\topbot{A}{B}\big)+h^{\bullet\bullet}\big(\topbot{C}{D}\big)~=~
h^{\bullet\bullet}\big(\topbot{A}{D}\big)+h^{\bullet\bullet}\big(\topbot{C}{B}\big)~
$$

\vspace{-20pt}
$$*\ \  *\ \  * $$

After immersing in the realm of toric geometry, we step back to discuss a more familiar class of compact Calabi-Yau manifolds realised as complete intersections of hypersurfaces in products of projective spaces \cite{Candelas:1987kf}. I will be using CICY as a short name for Calabi-Yau manifolds in this class. 

\begin{figure}[h]
\begin{center}
\includegraphics[width=6.2in]{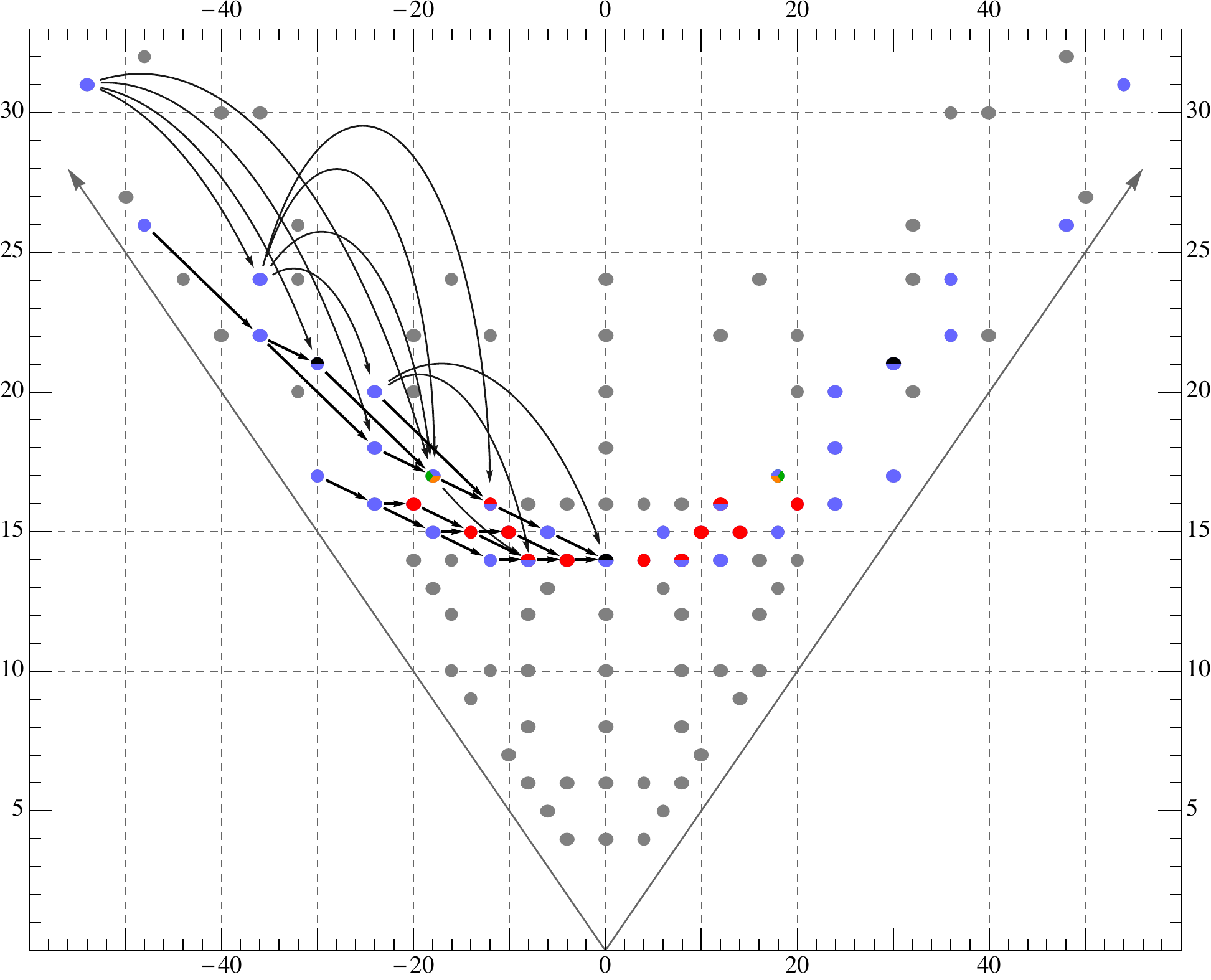}
\capt{6.5in}{Z3Webintro}{The web of $\IZ_3$ quotients of complete intersection Calabi-Yau manifolds. On the horizontal axis: the Euler number $\chi = 2\left( h^{1,1} - h^{2,1} \right)$.  On the vertical axis: the height $h^{1,1}+ h^{2,1}$.}
\end{center}
\end{figure}

In Chapter~\ref{Z3Quotients}, I discuss Calabi-Yau manifolds that are quotients of CICYs by freely acting discrete symmetry groups, in particular I discuss manifolds with fundamental group $\IZ_3$. As noted above, such manifolds provide an essential ingredient in heterotic model building. Thus this chapter  prepares the ground for the second part of the thesis. But there is a second rationale for including the discussion of CICY quotients, which has to do with the existence of yet another type of webs. 

It has been noted that CICY quotients with isomorphic fundamental group form webs connected by conifold transitions. This observation was first made in \cite{Candelas:2008wb}, in which a large number of CICY quotients were constructed. The search for CICY manifolds admitting free linear group actions was completed by Braun \cite{Braun:2010vc} by means of an automated scan, leading to a complete classification. In this chapter, I discuss the $\IZ_3$-quotients that were missed in \cite{Candelas:2008wb}, compute their Hodge numbers and study the conifold transitions between the covering manifolds and also the conifold transitions between the quotients. As it turns out, the $\IZ_3$-quotients, including the new quotients, are connected by conifold transitions so as to form a single web, as illustrated in \fref{Z3Webintro}.
\vspace{-8pt}

$$*\ \  *\ \  * $$

In the second part of this thesis, I present a systematic and algorithmic method of constructing heterotic string models exhibiting many features of the Standard Model. The need for such methods is easily understood given the large number of properties that one would like to match to those of the (supersymmetric) Standard Model. These include: the gauge group, the particle spectrum, the existence of light Higgs doublets, the doublet-triplet splitting problem, proton stability, the structure of (holomorphic) Yukawa couplings, neutrino physics, R-parity.

It is incredibly difficult to fine tune any particular construction in order to meet all these requirements. The history of heterotic string phenomenology proves it in a clear way: the number of models constructed so far that have the correct particle spectrum (let alone issues such as proton stability) is indeed very small. The approach we take is that of a `blind' automated scan over a huge number of models; for the present scan this number is of order $\sim\!\!10^{40}$. The feasibility of this attempt, initiated in \cite{Anderson:2011ns, Anderson:2012yf} and developed in the present work, relies on the particular details of the construction used. 

The construction is based on Calabi-Yau manifolds realised as CICY quotients by freely acting discrete symmetries. The distinctive and key feature of the construction is the fact that the vector bundle is a direct sum of five holomorphic line bundles. On one hand, the split nature of the bundle makes the algorithmic implementation of the various consistency and physical constraints possible. On the other hand, it leads to a rich phenomenology, such that one can easily envisage situations in which all the physical requirements mentioned above are simultaneously satisfied. I explain this briefly below. 

The choice of the internal gauge field as a connection on a vector bundle realised as a sum of five line bundles leads to a GUT group $G = SU(5)\times S\left(U(1)^5\right)$. The additional $U(1)$ factors are generically Green-Schwarz anomalous. As such, the corresponding gauge bosons are super-massive, thus leading to no inconsistencies with the observed physics. However, these broken $U(1)$ symmetries remain as global symmetries. This is a crucial observation: the Lagrangian must be invariant under these global $U(1)$ transformations, which leads to constraints on the allowed operators in the 4 dimensional effective supergravity operators. In specific models, this provides a solution to well-known problems arising in supersymmetric GUT constructions, such as proton stability or R-parity conservation.  

The consistency requirements, such as the anomaly cancellation and the conditions imposed on the vector bundle by supersymmetry can be checked in a straightforward manner. Supersymmetry requires that the vector bundle is holomorphic (automatic, in the present case) and that it satisfies the hermitian Yang-Mills equations. By the Donaldon-Uhlenbeck-Yau theorem \cite{Donaldson1985, Uhlenbeck1986}, this is possible if and only if the vector bundle has vanishing slope and is poly-stable. In general, checking (poly)-stability is one of the most challenging tasks involved in heterotic string constructions. For sums of line bundles, however, it reduces to the question of deciding whether a set of quadratic equations (corresponding to the vanishing slope for each line bundle) have common solutions in a certain domain. 
The particle content of the effective supergravity theory is computed in terms of ranks of various cohomology groups of the vector bundle. In general, this is very difficult. But the easiest situation one can hope for is that of line bundles. In practice, we are able to compute line bundle cohomology in the vast majority of the cases.
In Chapter~\ref{LineBundles}, I present the details of the line bundle construction, as well as the algorithm used in the computer-based scan. This search has led to approximately $35,000$ $SU(5)$ GUT models having the right content to induce low-energy models with the precise matter spectrum of the MSSM, with one, two or three pairs of Higgs doublets and no exotic fields of any kind.

\vspace{10pt}
The scan presented in Chapter~\ref{LineBundles} was pushed until no further viable models could be found. More precisely, line bundles are classified by their first Chern classes. In the automated search, the number of viable models reached a certain saturation limit after repeatedly increasing the range of integers defining the line bundles. In Chapter~\ref{TQ1}, I address this question from two different perspectives for the particular case in which the Calabi-Yau manifold is a hypersurface embedded in a product of four $\IC\IP^1$ spaces. One of the arguments provided in this chapter relies on an explicit formula for computing line bundle cohomology on the tetraquadric manifold. 

In Chapter~\ref{ModuliSpace}, I explore the moduli space of non-Abelian bundles in which line bundle sums live. After choosing a particular line bundle sum leading to a viable GUT model, I study the class of bundle deformations obtained via the monad construction. I will address questions such as the stability of monad bundles and the resulting particle spectrum. 
The comparison between the distinguished Abelian model and its non-Abelian deformations is carried out both at the high energy (geometrical) level and at the GUT level. For the chosen line bundle model, the class of non-Abelian bundles that lead to consistent and viable compactifications has co-dimension one in K\"ahler moduli space. I will make this statement more precise in Chapter~\ref{ModuliSpace}. 

Finally, Chapter~\ref{Conclusion} contains a summary of the main results presented in this thesis, as well as a number of concluding remarks and directions for future work. 

\bigskip

The work presented in this thesis is drawn from four research papers. The material presented in Chapter 2 is based on:
\begin{itemize}
\item[$1.$] P.~Candelas, A.~Constantin, H.~Skarke, {\itshape An Abundance of K3 Fibrations from Polyhedra with Interchangeable Parts}, to appear in  Comm. Math. Phys., arXiv:1207.4792 [hep-th]  \cite{Candelas:2010ve}
\end{itemize}

\bigskip
Chapter 3 is based on:
\begin{itemize}
\item[$2.$]P.~Candelas, A.~Constantin, {\itshape Completing the Web of $\mathbb{Z}_3$-Quotients of Complete Intersection Calabi-Yau Manifolds}, Fortsch.~Phys.~60, No.~4, 345-369 (2012), arXiv:1010.1878 [hep-th] \cite{Candelas:2012uu}
\end{itemize}

\bigskip
Chapter 4 is based on:
\begin{itemize}
\item[$3.$] L.~Anderson, A.~Constantin, J.~Gray, A.~Lukas and E.~Palti, {\itshape A Comprehensive Scan for Heterotic SU(5) GUT models}, arXiv:1307.4787 [hep-th] \cite{Anderson:2013xka}
\end{itemize}

\bigskip
Finally, Chapters 5 and 6 are based on the following article, in preparation:
\begin{itemize}
\item[$4.$] E.~Buchbinder, A.~Constantin, A.~Lukas, {\itshape Heterotic Line Bundle Standard Models: a Glimpse into the Moduli Space.} \cite{Buchbinder}
\end{itemize}
}

\part{The Manifold}
\chapter{Elliptic $K3$ Fibrations}\label{ToricCY}
{\setstretch{1.52}

Calabi-Yau manifolds made their way into physics through the discovery of their relevance for string compactifications \cite{Candelas:1985en}. 
Few years later, a certain type of duality, known as `mirror symmetry' was conjectured in relation to Calabi-Yau compactifications \cite{Dixon:1987bg, Lerche:1989uy}. The conjecture emerged from the observation that exchanging the K\"ahler moduli and the complex structure moduli of a Calabi-Yau manifold corresponds to an exchange of generators in the supersymmetry algebra of the underlying world-sheet theory leading to equivalent quantum field theories. Since the physical theory does not distinguish between the two cases, it was conjectured that Calabi-Yau manifolds come in pairs with interchangeable Hodge numbers $h^{1,1}$ and $h^{1,2}$.

The first explicit construction of large classes of Calabi-Yau three-folds as complete intersections of hypersurfaces in products of projective spaces \cite{Candelas:1987kf} did not seem to support the mirror symmetry conjecture. Complete intersections in products of projective spaces have negative Euler number $\chi=2\left( h^{1,1}-h^{1,2}\right)$, thus one could find no mirror pairs within this context. Later on, the construction of Calabi-Yau three-folds as hypersurfaces in weighted $\IC\IP^4$ \cite{Candelas:1989hd, Kreuzer:1992np, Klemm:1992bx} provided a large number of mirror pairs; however, it did not exhibit a perfect mirror symmetry at the level of Hodge numbers. 

The situation was rectified with the manifestly mirror symmetric construction of Calabi-Yau manifolds as hypersurfaces in toric varieties due to Batyrev's work \cite{Batyrev:1993dm}. Such manifolds are defined using reflexive polytopes. Following Batyrev's construction, Kreuzer and Skarke devised an algorithm which enabled them to compile a complete list of reflexive polytopes in two, three and four dimensions \cite{Kreuzer:1992bi, Kreuzer:1995cd, Kreuzer:2000xy, Kreuzer:2000qv}. Two dimensional reflexive polytopes correspond to complex elliptic curves. There are 16 isomorphism classes of reflexive polytopes in two dimensions. In three dimensions there are 4,319 reflexive polytopes and they correspond to $K3$ manifolds. The list of four dimensional reflexive polytopes, corresponding to Calabi-Yau three-folds has an impressive length of 473,800,766. 

\begin{figure}[h]
\begin{center}
\includegraphics[width=6.5in]{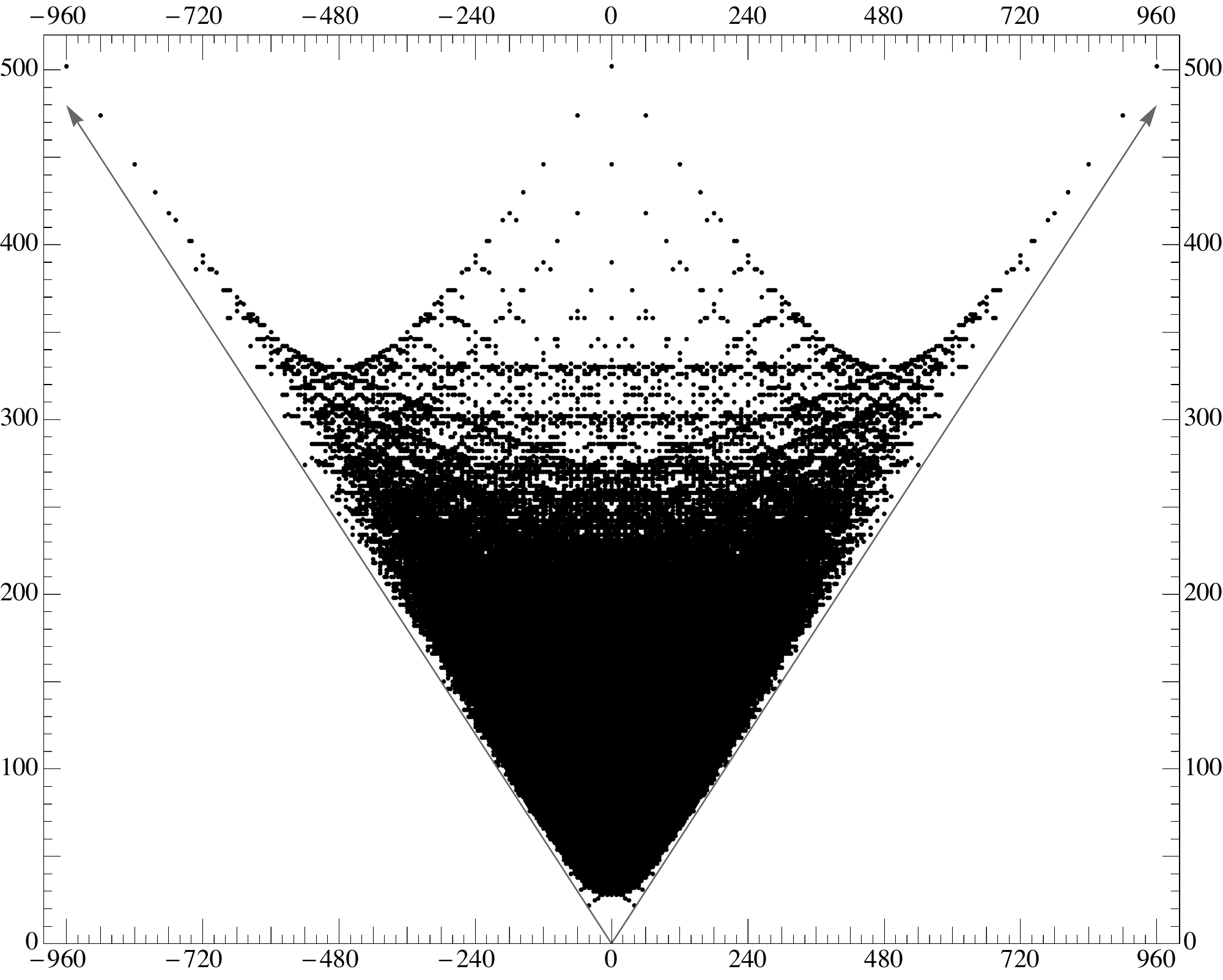}
\capt{6.5in}{BasicKSPlot}{The Hodge plot for the list or reflexive 4-polytopes. The Euler number
$\chi = 2\left( h^{1,1}-h^{1,2} \right)$ is plotted against the height $y=h^{1,1}+h^{1,2}$. The oblique axes correspond to $h^{1,1}=0$ and $h^{1,2}=0$.}
\end{center}
\end{figure}

The latter list is the subject of this chapter. The question which motivated the present work can be expressed as: {\itshape What kind of phenomenon gives rise to the symmetries present in the Hodge plot for the list of reflexive four dimensional polytopes?} In the rest of this chapter I will make this question more explicit, by pointing out a number of symmetries of the Hodge plot (see \fref{BasicKSPlot}). Then I will move on and present a few rudiments of toric geometry. Finally, I will present an answer to the above question at the level of lattice polytopes. An understanding of the physics associated to the topological transitions presented in this chapter is still missing and I hope it will be the subject of a future work.

\newpage
\section{Half-mirror Symmetry and Translation Vectors}\label{Sec2.1}
\vskip-10pt
The Kreuzer-Skarke list of four-dimensional reflexive polytopes \cite{Kreuzer} provides the largest class of Calabi-Yau threefolds that has been constructed explicitly. There are combinatorial formulas for the Hodge numbers $h^{1,1}$ and $h^{1,2}$ in terms of the polytopes \cite{Batyrev:1993dm}. By computing the Hodge numbers associated to the polytopes in the list, one obtains a list of 30,108 distinct pairs of values for $\hodgenos$.

\begin{figure}[h]
\begin{center}
\includegraphics[width=6.5in]{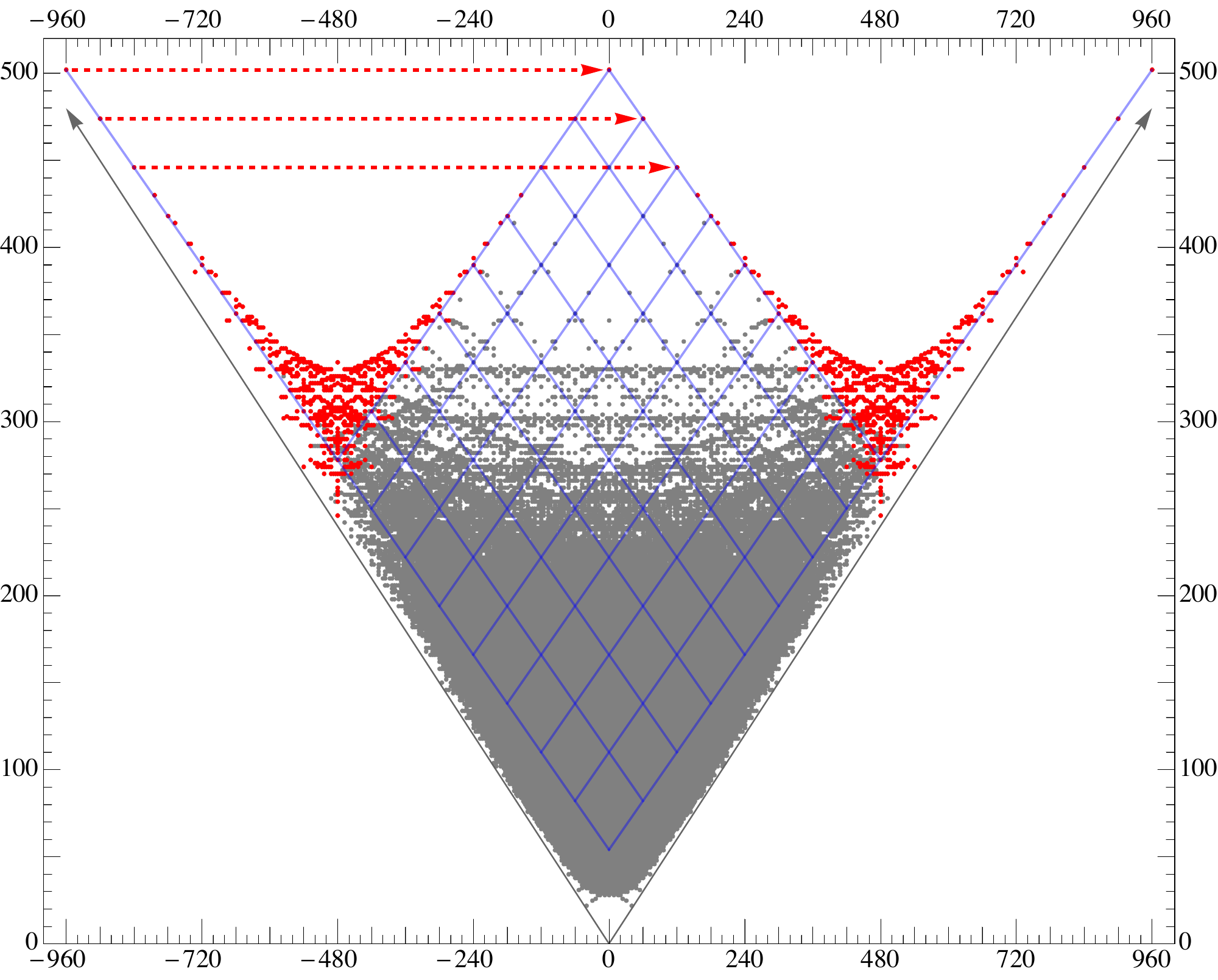}
\capt{6.2in}{YStructure}{The structure in red, on the left, contains points which have mirror images when reflected about the axis $\chi = -480$. The red arrows show that these points can be mapped into points corresponding to manifolds with positive Euler number by a change in Hodge numbers $\Delta\big(h^{1,1},h^{1,2} \big)=(240,-240)$, corresponding to $\D(\chi, y)=(960,0)$.}
\end{center}
\vskip-10pt
\end{figure}

These are presented in \fref{BasicKSPlot}, in which the Euler number $\chi = 2\left( h^{1,1}-h^{1,2} \right)$ is plotted against the height, $y=h^{1,1}+h^{1,2}$. The plot has an intriguing structure. One immediate feature of this plot, also evident from Batyrev's formulae, is the presence of mirror symmetry at the level of Hodge numbers: the Hodge numbers associated to a reflexive polytope are interchanged with respect to the dual polytope. This corresponds to the symmetry about the axis 
$\chi=0$.

Another striking feature is that both the left and the right hand sides of the plot contain structures symmetric about vertical lines corresponding to Euler numbers $\chi= \mp 480$. One can easily observe that the great majority of the points corresponding to manifolds with $\chi<-480$ have mirror images when reflected about the $\chi=-480$ axis. In \fref{YStructure}, the structure which exhibits this half-mirror symmetry is highlighted in red. Equivalently, one can observe that the red points, on the left, and only those, can be translated into other points of the plot, corresponding to manifolds with positive Euler number, by a change in Hodge numbers $\D(h^{1,1},h^{1,2} )=(240,-240)$ corresponding to $\D(\chi, y)=(960,0)$, as indicated by the red arrows in \fref{YStructure}. Together with mirror symmetry, this results in the symmetry described above.

\begin{figure}[h]
\begin{center}
\includegraphics[width=6.5in]{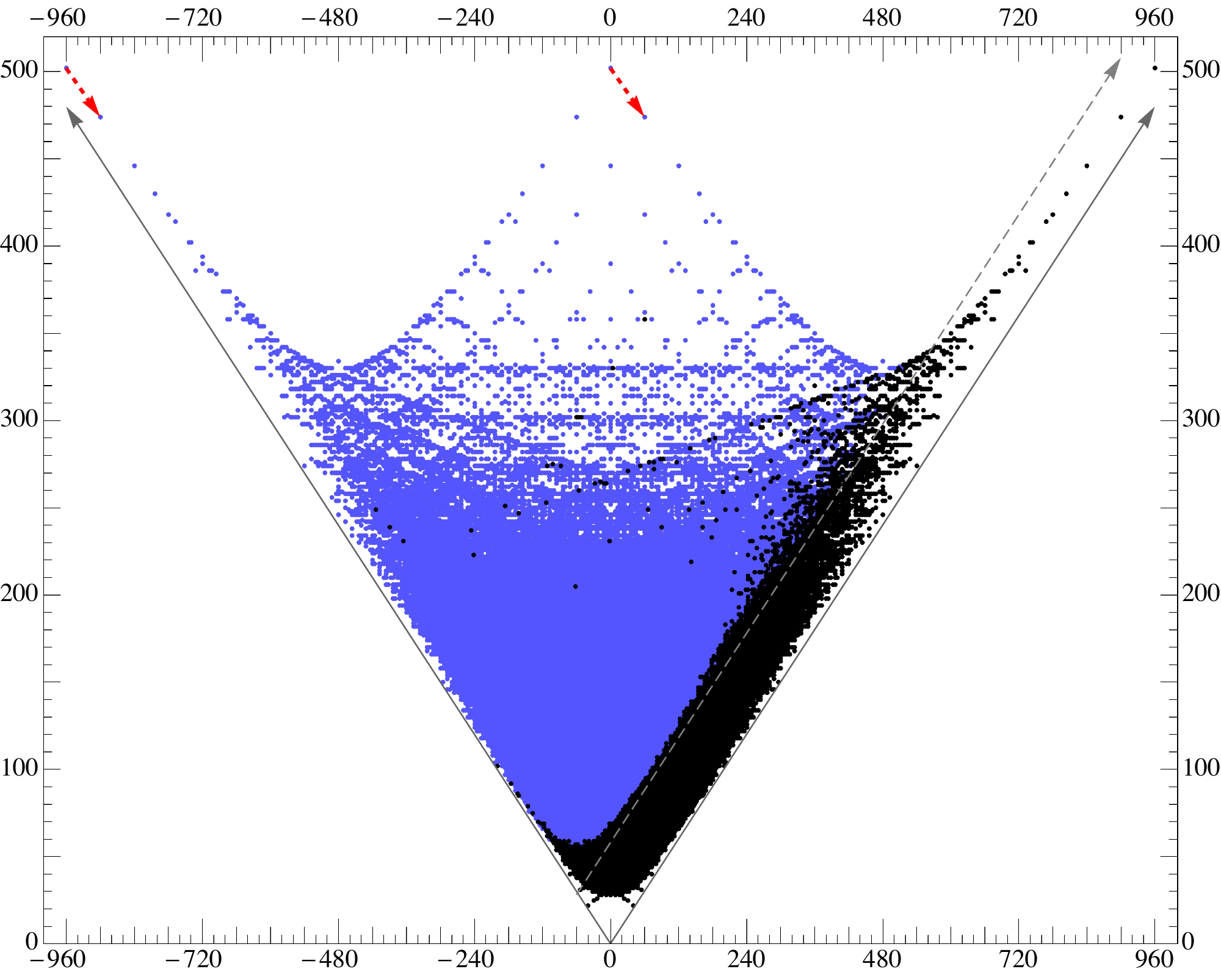}
\capt{5.8in}{BlueVectorPlot}{The blue points can be translated into other points of the plot by a change in Hodge numbers $\Delta\big(h^{1,1},h^{1,2} \big)=(1,-29)$. The red arrows illustrate this action on two pairs of points. The extra grey arrow corresponds to $h^{1,2} = 30$.}
\end{center}
\vskip-30pt
\end{figure}

Yet another intriguing feature is evident from \fref{YStructure}: a special role is played by  a vector $\Delta\big(h^{1,1},h^{1,2} \big)=(1,-29)$, corresponding to $\D(\chi,y)=(60,-28)$. This, together with its mirror, are the displacements corresponding to the blue grid. It is immediately evident that many points have a `right descendant' corresponding to these displacements. However, it is a fact that almost all points with $h^{1,2} \geq 30$ have such descendants. In \fref{BlueVectorPlot} the points  that have a right-descendant are coloured in blue.  Note also how these translations, together with their mirrors, account for the gridlike structure in the vicinity of the central peak of~the~plot.
\goodbreak
 
One can consider also `left-descendants', that is points of the plot that are displaced by the mirror vector, 
$\D(\chi,y)=(-60, -28)$, from a given point. There are very few points of the plot that do not have either a left or right descendant, as illustrated by \fref{NoDescendants}.

\begin{figure}[h]
\begin{center}
\includegraphics[width=6.5in]{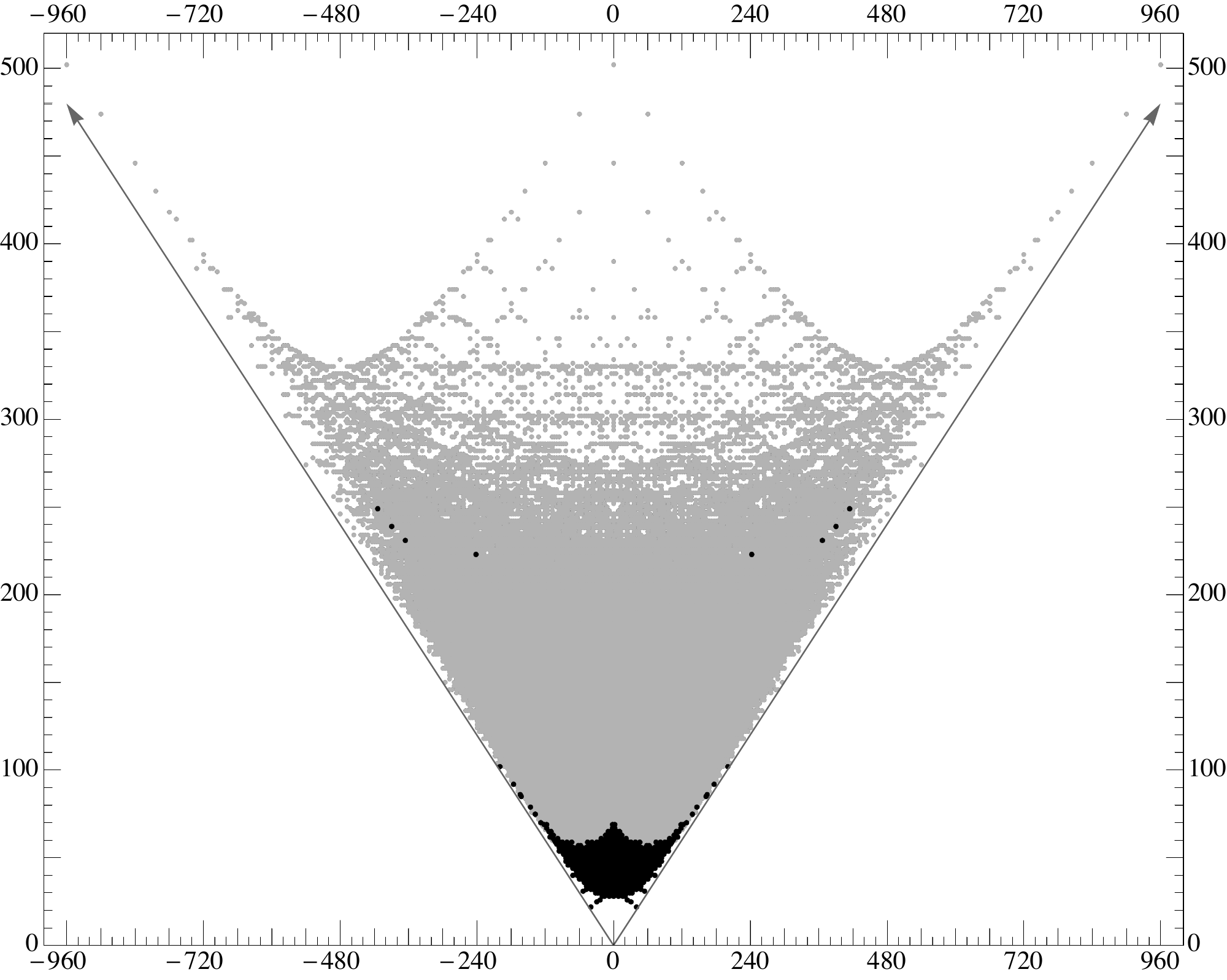}
\capt{6.3in}{NoDescendants}{The points in black are the only ones that have neither a left, nor a right descendant.}
\end{center}
\vskip-14pt
\end{figure}

The scope of this chapter is to study the structures discussed above. Our starting point is the observation that most of the points making up the red structure in \fref{YStructure} 
correspond to Calabi-Yau threefolds fibered by $K3$ surfaces, which are themselves elliptically fibered. This nested fibration structure is visible in the reflexive polytopes which provide the toric description of these Calabi-Yau manifolds. Let me briefly discuss the relevant elements of toric geometry involved in this description.

\newpage
\section{The Language of Toric Geometry}

\subsection{Reflexive polytopes and toric Calabi-Yau hypersurfaces} Let $N,M\simeq \mathbb Z^n$ be two dual lattices of rank $n$ and let 
$\langle \cdot ,\cdot \rangle:M\times N\rightarrow\mathbb Z$ 
denote the natural pairing. Define the real extensions of $N$ and $M$ as $N_{\mathbb R}:=N\otimes\, \mathbb R$ and $M_{\mathbb R}:=M\otimes\, \mathbb R$. A~polytope $\Delta \subset M_{\mathbb R}$ is defined as the convex hull of finitely many points in $M_{\mathbb R}$ (its vertices). The set of vertices of $\Delta$ is denoted by $\cV(\Delta)$ and its relative interior by $\text{int}(\Delta)$. A lattice polytope is a polytope for which $\cV(\Delta)\subset M$. Reflexivity of polytopes is a property defined for polytopes for which 
$\text{int}(\Delta)\cap M$ 
contains the origin. A lattice polytope is said to be reflexive if all its facets are at lattice distance 1 from the origin, that is there is no lattice plane parallel to any given facet, that lies between that facet and the origin.

The polar, or dual, polytope of a reflexive polytope is defined as the convex hull of inner normals to facets of~$\Delta$, normalised to primitive lattice points of $N$. Equivalently, a reflexive polytope $\Delta\subset M_{\mathbb R}$ is defined as a lattice polytope having the origin as a unique interior point, whose dual 
\begin{equation}
\SDelta = \{ y\in N_{\mathbb R}: \langle x,y\rangle \geq -1, \text{ for all } x \in \Delta \}
\end{equation}\label{dual}%
is also a lattice polytope. 

To any face $\theta$ of $\Delta$ one can assign a dual face $\theta^*$ of
$\SDelta$ as 
$$\theta^*=
\{ y\in\SDelta:\! \langle x,y\rangle = -1, \text{ for all } x \in \theta \}.$$
In this way a vertex is dual to a facet, an edge to a codimension 2 face and so on.
In particular, for 3-polytopes edges are dual to edges. 

One can construct a toric variety from the fan over a triangulation of the surface of $\SDelta$. A Calabi-Yau hypersurface in this toric variety can then be constructed as the zero locus of a polynomial whose monomials are in one-to-one correspondence with the lattice points of $\Delta$. This construction is described in the texts~\cite{0813.14039, Cox:2000vi, 1223.14001, Skarke:1998yk, Avram:1996pj, Bouchard:2006ah}. 


\subsection{A two dimensional example: the elliptic curve}
Instead of presenting in the abstract the construction of Calabi-Yau hypersurfaces in compact toric varieties, let me illustrate this in the case of 1 dimensional Calabi-Yau manifolds by discussing the Weierstrass elliptic curve. 

All the information we need is given in \fref{EllipticCurve}, in which the lattices $N$ and $M$ overlap each other. The red spot corresponds to the common origin of the two lattices. The black triangle, with vertices $v_1=(1,0), v_2=(0,1)$ and $v_3=(-2,-3)$, corresponds to the boundary of $\SDelta$. The corresponding fan consists of a 0-dimensional cone (the origin), three 1-dimensional cones (the rays $v_1, v_2$ and $v_3$) and three two dimensional cones. The dashed purple triangle with vertices $w_1=(-1,1), w_2=(-1,-1)$ and $w_3=(2,-1)$ defines $\Delta$. The dashed black lines correspond to a maximal `triangulation' of the surface of $\SDelta$. Notice that in this case the surface of $\SDelta$ corresponds to the black triangle and there is no ambiguity involved in the triangulation. 

\vspace{10pt}
\begin{figure}[h]
\begin{center}
\includegraphics[width=4in]{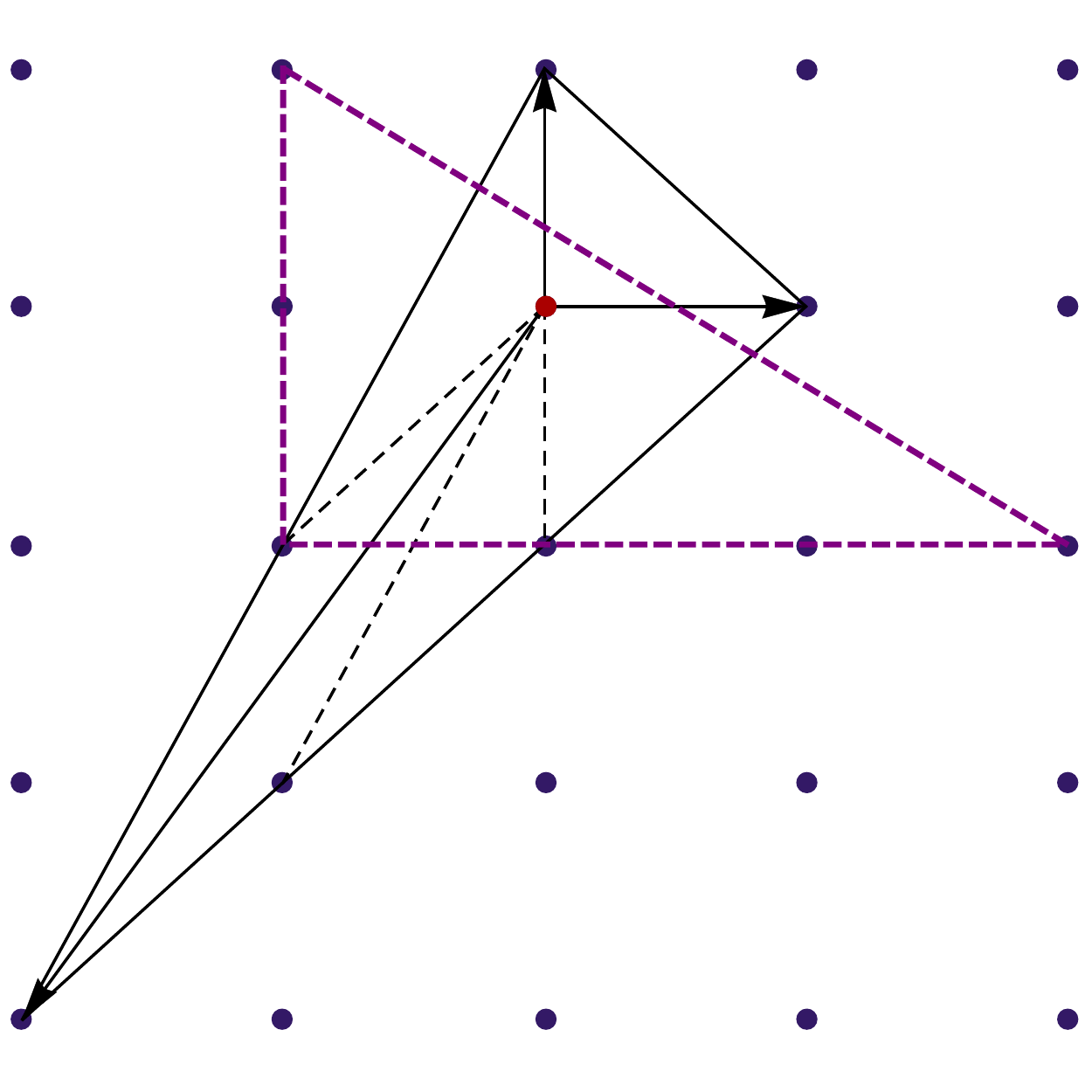}
\capt{5.8in}{EllipticCurve}{$\Delta$ and $\SDelta$ for the Weierstrass elliptic curve.}
\end{center}
\vskip-10pt
\end{figure}

We will construct the toric variety following Cox's approach, as an algebraic generalisation of complex weighted projective spaces. In this approach one assigns a homogeneous coordinate $x_i\in \IC$ to each vertex $v_i$ (here $1\leq i\leq 3$) and then identifies the points of $\IC^3\backslash \{0\}$ using the equivalence relation 
$$(x_1,x_2,x_3)\sim (\lambda^{q^1} x_1,\lambda^{q^2} x_2,\lambda^{q^3} x_3), \ \ \text{with }\lambda \in\IC^*$$

The weights $q_i$ can be read as the coefficients defining the linear relation $2v_1+3v_2+v_3=0$, chosen such that $\text{gcd}(q^1,q^2,q^3)=1$. The toric variety thus obtained is compact and corresponds to the weighted projective space $\IC\IP^3_{[2,3,1]}$. However, this variety is singular, since the action corresponding to $\lambda$ chosen to be equal to a root of unity of order 2 and, respectively 3, fixes the points $[1,0,0]$ and $[0,1,0]$. In toric geometry these singular points can be replaced by $\IC\IP^1$ spaces (`blown-up') by adding extra rays in the fan. In fact, in order to de-singularise completely the toric variety, one needs to add an extra ray for each point in the boundary of $\SDelta$ (excluding the vertices) as indicated by the black dashed lines in \fref{EllipticCurve}. 

Leaving aside the discussion of resolving singular points, let us mention the manner in which the Calabi-Yau hypersurface is constructed. In Cox's formulation of toric geometry, for each point $w\in \Delta \cap M$ one associates a monomial
$$m_w = x_1^{\langle v_1,w\rangle +1} x_2^{\langle v_2,w\rangle +1} x_3^{\langle v_3,w\rangle +1} $$
The zero locus of a polynomial constructed as a generic linear combination of these monomials defines a Calabi-Yau hypersurface $X$ embedded in a toric variety $V$. To see that this hypersurface is indeed Calabi-Yau one needs to realise that the defining polynomial is a section of the anticanonical bundle of the toric variety $K_V^{^{*}}$. Thus the normal bundle of $X$ is the restriction of $K_V^{^{*}}$ to the hypersurface. Then, using the normal bundle sequence and passing to the determinant bundles one arrives to the conclusion that the canonical bundle of $X$ is trivial, hence $X$ is Calabi-Yau.  

Since the singularities of the toric variety are point-like, a generic hypersurface will miss them, and thus a smooth Calabi-Yau hypersurface is obtained. Batyrev showed that generic hypersurfaces defined using reflexive $n$-dimensional polytopes, as above, are smooth up to $n=4$.

\subsection{Fibration structures}

In the following section, we will encounter Calabi-Yau three-folds that are $K3$ fibrations. So far, we became familiar with the fact that in toric geometry compact Calabi-Yau three-folds can be constructed from pairs $(\Delta, \SDelta) \subset M{\times}N$ of reflexive polytopes, where $M$ and $N$ are isomorphic to $\IZ^4$. Such a Calabi-Yau three-fold exhibits a fibration structure over $\IC\IP^1$, if  there exists a distinguished three dimensional sub-lattice $N_3\subset N$, such that $\SDelta_3= \SDelta \cap N_3$ is a three dimensional reflexive polytope. 

The sub-polytope $\SDelta_3$ corresponds to the $K3$ fiber and divides the polytope $\SDelta$ into two parts, a top and a bottom. The fan corresponding to the base space can be obtained by projecting the fan of the fibration along the linear space spanned by sub-polytope describing the fiber~\cite{Kreuzer:1997zg}. In the present case, $\SDelta$ a four-dimensional polytope and $\SDelta_3$ three dimensional, the fan defining the base space contains two opposite 1-dimensional cones and thus it always describes a $\IC\IP^1$. 

This description is dual to having a distinguished one dimensional sub-lattice $M_1 \subset M$, such that the projection of $\Delta$ along $M_1$ is $\Delta_3= (\SDelta_3)^{\!^*}$. The equivalence between the two descriptions has been proved in~\cite{Avram:1996pj}, and was expressed as `a slice is dual to a projection'. In the case where the mirror Calabi-Yau three-fold is a fibration over the mirror $K3$ it is possible to introduce distinguished three and, respectively, one dimensional sub-lattices $M_3$ and $N_1$, resulting in the splits $M = M_1 \oplus M_3$ and $N = N_1 \oplus N_3$.


\subsection{$K3$-fibrations in F-theroy/type IIA duality}
There has been a long standing interest in $K3$-fibered Calabi-Yau threefolds in string theory. $K3$~fibrations appear in a natural way in the study of four dimensional $\mathcal N=2$ heterotic/type IIA duality~\cite{Klemm:1995tj, Vafa:1995gm, Aspinwall:1995vk}. Toroidal compactifications of the strongly coupled heterotic string theory to six dimensions are dual to weakly coupled type IIA theory compactifications on $K3$ surfaces, in the sense that the moduli spaces of vacua for both sides match. In~\cite{Vafa:1995gm} it was noted that this duality can be carried over to four dimensions if the the two theories are fibered over $\mathbb P^1$, that is if the type IIA theory is compactified on a manifold which is a $K3$ fibration over $\mathbb P^1$ and the heterotic string is compactfied on $K3{\times} T^2$, which can be written as a $T^4$ fibration over $\mathbb P^1$. In this way, the six dimensional heterotic/type IIA duality can be used fiber-wise. 

In the same context, Aspinwall and Louis~\cite{Aspinwall:1995vk} showed that, after requiring that the pre-potentials in the two theories match, and assuming that the type IIA theory is compactified on a Calabi-Yau manifold, this manifold must admit a $K3$ fibration. At the time, several lists of $K3$ fibered Calabi-Yau threefolds have been compiled, first as hypersurfaces in weighted projective 4-spaces~\cite{Klemm:1995tj, Hosono:1996ua} and shortly after that by using the methods of toric geometry~\cite{Avram:1996pj}. The language of toric geometry was also used in~\cite{Candelas:1996su} where the authors noticed that the Dynkin diagrams of the gauge groups appearing in the type IIA theory can be read off from the polyhedron corresponding to the $K3$ fibered Calabi-Yau manifold used in the compactification. The singularity type of the fiber corresponds to the gauge group in the low-energy type IIA theory. In the case of an elliptically fibered $K3$ manifold, the Dynkin diagrams appear in a natural way~\cite{Perevalov:1997vw}.

\newpage
\section{Nested Elliptic-$K3$ Calabi-Yau Fibrations}
In this section I define the nested structures of reflexive polytopes which correspond to elliptic-$K3$ fibration structures of Calabi-Yau three-folds. The Hodge numbers associated with manifolds that exhibit such nested fibrations obey a certain additivity property, which applies (not exclusively) to tops and bottoms corresponding to a $K3$ slice of type $G_1{\times}G_2$, with Weierstrass elliptic fiber, for which the groups $G_1$ and $G_2$ are simply laced. 

\subsection{Elliptic-$K3$ fibrations}
This special type of fibration structure corresponds to Calabi-Yau manifolds which are $K3$ fibrations over $\mathbb P^1$ and for which the fiber is itself an elliptic fibration. Such manifolds appeared in~\cite{Candelas:1996su} in the discussion of heterotic/type IIA (F-theory) duality. In the toric language, such fibration structures are displayed in the form of nestings of the corresponding polytopes. 

A degeneration of the elliptic fibration may lead to a singularity of ADE type which can be resolved by introducing a collection of exceptional divisors whose intersection pattern is determined by the corresponding A, D or E type Dynkin diagram.
As the exceptional divisors correspond to lattice points of $\SDelta$, the group can be read off from the distribution of lattice points in the top and the bottom (above and below the polygon corresponding to the elliptic curve)~\cite{Candelas:1996su, Perevalov:1997vw}. 

The last polyhedron in \fref{polyselection} corresponds to an elliptically fibered $K3$ with a resolved $E_8$ singularity.
In this example, the bottom contains a single lattice point, which gives the Dynkin diagram for the trivial Lie group denoted by $\{1\}$, while the top corresponds to the extended Dynkin diagram for $E_8$. 

\vspace{12pt}
\begin{figure}[h]
\begin{center}
\framebox[6.5in][c]{\includegraphics[width=5.7in]{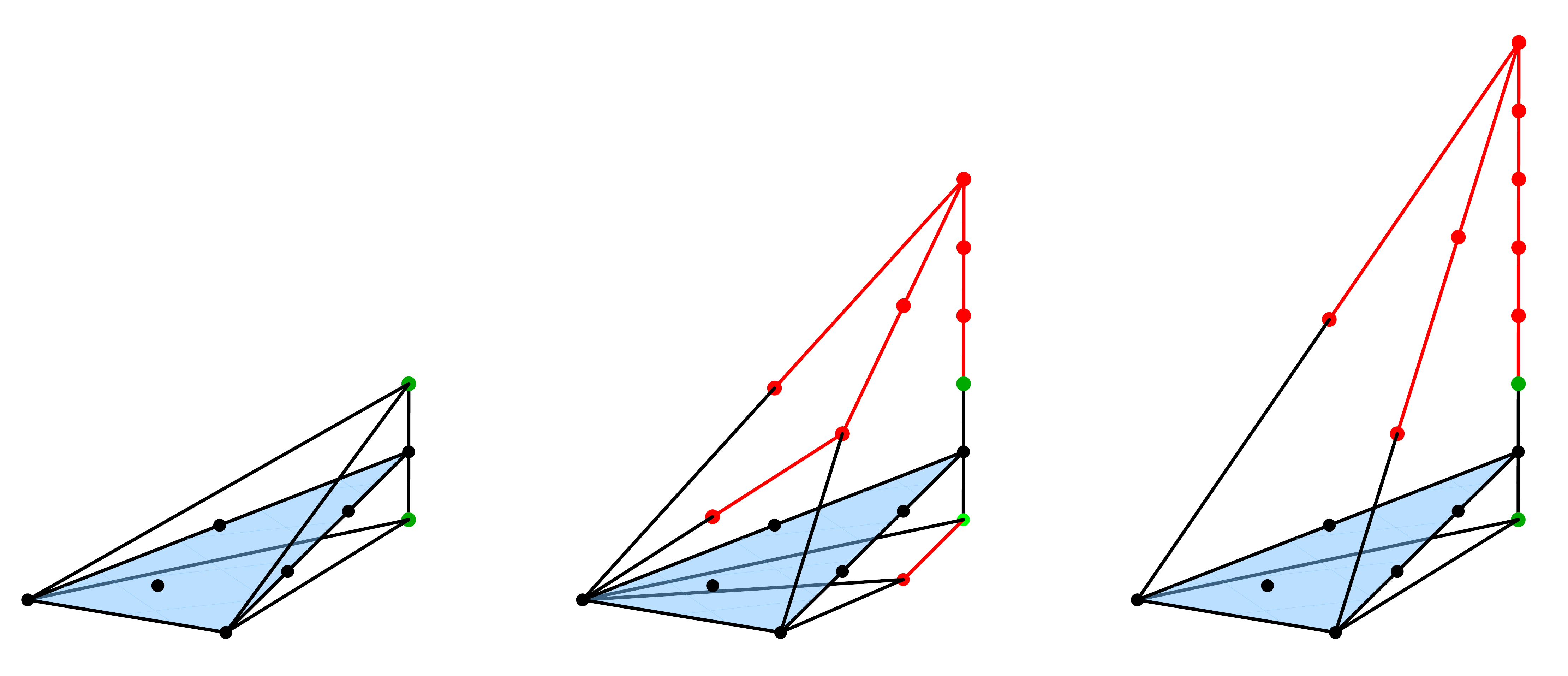}}
\capt{5.3in}{polyselection}{A selection of three-dimensional reflexive polytopes: the $\{1\}{\times}\{1\}$, $E_7{\times}SU(2)$~(self-dual) and $E_8{\times}\{1\}$~(self-dual) $K3$ polyhedra. The triangle corresponding to the elliptic fiber divides each polyhedron into a top and bottom.}
\end{center}
\end{figure}

\subsection{Composition of projecting tops and bottoms}\label{ComposingTops}
Above, we encountered the notion of a top as a lattice polytope as a part of a reflexive polytope lying on one side of a hyperplane through the origin whose intersection with the initial polytope is itself a reflexive polytope of lower-dimension. This definition was originally given in \cite{Candelas:1996su}.

A more general definition for a top as a lattice polytope with one facet through the origin and all the other facets at distance one from the origin is useful in the context of non-perturbative gauge groups~\cite{Candelas:1997pq}. All three-dimensional tops of this type were classified in~\cite{Bouchard:2003bu}. Implicit in this latter definition is the fact that, as before, the facet of an $n$-dimensional top that contains the origin is an $(n-1)$-dimensional reflexive polytope.

\vspace{12pt}
\begin{figure}[h]
\begin{center}
\framebox[6.5in][c]{\includegraphics[width=5.7in]{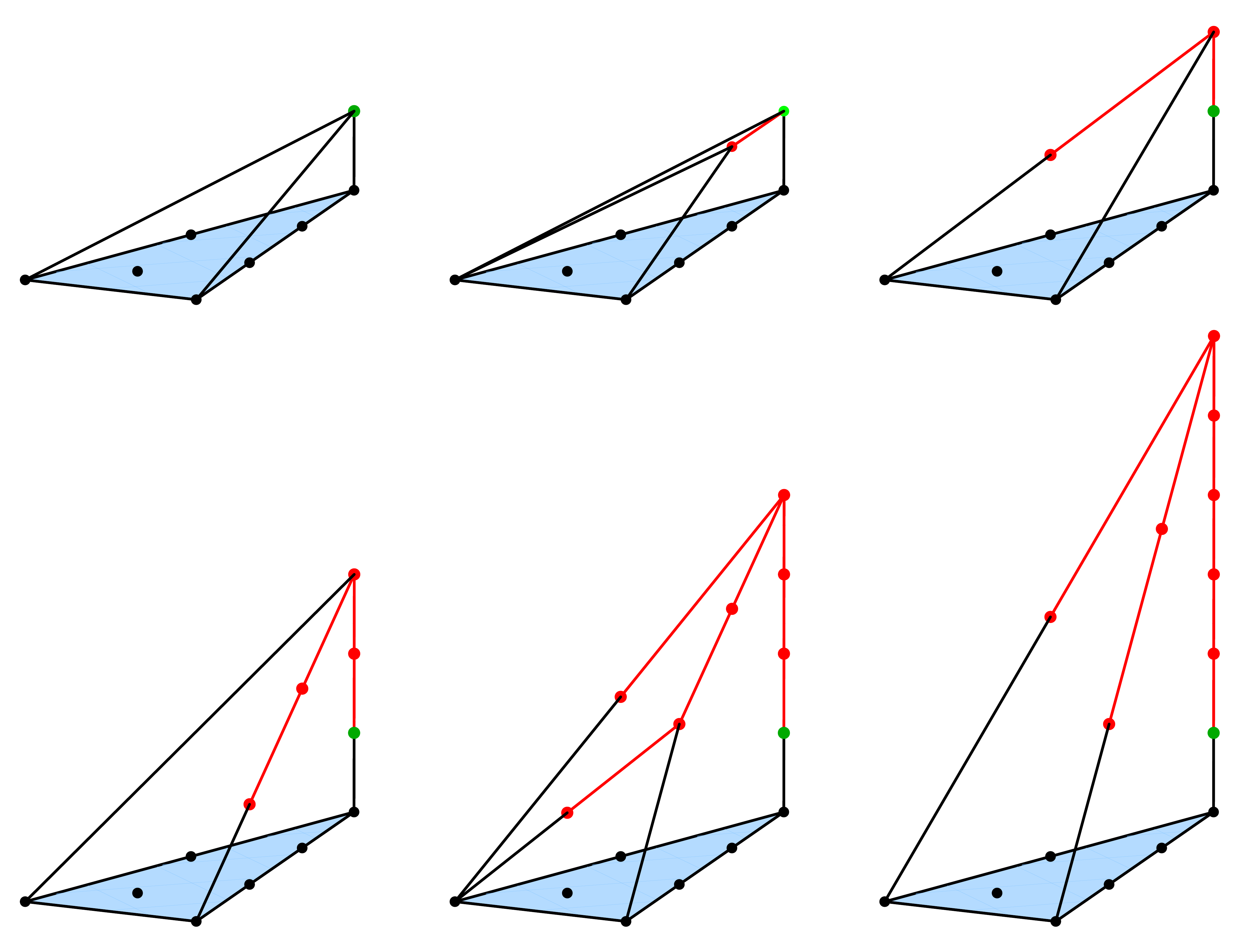}}
\capt{5.2in}{topselection}{A selection of $K3$ tops: the $\{1\}$, $SU(2)$, $G_2$, $F_4$, $E_7$ and $E_8$ tops.}
\end{center}
\end{figure}

In this section, we are interested in four-dimensional reflexive polytopes that encode $K3$ fibration structures such that the dual 4-polytope exhibits a fibration structure with respect to the dual $K3$ polytope. This means that in addition to a distinguished lattice vector $m\in M$ 
encoding the slice $\SDelta_3 = \{y\in\SDelta: \langle m,y\rangle=0\}$ of $\SDelta$, corresponding to the $K3$ polyhedron, there is a distinguished lattice vector $n\in N$ encoding the dual slice. As shown in~\cite{Avram:1996pj},~$\SDelta$~must then project to $\SDelta_3$ under $\pi_n: N\to N_3\simeq N/N_1$, where $N_1$ is the one-dimensional 
sublattice of $N$ that is generated by $n$. This implies, of course, that both the top and the bottom resulting from the slicing must project to $\SDelta_3$. A top and a bottom that project to~$\SDelta_3$ along the same direction $n$ can always be combined into a reflexive polytope, as shown below\footnote{The content of the following lemmata is, in a great measure, due to Harald Skarke.}. In the following, I will express this fact in terms of a notation where a top is indicated by a bra~$\Top{A}$, a bottom by a ket $\Bot{B}$ (to be interpreted as the reflection of $\Top{B}$ through $N_3$), and the resulting reflexive polytope $\SDelta=\Top{A}\cup\Bot{B}$ by $\topbot{A}{B}$. 

\vspace{30pt}

{\bf Lemma 1.} {Let $M$, $N$ be a dual pair of four-dimensional lattices splitting as $N=N_1\oplus N_3$, $M=M_1\oplus M_3$, with $N_1$ generated by  $n\in N$, $M_1$ generated by $m\in M$ and $\langle m,n\rangle=1$. Denote by $\pi_n:N\to N_3$, $\pi_m: M\to M_3$ the projections along $n$ and $m$, respectively. Let $\Delta_3$ and $\SDelta_3$ be a pair of dual polytopes that are reflexive with respect to $M_3$ and $N_3$, respectively.

Let $\Top{A}\subset N_\IR$ be a top over $\SDelta_3$, that is a lattice polytope satisfying $\langle m,y\rangle\ge 0,\, \forall y\in\Top{A}$, such that the facet saturating the inequality is $\SDelta_3$. Let $\{u_i\in M\}$ be a set of lattice vectors, such that $\langle u_i,y\rangle\ge -1,\,\, \forall y\in \Top{A}$. Define a lattice polytope $\Bot{\widetilde A}\subset M_\IR$:
\begin{center}
$\Bot{\widetilde A} = \{x\in M_\IR: ~\langle x,n\rangle\le 0,~\langle x,y\rangle\ge -1 ~~\forall\,y \in \Top{A} \}$
\end{center}
and similarly define $\Top{\widetilde {\widetilde A}}$. Further, assume that $\pi_n\Top{A}=\SDelta_3$. Then:

\begin{itemize}
\item[{\itshape a)}] the polytope $\Bot{\widetilde A}$ is a bottom under $\Delta_3$ with $\pi_m\Bot{\widetilde A}=\Delta_3$,
\item[{\itshape b)}] $\Top{\widetilde {\widetilde A}}=\Top{A}$, and
{\item[{\itshape c)}] the union $\topbot{A}{B}$ of $\Top{A}$ and any bottom $\Bot{B}$ with 
$\pi_n\Bot{B}=\SDelta_3$ is a reflexive polytope\\[5pt] with 
$\big(\topbot{A}{B}\big)^{{\!*}}=\topbot{\widetilde B}{\widetilde A}$.}
\end{itemize}
}

\vskip 20pt
{\bf Proof.} The four-dimensinoal dual of $\SDelta_3$ (in the sense of definition \ref{dual}) is the infinite prism $P^M = \IR \times \Delta_3\subset M_\IR$. Define the semi-infinite prism $P^M_{\le }=\IR_{\le }\times \Delta_3$, such that $\langle x,n\rangle\le 0$ for $x\in P^M_{\le }$. Analogously, define $P^M_{\ge }$, $P^N_{\ge }$ and $P^N_{\le }$. Then $\Bot{\widetilde A} = \Top{A}^*\cap P^M_{\le }$, where $\Top{A}^*$ is the unbounded polyhedron resulting from applying (\ref{dual}) to $\Top{A}$. The projection condition ensures that $P^M_{\ge} \subset \Top{A}^*$ and hence $\Top{A}^* = \Bot{\widetilde A} \cup P^M_{\ge }$. 

\begin{itemize}
\item[{\itshape a)}] The polytope $\Bot{\widetilde A}$ contains $\Delta_3$ as a consequence of its definition 
and the fact that $\Top{A}$ projects to $\SDelta_3$.
The facets of $\Bot{\widetilde A}$ are those of a bottom by definition, and 
its vertices are lattice points since they are either the $u_i$ or vertices of 
$\Delta_3$. 
The bottom $\Bot{\widetilde A} = \Top{A}^*\cap P^M_{\le }$ projects to $\Delta_3$ since 
$P^M_{\le }$ projects to $\Delta_3$.
\item[{\itshape b)}] $\Top{\widetilde {\widetilde A}}=\Bot{\widetilde{A}}^*\cap P^N_\ge = \left(\Top{A}^* \cap P^M_\le \right)^*\cap P^N_\ge = \left((\Top{A}^*)^*\cup P^N_\le\right)\cap P^N_\ge\,=\,\big(\Top{A}\cup P^N_{\le }\big)\cap P^N_{\ge }
\,=\,\Top{A}$
\item[{\itshape c)}] By construction, $\topbot{A}{B}$ is bounded by facets of the type 
$\langle u_i,y\rangle = -1$ and has vertices in~$N$. Convexity is a 
consequence of the projection conditions. The dual is given by
$$\big(\topbot{A}{B}\big)^*=\Top{A}^*\cap \Bot{B}^*
= \big(\Bot{\widetilde A} \cup P^M_{\ge }\big)\cap \big(\Top{\widetilde B}\cup P^M_{\le }\big)=
\topbot{\widetilde B}{\widetilde A}.$$
\hfill$\Box$
\end{itemize}

\subsection{An additivity lemma for the Hodge numbers}
The previous lemma taught us that whenever we have projections both at the $M$ and the $N$ lattice side, the top is determined by the dual bottom and vice versa. Also, a top and a bottom that project to $\SDelta_3$ along the same direction $n$ can always be combined into a reflexive polytope.

The following lemma shows that, under a specific assumption on the structure of $\SDelta_3$, this composition obeys additivity in the Hodge numbers of the resulting Calabi-Yau threefolds.

\vskip 20pt
{\bf Lemma 2.} {Let $\SDelta_3$ be a three-dimensional polytope that is reflexive with respect to the lattice $N_3$, with no edge $e^*$ such that both $e$ and $e^*$ have an interior lattice point. Let $N=N_1\oplus N_3$, where $N_1$ is generated by the primitive lattice vector $n$, and assume that $\Top{A}$ and $\Top{C}$ are tops and $\Bot{B}$ and $\Bot{D}$ are bottoms in $N_\IR$ that project to $\SDelta_3$ along $n$. 

Then the following relation holds:
\begin{equation}\label{HodgeFormula}
h^{\bullet\bullet}\big(\topbot{A}{B}\big)+h^{\bullet\bullet}\big(\topbot{C}{D}\big)~=~
h^{\bullet\bullet}\big(\topbot{A}{D}\big)+h^{\bullet\bullet}\big(\topbot{C}{B}\big)~
\end{equation}
where $h^{\bullet\bullet}$ stands for the Hodge numbers $\hodgenos$ of the Calabi-Yau hypersurface determined by the respective polytope.
}

\vskip 20pt
{\bf Proof.} The Hodge number $h^{1,1}$ is given~\cite{Batyrev:1993dm} by
$$ 
h^{1,1}~=~l(\SDelta)-5-\hskip-5pt\sum_{{\rm codim\, }\theta^* =1}l^*(\theta^*)+
		 \hskip-5pt\sum_{{\rm codim\, }\theta^* =2}l^*(\theta^*)l^*(\theta)~,   
$$
where $l(\SDelta)$ denotes the number of lattice points of $\SDelta$ and $l^*(\theta)$  denotes the number of interior lattice points of a face $\theta$.
\vspace{10pt}

This formula can be rewritten as
$$
h^{1,1} ~= -4 + \hskip-5pt\sum_{P\in\SDelta\cap N} \mathrm{mult}(P)
$$
where the sum runs over all the lattice points of $\SDelta$ and the multiplicities $\mathrm{mult}(P)$ are defined as
$$
\text{mult}(P)~= 
\begin{cases}
0 & \mbox{if } P \mbox{ is the interior point of } \SDelta \mbox{  or is interior to a facet, }\\[4pt] 
1 & \mbox{if } P \mbox{ is a vertex or interior to an edge of } \SDelta, \\[4pt]
\mbox{length}(\theta) & \mbox{if } P \mbox{ is interior to a 2-face } \theta^* \mbox{ of } \SDelta \mbox{ and } \theta \mbox{ is the dual edge of } \Delta, 
\end{cases}
$$
where the length of an edge is the number of integer segments, i.e.~$\text{length}(\theta) = l^*(\theta) +1$.

\vskip30pt
In the following we will see that the contributions of any lattice point $P$ add up to the same value for the different sides of the Hodge number relation. We distinguish the following cases:
\begin{itemize}
\item[{\itshape A)}] The case $\langle n,P\rangle \ne 0$. Assume, without loss of generality, that $P\in\Top{A}$. If $P$ is a vertex, or is interior to either a facet or an edge, it will contribute the 
same to $h^{1,1}(\topbot{A}{B})$ as to $h^{1,1}(\topbot{A}{D})$.

Otherwise we use the decomposition $P=\pi_nP+\lambda\, n$ with $\lambda > 0$: if $P$ is interior to the face~$\theta^*$ then any point $Q\in \theta$ must satisfy 
$$
-1~=~\langle Q,P\rangle ~=~\langle Q,\pi_nP\rangle+\langle Q,\lambda n\rangle
~\ge~ -1 +\lambda \langle Q,n\rangle~,
$$
hence $\langle Q,n\rangle \le 0$, so all of $\theta$ lies in the bottom $\Bot{\widetilde A}$ determined by the top  $\Top{A}$ to which $P$ belongs. In other words, $P$'s contribution to 
$h^{1,1}(\topbot{A}{B})$ is again the same as its contribution to $h^{1,1}(\topbot{A}{D})$.

\item[{\itshape B)}] $P$ is a vertex of $\SDelta_3$. Then $P$ is a vertex or interior to an edge for each of the four polytopes $\topbot{A}{B}, \topbot{C}{D},\topbot{A}{D},\topbot{C}{B}$ occurring in the Hodge number relation, so $P$ contributes the same value of unity each~time.

\item[{\itshape C)}] $P$ is interior to an edge $e^*$ of $\SDelta_3$.
There are two possibilities:
\begin{itemize}
\item[a)] $e^*$ is an edge of $\SDelta$, in which case $P$ contributes 1 to  
$h^{1,1}$.
\item[b)] $P$ lies within a two-face of $\SDelta$ which is dual 
(in the four-dimensional sense) 
to \hbox{$e\subset \Delta_3\subset\Delta$}. 
By our assumptions, $e$ has length 1 so $P$ again~contributes~1.
\end{itemize}
\item[{\itshape D)}] $P$ is interior to a facet of $\SDelta_3$ that is dual to a vertex 
$v$ of $\Delta_3$. In $\SDelta$, the face $\theta^*$ to which $P$ is interior 
can be a facet dual to $v$, implying $\text{mult}(P)=0$, or a codimension 2-face dual to an edge $\theta$ that projects to $v$, in which case 
$\text{mult}(P)=\text{length}(\theta)$.
The length of such an edge is additive under the composition of tops with 
bottoms, hence the contribution of $P$ is additive again.
\end{itemize}

This shows additivity for $h^{1,1}$.  Additivity for $h^{1,2}$ follows from the compatibility of top-bottom composition with mirror symmetry for projecting tops and bottoms.

\hfill$\Box$
 
\vskip 30pt
The assumption on the dual pairs of edges of the $K3$ polytopes is necessary, as the following example shows. Consider the polytope associated with the gauge group $F_4{\times} G_2$, as in the following figure. This polyhedron is self-dual and possesses dual pairs of edges of length $>1$. 

\begin{figure}[h]
\begin{center}
\phantom{}
\vskip20pt
\framebox{\hskip30pt
\begin{minipage}[t]{4.5in}
 $\begin{array}{rrrr}
(&\hskip-10pt -2,&2, &~3)\\[3pt]
(&                   0,&-1, &0)\\[3pt]
(&                   0,&  0,&-1)\\[3pt]
(&                   3,&  2,&3)\end{array}$
\hfill 
\raisebox{-1in}{\includegraphics[width=5.7cm]{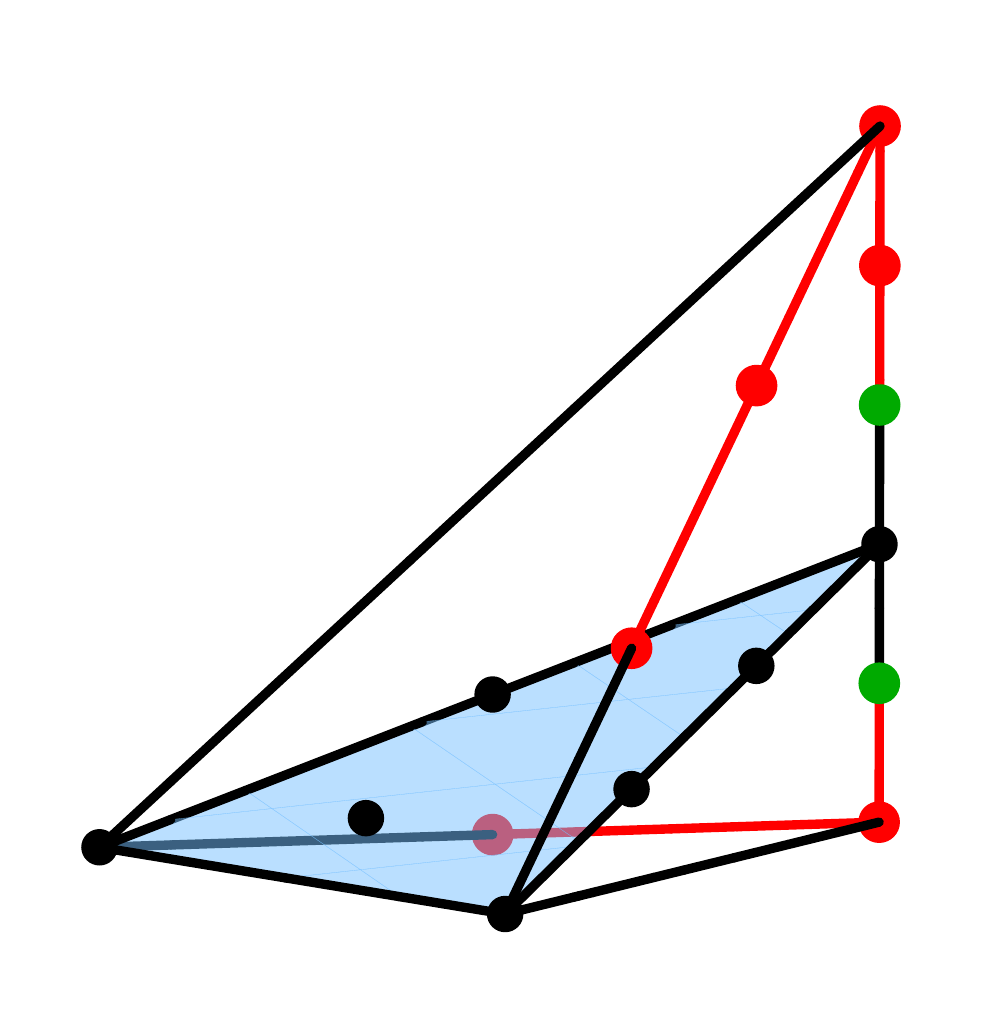}}
\hspace*{25pt}
\end{minipage}}
\capt{6.3in}{F4G2poly}{The polyhedron for an elliptically fibered $K3$ manifold of the type $F_4{\times} G_2$.}
\end{center}
\vskip -10pt
\end{figure}

A minimal extension of the $F_4{\times}G_2$ polyhedron to a four-dimensional top
\[
\Top{\text{min}} = \text{Conv}\Big(\big\{(0, -2, 2, 3),\, (0, 0, -1, 0),\, (0, 0,  0, -1),\, (0, 3, 2, 3),\, 
(1, 0, 0, 0)\big\}\Big)
\]
is easily seen to be dual to a prism shaped bottom
$\Bot{\text{prism}} = [-1,0]\times\SDelta_3$.

One finds the Hodge numbers 
\begin{equation*} 
\big(h^{1,1},h^{1,2} \big)\big(\topbot{\text{min}}{\text{min}}\big)=(21,45),~~~
\big(h^{1,1},h^{1,2} \big)\big(\topbot{\text{prism}}{\text{prism}}\big)=(45,21),
\end{equation*}
while
\begin{equation*}
\big(h^{1,1},h^{1,2} \big)\big(\topbot{\text{min}}{\text{prism}}\big)=
\big(h^{1,1},h^{1,2} \big)\big(\topbot{\text{prism}}{\text{min}}\big)=(31,31)
\end{equation*}
and we see that these Hodge numbers violate Formula the Hodge number relation.

\vskip20pt
For constructing this example we have used the fact that an edge in a toric diagram corresponding to a Lie group satisfies the assumption precisely if the Lie group is simply laced, i.e.~of ADE type~\cite{Perevalov:1997vw}. However, for applying Lemma 2 we do not necessarily require an elliptic fibration; conversely, an elliptic fibration structure with simply laced gauge groups is not sufficient since edges of the fiber polygon may violate the condition.

\newpage
\section{The $E_8{\times}\{1\}$ K3 Polytope and Its Web of Fibrations}

Let us now return to the questions raised in Section \ref{Sec2.1} in relation with the Hodge plot associated with the Kreuzer-Skarke list of 4-dimensional reflexive polytopes. I will explain how many of the intriguing features of the plot can be explained in terms of assembling reflexive polytopes that describe $K3$-fibered Calabi-Yau three-folds by mixing and matching tops over the same fiber in the spirit of Formula \eqref{HodgeFormula}. 

\vspace{21pt}
\begin{figure}[h]
\begin{center}
\includegraphics[width=6.5in]{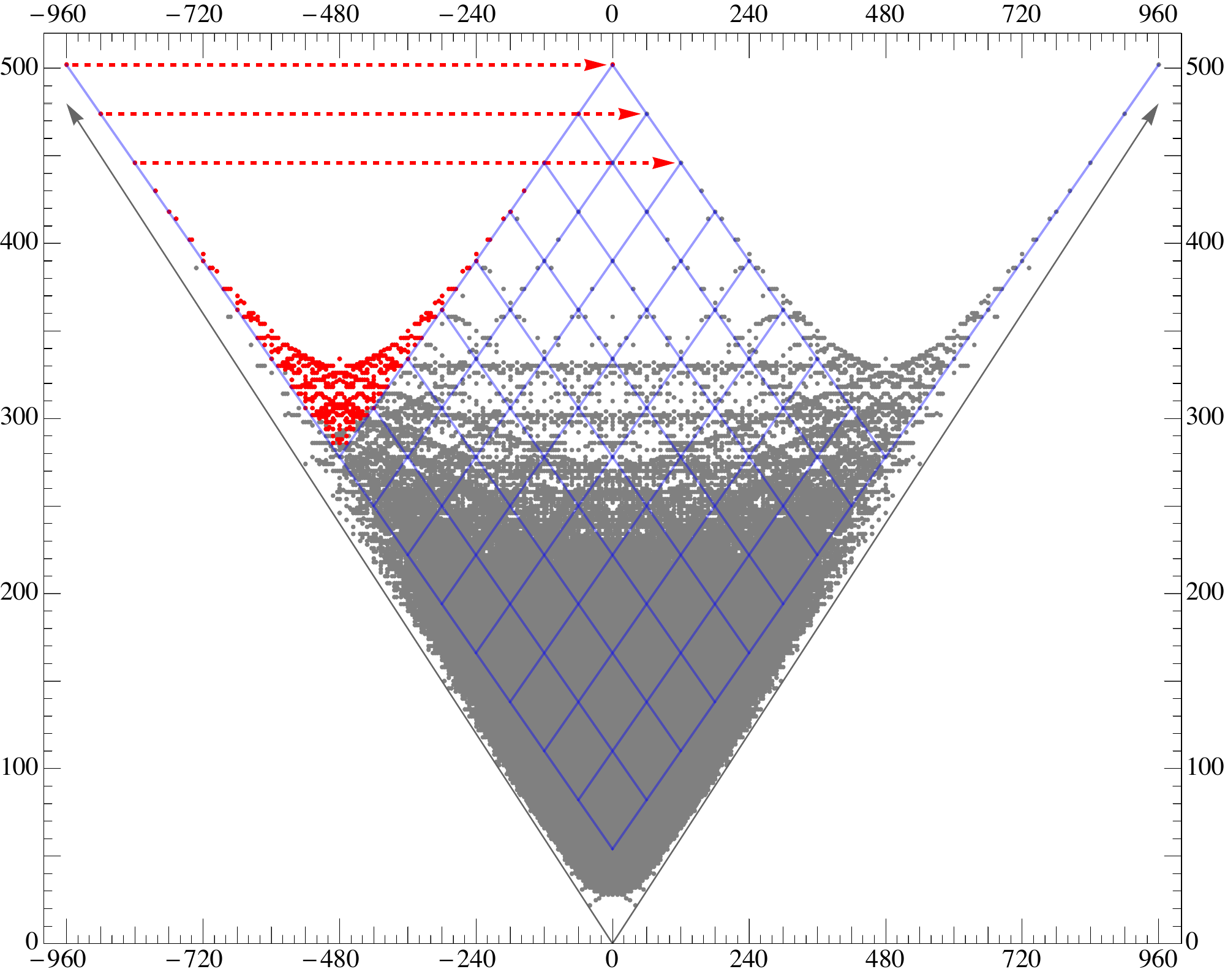}
\capt{6.5in}{BigBotPoints}{The 465 pairs of Hodge numbers that result from combining the 1263 possible $E_8{\times}\{1\}$ tops with the maximal bottom. Note that this structure is contained in, though is not identical to, the structure shown on the left in \fref{YStructure}. Exchanging the maximal bottom for the minimal bottom shifts the entire structure by the vector $\D(\chi,y)=(960,0)$, corresponding to the red arrows.}
\end{center}
\vskip-10pt
\end{figure}

Consider the mirror pair $\big(\cM_{11,491}$, $\cM_{491,11}\big)$ of Calabi-Yau three-folds occupying the top left and top right positions in the Hodge plot. Here, the subscripts refer to the Hodge numbers. The polytope $\Delta_{11,491}=\D^{^{\!\!*}}_{491,11}$ is actually the largest (that is, it has the largest number, 680, of lattice points) of all reflexive 4-polytopes and admits two distinct slicings along $K3$ polytopes. In the first case the $K3$ polytope is the largest reflexive 3-polytope (with 39 lattice points), which can be associated with a gauge group of either $E_8{\times} E_8$~\cite{Candelas:1996su} or $SO(32)$~\cite{Candelas:1997pq}; slicing along this $K3$ polytopes leads to the largest known gauge groups coming from F-theory compactification~\cite{Candelas:1997eh,Candelas:1997pq}.

\subsection{Translation vectors and the Hodge number formula}

For the present discussion, however, I will concentrate on the second case, where the $K3$ polytope is self-dual and corresponds to $E_8{\times}\{ 1\}$ (for the classification of elliptic-$K3$ polyhedra by Lie groups see, for example,~\cite{Candelas:1996su, Bouchard:2003bu}).  Moreover, the top and the bottom are the same and they are the largest among all the available tops and bottoms for that $K3$ manifold.  The mirror is also an elliptic $K3$ fibration. Its polytope is divided into a top and a bottom by the same slice, though now this top (which is again the same as the bottom) is the smallest  among all the available tops and~bottoms. By taking arbitrary tops, from the ones which fit this $K3$ slice, together with the biggest bottom we are able to obtain many elliptically fibered \cys . These give rise to the red points that form the $V$-shaped structure on the left of \fref{BigBotPoints}. Exchanging the maximal bottom with the smallest one, while keeping the top fixed, shifts the Hodge numbers by $\Delta\big(h^{1,1},h^{1,2} \big)=(240,-240)$,  in terms of the coordinates of the plot this is $\D(\chi,y)=(960,0)$, corresponding to the red arrow of~the~figure.

There are 1,263 different tops, and so also 1,263 bottoms that project onto this $K3$ slice, and these can be attached along the $K3$ polytope to obtain $1263{{\times}}1264/2 = 798,216$ \cys which are elliptic-$K3$ fibrations. This already gives us the largest collection of elliptic-$K3$ fibrations known hitherto. There are many different $K3$ polyhedra to which this construction can be applied, only a few of which we will discuss here.

The Hodge numbers of the manifolds obtained by combinations of the 1263 tops and bottoms mentioned above are related by Formula \eqref{HodgeFormula}. The 1,263 tops, when combined with the maximal bottom give rise to a set of 465 pairs of Hodge numbers. These are the points  shown in red in \fref{BigBotPoints}. I will refer to this set of points as the $V$-structure. The relation \eqref{HodgeFormula} has an important consequence for these points. Consider any vector taking the pair (11,491) to one of the remaining 464 points. The additive property of the Hodge numbers \eqref{HodgeFormula} ensures that we may translate the entire \hbox{$V$-structure} by each of these 464 vectors. Translating the $V$-structure by these vectors explains much of the repetitive structure associated with the blue grid~of~\fref{BigBotPoints}. It also enables us to calculate the Hodge numbers of the resulting Calabi-Yau manifolds. In this way, we find 16,148 distinct pairs of Hodge numbers. The result of performing all 464 translations on the $V$-structure is~shown~in~\fref{AllE8TimesSU1}.

The vectors with which our considerations began are included in the $V$-structure translations. The vector $\D\hodgenos=(240,-240)$ arises as the difference between $\hodgenos=(11,491)$ and 
$\hodgenos=(251,251)$. We know that $\D\hodgenos=(1,-29)$ appears among the translations, in fact the vectors $\D\hodgenos=k{\times}(1,-29)$ appear  for $1\leq k\leq 7$. 
Finally, the blue grid~of~\fref{BigBotPoints} closes due to the fact that the horizontal shift $\Delta\big(h^{1,1},h^{1,2} \big)=(240,-240)$ is related to the 
$\D(h^{11},h^{21})=(1,-29)$ shift and its mirror reflection by 
$$
(240,-240) ~=~ 8{\times}(1,-29) - 8{\times}(-29,1)~.
$$
Note, however, that the vector $8{\times}(1,-29)$ is not, itself, a $V$-structure translation. 
\begin{figure}[!t]
\begin{center}
\includegraphics[width=6.5in]{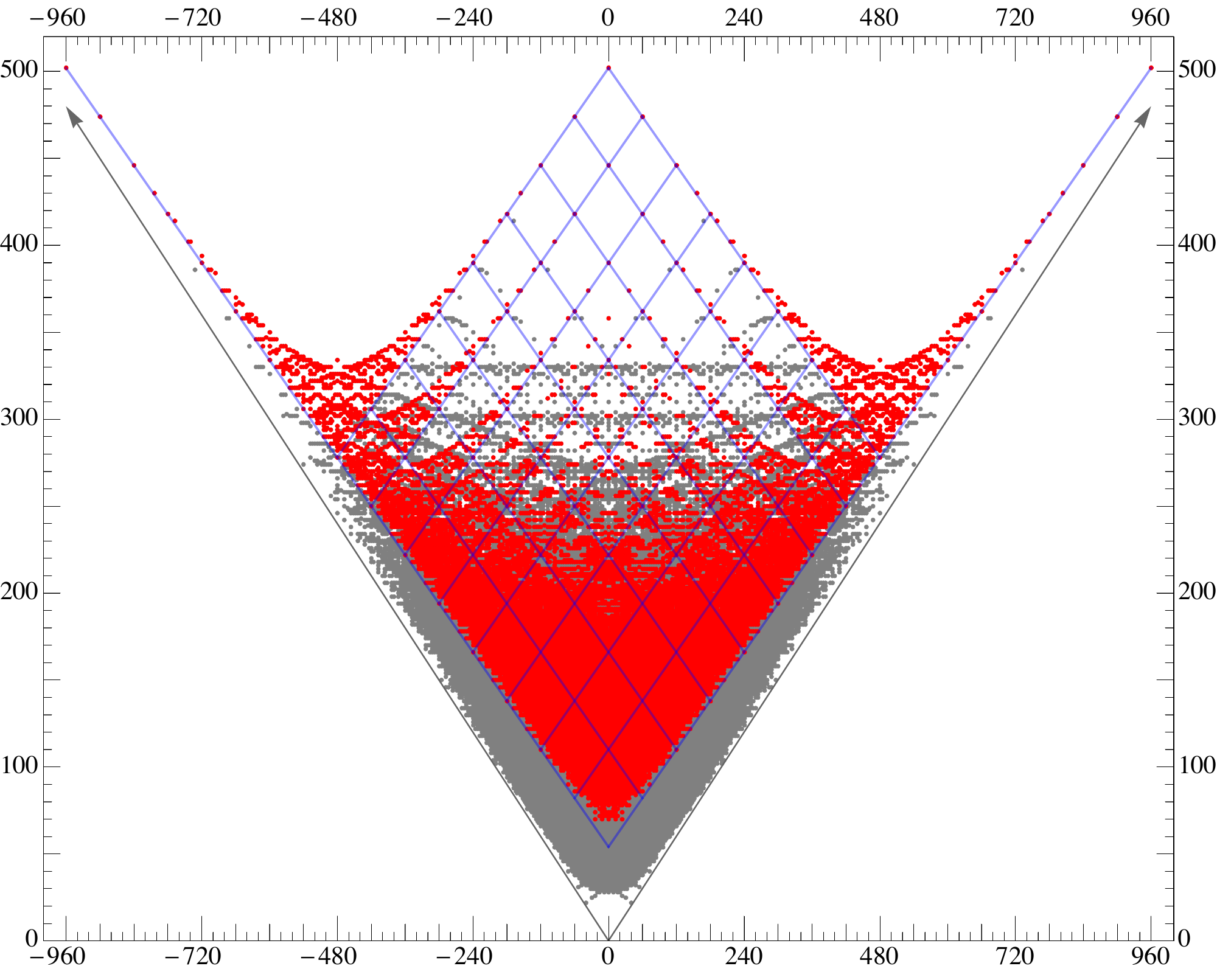}
\capt{5.5in}{AllE8TimesSU1}{The Hodge plot showing, in red, all $E_8{\times}\{1\}$ $K3$ fibrations for which the fiber is both a slice and a projection in the four-dimensional reflexive polytope.}
\end{center}
\end{figure}

\subsection{The web of $E_8{\times} \{1\}$ elliptic $K3$ fibrations}\label{E81}
So far, I have not mentioned the manner in which the complete set of tops over the $E_8{\times} \{1\}$ $K3$ fiber was obtained. A direct approach to this question would be to construct all possible tops from scratch as convex lattice polytopes with one facet containing the origin and all the other facets at lattice distance one from the origin. I will present below a method which avoids this computationally challenging task, by searching for $E_8{\times} \{1\}$ $K3$ fibrations through the list of reflexive 4-polytopes. 

The polytope $\SDelta_{E_8{\times}\{1\}}$ is isomorphic to the convex hull of the vertices
$$
\begin{array}{rrrr} 
(&\hskip-10pt -1,& 2,& 3)\\[2pt]
(&                   0,& -1,& 0)\\[2pt]
(&                   0,& 0, &-1)\\[2 pt]
(&                   6,& 2, & 3)\rlap{~.}
\end{array}
$$

This 3-polytope is shown in \fref{polyselection}. By extending the lattice into a fourth dimension and adding the point $(1, 6, 2, 3)$, one obtains a top, denoted by  $\Top{\text{min}}$. Similarly, adding the point $(-1, 6, 2, 3)$ results in the bottom $\Bot{\text{min}}$.
Adding both points gives the reflexive polytope $\topbot{\text{min}}{\text{min}}$, whose vertices are listed in Table \ref{E8SU1vertices}, along with the vertices for the dual~polytope $\topbot{\text{max}}{\text{max}}$.

\begin{table}[H]
\vspace{14pt}
\def\str{\varstr{12pt}{6pt}}
\begin{center}
\begin{tabular}{| >{$~~} r <{~~$} | >{$~~} r <{~~$} | >{$~~} r <{~~$} |}
\hline
\varstr{16pt}{8pt} \topbot{\text{min}}{\text{min}}^{11,491}
& \topbot{\text{max}}{\text{max}}^{491,11} & \topbot{\text{min}}{\text{max}}^{251,251} \\
\hline\hline
\varstr{14pt}{6pt} (-1,\ \ \ 6,\ \ \ 2,\ \ \ 3) & (-42,\ \ \ 6,\ \ \ 2,\ \ \ 3) & (-42,\ \ \ 6,\ \ \ 2,\ \ \ 3) \\
\str (\ \ 0,\ -1,\ \ \ 2,\ \ \ 3)     & (\hskip14pt 0,\  -1,\ \ \ 2,\ \ \ 3)       & (\hskip14pt 0,\ -1,\ \ \ 2,\ \ \ 3) \\
\str (\ \ 0,\ \ \ 0,\  -1,\ \ \ 0)    & (\hskip14pt 0,\ \ \ 0,\  -1,\ \ \ 0)       & (\hskip14pt 0,\ \ \ 0,\  -1,\ \ \ 0) \\
\str (\ \ 0,\ \ \ 0,\ \ \ 0,\  -1)    & (\hskip14pt 0,\ \ \ 0,\ \ \ 0,\  -1)       & (\hskip14pt 0,\ \ \ 0,\ \ \ 0,\  -1) \\
\varstr{12pt}{8pt} 
(\hskip9pt 1,\ \ \ 6,\ \ \ 2,\ \ \ 3)  & (\hskip9.5pt 42,\ \ \ 6,\ \ \ 2,\ \ \ 3)   & (\hskip16pt 1,\ \ \ 6,\ \ \ 2,\ \ \ 3) \\
\hline \hline
\end{tabular}
\capt{5.0in}{E8SU1vertices}{The minimal, the maximal and the maximal self-dual reflexive polytopes containing the $E_8{\times}\{1\}$ $K3$~polyhedron as a slice and as a projection.}
\end{center}
\end{table}
The Calabi-Yau threefold $\cM_{11,491}$ is determined by $\SDelta_{11,491}\cong\topbot{\text{min}}{\text{min}}$ and $\Delta_{11,491}\cong\topbot{\text{max}}{\text{max}}$. Combining $\Top{\text{min}}$ with $\Bot{\text{max}}$
results in a self-dual reflexive polytope $\topbot{\text{min}}{\text{max}}$ whose vertices are listed in the third column of Table~\ref{E8SU1vertices}, corresponding to the self-mirror threefold $\cM_{251,251}$. This manifold with vanishing Euler number is indicated in \fref{BasicKSPlot} by the topmost point lying on the axis 
$\chi=0$. 

The manifolds $\cM_{11,491}$ and $\cM_{251,251}$ are related by what we called `half-mirror symmetry' in the introduction. In fact, `half-mirror symmetry' corresponds to a combination of mirror symmetry with replacing $\Top{\text{max}}$ by $\Top{\text{min}}$ or $\Bot{\text{max}}$ by 
$\Bot{\text{min}}$.

\subsection{Searching for $E_8{\times} \{ 1\}$ elliptic $K3$ fibrations} Adding points to the polytope $\SDelta_{11,491}\simeq\topbot{\text{min}}{\text{min}}$ will not decrease $h^{1,1}$. This can be seen as follows.   For the smooth fibration $\cM_{11,491}$, the Picard number $h^{1,1} = 11$ comes from the 10 toric divisors in the $K3$, as well as from the generic fiber (the $K3$ itself). Enlarging the top/bottom corresponds to blowing up the points $z = 0$ or $z = \infty$ of the $\IC\IP^1$ that is the base of the fibration. These blowups take place separately at the two distinguished points and add up to 240 exceptional divisors at each of them. Denote by $N_{\text{top}}$ and $N_{\text{bottom}}$ the number of exceptional divisors resulting from adding points in the top 
$\Top{\text{min}}$ and bottom $\Bot{\text{min}}$ respectively. Then we have $h^{1,1} = 11 + N_{\text{bottom}} + N_{\text{top}}$ with $0\leq N_{\text{top}}, N_{\text{bottom}}\leq 240$. The maximal bottom corresponds to $h^{1,1} = 251 + N_{\text{top}}$. 

This argument tells us that the reflexive polytopes containing the maximal bottom and an arbitrary top are characterised by $h^{1,1}\geq 251$ and positive Euler number. The Kreuzer-Skarke list contains 2,219 polytopes that pass both of these requirements. It is possible to identify the polytopes that contain a maximal bottom by searching for a distinguished 3-face containing 4 vertices and 24 points. In the representation 
of \tref{E8SU1vertices}
the facet in question has vertices
$$ 
\big\{(-42, 6, 2, 3),\ (0, -1, 2, 3),\ (0, 0, -1, 0),\ (0, 0, 0, -1)\big\}~.
$$

This facet is isomorphic to the 3-polytope describing the $E_8\times \{1\}$ $K3$ fiber. It is also important to note that this facet is not orthogonal to the hyperplane determined by the 3-polytope associated with the $K3$. As such, since we are searching for polytopes which contain the $K3$ polytope both as a slice and as a projection (corresponding to $K3$ fibrations for which the mirror image is also a $K3$ fibration), this facet cannot extend into the top half. This means that, in order to find all reflexive polytopes containing the maximal bottom, it is enough to search for those reflexive polytopes which contain the distinguished facet and then check that this facet indeed belongs to a maximal bottom. 

The search yields a list of 1,263 reflexive polytopes, and thus an equal number of distinct tops, which are available at the URL given in Section~\ref{webaddress}. The Hodge numbers associated with these reflexive polytopes (465 distinct Hodge pairs) are shown in red in \fref{BigBotPoints}.  

\subsection{Generating all $E_8{\times} \{1\}$ elliptic $K3$ fibrations} 
It is now easy to generate a full list of polytopes corresponding to Calabi-Yau threefolds exhibiting the $E_8{\times} \{1\}$ $K3$ fibration structure discussed above. Indeed, this can be realised by taking all possible combinations $\topbot{A}{B}$, with $\Top{A}$ and $\Bot{B}$ being tops and bottoms from the previous list, glued along the polyhedron corresponding to the $K3$ fiber. 
There are 798,216 such reflexive polytopes. The 16,148 distinct Hodge pairs associated with these polytopes are shown in red in \fref{AllE8TimesSU1}. Prior to having a proof of \eqref{HodgeFormula}, I have checked the following, equivalent, relation for each of the combinations:
$$
h^{\bullet\bullet}\big(\topbot{A}{B}\big)~=~ h^{\bullet\bullet}\big(\topbot{A}{\text{max}}\big)+
h^{\bullet\bullet}\big(\topbot{\text{max}}{B}\big)-h^{\bullet\bullet}\big(\topbot{\text{max}}{\text{max}}\big)~.
$$

\vskip 40pt

\section{Other $K3$ Polytopes}
There are many types of elliptically fibered $K3$ polytopes. These are associated with groups $G_1{\times}G_2$ which describe the way in which the elliptic fiber degenerates along the base, according to the ADE classification of singularities. The elliptic fibration structure manifests itself as a slice along a reflexive polygon, with corresponding three-dimensional tops and bottoms to which the gauge groups can be associated, as first observed 
in~\cite{Candelas:1996su}.

\begin{figure}[h]
\begin{center}
\includegraphics[width=6.5in]{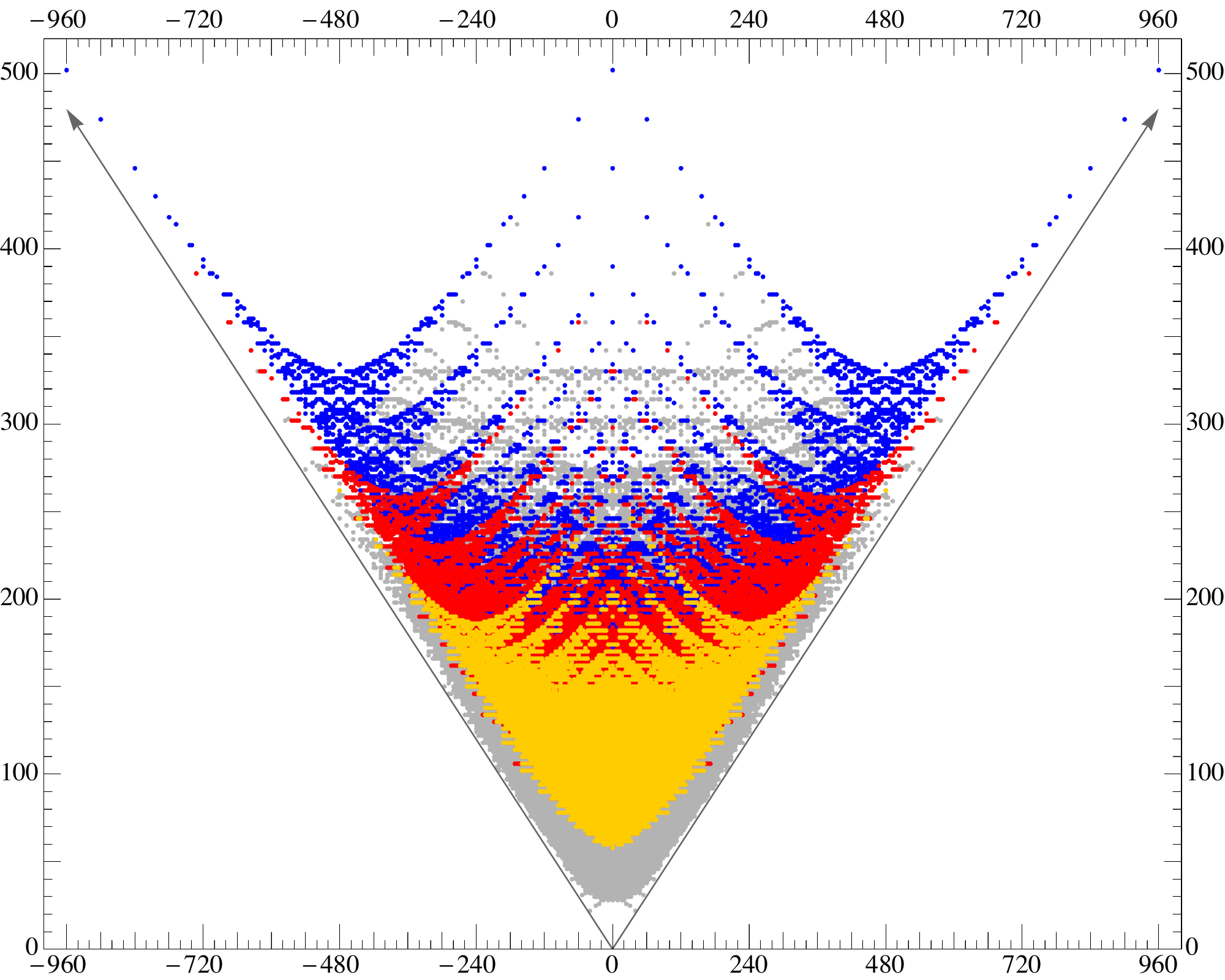}
\capt{5.3in}{RBGplot}{Hodge plot showing in blue all $E_8{\times} \{1\}$ $K3$ fibrations; in red all $E_7{\times} \{1\}$ and $E_8{\times} SU(2)$ $K3$ fibrations; and in yellow: all $E_7{\times} SU(2)$ $K3$ fibrations.}
\end{center}
\vskip20pt
\end{figure}

The complete lists of tops for any of the 16 reflexive polygons can be found in~\cite{Bouchard:2003bu}. In the situations presented here, the reflexive polygon is always the triangle corresponding to a Weierstrass model. Moreover, in the following discussion, both the $K3$ polyhedron and its dual are sub-polytopes of the polytope associated with $E_8\times E_8$. As such, the dual 3-polytope gives rise to $\widetilde G_1{\times}\widetilde G_2$,
where $\widetilde G_1$ is the commutant of $G_1$ in $E_8$, and likewise for $G_2$. 

In particular, if $G_2$ is the commutant of $G_1$ in $E_8$, the $K3$ polyhedron is self-dual. The class of elliptic-$K3$ fibered Calabi-Yau manifolds discussed in the previous section belongs to this category and corresponds~to~$(G_1,G_2)=(E_8,\{1\})$. 

I have also studied the case $(G_1,G_2)=(E_7, SU(2))$ which is very similar to the previous one and which gives rise to $725,410$ polyhedra and $7,929$ pairs of Hodge numbers. The $E_7{\times}SU(2)$ structure corresponds to the yellow points in \fref{RBGplot}. While similar to the $E_8\times\{1\}$ case, this structure is not immediately apparent in the original Hodge plot of \fref{BasicKSPlot} owing to the fact it appears in the region of the plot that is very dense.  

\goodbreak
The construction can be generalised to slices corresponding to non-self-dual $K3$ polyhedra. Below I discuss the case in which the $K3$ polyhedron has $(G_1,G_2)=(E_7, \{1\})$ and the corresponding dual $K3$ polyhedron has $(G_1,G_2)=(SU_2, E_8)$. The $E_7\times \{1\}$ structure and its mirror are presented in \fref{E7E8str}. The top part of these structures, which is not overlaid by the $E_7\times SU(2)$ structure corresponds to the red points in~\fref{RBGplot}. The $E_7\times \{1\}$ structure and its mirror are particularly interesting as they pick up some of the points of the structure presented in \fref{YStructure} that lie to the left of the $V$-structure.

\subsection{The web of $E_7{\times} SU(2)$ elliptic $K3$-fibrations}\label{E71}

For the elliptic $K3$ surface with degenerate fibers of the type $E_7 {\times} SU(2)$ we consider the polyhedron given by the vertices shown in \fref{K3polyhedron2}. As before, we extend this polyhedron to a 4-dimensional reflexive polytope by adding two points, above and below the point $(0,4,2,3)$. The resulting polytope, as well as its dual are given in~Table~\ref{E7SU2vertices}.
\vskip20pt
\begin{figure}[H]
\begin{center}
\framebox{\hskip30pt
\begin{minipage}[t]{4.5in}
$\begin{array}{rrrr}
(&\hskip-10pt -1,& 1,& 2)\\
(&\hskip-10pt -1,& 2,& 3)\\
(&                   0,& -1,& 0)\\
(&                   0,&  0,&-1)\\
(&                   0,&  2,& 3)\\
(&                   2,&  0,& 1)\\
(&                   4,&  2,& 3)\end{array}$
\hfill
\raisebox{-1.1in}{ \includegraphics[width=5.cm]{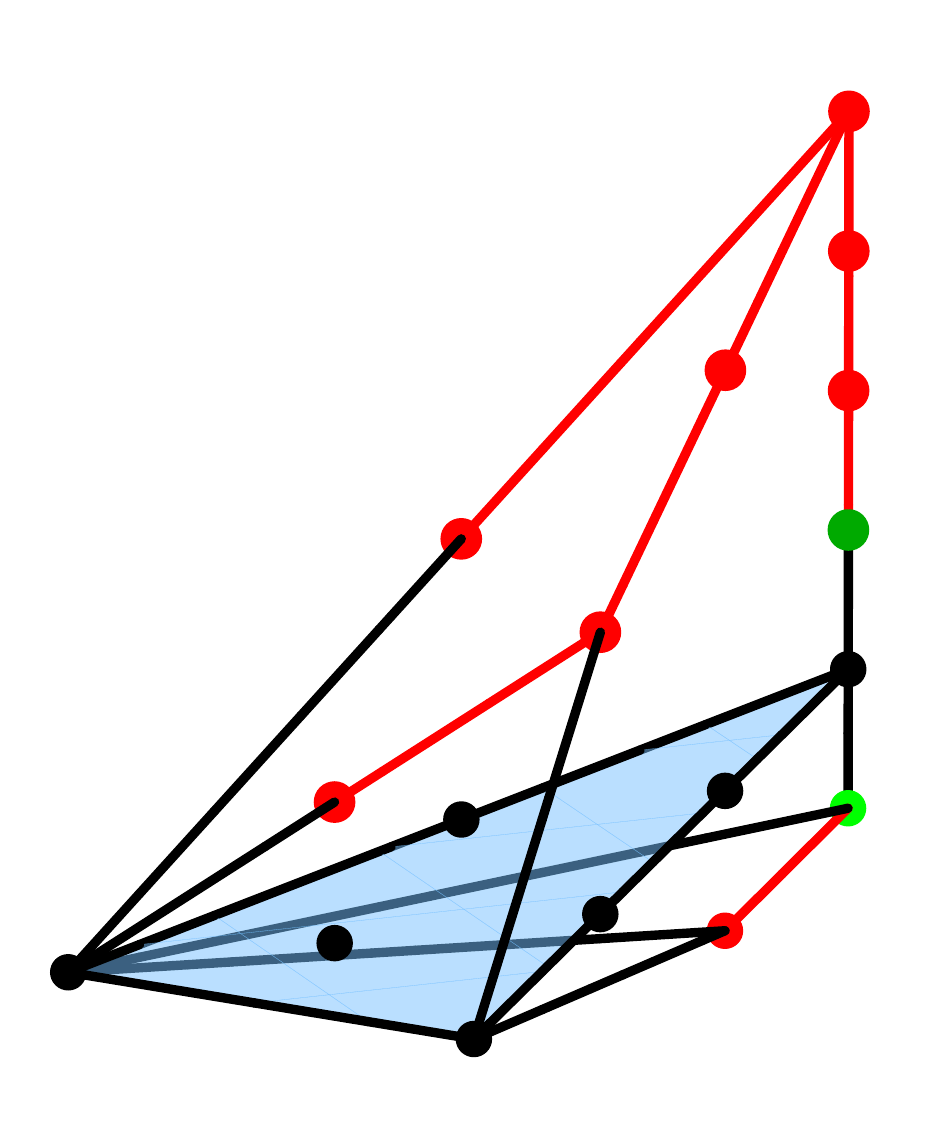}}
\hspace*{25pt}
\end{minipage}}
\capt{6.3in}{K3polyhedron2}{Elliptically fibered $K3$ manifold of the type $E_7{\times} SU(2)$.}
\end{center}
\end{figure}
\vfill
\begin{table}[H]
\def\str{\varstr{12pt}{6pt}}
\begin{center}
\begin{tabular}{| >{$~~} r <{~~$} | >{$~~} r <{~~$} |}
\hline
\varstr{16pt}{8pt} 
\topbot{\text{min}}{\text{min}}^{11,251} & \topbot{\text{max}}{\text{max}}^{251,11} \\
\hline\hline
\varstr{14pt}{6pt} (-1,\ \ \ 4,\ \ \ 2,\ \ \ 3) & (-22,\ \ \ 4,\ \ \ 2,\ \ \ 3) \\
\str (\ \ 0,\ -1,\ \ \ 1,\ \ \ 2) & (\hskip13.5pt 0,\ -1,\ \ \ 1,\ \ \ 2) \\
\str (\ \ 0,\ -1,\ \ \ 2,\ \ \ 3) & (\hskip13.5pt 0,\ -1,\ \ \ 2,\ \ \ 3) \\
\str (\ \ 0,\ \ \ 0,\ -1,\ \ \ 0) & (\hskip13.5pt 0,\ \ \ 0,\ -1,\ \ \ 0) \\
\str (\ \ 0,\ \ \ 0,\ \ \ 0,\  -1) & (\hskip13.5pt 0,\ \ \ 0,\ \ \ 0,\  -1) \\
\str (\ \ 0,\ \ \ 0,\ \ \ 2,\ \ \ 3) & (\hskip13.5pt 0,\ \ \ 0,\ \ \ 2,\ \ \ 3) \\
\str (\ \ 0,\ \ \ 2,\ \ \ 0,\ \ \ 1) & (\hskip13.5pt 0,\ \ \ 2,\ \ \ 0,\ \ \ 1) \\
\varstr{12pt}{8pt} (\hskip7.5pt 1,\ \ \ 4,\ \ \ 2,\ \ \ 3) & (\hskip8pt 22,\ \ \ 4,\ \ \ 2,\ \ \ 3) \\
\hline \hline
\end{tabular}
\capt{4.5in}{E7SU2vertices}{The minimal and the maximal reflexive polytopes containing the $E_7\times SU(2)$ $K3$~polyhedron as a slice and as a projection.}
\end{center}
\end{table}

\vspace{12pt}
\begin{figure}[h]
\begin{center}
\includegraphics[width=6.5in]{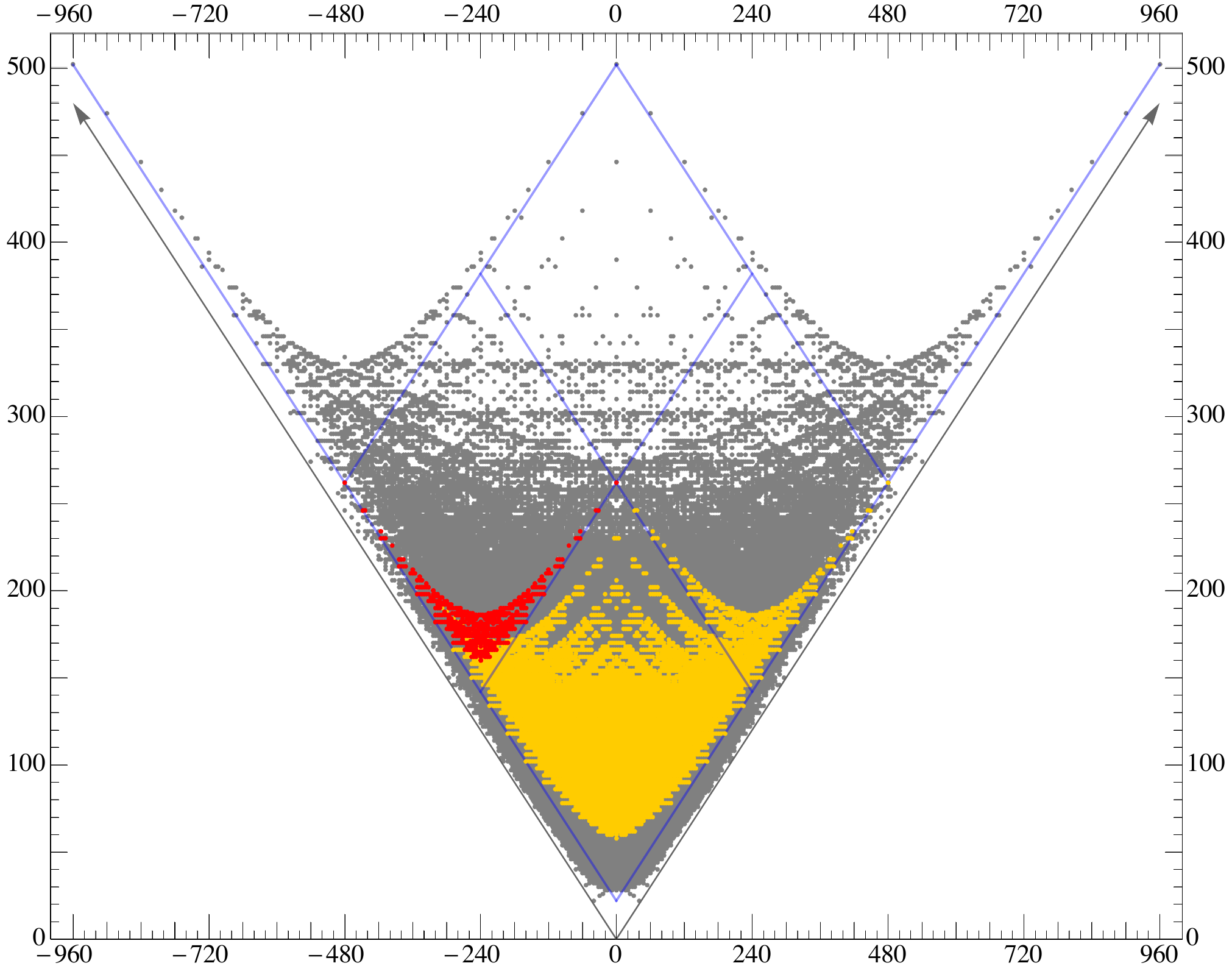}
\capt{6.2in}{AllE7TimesSU2}{Hodge plot showing all $E_7{\times}SU(2)$ $K3$ fibrations for which the fiber is both a slice and a projection in the four-dimensional reflexive polytope. The points in red form the $V$-region, corresponding to taking, for $\D$, all possible tops together with the maximal bottom.}
\end{center}
\end{figure}

As before, we'll be interested in finding all the reflexive polytopes which contain the $K3$ polyhedron of type $E_7{\times} SU(2)$ both as a slice and as a projection. The search for all the tops and bottoms that can be joined with the $K3$ polyhedron, performed in a way similar to the $E_8{\times} \{1\}$ case results in a list of $1204$ different tops and so also in $1204$ different bottoms. As in the previous case the set of polytopes
$\topbot{A}{\text{max}}$, as $\Top{A}$ ranges over all tops, gives rise to a $V$-structure, shown as the region formed by the red points on the left of \fref{AllE7TimesSU2} that is bounded by the blue lines. Within this region, the polytopes of greatest height are $\D=\topbot{\text{max}}{\text{max}}^{11,251}$ and $\D=\topbot{\text{min}}{\text{max}}^{131,131}$. The analogue of the vector that was $\D\hodgenos=(1,-29)$, for the case $E_8{\times}\{1\}$, is now $\D\hodgenos=(0,-16)$ and this vector, together with its mirror, determine the slope of the bounding lines. The analogue of the vector, that was previously, $\D\hodgenos=(240,-240)$ is now $\D\hodgenos=(120,-120)$.

By combining tops and bottoms from these lists, we obtain a number of $725,410$ polytopes which correspond to Calabi-Yau manifolds which are elliptic $K3$ fibrations of the type $E_7{\times} SU(2)$. Associated with these manifolds, there are $7,929$ distinct Hodge number pairs, indicated by the coloured points in \fref{AllE7TimesSU2}. The additive property for the Hodge numbers holds also in this case. It is interesting to note the similarity between the plots in \fref{AllE8TimesSU1} and \fref{AllE7TimesSU2}. The particular shape of the structures present in these plots seems to be a generic feature of the webs of elliptic $K3$ fibered Calabi-Yau manifolds with a self-dual $K3$ manifold.

\subsection{The web of $E_7{\times} \{1\}$ and $E_8{\times} SU(2)$ elliptic $K3$-fibrations} 

In the case when the $K3$ manifold is not self-dual, one needs to consider two webs at a time. For example, an elliptic $K3$ fibration of the type $E_7{\times} \{1\}$ is dual to an elliptic $K3$ fibration of the type $E_8{\times} SU(2)$. These polyhedra appear in \fref{K3polyhedron3}. 
\vskip-1pt
\begin{figure}[h]
\vskip20pt
\begin{center}
\framebox[5.8in][c]{
$\begin{array}{cc}
\hspace{-30pt}  
\includegraphics[width=3.7cm]{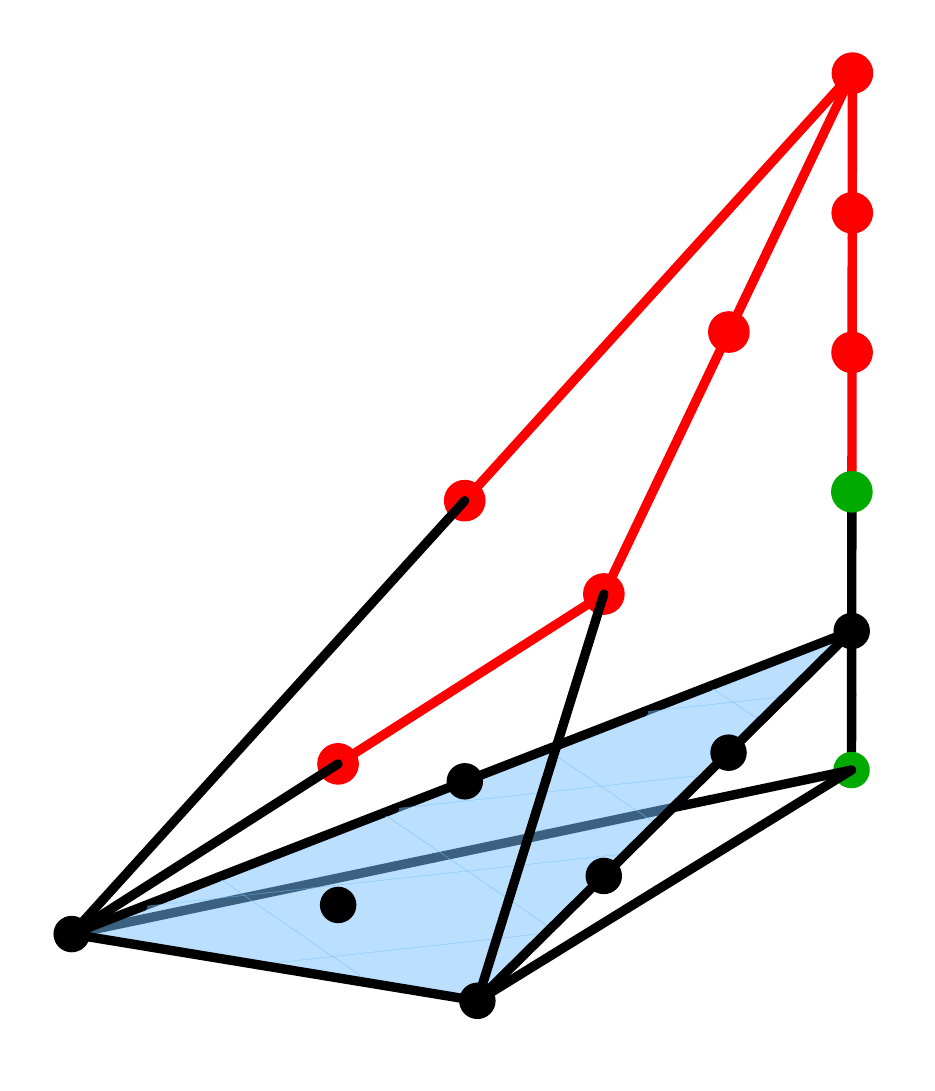}&\hspace{50pt}
\includegraphics[width=3.7cm]{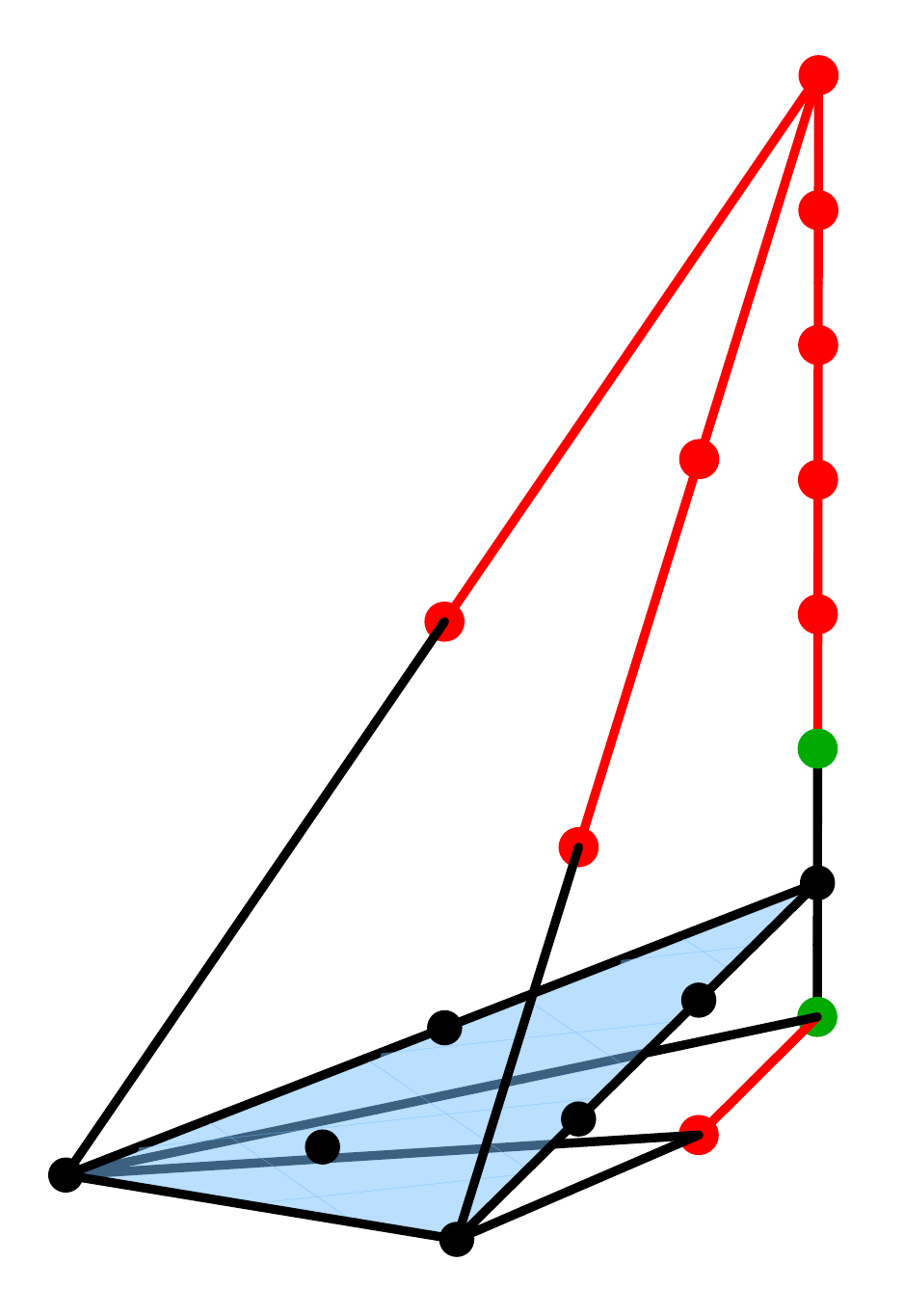}
\end{array}$
}
\capt{6.5in}{K3polyhedron3}{Elliptically fibered $K3$ manifolds of type $E_7{\times} \{1\}$ (left) and of type $E_8{\times} SU(2)$ (right).}
\end{center}
\end{figure}

\vspace{21pt}
%
%
%
\begin{table}[h]
\def\str{\varstr{12pt}{6pt}}
\begin{center}
\begin{tabular}{| >{$~} r <{~$} | >{$~} r <{~$} |  >{$~} r <{~$} |  >{$~} r <{~$} |}
\hline
\varstr{19pt}{12pt}
\topbot{\text{min}}{\text{min}}^{10,376}_{E_7\times \{1\}} 
& \topbot{\text{max}}{\text{max}}^{318,12}_{E_7\times \{1\}}
  & \topbot{\text{min}}{\text{min}}^{12,318}_{E_8{\times} SU(2)} 
     & \topbot{\text{max}}{\text{max}}^{376,10}_{E_8{\times} SU(2)} \\
\hline\hline
\varstr{14pt}{6pt} (-1,\ \ \ 4,\ \ \ 2,\ \ \ 3) & (-30,\ \ \ 4,\ \ \ 2,\ \ \ 3) 
                 &(-1,\ \ \ 6,\ \ \ 2,\ \ \ 3) & (-30,\ \ \ 6,\ \ \ 2,\ \ \ 3) \\
\str (\ \ 0,\ -1,\ \ \ 2,\ \ \ 3) & (-14,\ \ \ 2,\ \ \ 0,\ \ \ 1) 
                 & (\ \ 0,\ -1,\ \ \ 1,\ \ \ 2) & (\hskip4.5pt -2,\ -1,\ \ \ 2,\ \ \ 3) \\
\str (\ \ 0,\ \ \ 0,\ -1,\ \ \ 0) & (\hskip13.5pt 0,\ -1,\ \ \ 2,\ \ \ 3) 
                 & (\ \ 0,\ -1,\ \ \ 2,\ \ \ 3)  &  (\hskip13.5pt 0,\ -1,\ \ \ 1,\ \ \ 2) \\
\str (\ \ 0,\ \ \ 0,\ \ \ 0,\  -1)  &  (\hskip13.5pt 0,\ \ \ 0,\ -1,\ \ \ 0) 
                 & (\ \ 0,\ \ \ 0,\ -1,\ \ \ 0)  & (\hskip13.5pt 0,\ \ \ 0,\ -1,\ \ \ 0) \\
\str (\ \ 0,\ \ \ 2,\ \ \ 0,\ \ \ 1) & (\hskip13.5pt 0,\ \ \ 0,\ \ \ 0,\  -1) 
                 & (\ \ 0,\ \ \ 0,\ \ \ 0,\  -1) & (\hskip13.5pt 0,\ \ \ 0,\ \ \ 0,\  -1) \\ 
\str (\ \ 1,\ \ \ 4,\ \ \ 2,\ \ \ 3) & (\hskip8pt 14,\ \ \ 2,\ \ \ 0,\ \ \ 1) 
                 & (\hskip9.5pt 1,\ \ \ 6,\ \ \ 2,\ \ \ 3) & (\hskip13.5pt 2,\ -1,\ \ \ 2,\ \ \ 3) \\
\varstr{12pt}{8pt} & (\hskip8pt 30,\ \ \ 4,\ \ \ 2,\ \ \ 3)&& (\hskip9.5pt 30,\ \ \ 6,\ \ \ 2,\ \ \ 3) \\
\hline \hline
\end{tabular}
\capt{6in}{E7SU1vertices}{Minimal and maximal reflexive polytopes, $\D^*$, with a $K3$ slice of type $E_7{\times}\{1\}$  (first two columns) or an $E_8\times SU(2)$ slice (last two columns).}
\end{center}
\end{table}

The minimal and maximal 4D reflexive extensions of the $E_7{\times} \{1\}$ and the $E_8{\times} SU(2)$ $K3$~polyhedra are listed in Tables \ref{E7SU1vertices}.
The combined web of $E_7{\times} \{1\}$ and $E_8{\times} SU(2)$  contains $14,356$ distinct Hodge number pairs. This web is indicated by the coloured structure in \fref{E7E8str}. The red and the blue points correspond to $K3$ fibrations of the $E_7\times \{1\}$ and $E_8\times SU(2)$ type, respectively. The purple points correspond to fibrations of both types. Note the similarity between the structure formed by the purple points and the previous webs associated to self-dual $K3$ polyhedra.

\begin{figure}[h]
\vskip10pt
\begin{center}
\includegraphics[width=6.5in]{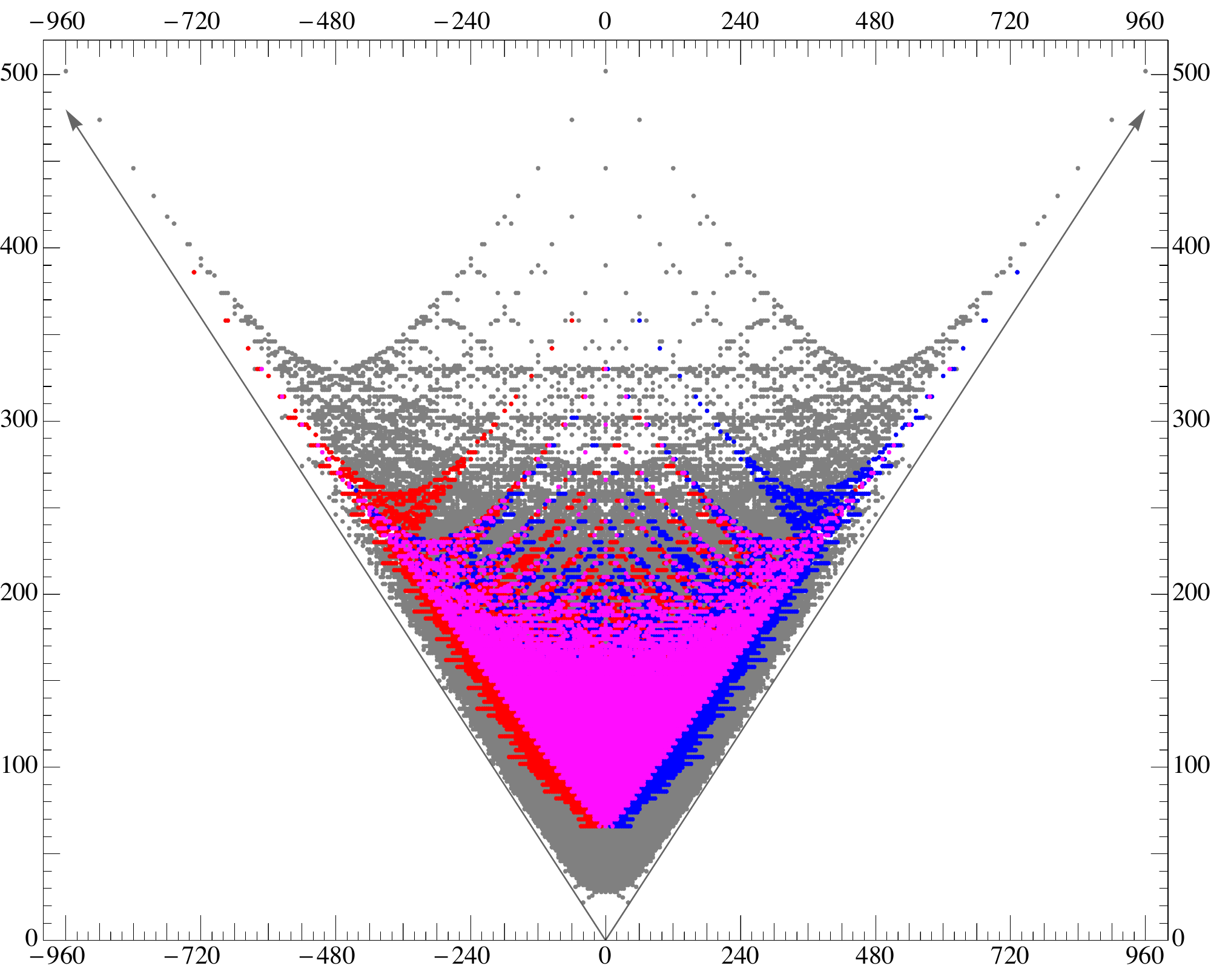}
\capt{5.0in}{E7E8str}{Hodge plot showing the $E_7{\times} \{1\}$ and $E_8{\times} SU(2)$ web.}
\end{center}
\vskip-10pt
\end{figure}

\subsection{Lists of tops\label{webaddress}}
In the course of this work I have compiled lists of all tops for $E_8{\times}\{1\}$, $E_7{\times}SU(2)$ and $E_8{\times}SU(2)$ fibrations. The tops for $E_7{\times}\{1\}$ fibrations can be recovered from those for $E_8{\times}SU(2)$ by computing dual polytopes. For reasons of length, I will not give these lists here; these can be found at \newline {\tt http://hep.itp.tuwien.ac.at/\raisebox{-7pt}{\Large\textasciitilde}skarke/NestedFibrations/}~.

\newpage
\section{Outlook}\label{Outlook}
We have seen that
the intricate structure of the upper region of the Hodge plot associated with the 
list of reflexive 4-polytopes can be largely understood as an overlap of webs of $K3$ fibrations. The pattern formed by the points of each web resembles that of the entire plot. Although very intricate, this pattern has a very regular structure, being formed by replicating a certain substructure many times. These intricately self-similar nested patterns within patterns give to the Hodge plot the appearance of a fractal. 
\begin{figure}[h]
\vskip20pt
\begin{center}
\includegraphics[width=6.2in]{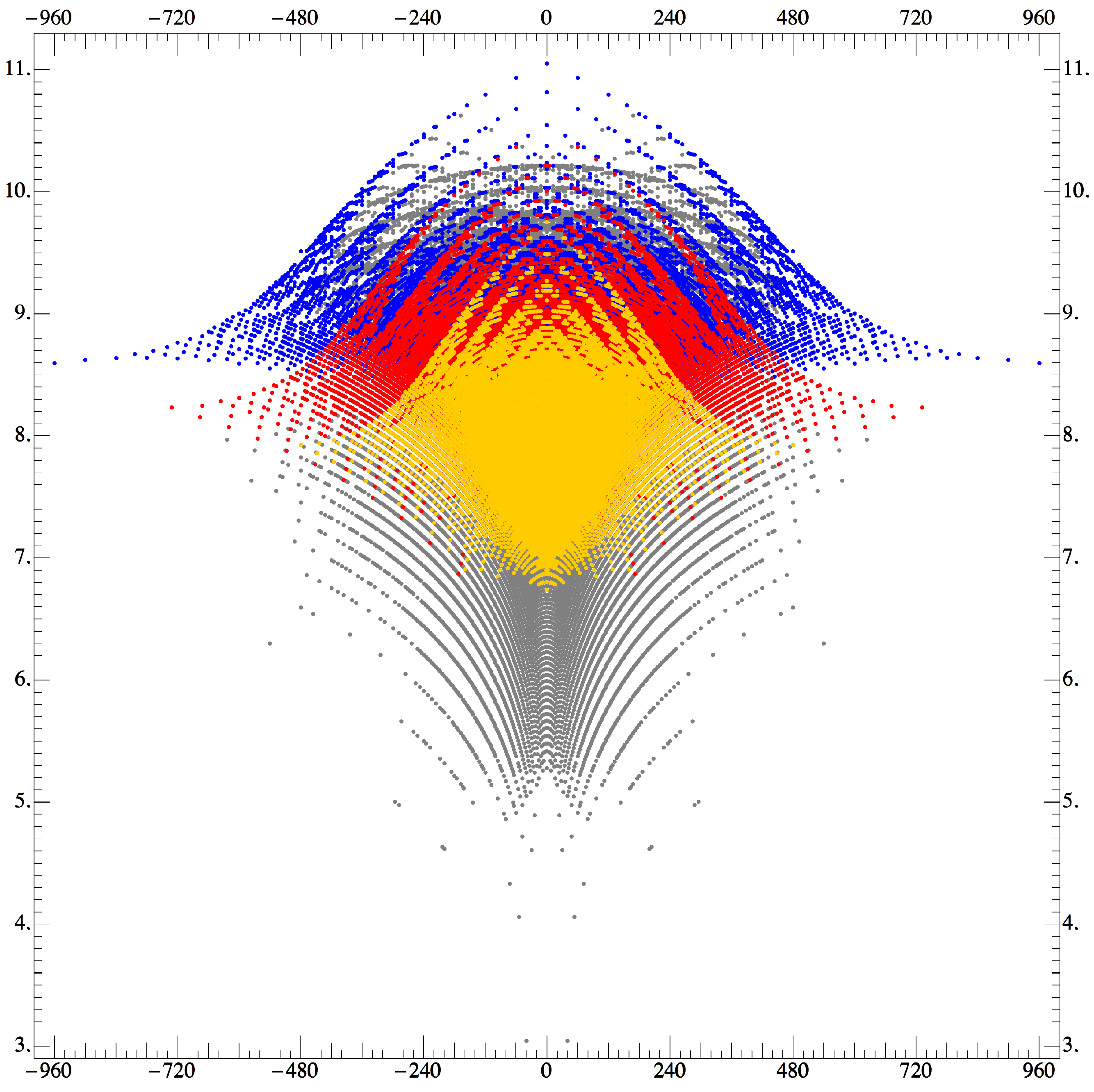}
\capt{5.5in}{NewKSPlot2}{Hodge plot showing in blue all $E_8\times \{1\}$ $K3$ fibrations; in red all $E_7\times \{1\}$ and $E_8\times SU(2)$ $K3$ fibrations; in green: all $E_7{\times}SU(2)$ $K3$ fibrations. Horizontal axis: the Euler number. Vertical axis: $\log h^{1,1}{+}\log h^{1,2}$.}
\end{center}
\vskip-10pt
\end{figure}
The plot contains many such webs, according to the different types of $K3$ manifolds. In this paper we have considered only three. Despite making this very restricted choice, we obtain a great number of $K3$ fibrations corresponding to $20,947$ distinct Hodge pairs, out of a total of $30,108$ Hodge~pairs. 

The plots in \fref{RBGplot} and \fref{NewKSPlot2} display the three overlapping webs discussed in this paper. The second of these plots uses a different coordinate system, $\log h^{1,1}+\log h^{1,2}$ against the Euler number. The most striking feature of this plot is the fact that the webs seem to separate.  

\vskip 20pt
One can continue to study the Hodge plot by trying to identify manifolds corresponding to Hodge numbers that do not belong to one of the webs identified so far. This is, in fact, the way in which we found the $E_7{\times} \{1\}$ web as the structure providing the first (gray) points in the top left corner of \fref{AllE8TimesSU1} that cannot be explained by the $E_8{\times} \{1\}$ web (red points). Similarly, the first gray point in the top left corner of \fref{RBGplot} corresponds to $h^{1,1}=9$, $h^{1,2}=321$. There are three polytopes $\SDelta$ giving rise to these Hodge numbers. All of them are fibrations of the $\topbot{\text{min}}{\text{min}}$ type over elliptic $K3$ manifolds, with the $K3$ polytopes $\SDelta_3$ being of the types $F_4{\times} \{1\}$, $E_6{\times} \{1\}$ and $E_6'{\times} \{1\}$, respectively; the 3-dimensional tops of $E_6$ and $E_6'$ type both correspond to the Lie group $E_6$, but the $E_6'$ diagram has one lattice point more than the $E_6$ diagram.

One could also ask about the gray dots remaining in the upper central portion of \fref{RBGplot}? It turns out that at least the first few of these correspond to cases where either the manifold or its mirror is a fibration over an elliptic $K3$ of the $E_8{\times} \{1\}$ type; remember that for the models of studied above, both the manifold and its mirror were of this type. In other words, now either slicing or projecting gives the last polyhedron of \fref{polyselection}, but the $E_8\times\{1\}$ polyhedron is not both a slice and a projection. 

A different approach to identifying structures in the set of toric Calabi-Yau hypersurfaces is to work directly with the polytopes. The classification of all reflexive 4-polytopes used the fact that there is a set of only 308 maximal polytopes (listed in Appendix A of \cite{Kreuzer:2000xy}) that contain all reflexive polytopes as subpolytopes, possibly on sublattices. Their duals are minimal polytopes in the sense that any reflexive polytope can be obtained from a minimal one by adding lattice points. It turns out that most of the minimal polytopes exhibit nested fibration structures, typically with Weierstrass elliptic fibers \cite{Candelas:2012uu}. This fits nicely with recent observations~\cite{Taylor:2012dr} about the connections between such fibrations and the structure of the Hodge plot.

}
\chapter{Discrete Calabi-Yau Quotients}\label{Z3Quotients}
{\setstretch{1.57}

\section{Introduction and Generalities}		


Non-simply connected Calabi-Yau threefolds have played an important role in the compactification of the heterotic string theory for a long time \cite{Candelas:1985en, Yau:1986gu, Tian:1978, Greene:1986jb, Greene:1986bm, Braun:2005ux, Braun:2005bw, Braun:2005nv, Bouchard:2005ag, Anderson:2007nc, Anderson:2008uw, Anderson:2009mh, Braun:2011ni}. Most of the known examples of such manifolds are quotients of complete intersection Calabi-Yau (CICY) manifolds by freely acting discrete symmetries. 

The interest in smooth quotients of CICY manifolds was renewed with the observation, made in \cite{Candelas:2007ac}, that there is an interesting corner in the string landscape, corresponding to Calabi-Yau threefolds with small Hodge numbers. Subsequently, this corner was populated with some 80 new manifolds \cite{Candelas:2008wb}, constructed either as free or resolved quotients of CICY manifolds. The observation was made also that quotients with isomorphic fundamental groups form webs connected by conifold transitions. The search for CICY manifolds admitting free linear group actions was completed, for the configurations of the CICY list, by Braun \cite{Braun:2010vc} by means of an automated scan, leading to a classification of all linear actions of discrete groups on the CICY manifolds constructed in \cite{Candelas:1987kf}. Together with many new examples of free quotients this search revealed also a new manifold with Euler number $-6$, leading to a physical model with three generations of particles interacting according to the  gauge group of the Standard Model. This manifold \cite{Braun:2009qy, Braun:2011ni}, realized as a quotient by a group of order 12, enjoys the remarkable property of having the smallest possible Hodge numbers for a manifold for which three generations of particles is achieved via the standard embedding, namely $\hodgenos = (1,4)$. 

It would be both interesting and important to give a detailed account of all the new manifolds and symmetries uncovered by \cite{Braun:2010vc}. The aim of this chapter, however, is more modest: I will give instead a detailed description of the six new $\IZ_3$ quotients appearing in the list \cite{Braun:2010vc}, which were missed in \cite{Candelas:2008wb}. This more modest goal is nevertheless worthwhile since the new $\IZ_3$ manifolds fit into the web of $\IZ_3$ manifolds in an interesting way, as may be seen by referring to \fref{Z3Web}, which shows this web, with the six new quotient manifolds indicated by red dots. 
\begin{figure}[ht]
\begin{center}
\includegraphics[width=6.5in]{CYLandscape3.pdf}
\vskip12pt
\framebox[5.0in]{\parbox{5.0in}{\vspace{7pt}
\hspace*{20pt}\includegraphics[width=7pt]{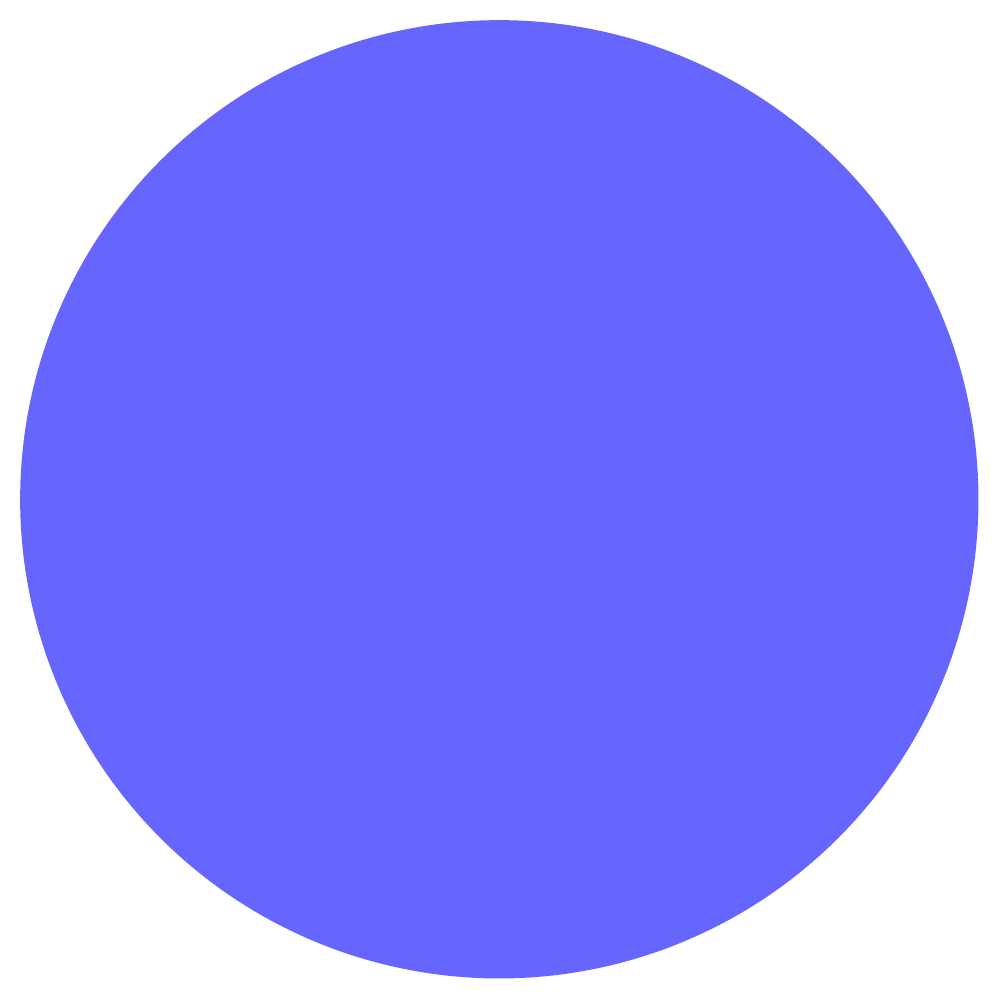}~~
\includegraphics[width=7pt]{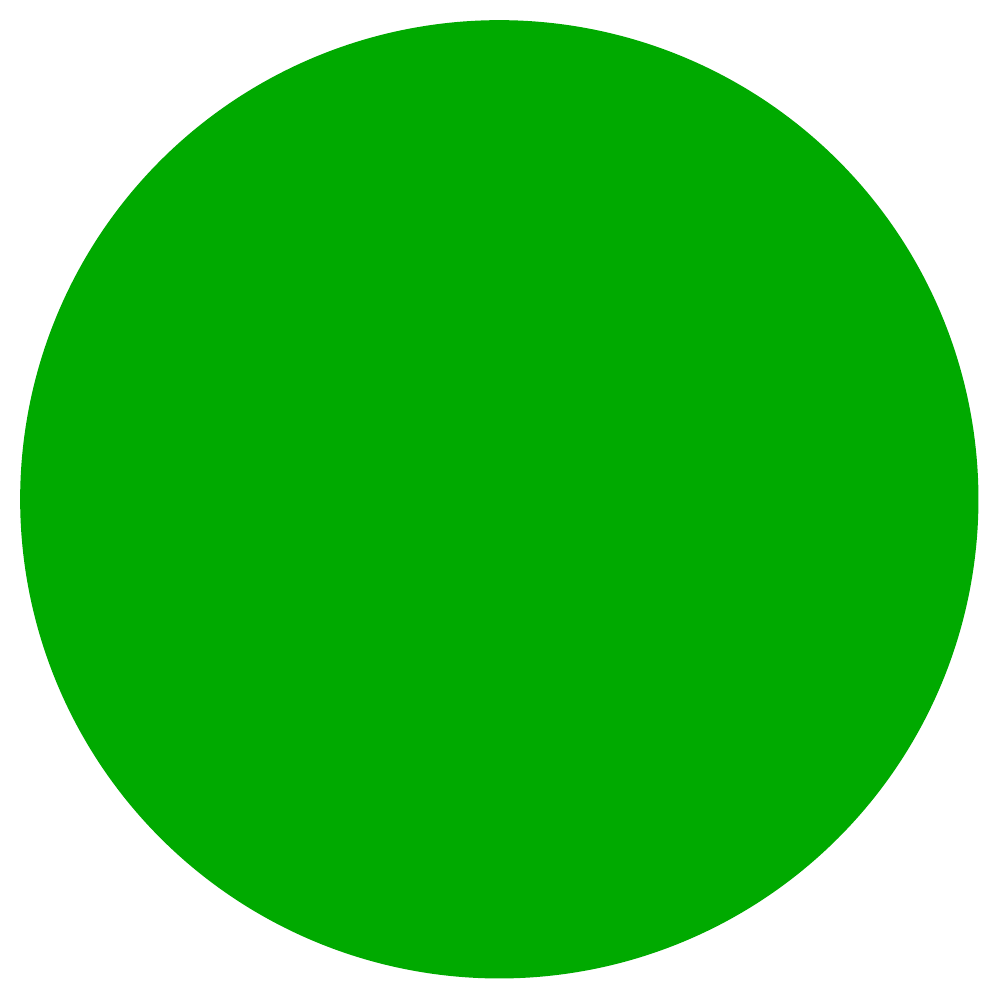}~~
\includegraphics[width=7pt]{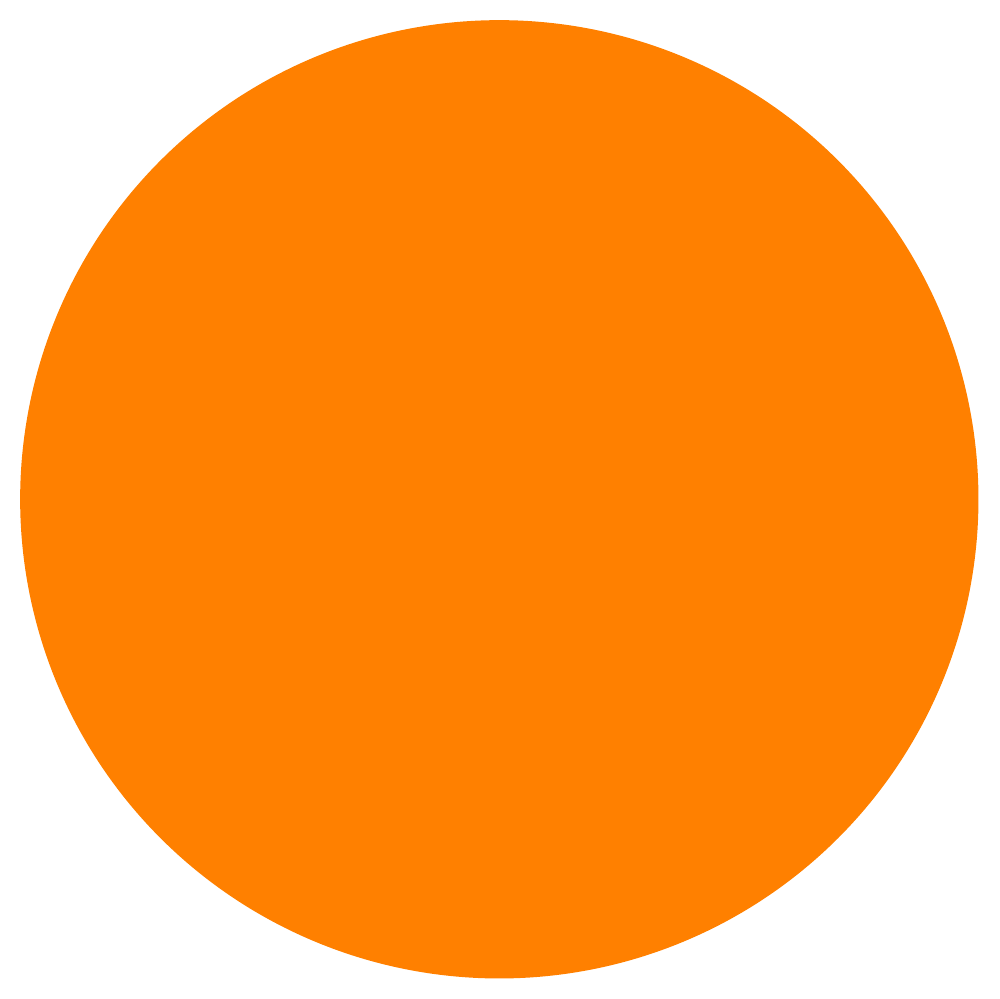}~~
\includegraphics[width=7pt]{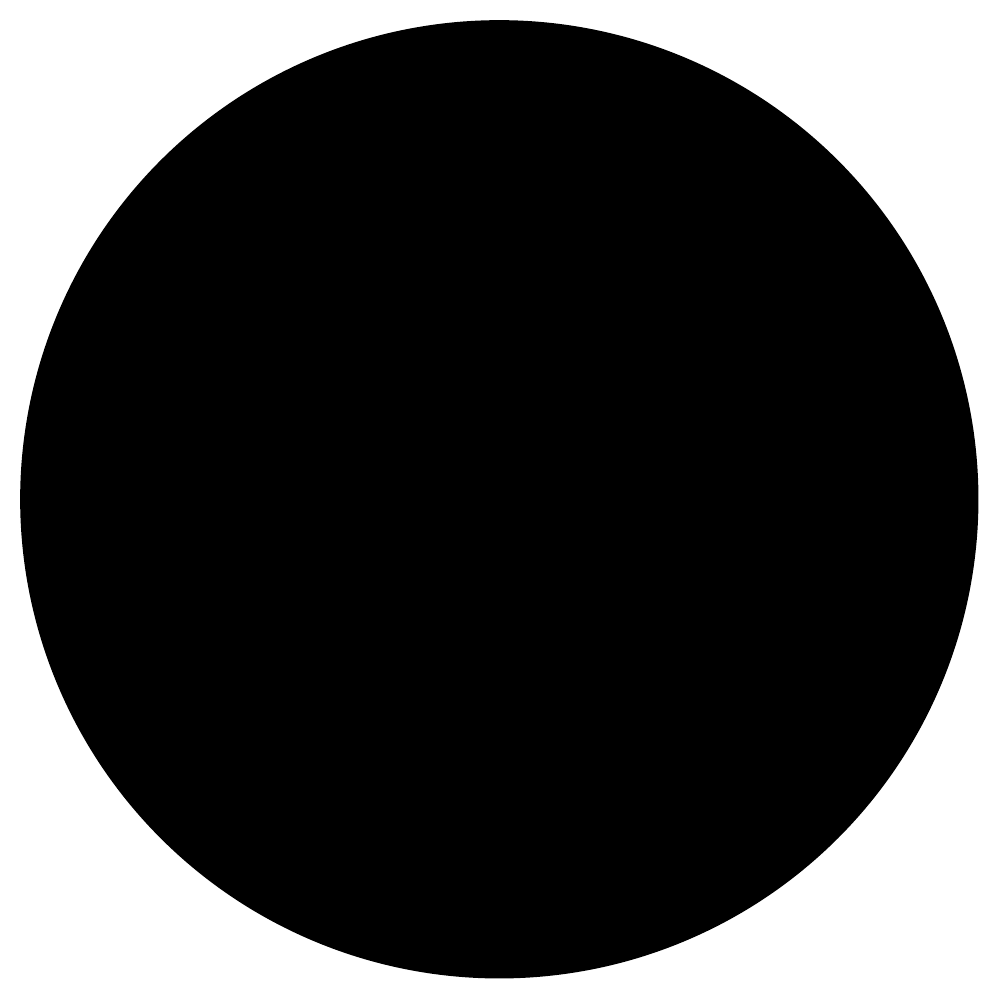}~~%
Old $\IZ_3$-free quotients of  CICY manifolds \cite{Candelas:2008wb, Candelas:2007ac}. \\ 
\hspace*{76pt}\includegraphics[width=7pt]{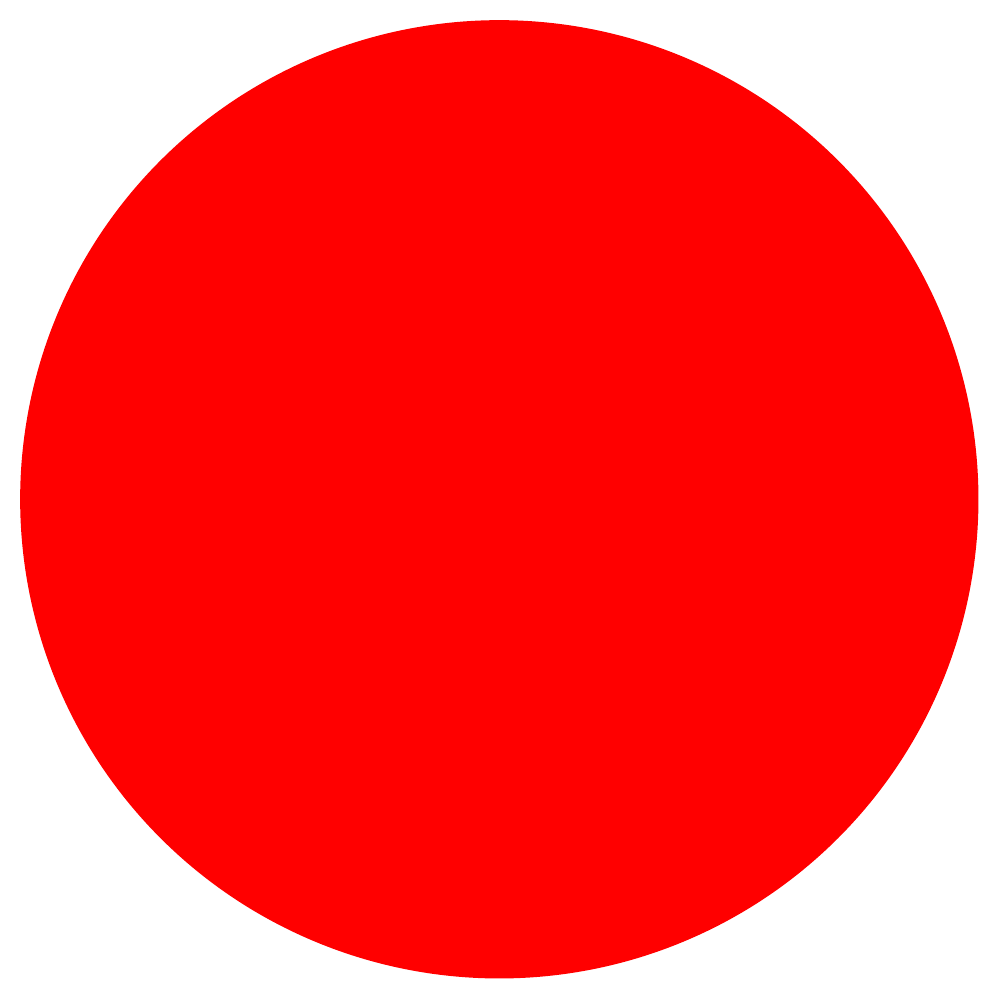}~~New $\IZ_3$-free quotients of CICY manifolds from \cite{Braun:2010vc}.\\
\hspace*{34.5pt}
\includegraphics[width=7pt]{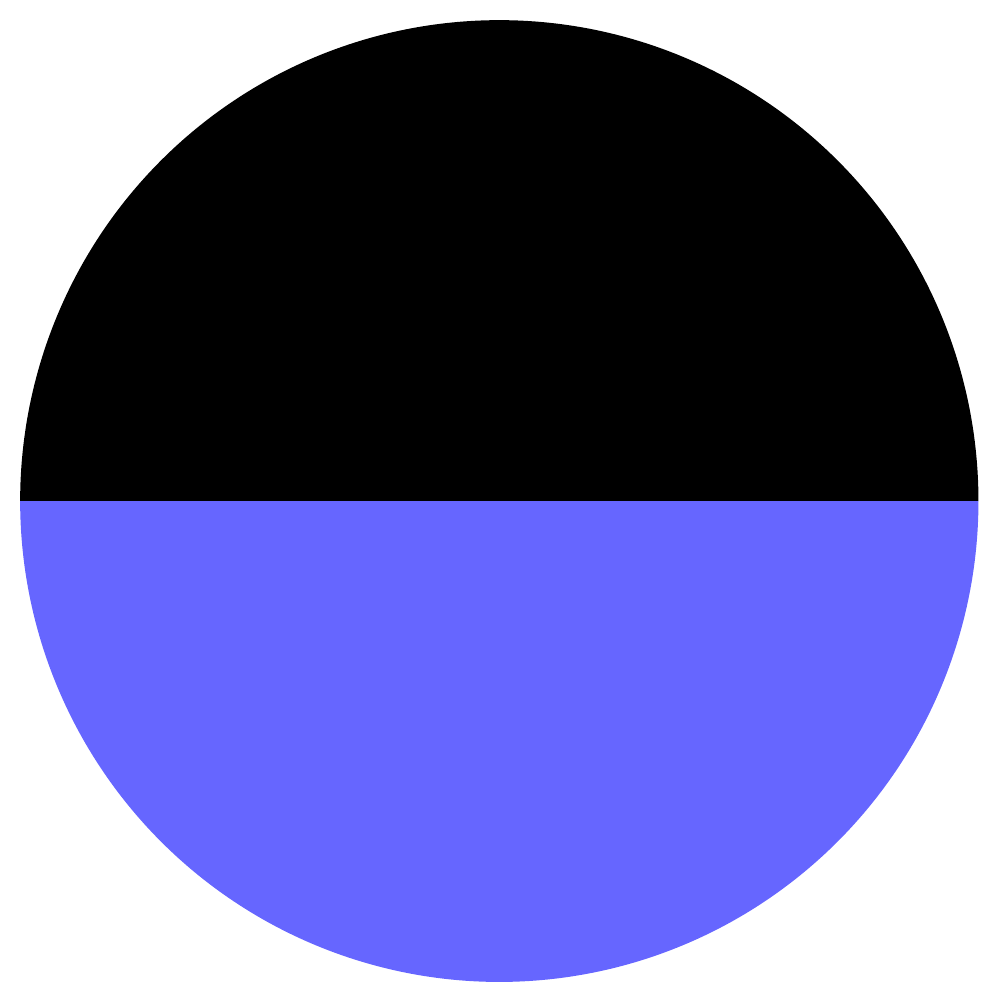}~~
\includegraphics[width=7pt]{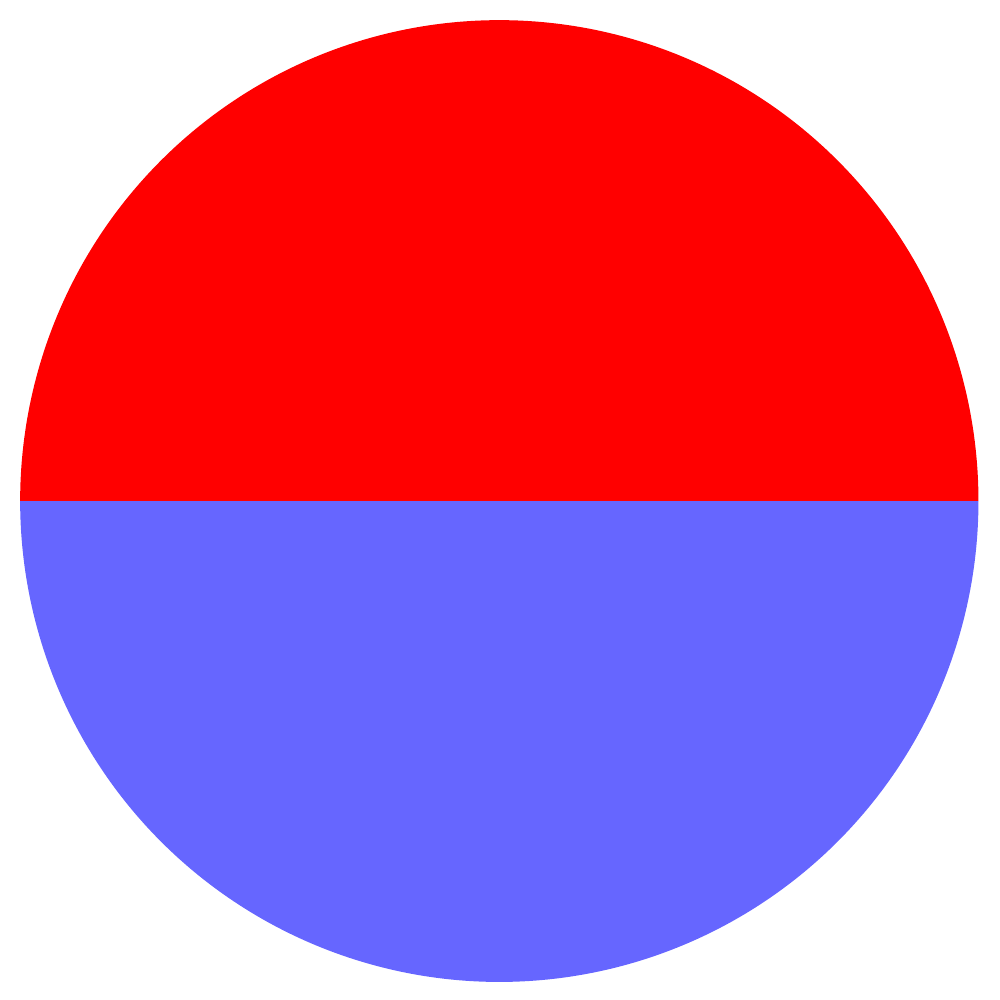}~~
\includegraphics[width=7pt]{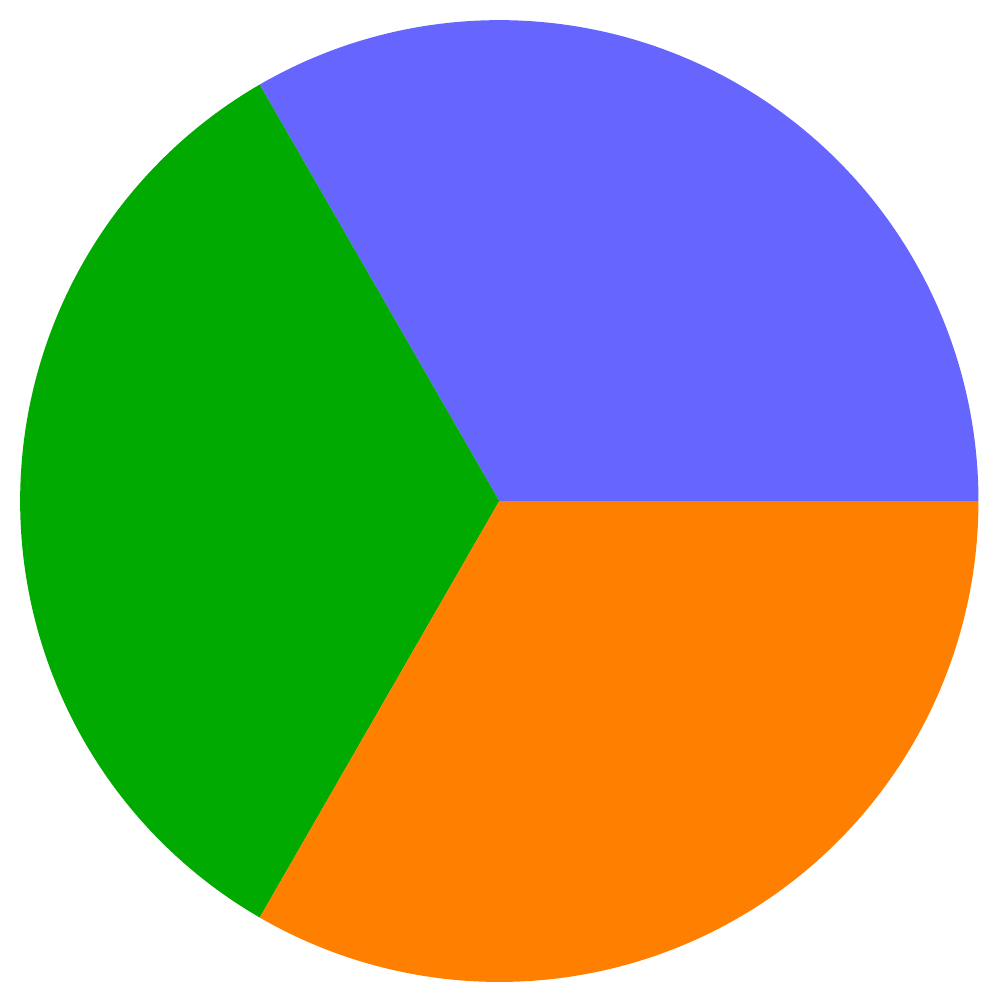}~~Multiply-occupied sites.\\[-4pt]
}}
\vskip3pt
\capt{5.5in}{Z3Web}{The web of $\IZ_3$ quotients of CICY manifolds. On the horizontal axis: the Euler number $\chi = 2\left( h^{1,1} - h^{2,1} \right)$.  On the vertical axis: the height $h^{1,1}+ h^{2,1}$.}
\end{center}
\vspace{-20pt}
\end{figure}

In this chapter, I will present the symmetries of the manifolds in an explicit and straightforward manner. While the group actions are given unambiguously in the results of \cite{Braun:2010vc} the form in which these are given are those found by the computer programme and, while equivalent, take a very different form from that which one would naturally write. I will also compute the Hodge numbers for each quotient. These are not given by the results of \cite{Braun:2010vc}. On computing these, certain regularities become apparent, as we see from 
\fref{Z3Web}. In particular five of the new quotients are seen to fit, for example, into the double sequence of \tref{doublesequenceIntro} for which the horizontal arrows correspond to $\D(h^{11},h^{21})=(1,-1)$ and the vertical arrows correspond to $\D(h^{11},h^{21})=(1,-2)$. This double sequence corresponds to the structure, evident in \fref{Z3Web}, whose top left member has coordinates $(-30,17)$. Other sequences will appear later. I have drawn \fref{Z3Web} so as to emphasise three sequences, which largely correspond to the straight arrows. These sequences are drawn with heavier lines though all the arrows represent conifold transitions.

In addition to computing the Hodge numbers, I will present in Section \ref{secWeb}, a study of the conifold transitions between the covering manifolds and also the conifold transitions between the quotients. I find, as in~\cite{Candelas:2008wb}, that the $\IZ_3$-quotients, including the new quotients, are connected by conifold transitions so as to form a single web, as one can see from \fref{Z3Web}. 

I will follow the conventions of \cite{Candelas:2008wb} and use the techniques discussed in Section 1 of that paper. Below, I summarise the most important aspects concerning the construction of smooth CICY quotients (see also \cite{Hubsch}). 
\begin{table}[ht]
\begin{center}
\framebox[6.0in]{
\begin{tabular}{c}
\begin{tikzpicture}[scale=1.2]
\clip (-2.5, -.4) rectangle (9.55,5.7);
\def\nodeshadowed[#1]#2;{\node[scale=1.1,above,#1]{#2};}

\nodeshadowed [at={(-1,0 )},yslant=0.0]
{ {\small \textcolor{black} {$\mathbf{\left( \mathscr{X}^{6,24}/ \IZ_3 \right) ^{4,10}}$}} };
\nodeshadowed [at={(2,0 )},yslant=0.0]
{ {\small \textcolor{black} {$\mathbf{\color{red} \left( \mathscr{X}^{9,21}/ \IZ_3 \right) ^{5,9}}$ }} };
\nodeshadowed [at={(5,0 )},yslant=0.0]
{ {\small \textcolor{black} {$\mathbf{\color{red} \left( \mathscr{X}^{12,18}/ \IZ_3 \right) ^{6,8}}$ }} };
\nodeshadowed [at={(8,0 )},yslant=0.0]
{ {\small \textcolor{black} {$\mathbf{\left( \mathscr{X}^{15,15}/ \IZ_3 \right) ^{7,7}}$}} };
\nodeshadowed [at={(-1,1.5 )},yslant=0.0]
{ {\small \textcolor{black} {$\mathbf{\left( \mathscr{X}^{5,32}/ \IZ_3 \right) ^{3,12}}$ }} };
\nodeshadowed [at={(2,1.5 )},yslant=0.0]
{ {\small \textcolor{black} {$\mathbf{\color{red}  \left( \mathscr{X}^{8,29}/ \IZ_3 \right) ^{4,11}}$}} };
\nodeshadowed [at={(5,1.5 )},yslant=0.0]
{ {\small \textcolor{black} {$\mathbf{\color{red} \left( \mathscr{X}^{11,26}/ \IZ_3 \right) ^{5,10}}$}} };
\nodeshadowed [at={(-1,3 )},yslant=0.0]
{ {\small \textcolor{black} {$\mathbf{\left( \mathscr{X}^{4,40}/ \IZ_3 \right) ^{2,14}}$}} };
\nodeshadowed [at={(2,3 )},yslant=0.0]
{ {\small \textcolor{black} {$\mathbf{\color{red} \left(  \mathscr{X}^{7,37} / \IZ_3 \right) ^{3,13}}$}} };
\nodeshadowed [at={(-1,4.5 )},yslant=0.0]
{ {\small \textcolor{black} {$\mathbf{\left( \mathscr{X}^{3,48}/ \IZ_3 \right) ^{1,16}}$}} };

\draw[very thick,blue,->] (-1, 4.5) -- (-1,3.7);
\draw[very thick,blue,->] (-1, 3.) -- (-1,2.2);
\draw[very thick,blue,->] (-1, 1.5) -- (-1,.7);
\draw[very thick,blue,->] (2, 3.) -- (2,2.2);
\draw[very thick,blue,->] (2, 1.5) -- (2,.7);
\draw[very thick,blue,->] (5, 1.5) -- (5,.7);

\draw[very thick,blue,->] (-.1, 3.3) -- (.8, 3.3);
\draw[very thick,blue,->] (-.1, 1.8) -- (.8, 1.8);
\draw[very thick,blue,->] (-.1, .3) -- (.8, .3);
\draw[very thick,blue,->] (2.9, 1.8) -- (3.74, 1.8);
\draw[very thick,blue,->] (2.9, .3) -- (3.74, .3);
\draw[very thick,blue,->] (5.9, .3) -- (6.8, .3);

\end{tikzpicture}
\end{tabular}}

\vskip 8pt
\framebox[6.0in]{\parbox{5.5in}{\vspace{-8pt}
\begin{center}
 For the covering spaces:   
$\Delta_{\raisebox{-3pt}{$\scriptstyle{}\hskip-2pt\color{blue}\rightarrow $}} \hodgenos =\left(3, -3\right)
\,;~\Delta_{\color{blue}\downarrow}  \hodgenos = \left( 1, -8 \right) $ \\[4pt]
For the quotient spaces: 
$\Delta_{\raisebox{-3pt}{$\scriptstyle{}\hskip-2pt\color{blue}\rightarrow $}} \hodgenos =\left(1, -1\right)
\,;~\Delta_{\color{blue}\downarrow}  \hodgenos = \left( 1, -2 \right) $ 
\end{center}
\vspace{-8pt}}}
\capt{6.0in}{doublesequenceIntro}{The first sub-web of $\IZ_3$-free CICY quotients. In red, the new quotients.}
\end{center}
\end{table}
\newpage
\subsection{Important aspects of CICY quotients}
\vskip -8pt
Let $\mathscr{X}$ be a Calabi-Yau manifold and $G \times {\mathscr X} \rightarrow {\mathscr X}$ an action of the finite group $G$ on ${\mathscr X}$. If $G$ acts freely and holomorphically, ${\mathscr X} / G$ will inherit the structure of a complex manifold from $\mathscr X$. Moreover, ${\mathscr X} / G$ inherits a nowhere vanishing holomorphic $n$-form, where $n$ is the complex dimension of $\mathscr X$, and consequently it is Calabi-Yau. 

Also, note that $\mathscr X$ is a covering space for the quotient ${\mathscr X} / G$, and thus the quotient ${\mathscr X} / G$ is multiply connected, unless $G$ is trivial and $\mathscr X$ simply connected.  In particular, if $\mathscr X$ is simply connected, the fundamental group of ${\mathscr X} / G$ is isomorphic to $G$. Since CICY manifolds are simply connected, all the $\IZ_3$-quotients discussed here will have fundamental group $\IZ_3$. 

Furthermore, the order of the group $G$ divides the following indices of $\mathscr X$: the Euler characteristic, the holomorphic Euler characteristic, the Hirzebruch signature and the index of the Dirac operator. These divisibility properties follows by expressing the indices as integrals of densities over the manifold $\mathscr X$. This is an important necessary condition for the existence of a free group action since the order of the group must divide the GCD of the four indices.

If $\mathscr X$ is a CICY manifold embedded in a product of projective spaces $\mathscr A = \IP^{n_1}\times \dots \times \IP^{n_m}$ and the holomorphic action $G \times {\mathscr X} \rightarrow {\mathscr X}$ comes from an action of $G$ on the ambient space $\mathscr A$, then $G$ must preserve the projectivity of the homogeneous coordinates: 
\begin{align*}
\left( \left[ x^0_1: \dotsc :x^{n_1}_1\right], \dots, \left[x^0_m: \dotsc : x^{n_m}_m\right]  \right)  \!\xmapsto {g\in G}  &~g\!\cdot\! \left( \left[ x^0_1: \dotsc : x^{n_1}_1\right], \dots, \left[x^0_m: 
\dotsc : x^{n_m}_m\right]  \right)\\[3pt]
 \cong  &~g\!\cdot \!\left( \left[ \lambda_1 x^0_1: \dotsc : \lambda_1 x^{n_1}_1\right], \dots, \left[\lambda_m x^0_m : \dotsc :\lambda_m x^{n_m}_m\right]  \right) 
\end{align*} 

This condition is clearly satisfied by all linear action of $G$. But in general, there may exist also nonlinear projective actions. For example, in a previous classification \cite{Bouchard:2007mf} of quotients of the split bicubic manifold given by the configuration matrix
$$
{\mathscr X}^{19, 19}~=~~
\cicy{\IP^1 \\ \IP^2\\ \IP^2}
{ 1& 1\\ 3 & 0\\ 0& 3 \\}_{0}^{19,19}
$$ 
the largest symmetry group had order 9. In \cite{Braun:2009qy}, however, it was shown that the split bicubic manifold admits free linear actions of two groups of order 12, when the manifold is represented as a complete intersection embedded in a product of seven $\IP^1$ spaces.  To my knowledge, nonlinear actions on CICY manifolds have not been studied systematically. The nonlinear actions that are currently known are all related to linear actions on equivalent CICY configurations with larger ambient spaces.

The quotients considered in the following sections will always come from linear projective actions. These are, in general, combinations of  internal actions on the coordinates of an individual projective space $\IP^n$, and permutations of the projective spaces, as they occur in~$\mathscr A$. 

The list of \cite{Braun:2010vc} records all free linear actions on complete intersection Calabi-Yau manifolds. More precisely, it indicates what symmetries exist for each of the 7890 classes of CICY manifolds constructed in \cite{Candelas:1987kf}. The list of CICY manifolds is available on the Calabi-Yau home page \cite{cicylist1} or at \cite{cicylist2}, the latter having appended Hodge numbers and other topological indices, as well as the free actions of finite groups found in \cite{Braun:2010vc}.  

For the purpose of this discussion, the kind of information needed from \cite{Braun:2010vc} is simply the indication that a certain subclass of the CICY deformation class $\mathscr X$ admits a $\IZ_3$ symmetry. Using this, I reconstruct the action by identifying the $\IZ_3$ invariant polynomials whose complete intersection define $\mathscr X$. Further, I check that this action is free. Finally, I compute the Hodge numbers for the quotients using the methods of \cite{Candelas:2008wb}, as follows: 

The Hodge number $h^{1,1}$ of K\"ahler structure parameters for the quotient ${\mathscr X}/G$ is computed by finding a representation in which $\mathscr X$ is embedded in a product of $h^{1,1}\left( \mathscr X\right)$ projective spaces. In this case, the $h^{1,1}\left( {\mathscr X}\right)$ linearly independent forms in the cohomology group $H^{1,1}\left( \mathscr X\right)$ arise as pullbacks from the hyperplane classes of $\mathscr A$ and the action of $G$ on $H^{1,1}\left( \mathscr X\right)$ is determined by its action on the ambient space $\mathscr A$.  

The number $h^{2,1}\left( \mathscr{X}/G \right)$ of complex structure parameters will be computed by counting the independent parameters in the $G$-invariant polynomials defining $\mathscr X$. By a theorem 
from~\cite{Green:1987rw}, the method is guaranteed to work whenever the parameter count for the covering space $\mathscr{X}$ is equal to $h^{2,1}(\mathscr{X})$  and the diagram associated with the CICY manifold 
$\mathscr X$ cannot be disconnected by cutting a single leg. 

As noted above, the quotient $\mathscr{X} /G$ is a smooth manifold if and only if $G$ acts freely on $\mathscr X$ and the manifold $\mathscr X$ is smooth. In Section 2 we will check, in each case, that the intersection of $\mathscr X$  with the fixed point set of the action $G\times \mathscr{A} \rightarrow \mathscr A$ is empty. The smoothness assumption for $\mathscr X$ is equivalent to the condition that the $G$-invariant polynomials  
$\{p_j\}_{j= 1,..,K}$ 
defining $\mathscr X$ are transverse, in other words $dp_1\wedge \dots \wedge dp_K \neq 0$ on the intersection $\left\{ p_j = 0\right\}_{j= 1,..,K}\,$. We check that the equations 
$dp_1\wedge \dots \wedge dp_K = 0$ and $\left\{ p_j = 0\right\}_{j= 1,..,K}\,$ have no common solution by performing a Groebner basis calculation, as explained in \cite{Candelas:2008wb}. The computation is implemented in Mathematica 7.0 and uses the computer algebra system SINGULAR \cite{Greuel} by means of the STRINGVACUA package \cite{Gray:2008zs}.

In many situations, there exist several different embeddings of the same manifold $\mathscr X$. For the purpose of computing the number of K\"ahler structure parameters, we will consider the embedding of $\mathscr X$ in a product of $h^{1,1}\left( \mathscr X\right)$ projective spaces. However, for checking transversality, I will always prefer an embedding in a product of projective spaces with fewer factors, owing to the fact that the complexity of the Groebner basis calculation increases rapidly with the number of projective spaces.  

\subsection{Conifold transitions}

It is known \cite{Green:1988wa, Green:1988bp, Candelas:1989ug} that the parameter spaces of CICY threefolds intersect along loci corresponding to conifolds, and thus form a connected web. In particular, any configuration matrix, defining a particular smooth deformation class of CICY manifolds can be brought into the form of any other configuration through a sequence of operations known as `splittings' and `contractions'. For example, if $\cicy{\mathcal{A}\ }{\, \mathcal{M}\,; c\,}$ is the configuration matrix describing the class $\mathscr X$, the deformation class $\Xcheck$ obtained from $\mathscr X$ by `splitting' the last polynomial with a $\IP^n$ space is defined by the operation
\begin{equation*}
\mathscr{X} = \cicy{\mathcal{A}\,}{\, \mathcal{M}\,; c\,}\ \twoheadlongrightarrow\ \Xcheck ~=~ 
\cicy{\IP^n \\  \mathcal{A\ }}{\mathbf{0} & 1 & 1 & \dots & 1 \\ 
\mathcal{M} & c_1 & c_2 & \dots & c_{n+1}}
\end{equation*}
where $c$ is a column vector and $\sum_{i=1}^{n+1} c_i = c$. The reversed process is called a contraction (see, e.g.~\cite{Hubsch, Candelas:1989ug}). 

Geometrically, such operations correspond to conifold transitions. The number of nodes of the associated conifold 
$\Xsharp$ is given as half the difference of the Euler characteristics of $\Xsharp$ and  $\Xcheck$ \cite{Candelas:1989ug, Candelas:1989js}.  In Section \ref{secWeb} I prove that if both $\mathscr X$ and $\Xcheck$ admit $G$-free quotients, and the actions on $\mathscr X$ and $\Xcheck$ reduce to the same action on $\Xsharp$, then the conifold transition commutes with taking quotients. It is known that the set of all CICY's form a web connected by conifold transitions \cite{Hubsch}. The observation made in this chapter is that the set of $\IZ_3$ free quotients of CICY's is similarly connected and that the new $\IZ_3$ quotients fit into this web.  

}

\newpage
\section{The New $\IZ_3$ Quotients}
{\setstretch{1.52}
In this section, I give a detailed description of the six new $\IZ_3$-symmetric manifolds. For each of these new manifolds I present $\IZ_3$-covariant polynomials together with the group action. I check that the polynomials are transverse and that the group action is fixed point free and, finally, I compute the Hodge numbers for the quotient manifold. 
\vspace{10pt}
\subsection{The manifold $\mathscr{X}^{12,18}$ with quotient 
$\mathscr{X}^{6,8} = \mathscr{ X}^{12,18}/ \IZ_3$} \label{sec:1.1}

This class of manifolds is described by the following configuration matrix
\begin{equation*}\label{X12,18}
\mathscr{X}^{12,18}~=~~
\cicy{\IP^1 \\ \IP^1\\  \vrule height10pt width0pt depth8pt  \IP^1\\ \IP^1\\ \IP^1\\ \vrule height10pt width0pt depth8pt  \IP^1\\ \IP^2\\ \IP^2\\ \vrule height10pt width0pt depth8pt  \IP^2\\ \IP^2}
{ \one& 0& 0&~ 0& 0 & 0&~0&0& 0&~ \one& 0 \\
 0& \one& 0&~ 0 & 0 & 0 &~0&0& 0&~ \one& 0 \\
\vrule height10pt width0pt depth8pt  0& 0& \one&~ 0& 0&0&~0&0& 0&~ \one& 0 \\
 0& 0& 0&~ \one& 0&0&~0&0& 0&~ 0& \one \\
 0& 0& 0&~ 0& \one& 0&~ 0&0 & 0&~ 0& \one \\
\vrule height10pt width0pt depth8pt   0& 0& 0&~ 0& 0& \one &~0&0& 0&~ 0& \one \\
\one& 0& 0&~ \one&0 &0&~\one& 0& 0&~ 0&0 \\
0& \one& 0&~ 0& \one&0&~0& \one& 0&~ 0& 0 \\
\vrule height10pt width0pt depth8pt  0& 0& \one&~ 0& 0&\one&~0& 0& \one&~ 0&0 \\
0& 0& 0&~ 0& 0&0&~\one&\one& \one&~ 0& 0 \\}_{-12}^{12,18}
\end{equation*}

corresponding to the following equivalent diagrams
\vspace{-15pt}
\begin{flushright}
$$
{\includegraphics[width=2.4in]{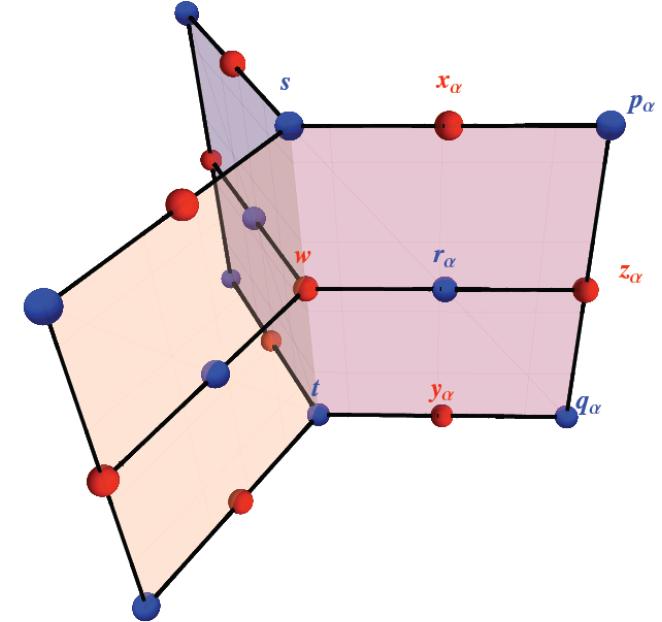}}
\hskip 32pt \lower -20pt\hbox{\includegraphics[width=2.4in]{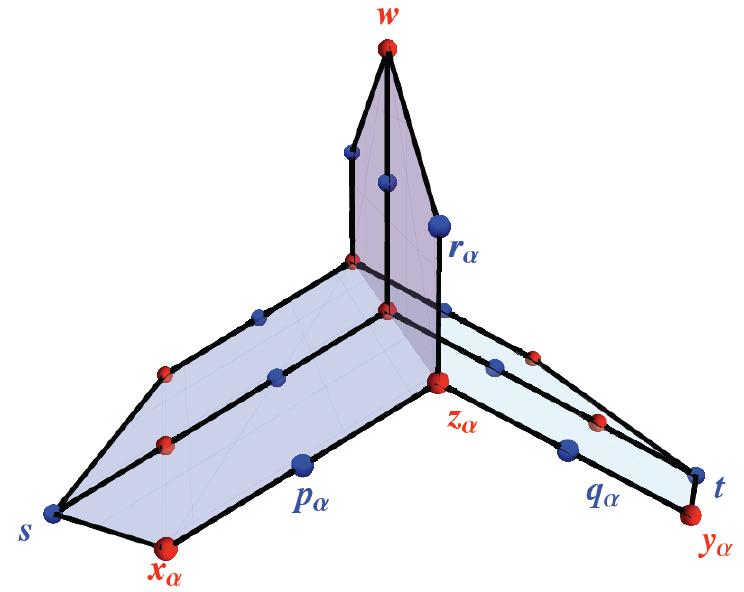}}
$$
\end{flushright}

Let us denote the coordinates of the first three $\IP^1$'s by  $x_{\alpha, j}$ and the coordinates of the remaining three $\IP^1$'s by $y_{\alpha, j}$. Here $\alpha$ labels the space and $j$ its coordinate, these labels are understood to take values in $\IZ_3$ and $\IZ_2$ respectively. Let $z_{\alpha, a}$ denote the coordinates of the first three $\IP^2$'s and $w_a$ the coordinates of the last $\IP^2$, where now the indices $\alpha$ and $a$ both take values in $\IZ_3$. 

Let us also denote by $ p^{\alpha}$ the first three polynomials, and by $q^{\alpha}$ the next three. Denote also by $s$ and $t$ the next two cubic polynomials in $\IP^1$ coordinates, and by $r^{\alpha}$ the last three cubic polynomials in $\IP^2$ coordinates. These polynomials have the general form: 
\beqnn
\begin{split}
p^{\alpha} &= \sum_{j,a} P^\a_{ja}\, x_{\a, j}\, z_{\alpha, a}~,~~~\,~~~~
q^{\alpha} = \sum_{j,a} Q^\a_{ja}\, y_{\alpha,j}\, z_{\alpha,a}\\[5pt]
s &= \sum_{i,j,k} S_{ijk}\, x_{0,i}\, x_{1,j}\, x_{2,k}~,~~~~
t = \sum_{i,j,k} T_{ijk}\, y_{0,i}\, y_{1,j}\, y_{2,k}\\[5pt]
&~~~~~~~~~~~~~~r^{\alpha} = \sum_{a,b} R^\a_{ab}\, z_{\alpha,a}\, w_{\a+b} 
\end{split}
\eeqnn

Counting all the terms that appear in the polynomials, there are 79 coefficients in total.  Of these, 
$18=6{\times}\left(4{-}1 \right)$ coefficients can be absorbed by coordinate redefinitions of the six $\IP^1$ spaces, $32 = 4{\times}\left(9{-}1 \right)$ by redefinitions of the coordinates of the four $\IP^2$ spaces and $11$ by rescalings of the polynomials.  

This parameter count shows that there are $79{-}61 = 18$ free parameters in the polynomials, which agrees with the dimension of the $(2,1)$--cohomology group for the manifold $\mathscr{X}^{12,18}$. Thus all deformations of the complex structure of $\mathscr{X}^{12,18}$ are parametrized by polynomials and so 
$h^{2,1}\left(\mathscr{X}^{12,18} / \IZ_3\right)$ can be found through a parameter count. 

The $\IZ_3$-action is generated by
\vspace{-20pt}
\begin{center}
\begin{minipage}{0.9\textwidth}
\begin{align*}
g:~x_{\alpha, j} & ~\to~  x_{\alpha+1, j}~,& y_{\alpha, j} &~\to~ y_{\alpha+1, j}~,& z_{\a,a} &~\to~ z_{\a +1, a} ~,& w_{a} &~\to~ w_{a+1} \\[5pt]
 p^{\alpha}&~\to~  p^{\alpha+1}~,& q^{\alpha}&~\to~  q^{\alpha+1}~,& r ^{\a}&~\to~ r^{\a+1}~, & s&~\to~ s~,~~~~~ t~\to~ t ~ . 
\end{align*}
\end{minipage}
\end{center}
and we require that the polynomials $p,\,q,\,r,\,s,\,t$ are covariant under this action. This implies that the coefficients $P^\a_{ja},\, Q^\a_{ja},\, R^\a_{ab}$ are independent of $\a$ and that $S_{ijk}$ and $T_{ijk}$ are invariant under cyclic permutations of indices. 

\vspace{12pt}
The action is fixed point free. In the embedding space, the fixed points of $g$ are of the form $x_{\a,j} = x_j$, $y_{\a,j} = y_j$, $z_{\a,a} = z_a$ and 
$w_a \in \{\left( 1,1,1 \right), \left(1, \o, \o^2 \right), \left( 1, \o^2, \o\right) \}$, where $\o$ is a nontrivial cube root of unity. The 5 independent polynomials $p,q,r,s,t$ become constraints on the coordinates of the product 
$\IP^1{\times} \IP^1{\times}\IP^2$ parametrized by $x_j, y_j, z_a$. As such, a fixed point is a simultaneous solution of 
$$
p(x_j, z_a) ~=~q(y_j, z_a) ~=~ r(z_a) ~=~ s(x_j) ~=~ t(y_j) ~=~ 0~.
$$  
In general, there are no such solutions. Suppose $x_j$ and $y_j$ are given. Then the constraints reduce to a system of three equations for $(z_0, z_1, z_2) \in \IP^2$ which for general coefficients have no solutions. 

\begin{table}
\begin{center}
\boxed{\hskip2pt
\begin{tabular}{cccc}
$\ \ $& & & $\ $\\
$\ \ $&$\cicy{\IP^1\\ \IP^1\\ \vrule height10pt width0pt depth8pt  \IP^1\\ \IP^1\\ \IP^1\\  \vrule height10pt width0pt depth8pt \IP^1\\ \IP^2\\ \IP^2\\ \vrule height10pt width0pt depth8pt  \IP^2\\ \IP^2\\ \IP^2\\ \IP^2}
{\one ~ 0 ~ 0 ~~ 0~ 0 ~ 0  ~~ 0 ~ 0 ~ 0 ~~ \one ~ 0 ~ 0~ ~ 0 ~ 0 ~ 0 \\
 0 ~ \one ~ 0 ~~ 0 ~ 0 ~ 0  ~~ 0 ~ 0 ~ 0~ ~ 0 ~ \one ~ 0~ ~ 0 ~ 0 ~ 0 \\
 \vrule height10pt width0pt depth8pt  0 ~ 0 ~ \one~ ~ 0 ~ 0 ~ 0  ~~ 0 ~ 0 ~ 0 ~~ 0 ~ 0 ~ \one ~~ 0 ~ 0 ~ 0  \\
 0 ~ 0 ~ 0 ~~ \one ~ 0 ~ 0 ~ ~ 0 ~ 0 ~ 0 ~~  0 ~ 0 ~ 0 ~~ \one ~ 0 ~ 0 \\
 0 ~ 0 ~ 0 ~~ 0 ~ \one ~ 0 ~ ~ 0 ~ 0 ~ 0 ~~  0 ~ 0 ~ 0~ ~ 0 ~ \one ~ 0  \\
 \vrule height10pt width0pt depth8pt  0 ~ 0 ~ 0 ~~ 0 ~ 0 ~ \one ~ ~ 0 ~ 0 ~ 0~ ~  0 ~ 0 ~ 0~ ~ 0 ~ 0 ~ \one  \\
 
\one~ 0 ~ 0~ ~ \one ~ 0 ~ 0~ ~ \one ~ 0 ~ 0 ~~ 0 ~ 0 ~ 0 ~~ 0 ~ 0 ~ 0  \\
 0 ~ \one ~ 0~ ~ 0 ~ \one ~ 0~ ~ 0 ~ \one ~ 0~ ~ 0 ~ 0 ~ 0~ ~ 0 ~ 0 ~ 0\\
\vrule height10pt width0pt depth8pt   0 ~ 0 ~ \one~ ~ 0 ~ 0 ~  \one~ ~ 0 ~ 0 ~ \one ~~0 ~ 0 ~ 0~ ~ 0 ~ 0 ~ 0 \\
 0 ~ 0 ~ 0 ~~ 0 ~ 0 ~ 0 ~~ \one ~ \one ~ \one~ ~  0 ~ 0 ~ 0~~ 0 ~ 0 ~ 0 \\
 0 ~ 0 ~ 0 ~~ 0 ~ 0 ~ 0 ~ ~0 ~ 0 ~ 0 ~ ~ \one ~ \one ~ \one~ ~ 0 ~ 0 ~ 0\\
 0 ~ 0 ~ 0 ~~ 0 ~ 0 ~ 0 ~ ~0 ~ 0 ~ 0 ~~ 0 ~ 0 ~ 0 ~~ \one ~ \one ~ \one}_{-12}^{12,18}$
&\hskip 21pt\lower 94pt\hbox{\includegraphics[width=2.4in]{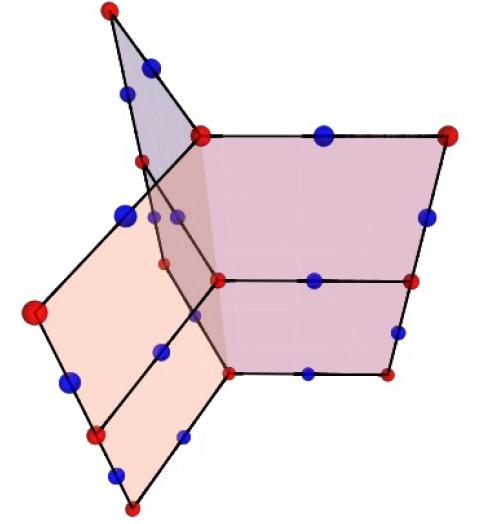}}& $\ $\\
$\ \ $& & & $\ $
\end{tabular}}
\vskip5pt
\capt{5.5in}{tab(12,18)Extended}{The matrix and diagram for the extended representation of 
$\mathscr{X}^{12,18}$ for which all 12 K\"ahler forms are represented by ambient spaces. The three vanes are identified by the $\IZ_3$ action, leaving only $6$ independent hyperplane classes.} 
\end{center}
\vspace{12pt}
\end{table}

\vspace{12pt}
The parameter count for the quotient manifold goes as follows. There are $29$ terms in the $\IZ_3$ covariant form of the polynomials. Since the $\IZ_3$ action identifies the first three $\IP^1$ spaces among themselves, and similarly for the last three $\IP^1$ spaces, the number of coordinate redefinitions associated with these spaces is reduced to $6$. Also, there are only $8$ coordinate redefinitions associated with the first three $\IP^2$ spaces and 2 associated with redefinitions in the last $\IP^2$ space, which, up to scale transformations have the form: 
$$
w_{a} \to \sum_{b}\Xi_{b}\, w_{a+b}~. 
$$
The number of scale transformations for the polynomials is now 5. Summing up, the number of free parameters that describe the $\IZ_3$-symmetric class of manifolds is $29-21 = 8$, so that
$h^{2,1}\left(\mathscr{X}^{12,18}/ \IZ_3\right)  =  8$. 

The dimension of the $(1,1)$--cohomology group can be obtained by considering an extended representation of the manifold $\mathscr{X}^{12,18}$ embedded in a product of $12$ projective spaces, providing the $12$ independent $(1,1)$--forms in $H^{1,1}\left( \mathscr{X}^{12,18}\right)$. In the $\IZ_3$ quotient only $6$ of these $(1,1)$--forms remain independent (see Table \ref{tab(12,18)Extended}). 

I have checked that the polynomial constraints are transverse. Since the action of $g$ is fixed point free, we obtain a smooth quotient manifold.  The new manifold 
$\mathscr{X}^{6,8} = \mathscr{X}^{12,18}/ \IZ_3$ has fundamental group $\IZ_3$ and 
$\hodgenos = (6,8)$.
%
%
\subsection{The manifold $\mathscr{X}^{9, 21}$ with quotient 
$\mathscr{X}^{5,9} = \mathscr{X}^{9,21}/ \IZ_3$} \label{sec:1.2}

The manifolds in this class are described by the configuration matrix
$$
\mathscr{X}^{9,21}~=~~
\cicy{\IP^1 \\ \IP^1\\ \vrule height10pt width0pt depth8pt  \IP^1\\ \IP^2\\ \IP^2\\ \vrule height10pt width0pt depth8pt  \IP^2\\ \IP^2\\ \IP^2}
{ \one& 0& 0& ~0& 0& 0&~0& 0& 0&~ \one \\
 0& \one& 0& ~0& 0&0&~0& 0& 0&~ \one \\
\vrule height10pt width0pt depth8pt  0& 0& \one&~ 0& 0&0&~0&0& 0&~ \one \\
\one & 0& 0&~ \one& 0&0&~\one& 0& 0&~ 0 \\
 0& \one & 0&~ 0& \one& 0&~ 0& \one& 0&~ 0 \\
 \vrule height10pt width0pt depth8pt 0& 0& \one&~ 0& 0& \one &~0&0& \one&~ 0 \\
0& 0& 0 &~ \one& \one &\one &~ 0& 0& 0&~0 \\
0& 0& 0&~ 0& 0&0&~\one& \one& \one&~ 0 \\}_{-24}^{9,21}
$$
corresponding to the equivalent diagrams
\vspace{-16pt}
\begin{flushright}
$$
{\includegraphics[width=2.4in]{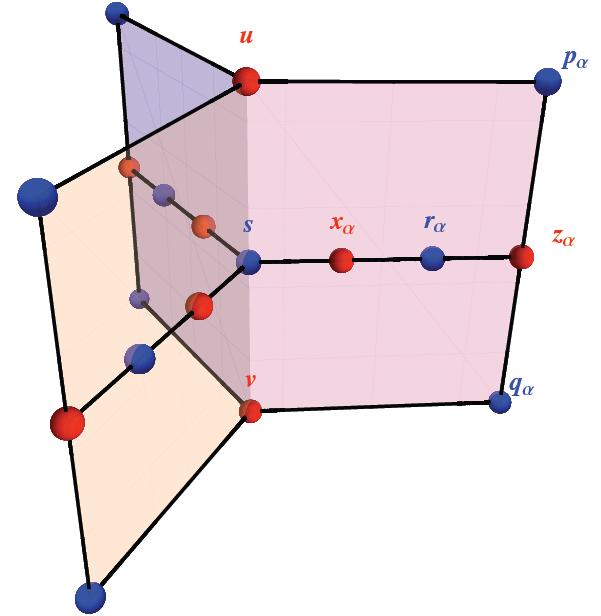}}
\hskip 32pt \lower -14pt\hbox{\includegraphics[width=2.4in]{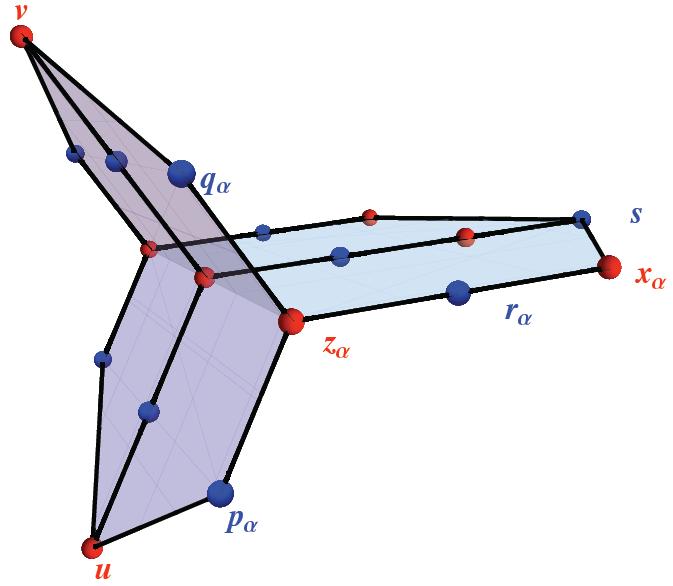}}
$$
\end{flushright}
\vspace{-5pt}
}

{\setstretch{1.57}
Denote the coordinates of the three $\IP^1$'s by  $x_{\alpha, j}$. As before, $\alpha$ labels the space and $j$ its coordinate, and they are understood to take values in $\IZ_3$ and $\IZ_2$ respectively. Let $z_{\alpha, a}$ denote the coordinates of the first three $\IP^2$'s, while $u_a$ and $v_a$ denote the coordinates of the last two~$\IP^2$'s.  Denote also by $ r^{\alpha}$ the first three polynomials, by $p^{\alpha}$ the next three, and  the last three by 
$q^{\a}$. Denote by $s$ the cubic polynomial in the $\IP^1$ coordinates. The polynomials can be expressed in the general form  
\vspace{-20pt}
\begin{center}
\begin{minipage}{0.7\textwidth}
\begin{align*}
p^{\alpha} ~&=~ \sum_{a,b} P^\a_{ab}\, z_{\a, a}\, u_{\alpha+b}~, & q^{\alpha} ~&=~ \sum_{a,b} Q^\a_{ab}\, z_{\alpha,a}\, v_{\alpha+b} \\[8pt]
r^{\alpha} ~&=~ \sum_{j,a} R^\a_{ja}\, x_{\alpha,j}\, z_{\a,a}~, & s ~&=~ \sum_{i,j,k} S_{ijk}\, x_{0,i}\, x_{1,j}\, x_{2,k}
\end{align*}
\end{minipage}
\end{center}

The parameter count is similar to the previous situation. There are 80 terms in the polynomials, 49 coordinate redefinitions, and 10 scale transformations for the polynomials. This gives a total of $80-59 = 21$ free parameters in the polynomials, equal to the dimension of $H^{2,1}\left(\mathscr{X}^{9,21} \right)$. So again, we can find the value of $h^{2,1}$ for the quotient manifold by counting the free parameters of the covariant polynomials after the $\IZ_3$ identifications. 

\vspace{30pt}
The $\IZ_3$-action is generated by
\vspace{-25pt}
\begin{center}
\begin{minipage}{0.8\textwidth}
\begin{align*}
g:~x_{\alpha, j} & ~\to~  x_{\alpha+1, j}~,&  z_{\a,a} &~\to~ z_{\a +1, a} ~,& u_{a} &~\to~ u_{a+1}~,& v_{a} &~\to~ v_{a+1} \\[5pt]
 p^{\alpha}&~\to~  p^{\alpha+1}~,& q^{\alpha}&~\to~  q^{\alpha+1}~,& r ^{\a}&~\to~ r^{\a+1}~, & s&~\to~ s~. 
\end{align*}
\end{minipage}
\end{center}
\vspace{-5pt}

As before, in order to have $g$-covariant polynomials, we require that $P^\a_{ab},\, Q^\a_{ab},\, R^\a_{ja}$ are independent of $\a$ and that $S_{ijk}$ is invariant under cyclic permutations of indices. 

\begin{table}[h]
\begin{center}
\vspace{20pt}
\boxed{\begin{tabular}{cc}
 & \\
\lower55pt\hbox{\includegraphics[width=2.4in]{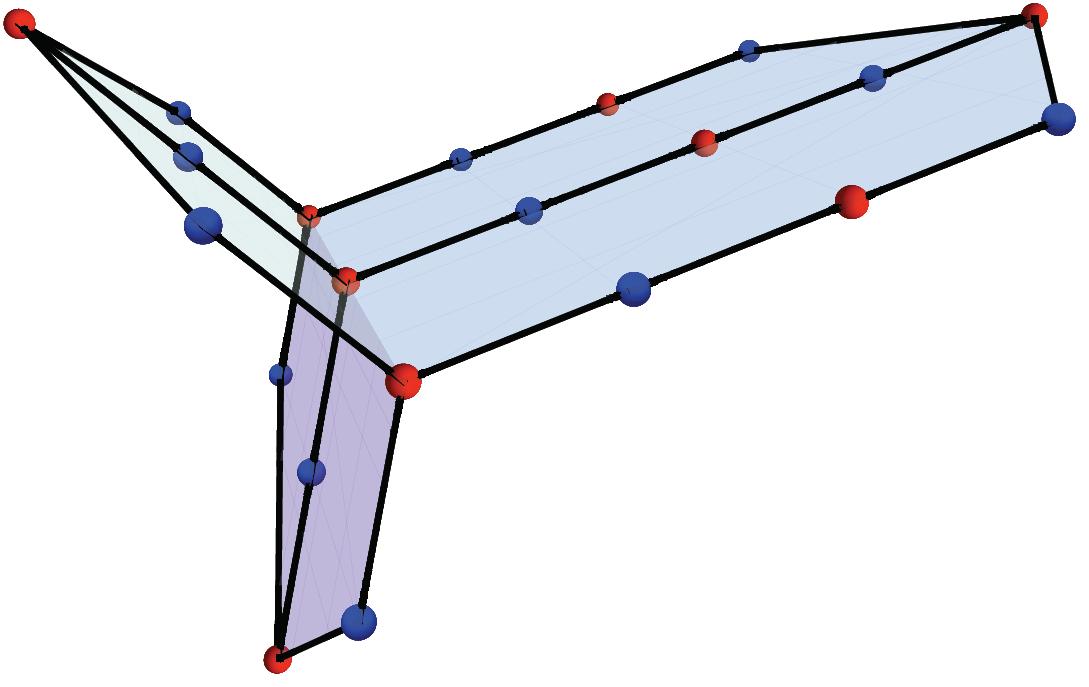}}&
\hskip .1in $\cicy{\IP^1 \\ \IP^1\\ \vrule height10pt width0pt depth8pt  \IP^1\\ \IP^2\\ \IP^2\\ \vrule height10pt width0pt depth8pt  \IP^2\\ \IP^2\\ \IP^2\\ \IP^2}
{ \one& 0& 0& ~0& 0& 0&~0& 0& 0&~ \one& 0 & 0 \\
 0& \one& 0& ~0& 0&0&~0& 0& 0&~ 0 &\one& 0 \\
\vrule height10pt width0pt depth8pt  0& 0& \one&~ 0& 0&0&~0&0& 0&~ 0 &0&\one \\
\one & 0& 0&~ \one& 0&0&~\one& 0& 0&~ 0 & 0 & 0\\
 0& \one & 0&~ 0& \one& 0&~ 0& \one& 0&~ 0 & 0 &0\\
 \vrule height10pt width0pt depth8pt 0& 0& \one&~ 0& 0& \one &~0&0& \one&~ 0 & 0 &0 \\
0& 0& 0 &~ \one& \one &\one &~ 0& 0& 0&~0 & 0 & 0 \\
0& 0& 0&~ 0& 0&0&~\one& \one& \one&~ 0 & 0 & 0\\
0& 0& 0&~ 0& 0&0&~0&0& 0&~ \one& \one & \one \\}_{-24}^{9,21}$\\
 & 
 \vspace{-4pt}
\end{tabular}}
\parbox{5.5in}{\capt{5.5in}{tab(9,21)Extended}{The matrix and diagram for the extended representation of $\mathscr{X}^{9,21}$, for which all 9 K\"ahler forms are represented by ambient spaces. After the $\IZ_3$ identification, only $5$ of the $(1,1)$--forms remain independent.}}
\end{center}
\end{table}

\vspace{30pt}
The action is fixed point free. Fixed points of $g$, in the embedding space, are of the form $x_{\a,j} = x_j$, $z_{\a,a} = z_a$ and $u_a = \xi^a$, $v_a = \zeta^a$, where $\xi$ and $\zeta$ are cube roots of unity, $\xi^3 = \zeta^3 =1$. The 4 independent polynomials $p(z_a),~ q(z_a),~ r(x_j, z_a),~ s(x_j) = 0$ become constraints on the coordinates of the product $\IP^1{\times}\IP^2$ parametrized by $x_j, z_a$. For fixed $x_j$, the constraints reduce to a system of three equations for $(z_0, z_1, z_2) \in \IP^2$ which have no solutions  for general choices of the coefficients. 

The $\IZ_3$ covariant polynomials have $28$ coefficients in total. There are also $15$ coordinate redefinitions consistent with the $\IZ_3$ identifications and $4$ scale transformations of the polynomials. This leads to 
$h^{2,1}\left(  \mathscr{X}^{9,21} / \IZ_3 \right) = 9$. 

The dimension of the $(1,1)$--cohomology group can be obtained by considering the extended representation of the manifold $\mathscr{X}^{9,21}$ as shown in Table \ref{tab(9,21)Extended}. It follows that $h^{1,1} = 5$ for the quotient manifold. 

\vspace{8pt}
I have checked that the polynomials are transverse. Thus we obtain a smooth $\IZ_3$ quotient of a CICY manifold, $\mathscr{X}^{5,9} =   \mathscr{X}^{9,21} / \IZ_3$, with fundamental group isomorphic to $\IZ_3$ and $\hodgenos = (5,9)$.
 
\vspace{10pt}
\subsection{The manifold $\mathscr{X}^{11, 26}$ with quotient 
$\mathscr{X}^{5,10} =  \mathscr{X}^{11,26}/ \IZ_3$} \label{sec:1.3}

This class of manifolds is described by the configuration
$$
\mathscr{X}^{11,26}~=~~
\cicy{\IP^1 \\ \IP^1\\ \vrule height10pt width0pt depth8pt \IP^1\\ \IP^1\\ \IP^1\\\vrule height10pt width0pt depth8pt  \IP^1\\ \IP^2\\ \IP^2\\\vrule height10pt width0pt depth8pt  \IP^2}
{ \one& 0& 0&~ 0& 0 & 0&~ \one&0& 0  \\
 0& \one& 0&~ 0 & 0 & 0 &~ \one& 0& 0 \\
\vrule height10pt width0pt depth8pt  0& 0& \one&~ 0& 0&0&~ \one & 0& 0 \\
 0& 0& 0&~ \one& 0&0&~ 0 & \one& 0\\
 0& 0& 0&~ 0& \one& 0&~ 0 & \one& 0 \\
\vrule height10pt width0pt depth8pt   0& 0& 0&~ 0& 0& \one & ~ 0 & \one& 0 \\
\one& 0& 0&~ \one&0 &0&~ 0&0 & \one\\
0& \one& 0&~ 0& \one&0&~ 0& 0 & \one\\
\vrule height10pt width0pt depth8pt  0& 0& \one&~ 0& 0&\one&~ 0&0 & \one\\}_{-30}^{11,26}
$$

with equivalent diagrams
\vspace{-35pt}
\begin{flushright}
$$
{\includegraphics[width=2.4in]{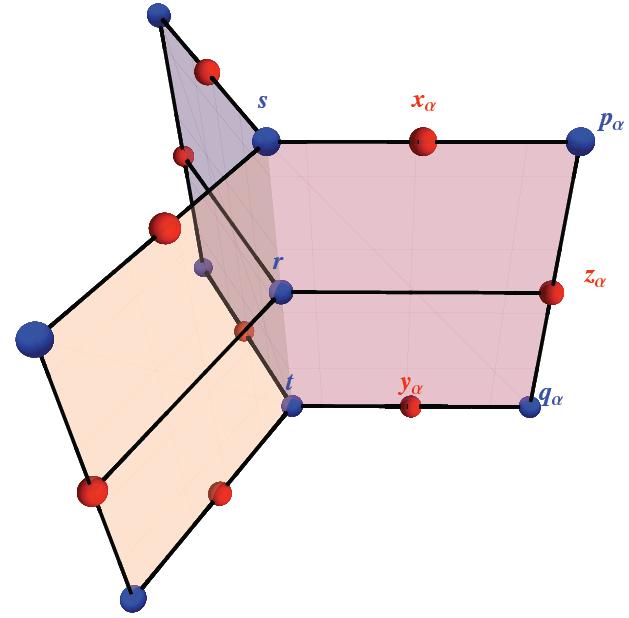}}
\hskip 32pt \lower -20pt\hbox{\includegraphics[width=2.4in]{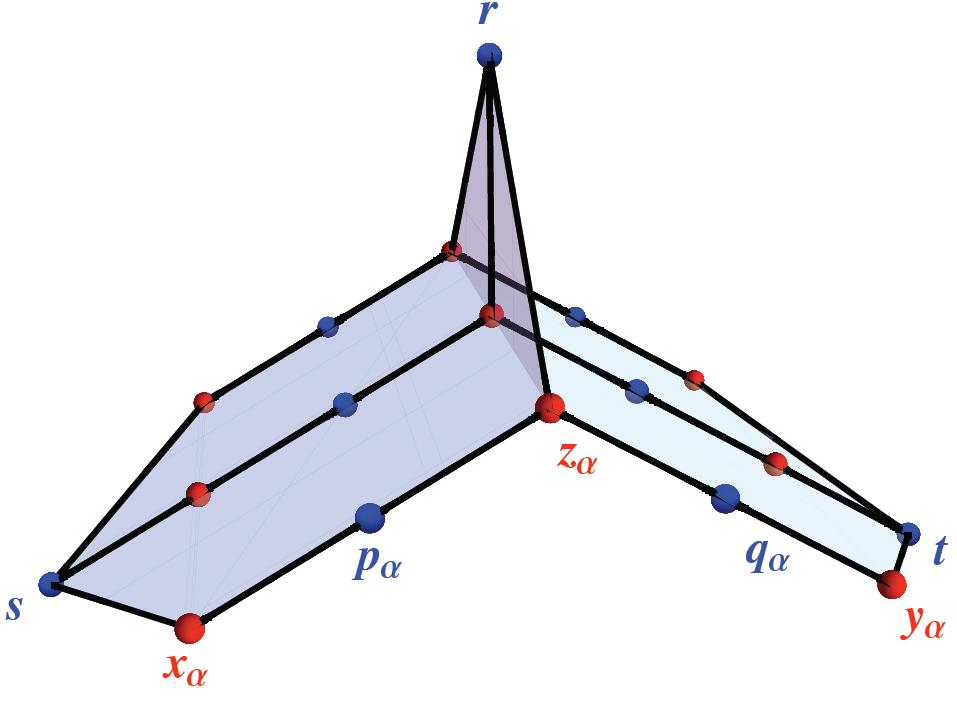}}
$$
\end{flushright}
\vspace{-7pt}

The manifold $\mathscr{X}^{11,26}$ is obtained by contracting the last $\IP^2$ in the configuration matrix of the manifold $\mathscr{X}^{12,18}$ discussed in Section \ref{sec:1.1}. In analogy with that case, we write down the following generic polynomials:
\begin{equation*}
\begin{split}
p^{\alpha} &= \sum_{j,a} P^\a_{ja}\, x_{\a, j}\, z_{\alpha, a}~,~~~~q^{\alpha}= \sum_{j,a} Q^\a_{ja}\, y_{\alpha,j}\, z_{\alpha,a}\\[8pt]
r = \sum_{a,b,c} R_{abc}\, z_{0,a}\, z_{1,b}& \, z_{2,c}~,~~~~s = \sum_{i,j,k} S_{ijk}\, x_{0,i}\, x_{1,j}\, x_{2,k}~,~~~~t = \sum_{i,j,k} T_{ijk}\, y_{0,i}\, y_{1,j}\, y_{2,k}
\end{split}
\end{equation*}

\begin{table}[h]
\begin{center}
\vspace{30pt}
\boxed{\hskip2pt
\begin{tabular}{cccc}
$\ \ $& & &$\ \ $ \\
$\ \ $&$\cicy{\IP^1\\ \IP^1\\ \vrule height10pt width0pt depth8pt  \IP^1\\ \IP^1\\ \IP^1\\  \vrule height10pt width0pt depth8pt \IP^1\\ \IP^2\\ \IP^2\\ \vrule height10pt width0pt depth8pt  \IP^2\\ \IP^2\\ \IP^2}
{\one ~ 0 ~ 0 ~~ 0~ 0 ~ 0  ~~ \one ~ 0 ~ 0~ ~ 0 ~ 0 ~ 0 ~~0\\
 0 ~ \one ~ 0 ~~ 0 ~ 0 ~ 0  ~~  0 ~ \one ~ 0~ ~ 0 ~ 0 ~ 0 ~~0\\
 \vrule height10pt width0pt depth8pt  0 ~ 0 ~ \one~ ~ 0 ~ 0 ~ 0  ~~ 0 ~ 0 ~ \one ~~ 0 ~ 0 ~ 0  ~~0\\
 0 ~ 0 ~ 0 ~~ \one ~ 0 ~ 0 ~ ~   0 ~ 0 ~ 0 ~~ \one ~ 0 ~ 0 ~~0\\
 0 ~ 0 ~ 0 ~~ 0 ~ \one ~ 0 ~ ~   0 ~ 0 ~ 0~ ~ 0 ~ \one ~ 0  ~~0\\
 \vrule height10pt width0pt depth8pt  0 ~ 0 ~ 0 ~~ 0 ~ 0 ~ \one ~ ~  0 ~ 0 ~ 0~ ~ 0 ~ 0 ~ \one ~~0 \\
 
\one~ 0 ~ 0~ ~ \one ~ 0 ~ 0 ~~ 0 ~ 0 ~ 0 ~~ 0 ~ 0 ~ 0  ~~1\\
 0 ~ \one ~ 0~ ~ 0 ~ \one ~ 0~ ~ 0 ~ 0 ~ 0~ ~ 0 ~ 0 ~ 0~~1\\
\vrule height10pt width0pt depth8pt   0 ~ 0 ~ \one~ ~ 0 ~ 0 ~  \one ~~0 ~ 0 ~ 0~ ~ 0 ~ 0 ~ 0 ~~1\\
 0 ~ 0 ~ 0 ~~ 0 ~ 0 ~ 0 ~~ \one ~ \one ~ \one~ ~  0 ~ 0 ~ 0~~ 0 \\
 0 ~ 0 ~ 0 ~~ 0 ~ 0 ~ 0 ~ ~0 ~ 0 ~ 0 ~ ~ \one ~ \one ~ \one~ ~ 0 }_{-30}^{11,26}$
&\hskip 21pt\lower 44pt\hbox{\includegraphics[width=2.4in]{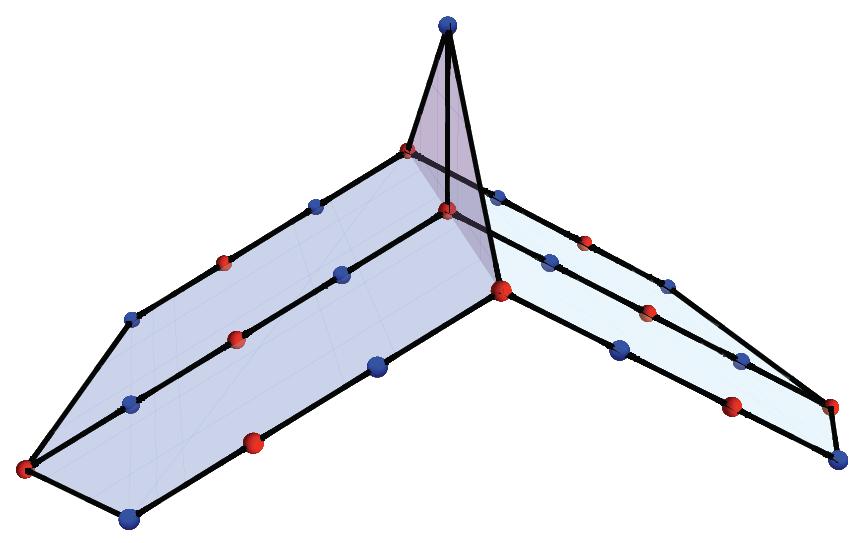}}&$\ \ $\\
$\ \ $& & &$\ \ $
\end{tabular}}
\vskip5pt
\capt{5.5in}{tab(11,26)Extended}{The matrix and diagram for the extended representation of 
$\mathscr{X}^{11,26}$ for which all 11 K\"ahler forms are represented by ambient spaces. After the $\IZ_3$ identification, only $5$ of the $(1,1)$--forms remain independent.} 
\end{center}
\vspace{20pt}
\end{table}

A parameter count shows that there are $79$ terms in this most general form of the polynomials. Of these, $42$ parameters can be eliminated by coordinate redefinitions, and $9$ by rescaling each of the polynomials. There are, however, two more degrees of freedom in the redefinition of $r$. These come from the following fact. For fixed $z$ coordinates, the polynomials $p^{\a} = 0$ can be regarded as a system of three equations linear in the $x$ coordinates. Using the freedom of coordinate redefinition, there are only three independent $x$ coordinates. If the equations $p^{\a} = 0$ have nontrivial solutions, we must have $\det\left.\left( \partial p / \partial x \right)\right|_z = 0$. But this is a trilinear polynomial in the $z$ coordinates and adding a multiple of it to $r$ is an allowed redefinition. Similarly, there is a second degree of freedom for $r$ associated with the vanishing of $\det\left.\left( \partial q / \partial y \right)\right|_z$. This count confirms the value of $h^{2,1} = 26 = 79 - 42 - 11$.

\vspace{20pt}
The $\IZ_3$-action is generated by
\begin{center}
\vspace{-30pt}
\begin{minipage}{0.9\textwidth}
\begin{align*}
g:~~x_{\alpha, j} & ~\to~  x_{\alpha+1, j}~,& y_{\alpha, j} &~\to~ y_{\alpha+1, j}~,& z_{\a, a} &~\to~ z_{\a, a+1}\\[5pt]
 p^{\alpha}&~\to~  p^{\alpha+1}~,& q^{\alpha}&~\to~  q^{\alpha+1}~,& r &~\to~ r~, & s&~\to~ s~,~~~~~~ t~\to~ t ~ . 
\end{align*}
\end{minipage}
\end{center}

Covariance under this action requires that $P^\a_{ab},\, Q^\a_{ab},\, R^\a_{ja}$ are independent of $\a$ and that $S_{ijk}$ and $T_{ijk}$ are invariant under cyclic permutations of indices. 
The action is again fixed point~free. 

\vspace{12pt}
After the $\IZ_3$ identification, there are 31 possible terms in the polynomials, 14 coordinate redefinitions and 7 degrees of freedom associated with redefinitions of polynomials. These give a total of 10 free parameters. Thus $h^{2,1} = 10$ for the quotient manifold. 

\vspace{12pt}
In order to find $h^{1,1}$, consider the extended representation of the manifold, obtained by splitting $s$ and $t$ with two $\IP^2$'s, as shown in Table \ref{tab(11,26)Extended}. In this way we see that $h^{1,1} = 5$ for the quotient manifold. We have also checked that the polynomials are transverse. Thus, we obtain a quotient 
$\mathscr{X}^{5,10} = \mathscr{X}^{11,26}/ \IZ_3$ with fundamental group $\IZ_3$. 


\newpage
\subsection{The manifold $\mathscr{X}^{8, 29}$ with quotient 
$\mathscr{X}^{4, 11} =  \mathscr{X}^{8,29}/ \IZ_3$} \label{sec:1.4}

This manifold can be obtained by contracting one of the last two $\IP^2$ spaces in the representation of $\mathscr{X}^{9,21}$  (see Section \ref{sec:1.2}). The configuration matrix is
\vskip 5pt
$$
\mathscr{X}^{8,29}~=~~
\cicy{\IP^1 \\ \IP^1\\  \vrule height10pt width0pt depth8pt  \IP^1\\ \IP^2\\ \IP^2\\ \vrule height10pt width0pt depth8pt  \IP^2\\ \IP^2}
{ \one& 0& 0&~ 0&0& 0&~ \one & 0\\
 0& \one& 0&~ 0&0& 0&~ \one & 0\\
\vrule height10pt width0pt depth8pt  0& 0& \one&~ 0&0& 0&~ \one & 0\\
\one& 0& 0&~ \one& 0& 0&~ 0 & \one\\
0& \one& 0&~ 0& \one& 0&~ 0 & \one\\
\vrule height10pt width0pt depth8pt  0& 0& \one&~0& 0& \one&~ 0 & \one\\
0& 0& 0&~\one&\one& \one&~ 0 & 0\\}_{-42}^{8,29}
$$
\vskip 5pt
with equivalent diagrams

\vspace{-21pt}
\begin{flushright}
$${\includegraphics[width=1.8in]{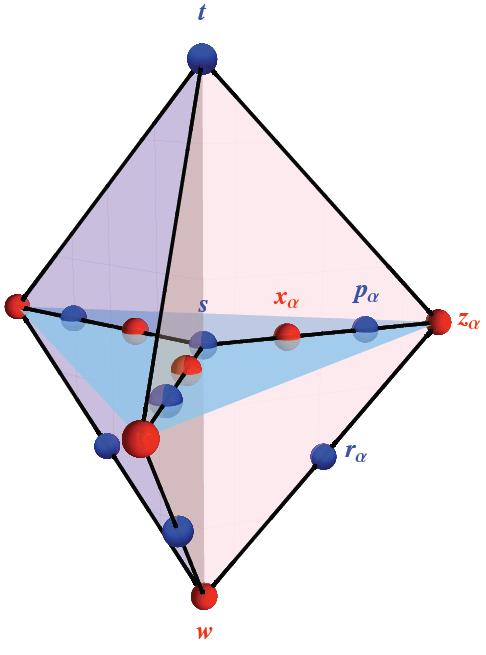}}
\hskip 52pt \lower -20pt\hbox{\includegraphics[width=2.4in]{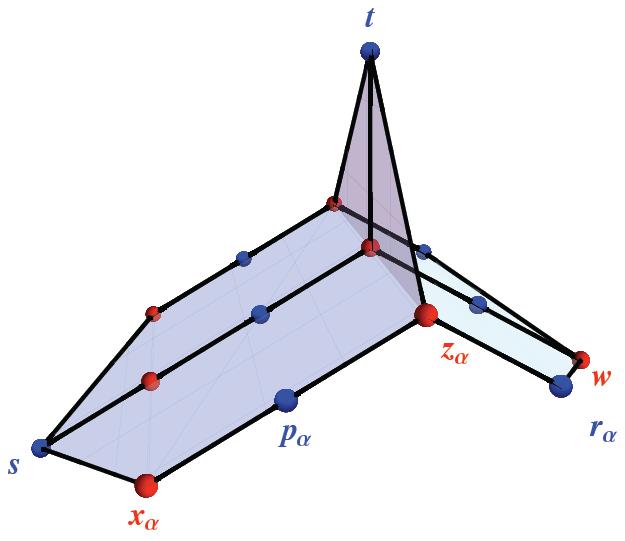}}
$$
\end{flushright}
\vspace{-8pt}
As before, take coordinates $x_{\a,j}$ for the three $\IP^1$ spaces, $z_{\a,a}$ for the first three $\IP^2$ spaces and $w_a$ for the last $\IP^2$ space. Denote by $p^\a$ the first three polynomials, by $r^\a$ the following three and by $s$ and $t$ the last two. The $\IZ_3$-action is generated by
\vspace{-20pt}
\begin{center}
\begin{minipage}{0.8\textwidth}
\begin{align*}
g:~x_{\alpha, j} & ~\to~  x_{\alpha+1, j}~,& z_{\alpha, a} &~\to~ z_{\alpha+1, a}~,& w_{a} &~\to~ w_{a+1}\\[5pt]
 p^{\alpha}&~\to~  p^{\alpha+1}~,& r &~\to~ r~, & s&~\to~ s~,~~~~~~~ t~\to~ t ~ . 
\end{align*}
\end{minipage}
\end{center}

The most general polynomials, covariant with the $g$ action 
\vspace{-10pt}
\begin{center}
\begin{minipage}{0.7\textwidth}
\begin{align*}
p^{\alpha} ~&=~ \sum_{j,a} P^\a_{ja}\, x_{\a, j}\, z_{\alpha, a}~, & r^{\a} ~&=~ \sum_{a,b} R^\a_{ab}\, z_{\a,a}w_{\a+b} \\[8pt]
s ~&=~ \sum_{i,j,k} S_{ijk}\, x_{0,i}\, x_{1,j}\, x_{2,k}~, & t ~&=~ \sum_{a,b,c} T_{abc}\, z_{0,a}\, z_{1,b}\, z_{2,c}
\end{align*}
\end{minipage}
\end{center}
where $S_{ijk}$ and $T_{abc}$ are cyclic in their indices and, for given $a,b,j$, the coefficients $P^\a_{ja}, R^\a_{ab}$ take the same values for all $\a$'s. It is easy to show that the action is fixed point free. 

A parameter count shows that there are $80$ terms in the most general form of the polynomials, before imposing covariance under the $\IZ_3$ action. Of these, $41$ parameters can be eliminated by coordinate redefinitions, $8$ by rescalings of the polynomials and $2$ more by redefinitions of $t$ in terms of $D_1 = \det \left.\partial\left(p; x\right) \right|_z$ and $D_2 = \det \left.\partial\left(r; w\right) \right|_z$. The parameter count confirms the value of $h^{2,1} = 29 = 80 - 41 - 10$. 

After the $\IZ_3$ identification, there are $30$ possible terms in the polynomials, $13$ coordinate redefinitions and $6$ degrees of freedom associated with redefinitions of polynomials. These give a total of $11$ free parameters, thus $h^{2,1} = 11$ for the quotient manifold. 

In order to find $h^{1,1}$, let us consider an extended representation of the manifold, obtained by splitting $s$ by means of a $\IP^2$, as shown in Table \ref{tab(8,29)Extended}. We see that the quotient 
manifold~has~\hbox{$h^{1,1} = 4$}. 

I have checked that the polynomials are transverse. Thus we obtain a smooth quotient of a CICY manifold  $\mathscr{X}^{4,11} = \mathscr{X}^{8,29}/ \IZ_3$  with fundamental group $\IZ_3$ and $\hodgenos = (4,11)$.

\begin{table}[ht]
\begin{center}
\boxed{\begin{tabular}{ccc}
& & $\ $ \\
\hskip .3in $\cicy{\IP^1 \\ \IP^1\\ \vrule height10pt width0pt depth8pt  \IP^1\\ \IP^2\\ \IP^2\\ \vrule height10pt width0pt depth8pt  \IP^2\\ \IP^2\\ \IP^2}
{ \one& 0& 0& ~0& 0& 0&~\one& 0& 0&~ 0 \\
 0& \one& 0& ~0& 0&0&~0& \one& 0&~ 0 \\
\vrule height10pt width0pt depth8pt  0& 0& \one&~ 0& 0&0&~0&0& \one&~ 0 \\
\one & 0& 0&~ \one& 0&0&~0& 0& 0&~ \one \\
 0& \one & 0&~ 0& \one& 0&~ 0& 0& 0&~ \one \\
 \vrule height10pt width0pt depth8pt 0& 0& \one&~ 0& 0& \one &~0&0& 0&~ \one \\
0& 0& 0 &~ \one& \one &\one &~ 0& 0& 0&~0 \\
0& 0& 0&~ 0& 0&0&~\one& \one& \one&~ 0 \\}_{-42}^{8,29}
$ &  \hskip .3in \lower 70pt \hbox{\includegraphics[width=2.4in]{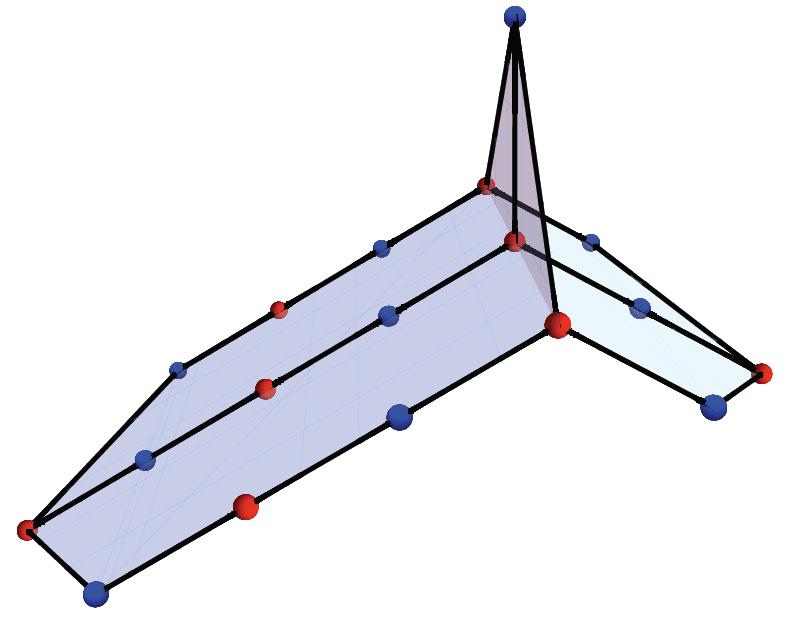}}& $\ $\\
 & & $\ $ 
\end{tabular}}
\capt{5.5in}{tab(8,29)Extended}{The matrix and diagram for the extended representation of 
$\mathscr{X}^{8,29}$ for which all 8 K\"ahler forms are represented by ambient spaces. After the $\IZ_3$ identification, only $4$ of the $(1,1)$--forms remain independent.} 
\end{center}
\end{table}

\pagebreak


\subsection{The manifold $\mathscr{X}^{7, 37}$ with quotient
 $\mathscr{X}^{3, 13} =  \mathscr{X}^{7,37}/ \IZ_3$} \label{sec:1.5}
\vspace{0pt}
This manifold can be obtained by contracting the last $\IP^2$ space of the previous 
manifold, $\mathscr{X}^{8,29}$. The configuration matrix is
$$
\mathscr{X}^{7,37}~=~~
\cicy{\IP^1 \\ \IP^1\\  \vrule height10pt width0pt depth8pt  \IP^1\\ \IP^2\\ \IP^2\\ \vrule height10pt width0pt depth8pt  \IP^2}
{ \one& 0& 0&~ \one & 0 & 0\\
 0& \one& 0&~  \one & 0 & 0\\
\vrule height10pt width0pt depth8pt  0& 0& \one&~ \one & 0 & 0\\
\one& 0& 0&~  0 & \one & \one\\
0& \one& 0&~  0 & \one & \one\\
\vrule height10pt width0pt depth8pt  0& 0& \one&~ 0 & \one & \one}_{-60}^{7,37}
$$
\vskip 5pt
with equivalent diagrams
\vspace{-35pt}
\begin{flushright}
$${\includegraphics[width=2.1in]{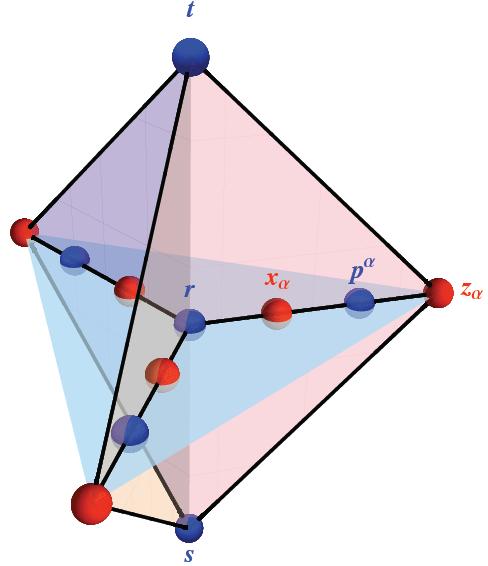}}
\hskip 52pt \lower 0pt\hbox{\includegraphics[width=2.4in]{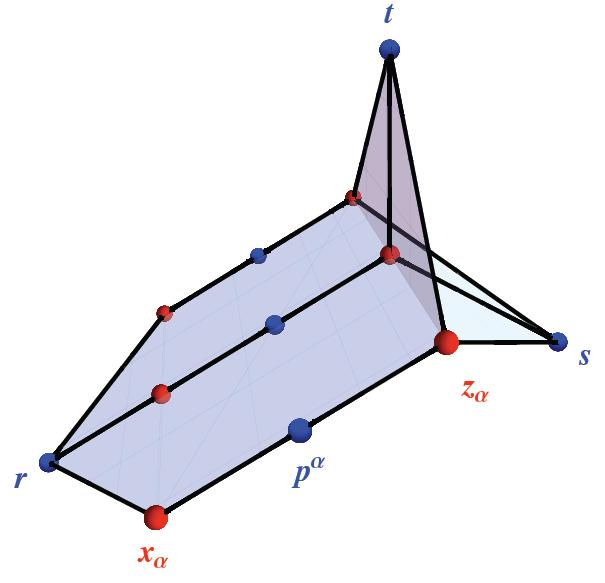}}
$$
\end{flushright}

As before, take coordinates $x_{\a,j}$ for the three $\IP^1$ spaces and $z_{\a,a}$ for the three $\IP^2$'s. Denote by $p^\a$ the first three polynomials, and by $r,s,t$  the last three. The $\IZ_3$ action is generated by 
\vspace{-30pt}
\begin{center}
\begin{minipage}{0.8\textwidth}
\begin{align*}
g:~x_{\alpha, j} & ~\to~  x_{\alpha+1, j}~,& z_{\alpha, a} &~\to~ z_{\alpha+1, a}\\[5pt]
 p^{\alpha}&~\to~  p^{\alpha+1}~,& r &~\to~ r~, & s~\to~ s~,~~~~~~~~t~\to~ t ~ . 
\end{align*}
\end{minipage}
\end{center}

The most general polynomials, covariant with the $g$ action are
\vspace{-10pt}
\begin{center}
\begin{minipage}{0.7\textwidth}
\begin{align*}
p^{\alpha} ~&=~ \sum_{j,a} P^\a_{ja}\, x_{\a, j}\, z_{\alpha, a}~,& r~&=~ \sum_{i,j,k} R_{ijk}\, x_{0,i}\, x_{1,j}\, x_{2,k}\\[8pt]
s ~&=~ \sum_{a,b,c} S_{abc}\, z_{0,a}\, z_{1,b}\, z_{2,c}~, & t ~&=~ \sum_{a,b,c} T_{abc}\, z_{0,a}\, z_{1,b}\, z_{2,c}
\end{align*}
\end{minipage}
\end{center}
where $R_{ijk}$, $S_{abc}$ and $T_{abc}$ are cyclic in their indices and the coefficients $P^\a_{ja}$ do not depend on $\a$'s. The search for fixed points is very similar to the previous cases and gives an empty result.

The parameter count shows that there are again $80$ terms in the most general form of the polynomials, before imposing covariance under the $\IZ_3$ action. Of these, $33$ parameters can be eliminated by coordinate redefinitions, $4$ by rescaling the polynomials $r$ and $p^\a$. The most general redefinitions for $t$ and $s$ involve $6$ parameters: 
$$ t~ \to~ c_{tt} t + c_{ts} s + c_{1} D~, ~~~~s ~\to~ c_{st} t + c_{ss} s + c_{2} D$$
where $D = \det \left.\partial\left(p; x\right) \right|_z$. Thus the number of free parameters is $80-33-10 = 37$, confirming the value of $h^{2,1}$.

In order to find $h^{1,1}$, let us consider an extended representation of the manifold, obtained by splitting $s$ with a $\IP^2$, as shown in Table \ref{tab(7,37)Extended}. It follows that the quotient manifold has 
$h^{1,1} = 3$. Thus, we obtain a quotient $\mathscr{X}^{3,13} = \mathscr{X}^{7,37}/ \IZ_3$. 

\begin{table}
\begin{center}
\boxed{\begin{tabular}{ccc}
& & $\ \ $\\
\hskip .3in $\cicy{\IP^1 \\ \IP^1\\ \vrule height10pt width0pt depth8pt  \IP^1\\ \IP^2\\ \IP^2\\ \vrule height10pt width0pt depth8pt  \IP^2\\ \IP^2}
{ \one& 0& 0& ~\one& 0& 0&~ 0 & 0\\
 0& \one& 0&~0& \one& 0&~ 0 & 0\\
\vrule height10pt width0pt depth8pt  0& 0& \one&~0&0& \one&~ 0 & 0\\
\one & 0& 0&~0& 0& 0&~ \one & \one\\
 0& \one & 0&~ 0& 0& 0&~ \one & \one \\
 \vrule height10pt width0pt depth8pt 0& 0& \one&~0&0& 0&~ \one & \one \\
0& 0& 0&~\one& \one& \one&~ 0 & 0\\}_{-60}^{7,37}
$ &  \hskip .5in \lower 75pt \hbox{\includegraphics[width=2.4in]{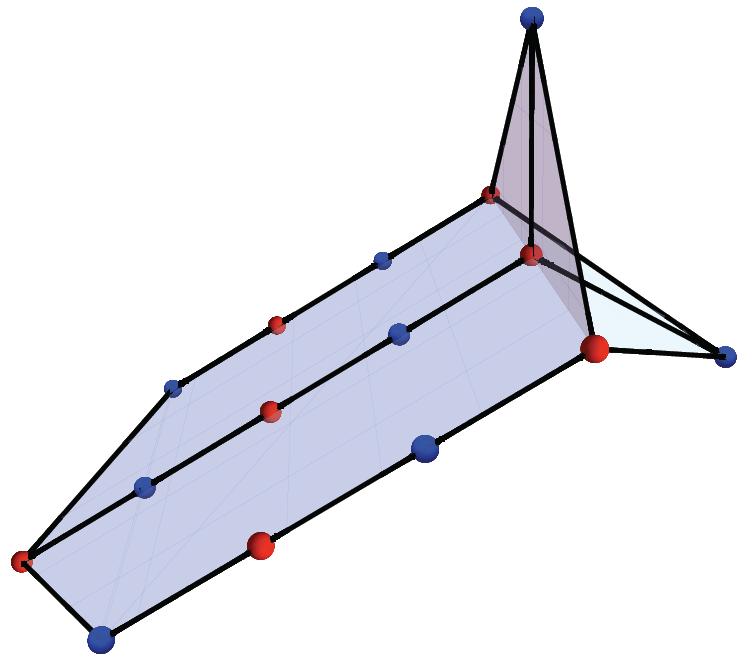}} & $\ $\\
& & $\ $
\end{tabular}}
\capt{5.5in}{tab(7,37)Extended}{The matrix and diagram for the extended representation of 
$\mathscr{X}^{7,37}$ for which all 7 K\"ahler forms are represented by ambient spaces. After the $\IZ_3$ identification, only $3$ of the $(1,1)$--forms remain independent.} 
\end{center}
\vspace{10pt}
\end{table}

I have checked that the polynomials are transverse. Thus we obtain a smooth Calabi-Yau manifold with fundamental group $\IZ_3$ and $\hodgenos = (3,13)$.


\subsection{The manifold $\mathscr{X}^{11,29}$ with quotient
$\mathscr{X}^{5, 11} = \mathscr{X}^{11,29}/ \IZ_3$} \label{sec:1.7}
\vskip -8pt
The configuration matrix for this manifold is 
$$
\mathscr{X}^{11,29}~=~~
\cicy{\IP^1 \\ \IP^1\\  \vrule height10pt width0pt depth8pt  \IP^1\\ \IP^3\\ \IP^2}
{ \one& \one&~ 0 & 0 & 0\\
 \one& \one& ~ 0 & 0 & 0\\
\vrule height10pt width0pt depth8pt  \one& \one&~ 0 & 0 & 0\\
0 &\one& ~ \one & \one & \one\\
0& 0&~  \one & \one & \one}_{-36}^{11,29}
\hskip0.35in
\lower0.34in\hbox{\includegraphics[width=2.5in]{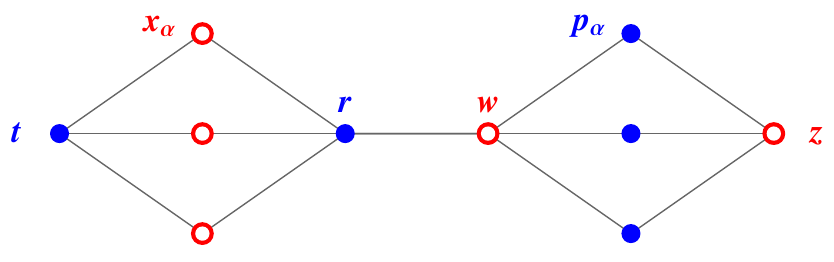}}
$$

Let $x_{\a,j}$ denote the homogeneous coordinates of the three $\IP^1$ spaces, $\left( w_0, w_a\right)$ those of the $\IP^3$ and $z_a$ the coordinates of  the $\IP^2$. As before, $\a$ and $a$ take values in $\IZ_3$ and $j$ in $\IZ_2$.  Also, denote by $t$ and $r$ the first two polynomials and by $p^\a$ the last three. 

We can write the most general form of the defining polynomials as
\begin{equation*}
\begin{split}
&t = \sum_{i,j,k} T_{ijk}\, x_{0,i}\, x_{1,j}\, x_{2,k} \ \ \ \ \  p^\a = \sum_{\a, a,b} P^\a_{ab} z_{\a + a} w_{\a + b} + w_0 \sum_{\a, a} \widetilde{P}^\a_a z_{\a+a} \\[5pt]
& r = \sum_{a, i, j, k} R_{a i j k} \, x_{0,i}\, x_{1,j}\, x_{2,k}\, w_a + w_0\sum_{ i, j, k} \widetilde{R}_{i j k} \, x_{0,i}\, x_{1,j}\, x_{2,k}\\
\end{split}
\end{equation*}
There are 76 terms in the polynomials, out of which 32 can be eliminated by redefinitions of coordinates, 1 by rescaling $t$, 5 by rescaling $r$ and adding to it polynomials of the form $C_q\, w_q\, t$ ($q\in \IZ_4$) and 9 by redefining the polynomials $p^{\a}$. This leaves 29 free parameters, which agree with the number of complex structure parameters. Note, however, that in this case we cannot rely on our previous method for computing $h^{2,1}$ for the quotient manifold, since the diagram associated with ${\mathscr X}^{11,29}$ does not fulfil the sufficient condition discussed in the introduction. Assuming, however, that the method is still valid, we can compute a value for $h^{2,1}\left( {\mathscr X}^{11,29} /\IZ_3\right)$ and compare it against the value obtained by expressing $h^{2,1}$ in terms of the Euler characteristic and $h^{1,1}$. 

The $\IZ_3$ action is generated by 
\vspace{-20pt}
\begin{center}
\begin{minipage}{0.8\textwidth}
\begin{align*}
g:~x_{\alpha, j} & ~\to~  x_{\alpha+1, j}~,& z_{\alpha} &~\to~ z_{\alpha+1} ~, &\left( w_0, w_{a}\right) & ~\to~ \left( w_0, w_{a+1}\right)\\[5pt]
 p^{\alpha}&~\to~  p^{\alpha+1}~,& r &~\to~ r~, & t & ~\to~ t ~ . 
\end{align*}
\end{minipage}
\end{center}
\vspace{-5pt}

The polynomials are covariant under this action if $T_{ijk}$ and $\widetilde{R}_{ijk}$ are cyclic in their indices, $R_{aijk} = R_{a+1, kji} = R_{a+2, jik}$ and $P^\a_{ab}, \widetilde{P}^\a_a$ do not depend on $\a$. This leaves a total of 28 terms in the polynomials, out of which 10 can be eliminated by coordinate redefinitions and 7 by redefinitions of the $\IZ_3$-covariant polynomials. The parameter count indicates that there are $28-17 =11$ complex structure parameters, a result which is consistent with the value $h^{1,1}\left( {\mathscr X}^{11,29} /\IZ_3 \right) = 5$, obtained from the extended representation in Table \ref{tab(11,29)Extended}. 

Fixed points of $g$, in the embedding space, are of the form $x_{\a,j} = x_j$, and $z_\a = \xi^\a$, $w_a = \zeta^\a$, where $\xi^\a$ and $\zeta^\a$ are cube roots of unity. The last three polynomials also determine the value of $w_0$, leaving two equations $t\left(x_j\right) = r\left(x_j\right) = 0$ which have no solutions in $\IP^1$, in general. Therefore, the action is fixed point free, on the manifold.

\begin{table}[H]
\begin{center}
\boxed{\begin{tabular}{cccc}
& & & \\
$\ $ & \hskip .0in \footnotesize $\cicy{\IP^1 \\ \IP^1\\ \vrule height10pt width0pt depth8pt   \IP^1\\ \IP^1\\ \IP^1\\ \vrule height10pt width0pt depth8pt  \IP^1\\ \IP^1\\ \IP^1\\ \vrule height10pt width0pt depth8pt  \IP^1\\  \IP^2\\ \IP^3}
{ \one& 0& 0&~ \one&0& 0&~ 0 & 0 & 0&~0& 0  \\
 0& \one& 0&~ 0& \one&  0&~ 0 & 0 & 0&~0 & 0\\
\vrule height10pt width0pt depth8pt  0& 0& \one&~ 0 & 0& \one&~ 0 & 0 & 0 &~0 & 0\\
  \one& 0&0& ~ 0& 0& 0&~ \one & 0 & 0&~0 & 0\\
  0& \one& 0&~ 0& 0& 0&~ 0 & \one & 0&~0 & 0\\
\vrule height10pt width0pt depth8pt    0& 0& \one &~0& 0& 0&~ 0 & 0 & \one &~0 & 0\\
0& 0& 0&~0& 0& 0&~0& 0& 0&~ \one & \one\\
0& 0& 0&~0& 0& 0&~0& 0& 0&~\one & \one\\
\vrule height10pt width0pt depth8pt  0& 0& 0&~0& 0& 0&~0& 0& 0&~ \one & \one\\
0& 0& 0&~\one&\one& \one&~0&0& 0 &~0 & 0\\
 0& 0& 0&~ 0& 0 & 0&~\one&\one& \one& ~\one& 0}_{-36}^{11,29}
$ &  \hskip-0.3in \lower21pt \hbox{\includegraphics[width=3.3in]{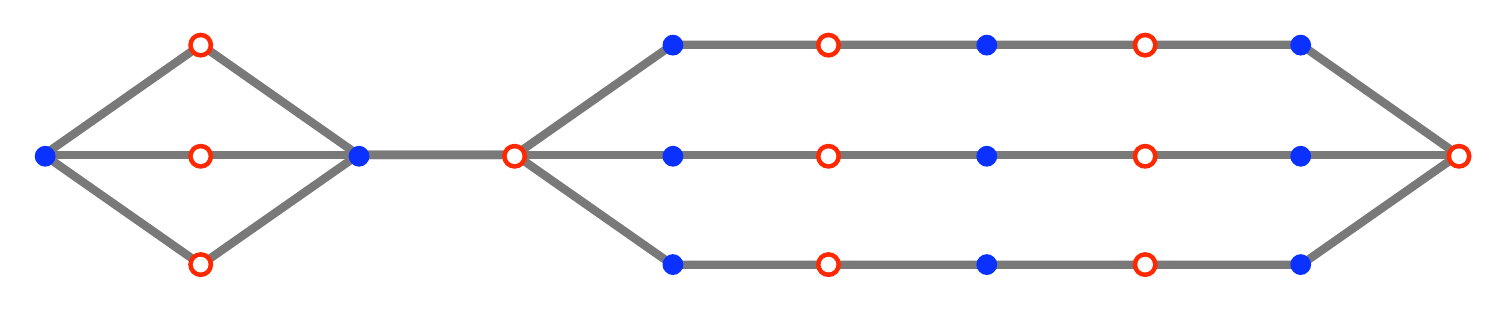}}\hspace{-0.2in} &\\
& & &  
\end{tabular}}
\capt{5.5in}{tab(11,29)Extended}{The matrix and diagram for the extended representation of $\mathscr{X}^{11,29}$ for which all 11 K\"ahler forms are represented by ambient spaces. After the $\IZ_3$ identification, only $5$ of the $(1,1)$--forms remain independent.} 
\end{center}
\vspace{-17pt}
\end{table}

There is another manifold for which the list \cite{Braun:2010vc} indicates a $\IZ_3$-free symmetry, namely the manifold, or rather the deformation class, given by the following configuration matrix and diagram: 
$$
\mathscr{X}^{11,29}~=~~
\cicy{\IP^2 \\   \IP^2\\ \IP^3\\ \IP^2}
{ \one& \one & \one&~ 0 & 0 & 0\\
 \one& \one& \one &  ~ 0 & 0 & 0\\
0& 0 &\one& ~ \one & \one & \one\\
0& 0& 0&~  \one & \one & \one}_{-36}^{11,29}
\hskip0.25in
\lower0.4in\hbox{\includegraphics[width=2.5in]{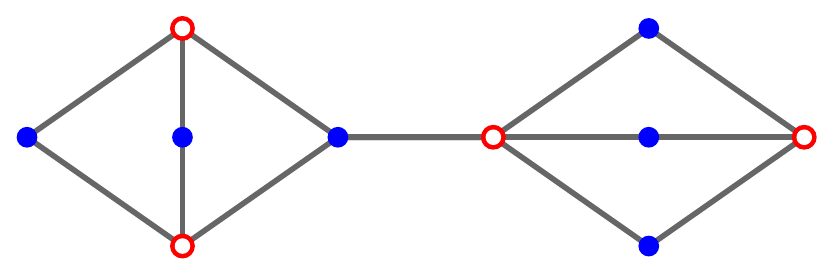}}
$$

Note, however, that this is only an alternative representation of the previous manifold, since
$$
\cicy{\IP^2 \\   \IP^2}
{ \one& \one\\
 \one& \one} \,\cong~
\cicy{\IP^1 \\   \IP^1\\ \IP^1}
{ \one\\ \one \\ \one} 
$$
are equivalent alternative representations of the del Pezzo surface $\mathrm{dP}_6$.


\newpage
\section{The Web of $\IZ_3$ Quotients}\label{secWeb}
\vskip -8pt
CICY threefolds form a web, connected by conifold transitions. In \cite{Candelas:2008wb}, it was suggested that a similar statement might hold for the webs of discrete quotients of CICY threefolds. The purpose of this section is to clarify this statement. 
Since conifold transitions do not affect the fundamental group, there will be several webs corresponding to different fundamental groups.\footnote{One can migrate between such webs via hyperconifold transitions \cite{Davies:2009ub, Davies:2011is, Davies:2011fr}} Here, we are mainly concerned with the web of smooth $\IZ_3$ quotients of CICY threefolds, having fundamental group $\IZ_3$. 

\subsection{Conifold transitions between smooth quotients of Calabi-Yau threefolds}
\vskip -8pt
Let $\Xcheck \to \Xsharp \to \mathscr{X}$ be a conifold transition between the 
\cy threefolds $\Xcheck$ and $\mathscr{X}$, where $\chi\left( {\Xcheck} \right) > \chi ( \mathscr{X} )$ and 
$\Xsharp$ is a conifold. Assume also that $\Xcheck$ and $\mathscr{X}$ both admit a free action of $G$. 

The argument\footnote{This argument is partly due to Rhys Davies.} presented below shows that, if the two actions of $G$ agree on $\Xsharp$, then the conifold transition commutes with taking quotients:
\beqnn
\begin{CD}
 \Xcheck   @>\wp >> \Xsharp @>\mathrm{deformation}>> \mathscr{X}\\
@V \mathrm{~} VV @VV \mathrm{~}V@VV \mathrm{~}V\\
\Xcheck / G @> \bar \wp  >> \Xsharp / G @>\mathrm{deformation}>> \mathscr{X} / G
\end{CD}
\eeqnn
where the map $\wp$ projects an entire $\IP^1\subset {\Xcheck}$ to each node of $\Xsharp$ and the deformation replaces each node of the conifold with a copy of $S^3$.  

The free $G$ action on $\Xcheck$ descends to a free action of $G$ on $\Xsharp$. Indeed, the conifold inherits a $G$-action from ${\Xcheck}$ via the projection $\wp: {\Xcheck} \to \Xsharp$. Moreover, the action of $G$ on the conifold is free, as we can see by the following argument. The only points which could be fixed by the group action are the conifold singularities. Assume this was the case for a node $p \in \Xsharp$. Then there exists an element 
$g\in G$ such that $g(p) = p$. But the $G$-action on $\Xsharp$ descends from the free $G$-action on $\Xcheck$, where $p$ is replaced by an $S^2$, and hence this implies the existence of a holomorphic map $g:S^2 \to S^2$, which being free, can have no fixed points, but there is no such holomorphic map. 
Thus, the action permutes the nodes of the conifold. In particular, the number of nodes of the conifold quotient $\Xsharp/ G$ reduces by a factor equal to the order of $G$. 

For all the quotients of CICY threefolds discussed above, in \cite{Candelas:2008wb} and \cite{Candelas:2007ac}, the group actions are given explicitly in terms of homogeneous coordinates of the embedding spaces. This makes it very easy to decide whether, in a conifold transition, the actions on the resolved spaces descend to the same action on the conifold. 

\vspace{-4pt}
\subsection{The web of $\IZ_3$ quotients	}
\vskip -8pt
The manifolds discussed in the previous section complete the web of smooth $\IZ_3$-free quotients of CICY threefolds in a nice way. Five of the new quotients can be organized in the double sequence that we have already seen in the introduction, but which we repeat here for completeness. The arrows of the table stand for conifold transitions, and the appended numbers are the respective Hodge numbers: 
\begin{table}[ht]
\vspace{-2pt}
\centering
\capt{6.0in}{doublesequenceBody}{The first double sequence of $\IZ_3$-free CICY quotients. In red, the new quotients.}
\vskip8pt
\framebox[6.0in]{\centering
\parbox{5.2in}{\vspace{4pt}
For the covering spaces:   
$\Delta_{\raisebox{-3pt}{$\scriptstyle{}\hskip-2pt\color{blue}\rightarrow $}} \hodgenos =\left(3, -3\right)
\,;~\Delta_{\color{blue}\downarrow}  \hodgenos = \left( 1, -8 \right) $ \\[4pt]
For the quotient spaces: 
$\Delta_{\raisebox{-3pt}{$\scriptstyle{}\hskip-2pt\color{blue}\rightarrow $}} \hodgenos =\left(1, -1\right)
\,;~\Delta_{\color{blue}\downarrow}  \hodgenos = \left( 1, -2 \right) $
\vspace{4pt}} }
\vskip8pt
\framebox[6.0in]{
\begin{tabular}{c}
\begin{tikzpicture}[scale=1.2]
\clip (-2.5, -.4) rectangle (9.55,5.7);
\def\nodeshadowed[#1]#2;{\node[scale=1.1,above,#1]{#2};}

\nodeshadowed [at={(-1,0 )},yslant=0.0]
{ {\small \textcolor{black} {$\mathbf{\left( \mathscr{X}^{6,24}/ \IZ_3 \right) ^{4,10}}$}} };
\nodeshadowed [at={(2,0 )},yslant=0.0]
{ {\small \textcolor{black} {$\mathbf{\color{red} \left( \mathscr{X}^{9,21}/ \IZ_3 \right) ^{5,9}}$ }} };
\nodeshadowed [at={(5,0 )},yslant=0.0]
{ {\small \textcolor{black} {$\mathbf{\color{red} \left( \mathscr{X}^{12,18}/ \IZ_3 \right) ^{6,8}}$ }} };
\nodeshadowed [at={(8,0 )},yslant=0.0]
{ {\small \textcolor{black} {$\mathbf{\left( \mathscr{X}^{15,15}/ \IZ_3 \right) ^{7,7}}$}} };
\nodeshadowed [at={(-1,1.5 )},yslant=0.0]
{ {\small \textcolor{black} {$\mathbf{\left( \mathscr{X}^{5,32}/ \IZ_3 \right) ^{3,12}}$ }} };
\nodeshadowed [at={(2,1.5 )},yslant=0.0]
{ {\small \textcolor{black} {$\mathbf{\color{red}  \left( \mathscr{X}^{8,29}/ \IZ_3 \right) ^{4,11}}$}} };
\nodeshadowed [at={(5,1.5 )},yslant=0.0]
{ {\small \textcolor{black} {$\mathbf{\color{red} \left( \mathscr{X}^{11,26}/ \IZ_3 \right) ^{5,10}}$}} };
\nodeshadowed [at={(-1,3 )},yslant=0.0]
{ {\small \textcolor{black} {$\mathbf{\left( \mathscr{X}^{4,40}/ \IZ_3 \right) ^{2,14}}$}} };
\nodeshadowed [at={(2,3 )},yslant=0.0]
{ {\small \textcolor{black} {$\mathbf{\color{red} \left(  \mathscr{X}^{7,37} / \IZ_3 \right) ^{3,13}}$}} };
\nodeshadowed [at={(-1,4.5 )},yslant=0.0]
{ {\small \textcolor{black} {$\mathbf{\left( \mathscr{X}^{3,48}/ \IZ_3 \right) ^{1,16}}$}} };

\draw[very thick,blue,->] (-1, 4.5) -- (-1,3.7);
\draw[very thick,blue,->] (-1, 3.) -- (-1,2.2);
\draw[very thick,blue,->] (-1, 1.5) -- (-1,.7);
\draw[very thick,blue,->] (2, 3.) -- (2,2.2);
\draw[very thick,blue,->] (2, 1.5) -- (2,.7);
\draw[very thick,blue,->] (5, 1.5) -- (5,.7);

\draw[very thick,blue,->] (-.1, 3.3) -- (.8, 3.3);
\draw[very thick,blue,->] (-.1, 1.8) -- (.8, 1.8);
\draw[very thick,blue,->] (-.1, .3) -- (.8, .3);
\draw[very thick,blue,->] (2.9, 1.8) -- (3.74, 1.8);
\draw[very thick,blue,->] (2.9, .3) -- (3.74, .3);
\draw[very thick,blue,->] (5.9, .3) -- (6.8, .3);

\end{tikzpicture}
\end{tabular}}
\vspace{-12pt}
\end{table}

Note that each horizontal arrow corresponds to the same change in the Hodge numbers, as does each vertical arrow. This is also the case for the Hodge numbers of the quotient manifolds in the following sequence, to which the sixth new manifold belongs: 

\begin{center}
\vspace{-16pt}
\begin{table}[H]
\centering
\capt{6.0in}{OneLineSequence}{Sequence of $\IZ_3$-free CICY quotients. In red, one of the new quotients.}
\vskip8pt
\framebox[6.0in]{\centering
\parbox{5.2in}{\centering\vspace{2pt}
For the quotient spaces: 
$\Delta_{\raisebox{-3pt}{$\scriptstyle{}\hskip-2pt\color{blue}\rightarrow $}} \hodgenos =\left(1, -2\right)$
\vspace{0pt}} }
\vskip8pt
\framebox[6.0in]{
\begin{tabular}{c}
\begin{tikzpicture}[scale=1.2]
\clip (-2.5, -.2) rectangle (9.55,1.1);
\def\nodeshadowed[#1]#2;{\node[scale=1.1,above,#1]{#2};}

\nodeshadowed [at={(-1,0 )},yslant=0.0]
{ {\small \textcolor{black} {$\mathbf{\left( \mathscr{X}^{8,35}/\IZ_3 \right)^{4,13} }$}} };
\nodeshadowed [at={(2,0 )},yslant=0.0]
{ {\small \textcolor{black} {$\mathbf{\color{red} \left( \mathscr{X}^{11,29}/\IZ_3 \right)^{ 5,11} }$}} };
\nodeshadowed [at={(5,0 )},yslant=0.0]
{ {\small \textcolor{black} {$\mathbf{\left( \mathscr{X}^{14,23}/\IZ_3 \right)^{ 6,9}}$ }} };
\nodeshadowed [at={(8,0 )},yslant=0.0]
{ {\small \textcolor{black} {$\mathbf{\left( \mathscr{X}^{19,19} /\IZ_3 \right)^{7,7} }$}} };

\draw[very thick,blue,->] (-.1, .3) -- (.74, .3);
\draw[very thick,blue,->] (2.9, .3) -- (3.74, .3);
\draw[very thick,blue,->] (5.9, .3) -- (6.77, .3);

\end{tikzpicture}
\end{tabular}}
\end{table}
\end{center}

There is yet a third sub-web of smooth $\IZ_3$-quotients which can be organized into a double sequence: 

\begin{table}[ht]
\vspace{-10pt}
\centering
\capt{6.0in}{doublesequenceAppendix}{The second double sequence of $\IZ_3$-free CICY quotients.}
\vskip8pt
\framebox[6.0in]{\centering
\parbox{5.2in}{\vspace{4pt}
For the covering spaces:   
$\Delta_{\raisebox{-3pt}{$\scriptstyle{}\hskip-2pt\color{blue}\rightarrow $}} \hodgenos =\left(3, -6\right)
\,;~\Delta_{\color{blue}\downarrow}  \hodgenos = \left( 1, -7 \right) $ \\[4pt]
For the quotient spaces: 
$\Delta_{\raisebox{-3pt}{$\scriptstyle{}\hskip-2pt\color{blue}\rightarrow $}} \hodgenos =\left(1, -2\right)
\,;~\Delta_{\color{blue}\downarrow}  \hodgenos = \left( 1, -5 \right) $
\vspace{4pt}} }
\vskip8pt
\framebox[6.0in]{
\begin{tabular}{c}
\begin{tikzpicture}[scale=1.2]
\clip (-1.3, 1.5) rectangle (8.35,5.4);
\def\nodeshadowed[#1]#2;{\node[scale=1.1,above,#1]{#2};}

\nodeshadowed [at={(.5,1.5 )},yslant=0.0]
{ {\small \textcolor{black} {$\mathbf{\left( \mathscr{X}^{3,39} / \IZ_3 \right) ^{ 3,15}} $}} };
\nodeshadowed [at={(3.5,1.5 )},yslant=0.0]
{ {\small \textcolor{black} {$\mathbf{\left( \mathscr{X}^{6,33} / \IZ_3 \right) ^{4,13}} $}} };
\nodeshadowed [at={(6.5,1.5 )},yslant=0.0]
{ {\small \textcolor{black} {$\mathbf{\left( \mathscr{X}^{9,27}/ \IZ_3 \right) ^{5,11}} $}} };
\nodeshadowed [at={(.5,3 )},yslant=0.0]
{ {\small \textcolor{black} {$\mathbf{\left( \mathscr{X}^{2,56} / \IZ_3 \right) ^{ 2,20}}$}} };
\nodeshadowed [at={(3.5,3 )},yslant=0.0]
{ {\small \textcolor{black} {$\mathbf{\left( \mathscr{X}^{5,50} / \IZ_3 \right) ^{ 3,18}} $}} };
\nodeshadowed [at={(.5,4.5 )},yslant=0.0]
{ {\small \textcolor{black} {$\mathbf{\left( \mathscr{X}^{1,73}/ \IZ_3 \right) ^{ 1,25}}$}} };

\draw[very thick,blue,->] (.5, 4.5) -- (.5,3.7);
\draw[very thick,blue,->] (.5, 3.) -- (.5,2.2);
\draw[very thick,blue,->] (3.5, 3.) -- (3.5,2.2);

\draw[very thick,blue,->] (1.4, 3.3) -- (2.3, 3.3);
\draw[very thick,blue,->] (1.4, 1.8) -- (2.3, 1.8);
\draw[very thick,blue,->] (4.4, 1.8) -- (5.24, 1.8);

\end{tikzpicture}
\end{tabular}}
\end{table}
\vspace{10pt}
The last two sequences above partially overlay each other in \fref{Z3Web}. However, they do not form a single structure as is seen by examining the Hodge numbers of the covering manifolds.

The sequences above are part of the connected web of conifold transitions between all the $\IZ_3$-free quotients of CICY threefolds, shown in Table \ref{TheWeb}. In order to save space, the chart contains only the Hodge numbers. For example, $\mathbf{\frac{3,48}{1,16}}~$ should be read as 
$\mathbf{\left( \mathscr{X}^{3,48}/ \IZ_3 \right) ^{1,16}}$.

\begin{center}
\begin{table}[H]
\begin{center}
\capt{6.0in}{TheWeb}{The web of $\IZ_3$-free CICY quotients.}
\vskip8pt
\framebox[6.0in]{
\begin{tabular}{c}
\begin{tikzpicture}[scale=.71]
\clip (0,-.3) rectangle (19.4,8.4);
\def\nodeshadowed[#1]#2;{\node[scale=.94,above,#1]{#2};}

\nodeshadowed [at={(.5,6 )},yslant=0.0]
{ {\small \textcolor{black} {$\mathbf{\dfrac{3,48}{1,16}}$}} };
\nodeshadowed [at={(2.5,6 )},yslant=0.0]
{ {\small \textcolor{black} {$\mathbf{\dfrac{4,40}{2,14}}$}} };
\nodeshadowed [at={(4.5,6 )},yslant=0.0]
{ {\small \textcolor{black} {$\mathbf{\dfrac{5,32}{3,12}}$}} };
\nodeshadowed [at={(6.5,6 )},yslant=0.0]
{ {\small \textcolor{black} {$\mathbf{\dfrac{6,24}{4,10}}$}} };
\nodeshadowed [at={(2.5,4 )},yslant=0.0]
{ {\small \textcolor{red} {$\mathbf{\dfrac{7,37}{3,13}}$}} };
\nodeshadowed [at={(4.5,4 )},yslant=0.0]
{ {\small \textcolor{red} {$\mathbf{\dfrac{8,29}{4,11}}$}} };
\nodeshadowed [at={(6.5,4 )},yslant=0.0]
{ {\small \textcolor{red} {$\mathbf{\dfrac{9,21}{5,9}}$}} };
\nodeshadowed [at={(4.5,2 )},yslant=0.0]
{ {\small \textcolor{red} {$\mathbf{\dfrac{11,26}{5,10}}$}} };
\nodeshadowed [at={(6.5,2 )},yslant=0.0]
{ {\small \textcolor{red} {$\mathbf{\dfrac{12,18}{6,8}}$}} };
\nodeshadowed [at={(6.5,0 )},yslant=0.0]
{ {\small \textcolor{black} {$\mathbf{\dfrac{15,15}{7,7}}$}} };

\nodeshadowed [at={(8.5,6 )},yslant=0.0]
{ {\small \textcolor{black} {$\mathbf{\dfrac{3,48}{3,18}}$}} };
\nodeshadowed [at={(8.5,4 )},yslant=0.0]
{ {\small \textcolor{black} {$\mathbf{\dfrac{6,33}{4,13}}$}} };
\nodeshadowed [at={(8.5,2 )},yslant=0.0]
{ {\small \textcolor{black} {$\mathbf{\dfrac{9,21}{5,9}}$}} };

\nodeshadowed [at={(10.5,6 )},yslant=0.0]
{ {\small \textcolor{black} {$\mathbf{\dfrac{8,35}{4,13}}$}} };
\nodeshadowed [at={(10.5,4 )},yslant=0.0]
{ {\small \textcolor{red} {$\mathbf{\dfrac{11,29}{5,11}}$}} };
\nodeshadowed [at={(10.5,2 )},yslant=0.0]
{ {\small \textcolor{black} {$\mathbf{\dfrac{14,23}{6,9}}$}} };
\nodeshadowed [at={(10.5,0 )},yslant=0.0]
{ {\small \textcolor{black} {$\mathbf{\dfrac{19,19}{7,7}}$}} };

\nodeshadowed [at={(12.5,6 )},yslant=0.0]
{ {\small \textcolor{black} {$\mathbf{\dfrac{2,83}{2,29}}$}} };
\nodeshadowed [at={(12.5,4 )},yslant=0.0]
{ {\small \textcolor{black} {$\mathbf{\dfrac{5,59}{3,21}}$}} };
\nodeshadowed [at={(12.5,2 )},yslant=0.0]
{ {\small \textcolor{black} {$\mathbf{\dfrac{8,44}{4,16}}$}} };

\nodeshadowed [at={(14.5,6 )},yslant=0.0]
{ {\small \textcolor{black} {$\mathbf{\dfrac{3,39}{3,15}}$}} };
\nodeshadowed [at={(16.5,6 )},yslant=0.0]
{ {\small \textcolor{black} {$\mathbf{\dfrac{2,56}{2,20}}$}} };
\nodeshadowed [at={(18.5,6 )},yslant=0.0]
{ {\small \textcolor{black} {$\mathbf{\dfrac{1,73}{1,25}}$}} };
\nodeshadowed [at={(14.5,4 )},yslant=0.0]
{ {\small \textcolor{black} {$\mathbf{\dfrac{6,33}{4,13}}$}} };
\nodeshadowed [at={(16.5,4 )},yslant=0.0]
{ {\small \textcolor{black} {$\mathbf{\dfrac{5,50}{3,18}}$}} };
\nodeshadowed [at={(14.5,2 )},yslant=0.0]
{ {\small \textcolor{black} {$\mathbf{\dfrac{9,27}{5,11}}$}} };

\draw[very thick,blue,->] (1.38,6.80) -- (1.82,6.80);
\draw[very thick,blue,->] (3.38,6.80) -- (3.82,6.80);
\draw[very thick,blue,->] (5.38,6.80) -- (5.82,6.80);
\draw[very thick,blue,->] (3.38,4.80) -- (3.82,4.80);
\draw[very thick,blue,->] (5.38,4.80) -- (5.82,4.80);
\draw[very thick,blue,->] (5.38,2.80) -- (5.82,2.80);
\draw[very thick,blue,->] (2.5,5.9) -- (2.5,5.46);
\draw[very thick,blue,->] (4.5,5.9) -- (4.5,5.46);
\draw[very thick,blue,->] (6.5,5.9) -- (6.5,5.46);
\draw[very thick,blue,->] (4.5,3.9) -- (4.5,3.46);
\draw[very thick,blue,->] (6.5,3.9) -- (6.5,3.46);
\draw[very thick,blue,->] (6.5,1.9) -- (6.5,1.46);

\draw[very thick,blue,->] (7.82,6.80) -- (7.38,6.80);
\draw[very thick,blue,->] (7.82,4.80) -- (7.38,4.80);
\draw[very thick,blue,->] (7.82,2.80) -- (7.38,2.80);
\draw[very thick,blue,->] (8.5,5.9) -- (8.5,5.46);
\draw[very thick,blue,->] (8.5,3.9) -- (8.5,3.46);

\draw[very thick,blue,->] (11.82,6.80) -- (11.38,6.80);
\draw[very thick,blue,->] (11.82,4.80) -- (11.38,4.80);
\draw[very thick,blue,->] (11.82,1.87) -- (11.38,1.44);
\draw[very thick,blue,->] (13.38,6.80) -- (13.82,6.80);
\draw[very thick,blue,->] (13.38,4.80) -- (13.82,4.80);
\draw[very thick,blue,->] (13.38,2.80) -- (13.82,2.80);
\draw[very thick,blue,->] (10.5,5.9) -- (10.5,5.46);
\draw[very thick,blue,->] (10.5,3.9) -- (10.5,3.46);
\draw[very thick,blue,->] (10.5,1.9) -- (10.5,1.46);

\draw[very thick,blue,->] (15.82,6.80) -- (15.38,6.80);
\draw[very thick,blue,->] (15.82,4.80) -- (15.38,4.80);
\draw[very thick,blue,->] (17.82,6.80) -- (17.38,6.80);
\draw[very thick,blue,->] (12.5,5.9) -- (12.5,5.46);
\draw[very thick,blue,->] (12.5,3.9) -- (12.5,3.46);
\draw[very thick,blue,->] (14.5,5.9) -- (14.5,5.46);
\draw[very thick,blue,->] (14.5,3.9) -- (14.5,3.46);
\draw[very thick,blue,->] (16.5,5.9) -- (16.5,5.46);

\draw[very thick, blue,->] (12.2,7.46) ..  controls +(up: .7cm)  and +(up: 1.2cm)  .. (8.8,7.46);
\draw[very thick, blue,->] (12.2,5.46) ..  controls +(up: .7cm)  and +(up: 1.2cm)  .. (8.8,5.46);
\draw[very thick, blue,->] (12.2,3.46) ..  controls +(up: .7cm)  and +(up: 1.2cm)  .. (8.8,3.46);

\end{tikzpicture}
\end{tabular}}
\end{center}
\end{table}
\end{center}
\newpage
%
%
\newpage
\section{The List of Smooth $\IZ_3$ Quotients of CICYs}
In this section, we take a second look at the three sequences that were given previously with the aim of presenting the changing structure of the CICYs that appear. Thus we start with the sequences that were previously given and then give further tables that present the diagrams for the CICYs of the sequence.

\vspace{20pt}
\def\str{\vrule height14pt depth8pt width0pt}
\setlength{\doublerulesep}{3pt}
\vspace{-10pt}
%
%
\begin{center}
\begin{table}[H]
\begin{center}
\capt{6.0in}{List1}{The first sub-web of $\IZ_3$ quotients generated by 
$\mathscr{X}^{1,16} = \mathscr{X}^{3,48}/ \IZ_3$.}
\addtocounter{table}{-1}
\vskip10pt
\framebox[6.2in]{
\begin{tabular}{c}
\begin{tikzpicture}[scale=1.2]
\clip (-2.5, -.4) rectangle (9.55,5.7);
\def\nodeshadowed[#1]#2;{\node[scale=1.1,above,#1]{#2};}

\nodeshadowed [at={(-1,0 )},yslant=0.0]
{ {\small \textcolor{black} {$\mathbf{\left( \mathscr{X}^{6,24}/ \IZ_3 \right) ^{4,10}}$}} };
\nodeshadowed [at={(2,0 )},yslant=0.0]
{ {\small \textcolor{black} {$\mathbf{\color{red} \left( \mathscr{X}^{9,21}/ \IZ_3 \right) ^{5,9}}$ }} };
\nodeshadowed [at={(5,0 )},yslant=0.0]
{ {\small \textcolor{black} {$\mathbf{\color{red} \left( \mathscr{X}^{12,18}/ \IZ_3 \right) ^{6,8}}$ }} };
\nodeshadowed [at={(8,0 )},yslant=0.0]
{ {\small \textcolor{black} {$\mathbf{\left( \mathscr{X}^{15,15}/ \IZ_3 \right) ^{7,7}}$}} };
\nodeshadowed [at={(-1,1.5 )},yslant=0.0]
{ {\small \textcolor{black} {$\mathbf{\left( \mathscr{X}^{5,32}/ \IZ_3 \right) ^{3,12}}$ }} };
\nodeshadowed [at={(2,1.5 )},yslant=0.0]
{ {\small \textcolor{black} {$\mathbf{\color{red}  \left( \mathscr{X}^{8,29}/ \IZ_3 \right) ^{4,11}}$}} };
\nodeshadowed [at={(5,1.5 )},yslant=0.0]
{ {\small \textcolor{black} {$\mathbf{\color{red} \left( \mathscr{X}^{11,26}/ \IZ_3 \right) ^{5,10}}$}} };
\nodeshadowed [at={(-1,3 )},yslant=0.0]
{ {\small \textcolor{black} {$\mathbf{\left( \mathscr{X}^{4,40}/ \IZ_3 \right) ^{2,14}}$}} };
\nodeshadowed [at={(2,3 )},yslant=0.0]
{ {\small \textcolor{black} {$\mathbf{\color{red} \left(  \mathscr{X}^{7,37} / \IZ_3 \right) ^{3,13}}$}} };
\nodeshadowed [at={(-1,4.5 )},yslant=0.0]
{ {\small \textcolor{black} {$\mathbf{\left( \mathscr{X}^{3,48}/ \IZ_3 \right) ^{1,16}}$}} };

\draw[very thick,blue,->] (-1, 4.5) -- (-1,3.7);
\draw[very thick,blue,->] (-1, 3.) -- (-1,2.2);
\draw[very thick,blue,->] (-1, 1.5) -- (-1,.7);
\draw[very thick,blue,->] (2, 3.) -- (2,2.2);
\draw[very thick,blue,->] (2, 1.5) -- (2,.7);
\draw[very thick,blue,->] (5, 1.5) -- (5,.7);

\draw[very thick,blue,->] (-.1, 3.3) -- (.8, 3.3);
\draw[very thick,blue,->] (-.1, 1.8) -- (.8, 1.8);
\draw[very thick,blue,->] (-.1, .3) -- (.8, .3);
\draw[very thick,blue,->] (2.9, 1.8) -- (3.74, 1.8);
\draw[very thick,blue,->] (2.9, .3) -- (3.74, .3);
\draw[very thick,blue,->] (5.9, .3) -- (6.8, .3);

\end{tikzpicture}
\end{tabular}}
\end{center}
\end{table}
\end{center}
\vspace{-10pt}
\begin{center}
\begin{longtable}{|c|c|c|}

\hline \multicolumn{1}{|c|}{\str\textbf{CICY Manifold}}&  \multicolumn{1}{|c|}{\textbf{$\IZ_3-$Quotient}} & \multicolumn{1}{|c|}{\textbf{CICY Diagram}} \\ \hline 
\endfirsthead


\hline\multicolumn{1}{|c|}{\str\textbf{CICY Manifold}} &
\multicolumn{1}{|c|}{\textbf{$\IZ_3-$Quotient}} &
\multicolumn{1}{|c|}{\textbf{CICY Diagram}} \\ \hline 
\endhead

\hline\hline \multicolumn{3}{|r|}{{\str Continued on next page}} \\ \hline
\endfoot

\hline\hline\multicolumn{3}{|c|}{\str}\\ \hline
\endlastfoot

\hline
\hline
\multicolumn{3}{|l|}{\str\textbf{First Row}}\\
\hline
\vrule height28pt  width0pt depth20pt  $\mathscr{X}^{3,48}$ & $\mathscr{X}^{1,16}$ & \lower16pt \hbox{\includegraphics[width=.85in]{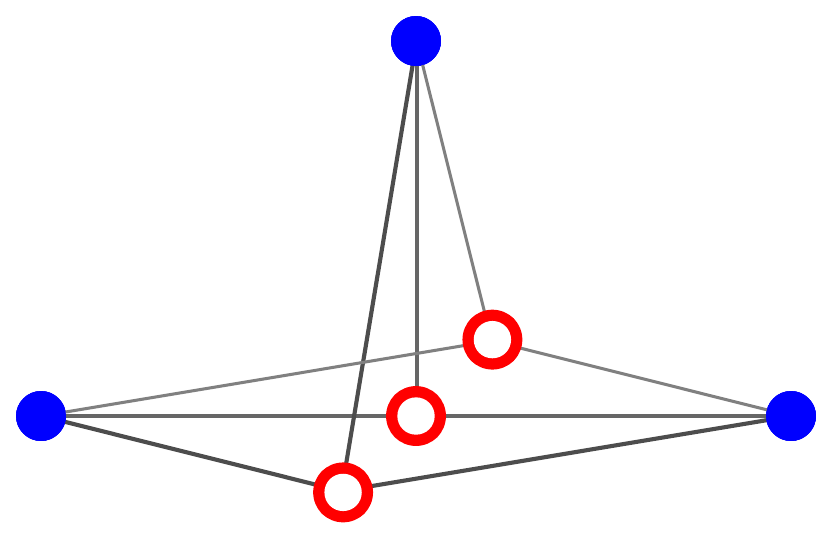}}  \\
\hline
\hline
\multicolumn{3}{|l|}{\str\textbf{Second Row}}\\
\hline
\vrule height30pt  width0pt depth20pt  $\mathscr{X}^{4,40}$ & $\mathscr{X}^{2,14}$ &\hskip .4in \lower16pt \hbox{\includegraphics[width=1.275in]{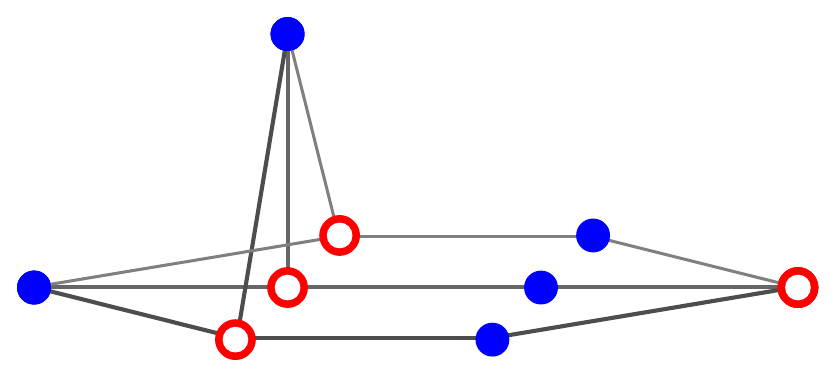}}  \\
\hline
\vrule height32pt  width0pt depth20pt  $\mathscr{X}^{7,37}$ & $\mathscr{X}^{3,13}$ &\hskip 1.2in \lower16pt \hbox{\includegraphics[width=2.125in]{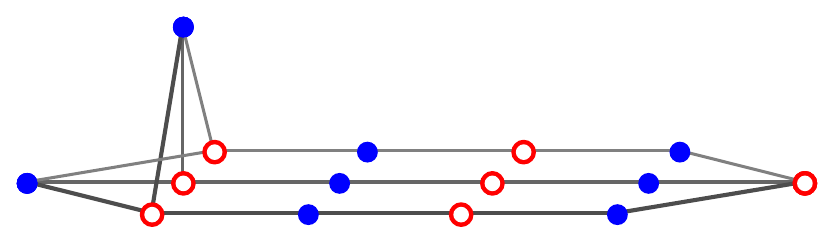}}  \\
\hline
\hline
\multicolumn{3}{|l|}{\str\textbf{Third Row}}\\
\hline
\vrule height30pt  width0pt depth20pt  $\mathscr{X}^{5,32}$ & $\mathscr{X}^{3,12}$ &\hskip 0in \lower16pt \hbox{\includegraphics[width=1.7in]{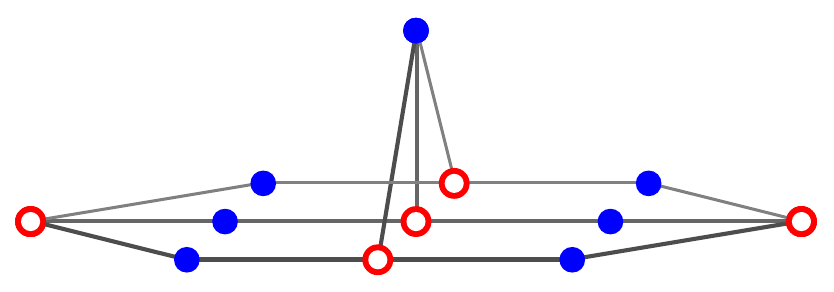}}  \\
\hline
\vrule height32pt  width0pt depth20pt  $\mathscr{X}^{8,29}$ & $\mathscr{X}^{4,11}$ &\hskip .8in \lower16pt \hbox{\includegraphics[width=2.55in]{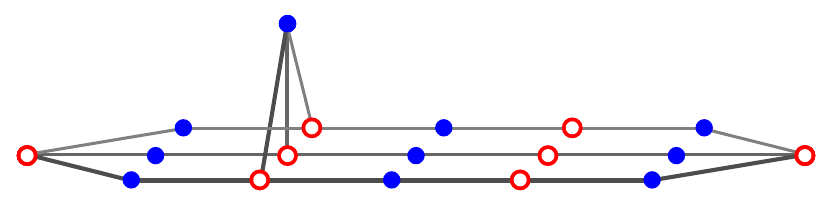}}  \\
\hline
\vrule height32pt  width0pt depth20pt  $\mathscr{X}^{11,26}$ & $\mathscr{X}^{5,10}$ &\hskip 0in \lower16pt \hbox{\includegraphics[width=3.4in]{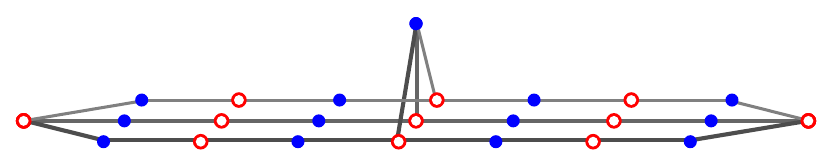}}  \\
\hline
\hline
\multicolumn{3}{|l|}{\str\textbf{Fourth Row}}\\
\hline
\vrule height61pt  width0pt depth20pt \lower -14pt \hbox{$\mathscr{X}^{6,24}$} & \lower -14pt \hbox{$\mathscr{X}^{4,10}$} &\hskip 0in \lower16pt \hbox{\includegraphics[width=1.7in]{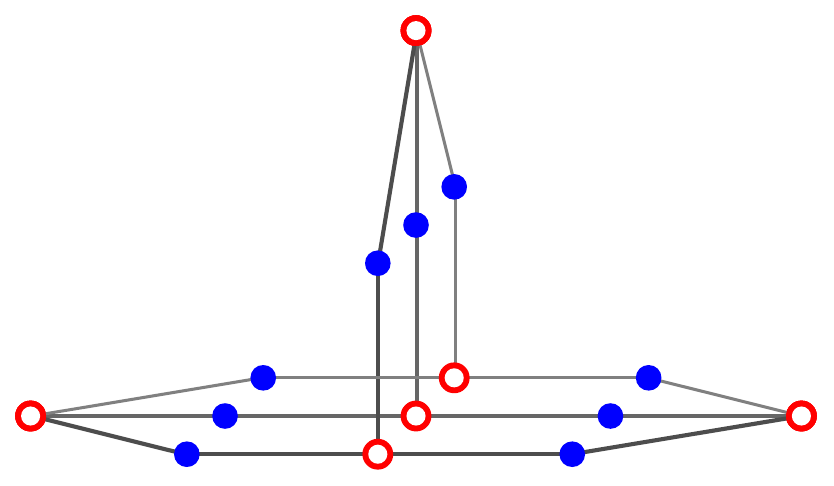}}  \\
\hline
\vrule height61pt  width0pt depth20pt  \lower -14pt \hbox{$\mathscr{X}^{9,21}$} & \lower -14pt \hbox{$\mathscr{X}^{5,9}$} &\hskip .8in \lower16pt \hbox{\includegraphics[width=2.55in]{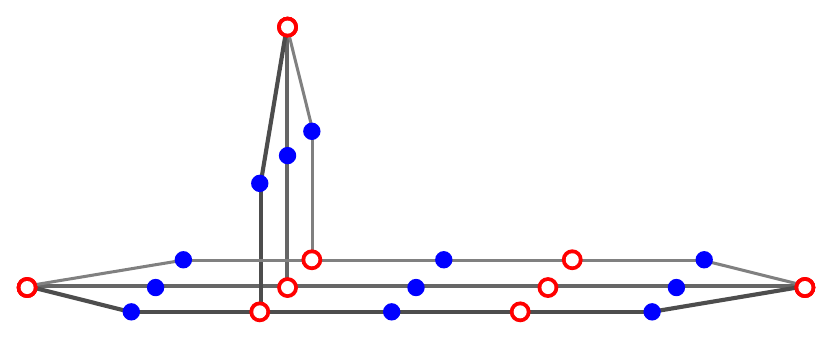}}  \\
\hline
\vrule height61pt  width0pt depth20pt  \lower -14pt  \hbox{$\mathscr{X}^{12,18}$} & \lower -14pt \hbox{$\mathscr{X}^{6,8}$} &\hskip 0in \lower16pt \hbox{\includegraphics[width=3.4in]{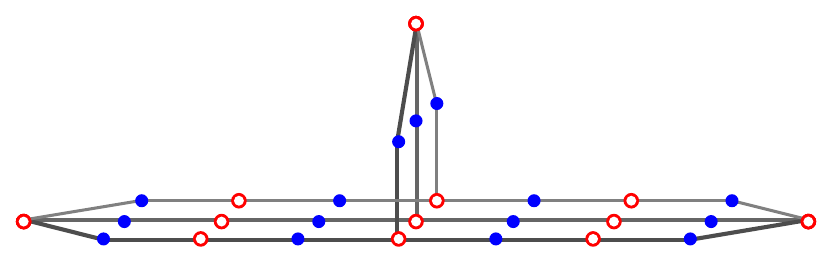}}  \\
\hline
\vrule height81pt  width0pt depth20pt  \lower -44pt \hbox{$\mathscr{X}^{15,15}$} & \lower -44pt \hbox{$\mathscr{X}^{7,7}$} &\hskip 0in \lower16pt \hbox{\includegraphics[width=3.4in]{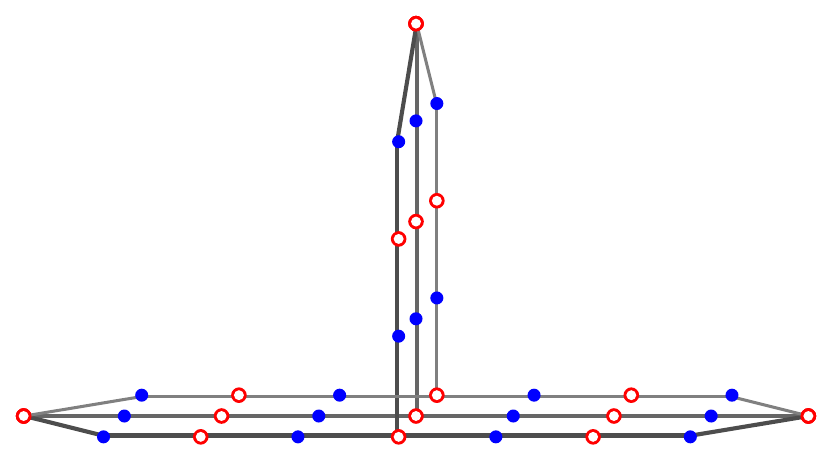}}  
\end{longtable}
\end{center}
%
%
\vspace{-60pt}
\begin{center}
\begin{table}[H]
\begin{center}
\capt{6.0in}{List3}{The second sub-web of $\IZ_3$ quotients generated by 
$\mathscr{X}^{1,25} = \mathscr{X}^{1,73}/ \IZ_3$.}
\addtocounter{table}{-1}
\vskip10pt
\framebox[6.2in]{
\begin{tabular}{c}
\begin{tikzpicture}[scale=1.2]
\clip (-1.3, 1.3) rectangle (8.35,5.5);
\def\nodeshadowed[#1]#2;{\node[scale=1.1,above,#1]{#2};}

\nodeshadowed [at={(.5,1.5 )},yslant=0.0]
{ {\small \textcolor{black} {$\mathbf{\left( \mathscr{X}^{3,39} / \IZ_3 \right) ^{ 3,15}} $}} };
\nodeshadowed [at={(3.5,1.5 )},yslant=0.0]
{ {\small \textcolor{black} {$\mathbf{\left( \mathscr{X}^{6,33} / \IZ_3 \right) ^{4,13}} $}} };
\nodeshadowed [at={(6.5,1.5 )},yslant=0.0]
{ {\small \textcolor{black} {$\mathbf{\left( \mathscr{X}^{9,27}/ \IZ_3 \right) ^{5,11}} $}} };
\nodeshadowed [at={(.5,3 )},yslant=0.0]
{ {\small \textcolor{black} {$\mathbf{\left( \mathscr{X}^{2,56} / \IZ_3 \right) ^{ 2,20}}$}} };
\nodeshadowed [at={(3.5,3 )},yslant=0.0]
{ {\small \textcolor{black} {$\mathbf{\left( \mathscr{X}^{5,50} / \IZ_3 \right) ^{ 3,18}} $}} };
\nodeshadowed [at={(.5,4.5 )},yslant=0.0]
{ {\small \textcolor{black} {$\mathbf{\left( \mathscr{X}^{1,73}/ \IZ_3 \right) ^{ 1,25}}$}} };

\draw[very thick,blue,->] (.5, 4.5) -- (.5,3.7);
\draw[very thick,blue,->] (.5, 3.) -- (.5,2.2);
\draw[very thick,blue,->] (3.5, 3.) -- (3.5,2.2);

\draw[very thick,blue,->] (1.4, 3.3) -- (2.3, 3.3);
\draw[very thick,blue,->] (1.4, 1.8) -- (2.3, 1.8);
\draw[very thick,blue,->] (4.4, 1.8) -- (5.24, 1.8);

\end{tikzpicture}
\end{tabular}}
\end{center}
\end{table}
\end{center}
\vspace{-15pt}
\begin{center}
\begin{longtable}{|c|c|c|}

\hline \multicolumn{1}{|c|}{\str\textbf{CICY Manifold}}&  \multicolumn{1}{|c|}{\textbf{$\IZ_3-$Quotient}} & \multicolumn{1}{|c|}{\textbf{CICY Diagram}} \\ \hline 
\endfirsthead


\hline \multicolumn{1}{|c|}{\str\textbf{CICY Manifold}} &
\multicolumn{1}{|c|}{\textbf{$\IZ_3-$Quotient}} &
\multicolumn{1}{|c|}{\textbf{CICY Diagram}} \\ \hline 
\endhead

\hline\hline \multicolumn{3}{|r|}{{\str Continued on next page}} \\ \hline
\endfoot

\hline\hline\multicolumn{3}{|c|}{\str}\\ \hline
\endlastfoot

\hline
\hline
\multicolumn{3}{|l|}{\str\textbf{First Row}}\\
\hline
\vrule height28pt  width0pt depth20pt  $\mathscr{X}^{1,73}$ & $\mathscr{X}^{1,25}$ & \lower4pt \hbox{\includegraphics[width=.85in]{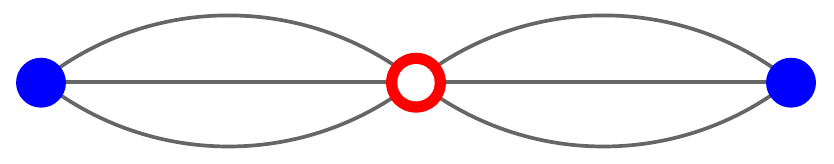}}  \\
\hline
\hline
\multicolumn{3}{|l|}{\str\textbf{Second Row}}\\
\hline
\vrule height32pt  width0pt depth20pt  $\mathscr{X}^{2,56}$ & $\mathscr{X}^{2,20}$ &\hskip .4in \lower10pt \hbox{\includegraphics[width=1.275in]{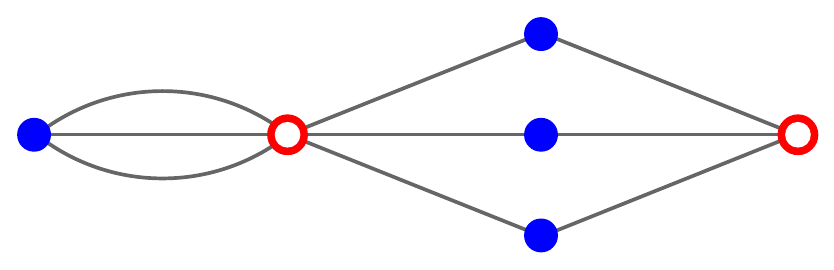}}  \\
\hline
\vrule height32pt  width0pt depth20pt  $\mathscr{X}^{5,50}$ & $\mathscr{X}^{3,18}$ &\hskip 1.20in \lower12pt \hbox{\includegraphics[width=2.125in]{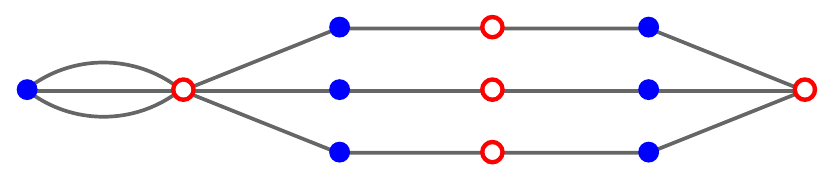}}  \\
\hline
\hline
\multicolumn{3}{|l|}{\str\textbf{Third Row}}\\
\hline
\vrule height30pt  width0pt depth20pt  $\mathscr{X}^{3,39}$ & $\mathscr{X}^{3,15}$ &\hskip .0in \lower12pt \hbox{\includegraphics[width=1.7in]{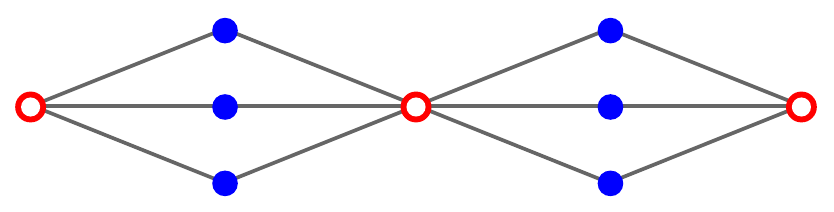}}  \\
\hline
\vrule height32pt  width0pt depth20pt  $\mathscr{X}^{6,33}$ & $\mathscr{X}^{4,13}$ &\hskip .8in \lower12pt \hbox{\includegraphics[width=2.55in]{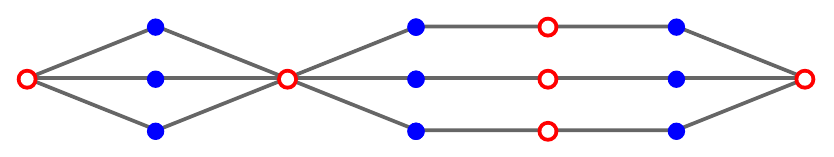}}  \\
\hline
\vrule height32pt  width0pt depth20pt  $\mathscr{X}^{9,27}$ & $\mathscr{X}^{5,11}$ &\hskip 0in \lower12pt \hbox{\includegraphics[width=3.4in]{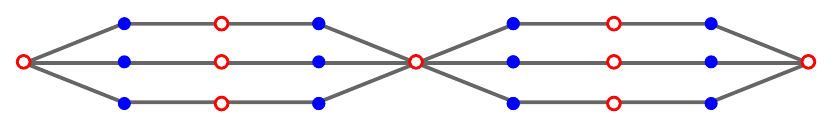}}  \\
\end{longtable}
\end{center}

%
%
\vspace{-60pt}
\begin{center}
\begin{table}[H]
\begin{center}
\capt{6.0in}{List2}{The third sub-web of $\IZ_3$ quotients generated by 
$\mathscr{X}^{2,29} = \mathscr{X}^{2,83}/ \IZ_3$.}
\vskip10pt
\framebox[6.2in]{
\begin{tabular}{c}
\begin{tikzpicture}[scale=1.2]
\clip (-2.5, -.2) rectangle (9.55,4.3);
\def\nodeshadowed[#1]#2;{\node[scale=1.1,above,#1]{#2};}

\nodeshadowed [at={(-1,3 )},yslant=0.0]
{ {\small \textcolor{black} {$\mathbf{\left( \mathscr{X}^{3,48}/\IZ_3 \right)^{3,18}}$}} };
\nodeshadowed [at={(2,3)},yslant=0.0]
{ {\small \textcolor{black} {$\mathbf{\left( \mathscr{X}^{6,33} /\IZ_3 \right)^{4,13}}$}} };
\nodeshadowed [at={(5,3)},yslant=0.0]
{ {\small \textcolor{black} {$\mathbf{\left( \mathscr{X}^{9,21} /\IZ_3 \right)^{5,9}}$}} };

\nodeshadowed [at={(-1,1.5 )},yslant=0.0]
{ {\small \textcolor{black} {$\mathbf{\left( \mathscr{X}^{2,83} /\IZ_3 \right)^{2,29}}$}} };
\nodeshadowed [at={(2,1.5 )},yslant=0.0]
{ {\small \textcolor{black} { $\mathbf{\left( \mathscr{X}^{5,59} /\IZ_3 \right)^{3,21}}$}} };
\nodeshadowed [at={(5,1.5 )},yslant=0.0]
{ {\small \textcolor{black} {$\mathbf{\left( \mathscr{X}^{8,44} /\IZ_3 \right)^{4,16}}$}} };

\nodeshadowed [at={(-1,0 )},yslant=0.0]
{ {\small \textcolor{black} {$\mathbf{\left( \mathscr{X}^{8,35}/\IZ_3 \right)^{4,13} }$}} };
\nodeshadowed [at={(2,0 )},yslant=0.0]
{ {\small \textcolor{black} {$\mathbf{\color{red} \left( \mathscr{X}^{11,29}/\IZ_3 \right)^{ 5,11} }$}} };
\nodeshadowed [at={(5,0 )},yslant=0.0]
{ {\small \textcolor{black} {$\mathbf{\left( \mathscr{X}^{14,23}/\IZ_3 \right)^{ 6,9}}$ }} };
\nodeshadowed [at={(8,0 )},yslant=0.0]
{ {\small \textcolor{black} {$\mathbf{\left( \mathscr{X}^{19,19} /\IZ_3 \right)^{7,7} }$}} };

\draw[very thick,blue,->] (-.1, 3.3) -- (.74, 3.3);
\draw[very thick,blue,->] (2.9, 3.3) -- (3.74, 3.3);
\draw[very thick,blue,->] (-.1, 1.8) -- (.74, 1.8);
\draw[very thick,blue,->] (2.9, 1.8) -- (3.74, 1.8);
\draw[very thick,blue,->] (-.1, .3) -- (.74, .3);
\draw[very thick,blue,->] (2.9, .3) -- (3.74, .3);
\draw[very thick,blue,->] (5.9, .3) -- (6.77, .3);

\draw[very thick,blue,->] (-1, 2.25) -- (-1, 2.95);
\draw[very thick,blue,->] (2, 2.25) -- (2, 2.95);
\draw[very thick,blue,->] (5, 2.25) -- (5, 2.95);

\draw[very thick,blue,->] (-1, 1.45) -- (-1, .75);
\draw[very thick,blue,->] (2, 1.45) -- (2, .75);
\draw[very thick,blue,->] (5.9, 1.45) -- (6.8, .75);
\end{tikzpicture}
\end{tabular}}
\end{center}
\end{table}
\end{center}
\begin{center}
\vspace{40pt}
\begin{longtable}{|c|c|c|}

\hline \multicolumn{1}{|c|}{\str\textbf{CICY Manifold}}&  \multicolumn{1}{|c|}{\textbf{$\IZ_3-$Quotient}} & \multicolumn{1}{|c|}{\textbf{CICY Diagram}} \\ \hline 
\endfirsthead
%
%
\hline \multicolumn{1}{|c|}{\str\textbf{CICY Manifold}} &
\multicolumn{1}{|c|}{\textbf{$\IZ_3-$Quotient}} &
\multicolumn{1}{|c|}{\textbf{CICY Diagram}} \\ \hline 
\endhead
\hline\hline \multicolumn{3}{|r|}{{\str Continued on next page}} \\ \hline
\endfoot
\hline\hline\multicolumn{3}{|c|}{\str}\\ \hline
\endlastfoot
\hline\hline
\multicolumn{3}{|l|}{\str\textbf{First Row}}\\
\hline
\vrule height28pt  width0pt depth20pt  $\mathscr{X}^{3,48}$ & $\mathscr{X}^{3,18}$ & \lower16pt \hbox{\includegraphics[width=.85in]{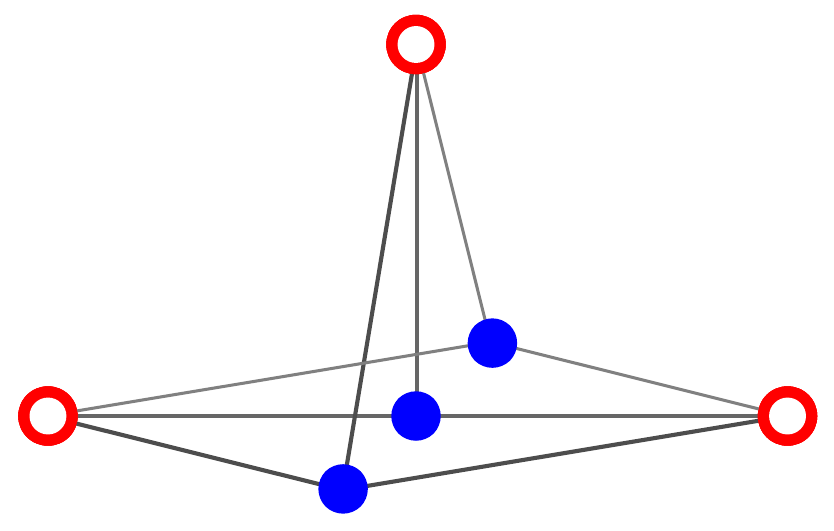}}  \\
\hline
\vrule height32pt  width0pt depth20pt  $\mathscr{X}^{6,33}$ & $\mathscr{X}^{4,13}$ &\hskip .8in \lower16pt \hbox{\includegraphics[width=1.7in]{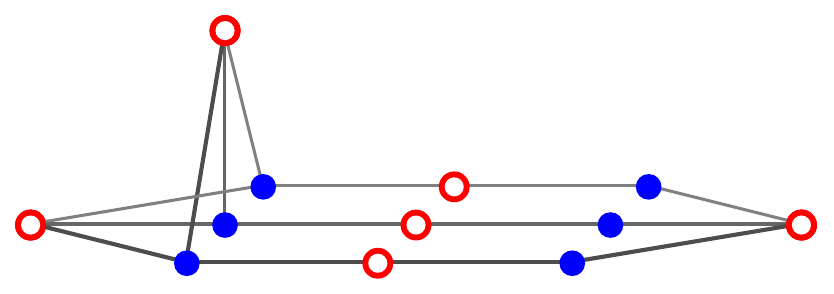}}  \\
\hline
\vrule height32pt  width0pt depth20pt  $\mathscr{X}^{9,21}$ & $\mathscr{X}^{5,9}$ &\hskip 0in \lower16pt \hbox{\includegraphics[width=2.55in]{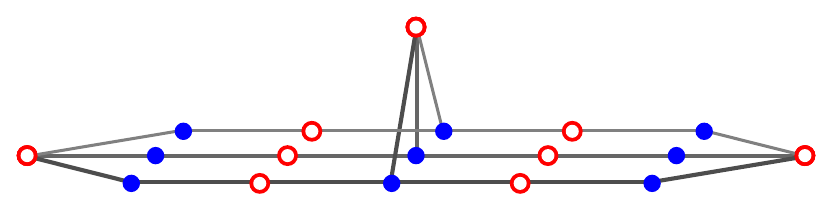}}  \\
\hline
\hline
\multicolumn{3}{|l|}{\str\textbf{Second Row}}\\
\hline
\vrule height30pt  width0pt depth20pt  $\mathscr{X}^{2,83}$ & $\mathscr{X}^{2,29}$ &\hskip 0in \lower 4pt \hbox{\includegraphics[width=.85in]{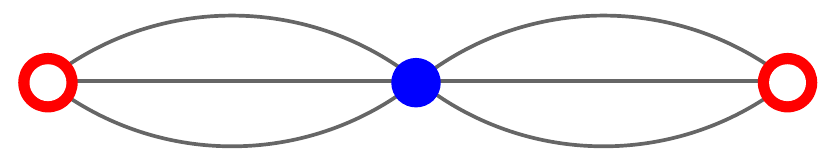}}  \\
\hline
\vrule height32pt  width0pt depth20pt  $\mathscr{X}^{5,59}$ & $\mathscr{X}^{3,21}$ &\hskip .8in \lower10pt \hbox{\includegraphics[width=1.7in]{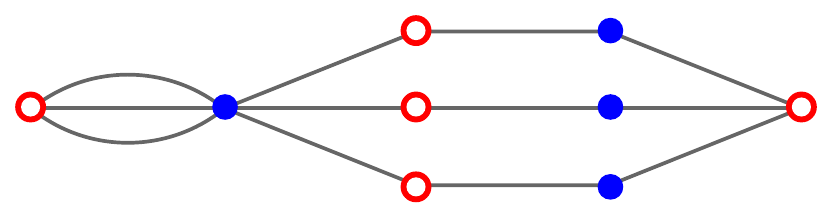}}  \\
\hline
\vrule height32pt  width0pt depth20pt  $\mathscr{X}^{8,44}$ & $\mathscr{X}^{4,16}$ &\hskip 0in \lower12pt \hbox{\includegraphics[width=2.55in]{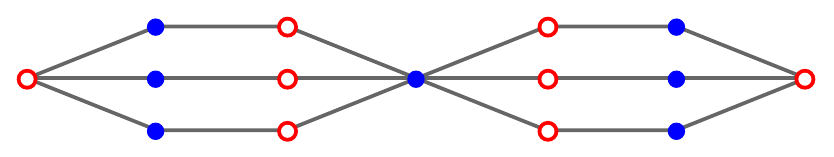}}  \\
\hline
\hline
\multicolumn{3}{|l|}{\str\textbf{Third Row}}\\
\hline
\vrule height30pt  width0pt depth20pt  $\mathscr{X}^{8,35}$ & $\mathscr{X}^{4,13}$ &\hskip 0in \lower12pt \hbox{\includegraphics[width=1.7in]{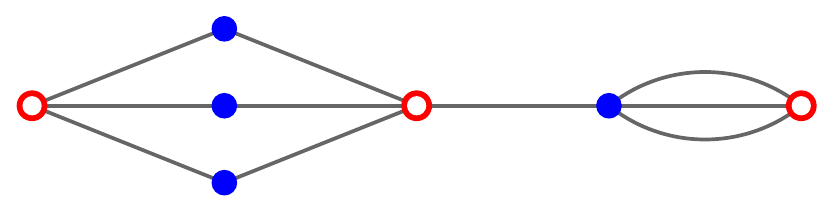}}  \\
\hline
\vrule height32pt  width0pt depth20pt  $\mathscr{X}^{11,29}$ & $\mathscr{X}^{5,11}$ &\hskip .8in \lower16pt \hbox{\includegraphics[width=2.55in]{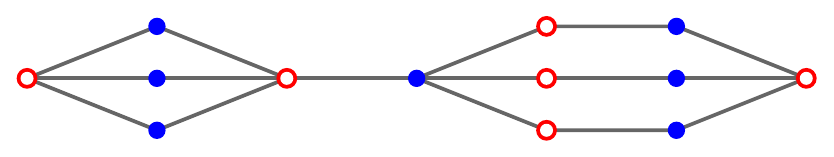}}  \\
\hline
\vrule height32pt  width0pt depth20pt  $\mathscr{X}^{14,23}$ & $\mathscr{X}^{6,9}$ &\hskip .8in \lower16pt \hbox{\includegraphics[width=2.5in]{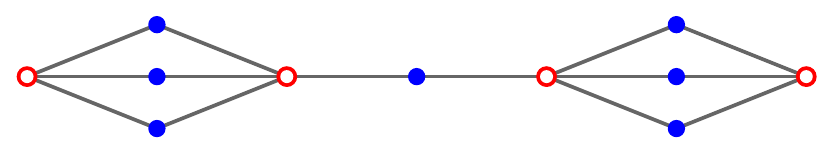}}  \\
\hline
\vrule height32pt  width0pt depth20pt  $\mathscr{X}^{19,19}$ & $\mathscr{X}^{7,7}$ &\hskip 0in \lower16pt \hbox{\includegraphics[width=3.4in]{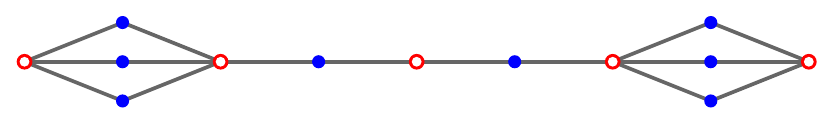}}  \\
\end{longtable}
\end{center}
}

\part{The Bundle}
\chapter{Line Bundle Models on CICY Quotients}\label{LineBundles}

\vspace{-21pt}
\section{Introduction}
\vspace{-8pt}
Heterotic string compactifications on Calabi-Yau threefolds provided one of the most promising approaches to string phenomenology for almost three decades. The ability to realise realistic particle physics models, affirmed for the first time in \cite{Gross:1984dd, Gross:1985fr, Gross:1985rr, Candelas:1985en}, has always been one of the strengths of this construction. For this purpose, several approaches have been proposed and used over the years: smooth Calabi-Yau compactifications based on the standard embedding \cite{Candelas:1985en, Greene:1986bm, Greene:1986jb, Braun:2009qy, Braun:2011ni}, non-standard embedding models \cite{Distler:1987ee, Distler:1993mk, Kachru:1995em, Braun:2005ux, Braun:2005bw, Braun:2005nv, Bouchard:2005ag, Blumenhagen:2006ux, Blumenhagen:2006wj, Anderson:2007nc, Anderson:2008uw, Anderson:2009mh}, models based on orbifolds \cite{Buchmuller:2005jr, Buchmuller:2006ik, Lebedev:2006kn, Lebedev:2007hv, Lebedev:2008un, Kim:2007mt, Nibbelink:2009sp, Blaszczyk:2009in, Blaszczyk:2010db, Kappl:2010yu}, on free fermionic strings \cite{Assel:2009xa, Christodoulides:2011zs, Cleaver:2011ir} and Gepner models \cite{GatoRivera:2009yt, GatoRivera:2010xn, Maio:2011qn}. 
In this chapter, I will present a recently developed approach to model building in the context of smooth Calabi-Yau compactifications of the heterotic string. This programme, aimed at achieving more detailed phenomenology than has to date been possible in this context, was initiated in the publications~\cite{Anderson:2011ns, Anderson:2012yf}. The history of string phenomenology suggests that it is difficult to fine tune any particular construction in order to simultaneously meet all the properties of the Standard Model. Instead, the approach taken here is that of a `blind' automated scan over a huge number of models; in the following sections I will describe a scan for which the number of models considered is of order $10^{40}$.  

\vspace{10pt}
What lies in front of the heterotic string model builder is  a set of highly non-trivial challenges that can be summarised in the following checklist:
\vspace{-10pt}
\begin{itemize}
	\item[1.] Construct a geometrical set-up, such that the 4-dimensional compactification of the $\cN=1$ supergravity limit of the heterotic string contains the symmetry $SU(3)\times SU(2)\times U(1)$ of the Standard Model of particle physics. This step is usually realised in two stages, by firstly breaking the $E_8$ heterotic symmetry to a Grand Unified Theory (GUT) group and then breaking the latter to the Standard Model gauge group (plus possibly $U(1)$ factors). This two-stage route can be shortcut, as in \cite{Blumenhagen:2006ux, Blumenhagen:2006wj}, though going through the GUT stage allows for more control over the following steps. In any case, in order to break the $E_8$ symmetry one needs to specify a VEV of the gauge connection on the internal (compact) 6-dimensional space $X$, or, equivalently, one needs to construct a vector bundle $V\longrightarrow X$. In order to preserve $\cN=1$ supersymmetry in four dimensions, this bundle has to be holomorphic and poly-stable.
	\item[2.] Derive the matter spectrum of the 4-dimensional theory. At low energy, the fermion fields transforming under the broken gauge group must be massless modes of the Dirac operator on the internal space $X$. The number of massless modes for a given representation is given by the dimension of certain bundle-valued cohomology groups on $X$. Such cohomology computations are generically difficult to perform. Ideally, one would like to obtain the matter spectrum of the minimally supersymmetric Standard Model (MSSM). In  practice, however, one often ends up with additional matter charged under the Standard Model gauge group (exotic fields) and with a number of scalar fields, the so-called moduli. By turning on VEVs for the moduli fields, one can deform the geometry of $X$ or $V$ or a combination thereof. Outside of the programme described in the following sections, only a small number of heterotic models that exhibit a realistic, exotic-free spectrum have been constructed so far \cite{Braun:2011ni, Braun:2005ux, Bouchard:2005ag, Anderson:2009mh}. 
	\item[3.] Constrain the resulting Lagrangian, in order to avoid well-known problems of supersymmetric GUT models, such as fast proton decay. For this purpose, additional discrete or continuous symmetries derived from the compactification set-up may be helpful in order to constrain the undesired operators. This is a clear advantage of string-derived models over traditional GUTs. 
	\item[4.] Derive information about the detailed properties of the model, such as the superpotential, the holomorphic Yukawa couplings, fermion mass-terms and $\mu$-terms. Such holomorphic quantities can usually be understood using techniques from algebraic geometry.
	\item[5.] Compute the physical Yukawa couplings. The physical Yukawa couplings consist of holomorphic superpotential terms times a non-holomorphic prefactor, whose computation requires the explicit knowledge of the metric on $X$ and the gauge connection on the vector bundle $V$. For the case when $X$ is Calabi-Yau, Yau's proof \cite{Yau:1978} guarantees the existence of a Ricci-flat metric, while for poly-stable vector bundles on Calabi-Yau manifolds, the Donaldson-Uhlenbeck-Yau theorem \cite{Donaldson1985, Uhlenbeck1986} guarantees the existence of a Hermitian Yang-Mills connection. However, except in very special cases, these quantities are not known analytically. So far, one can approach this differential geometric problem only numerically \cite{Douglas:2006hz,Anderson:2010ke,Anderson:2011ed}. 
	\item[6.] Stabilize the moduli and break supersymmetry. Recently, some progress has been made by including the effect of the $E_8\times E_8$ bundle flux~\cite{Anderson:2011cza, Anderson:2011ty, Anderson:2013qca}. 
	\item[7.] Compute soft-breaking parameters.
\end{itemize}

\vspace{0pt}
Every phenomenological requirement from the above list leads to a substantial reduction in the number of viable models. It is, therefore, crucial to start with a large number of models, if one hopes to retain a realistic model in the end. In this chapter I will concentrate on precisely this task and obtain, within a certain class of constructions, the largest possible set of models after the first two steps.

\vspace{10pt}
The history of heterotic string compactifications can be largely understood by looking at the types of poly-stable holomorphic vector bundles that have been the focus of different studies. In the early days of the subject, researchers largely concentrated on small deviations from the ``standard embedding", where the vector bundle $V$ was taken to be a holomorphic deformation of the tangent bundle \cite{Greene:1986bm, Greene:1986jb}. Such work has been continued to the current day with the first exact MSSM being produced from such an approach relatively recently \cite{Braun:2011ni}. In the 1990's and later more general poly-stable holomorphic vector bundles, or ``non-standard embeddings", began to be considered in ernest \cite{Distler:1987ee, Distler:1993mk, Kachru:1995em, Braun:2005ux, Braun:2005bw, Braun:2005nv, Bouchard:2005ag, Blumenhagen:2006ux, Blumenhagen:2006wj, Anderson:2007nc, Anderson:2008uw, Anderson:2009mh}. These gauge bundles were typically taken to have structure groups $SU(3)$, $SU(4)$ or $SU(5)$ leading to an $E_6$, $SO(10)$ or $SU(5)$ GUT group, respectively.

Recently, a new approach to building heterotic models on smooth Calabi-Yau threefolds has been advanced~\cite{Anderson:2011ns,Anderson:2012yf}. 
In this approach, the vector bundles in consideration are chosen to be direct sums of holomorphic line bundles. Such abelian gauge configurations lead to GUT groups which naively include additional~$U(1)$ factors. However, these extra $U(1)$ symmetries are frequently broken, in addition to other effects, by the Green-Schwarz mechanism, thus leading to no phenomenological inconsistencies. 

There are several advantages to working with sums of line bundles, as opposed to irreducible vector bundles. Firstly, such configurations are relatively simple to deal with from a computational point of view and, as a result, vastly greater numbers of models can be considered as compared to other approaches, such as \cite{Anderson:2007nc, Anderson:2008uw, Anderson:2009mh}. Secondly, although broken at a high scale, the additional~$U(1)$ symmetries remain as global symmetries and, as such, can severely constrain the Lagrangian of these models giving more information about the superpotential, and in particular the K\"ahler potential, than it is usually available in smooth heterotic constructions. Finally, although line bundle sums often represent special loci in the moduli space of vector bundles of a given topology, one can move away from the `split' locus by turning on VEVs for certain bundle moduli, thus reaching non-abelian bundles. As such, these simple configurations provide a computationally accessible window into an even bigger moduli space of heterotic compactifications.

\vspace{10pt}
The work presented in this chapter focuses on the description of an efficient algorithm used in order to perform a comprehensive scan over a large set of line bundle sums. In fact, for each Calabi-Yau manifold considered, the scan could be run by increasing the range of integers defining the first Chern class of the line bundles, until no more models were found. This amounted to a set of roughly $10^{40}$ different compactifications of heterotic string theory being investigated. The scan ran on a computer cluster over a period of seven months. The data set of GUT models resulted from this search comprises $34,989$ GUT models which will then lead to several orders of magnitude more heterotic standard models once the GUT group is broken by adding Wilson lines. Given the size and extra technical complications resulting from dealing with such huge numbers of heterotic compactifications, the detailed analysis of  incorporating the effects of the Wilson line breaking will not be included here. This represents in and of itself a huge task in computational algebraic geometry, which will be the subject of future work.

The table below presents a statistics on the total number of consistent GUT models which have resulted from the automated search discussed in Section \ref{sec:alg}. The first column counts $SU(5)$~GUT models having the correct chiral asymmetry, which can, however, suffer from the presence of $\mathbf{\overline{10}}$ multiplets or the absence of $\mathbf{5}-\bar{\mathbf{5}}$ pairs 
 to accomodate the Higgs content of the standard model. In the second column we eliminate those models that contain $\mathbf{\overline{10}}$ anti-family matter. This step relies on computations of line bundle cohomology groups, which we are able to perform in 94\% of all cases. The number in parentheses indicates the GUT models for which we could not decide upon the presence of $\mathbf{\overline{10}}$ multiplets. Similarly, in the third column we select from the $44343$ models that definitely have no anti-families, those which contain at least one $\mathbf{5}-\bar{\mathbf{5}}$ pair to contain MSSM Higgs fields.

\vspace{0pt}
\begin{footnotesize}
\begin{longtable}{| c || c | c | c |c|}
\captionsetup{width=14cm}
\caption{\label{basicstat}\it  Statistics on the number of models:}\\
\hline

\myalign{| c||}{\varstr{21pt}{16pt}$\ \ \ \ $ All $\left( X, |\Gamma|\right) $ $\ \ \ \ $} &
\myalign{m{2.2cm}|}{$\ $GUT models} &
\myalign{m{3.5cm}|}{ $\ \ \ \ $ no $ \overline{\mathbf{10}}$ multiplets$\ \ \ $ }&
\myalign{m{3.5cm}|}{$\ \ \ \ \ \ \ $ no $ \overline{\mathbf{10}}\,$s  and $\ \ \ \ \ \ \ $ at least one $\mathbf{5}-\overline{\mathbf{5}}$ pair}
\\ \hline
\varstr{14pt}{8pt} Total & 63325 & 44343 (3606) & 34989 (5291) 
\\ \hline
\end{longtable}
\end{footnotesize}

In the next section I will give a brief overview of the line bundle construction. For a more detailed account of the construction, the reader is directed towards the original papers \cite{Anderson:2011ns, Anderson:2012yf}, as well as Ref.~\cite{Anderson:2013xka}. For a general treatment of heterotic string compactifications, see the classic textbook by Green, Schwarz and Witten \cite{GSW}, the excellent introduction to string phenomenology by Iba\~nez and Uranga \cite{Ibanez:2012zz}, as well as other review articles \cite{Distler:1987ee, Quevedo:1997uy, Kachru:1997pc, He:2010uj} The reader interested to find out more about Calabi-Yau manifolds is invited to consult  H\"ubsch's physics-oriented compendium \cite{Hubsch}, or towards the more mathematically-flavoured texts by Joyce \cite{Joyce, Joyce:2001xt}. 

After pointing out the main elements of the line bundle construction and discussing the resulting GUT theory, in Section \ref{Sec3} I will define the class of manifolds under consideration, that is Calabi-Yau manifolds realised as complete intersections of hypersurfaces in products of projective spaces that admit free actions of finite groups. The following section contains a discussion of the requirements imposed on line bundle sums, succeeded, in Section \ref{sec:alg} by an outline of the algorithm used in the automated scan. In Section~\ref{sec:results} I will present the number of viable models obtained for each of the studied manifolds, noting that in all the cases we reach a limit beyond which no realistic line bundle vacua exist. 


\vspace{10pt}

\section{Overview of the Construction}


\subsection{Heterotic line bundle compactifications}

Schematically, the construction and analysis of heterotic string line bundle models can be broken up into three steps.
 \begin{itemize}
\item[$1.$]  In the first step, a solution to the 10 dimensional supergravity limit of the $E_8\times E_8$ heterotic string is obtained by specifying several geometrical elements. Firstly, we compactify the 10 dimensional space-time of the heterotic string on a smooth Calabi-Yau threefold $X$. Over this manifold we specify a poly-stable holomorphic vector bundle $V$ with structure group $H \subset E_8 \times E_8$ which describes the gauge field expectation values in the supergravity solution. The possible choices of $V$ over a given $X$ are restricted by several consistency requirements as described in Sections \ref{gut_group} and \ref{bundle_sec}. In the line bundle construction, the vector bundle is taken to be a direct sum of five holomorphic line bundles  $$V = \bigoplus_a L_a$$
As discussed in Section~\ref{gut_group}, I will choose the five line bundles $L_a$ such that the structure group $H \subset E_8$ is Abelian and of the form $H =S\left(U(1)^5\right)\cong U(1)^4$. 

If the background derived in this first step was used to dimensionally reduce the heterotic string theory to obtain an $\cN=1$ four dimensional supergravity without further modification, then the result would be a supersymmetric GUT. The gauge group, $G$, seen in four dimensions would, naively, be the commutant of $H$ inside $E_8$. For the line bundle models mentioned above, this leads to a GUT group
\vspace{-8pt}
 $$G = SU(5)\times S\left(U(1)^5\right)$$ 
\vspace{-28pt}

However, the additional $U(1)$ factors are generically Green-Schwarz anomalous and thus the associated gauge bosons often obtain St\"uckelberg mass terms which are close to the compactification scale in magnitude. 

\item[$2.$] In the second step, Wilson lines are added on the Calabi-Yau in such a way as to break the GUT group described above, down to that of the Standard Model. Adding such structure to the compactification is only possible if $X$ is not simply connected. Most standard constructions of Calabi-Yau threefolds lead to manifolds for which $\pi_1(X)=0$. Fortunately this situation can be resolved by quotienting a monifold $X$ obtained from one of the usual constructions by a freely acting discrete symmetry $\Gamma$. The fundamental group of the resulting smooth quotient manifold $\hat{X}=X/\Gamma$ is non-trivial, and in fact is isomorphic to $\Gamma$. 

The vector bundle $V$ constructed in step 1 must be consistent with this quotienting procedure. We must ensure that our bundle $V\rightarrow X$ descends to a well defined vector bundle $\hat{V} \rightarrow \hat{X}$. This is only the case if $V$ admits an equivariant structure under the symmetry $\Gamma$. Indeed, the set of vector bundles on $\hat{X}$ is in one-to one correspondence with the set of equivariant vector bundles on $X$.

The heterotic theory is then compactified to four dimensions on this new quotiented configuration including a non-trivial Wilson line. The gauge group obtained in four dimensions is then the commutant of the structure group of the flat bundle associated to the Wilson line inside $G$. This result is corrected as described in the first step by the Green-Schwarz mechanism. If the configuration is chosen correctly this can lead to the standard model gauge group $G_{SM}$ in four dimensions. The matter content must be computed by the usual techniques of dimensional reduction - including the effects of the Wilson line. One wishes to obtain examples where the resulting four dimensional standard model charged matter is exactly that of the MSSM.

\item[$3.$] As a final step in analysing a heterotic line bundle standard model, one can use global remnants of the additional $U(1)$ four dimensional gauge symmetries which are broken by the Green-Schwarz mechanism to constrain the operators present in the four dimensional Lagrangian. This allows a degree of analytical control over the low energy theory associated to these models which is unusual in the context of smooth Calabi-Yau reductions - in particular with regards to the K\"ahler potential for matter fields. In specific models, these symmetries can forbid operators in the four dimensional theory whose presence can be problematic for issues such as proton stability. 
\end{itemize}

In this chapter, I will focus on presenting the results obtained after pursuing the avenue described in the first step above. In the rest of this section I will describe in more detail the GUT gauge group and particle spectrum that are obtained in such constructions.

\vspace{20pt}
\subsection{The GUT gauge group}\label{gut_group}
As discussed above, the GUT gauge group $G$ is given by the commutant in $E_8$ of the structure group $H$ of the holomorphic vector bundle $V$. Thus, $G$ is the maximal subgroup of $E_8$, such that $H \times G \subset E_8$. This section addresses the question of  determining the structure group of a sum of line bundles. For generic bundles, the problem of determining the structure group cannot be solved without explicitly knowing the connection or the transition functions. However, in the present case, the topology of the bundle entirely specifies the structure group. 

The structure group of a single line bundle $L$ is $U(1)$. For a sum of five line bundles $V=\bigoplus_{a=1}^{5} L_a$, one would naively assign a structure group $H=U(1)^5$, or, if the first Chern class of $V$ vanishes, $H=S\left(U(1)^5\right) \cong U(1)^4$. The maximal subgroup of $E_8$ which commutes with $S\left( U(1)^5\right)$ is $SU(5)\times S\left(U(1)^5\right)\cong SU(5)\times U(1)^4$. This can be obtained via the sequence of embeddings $S\left( U(1)^5\right)\times S\left( U(1)^5\right) \subset SU(5)$ and $SU(5)\times SU(5) \subset E_8$ As explained below, the Green-Schwarz mechanism guarantees that the gauge bosons associated with the extra $U(1)$ symmetries are generically massive and thus these symmetries are explicitly broken. This leaves us with an $SU(5)$ GUT symmetry with additional global $U(1)$s, a situation which is very appealing from a phenomenological perspective. 

However, one should be careful about a number of topological elements which can modify the structure group of a sum of five line bundles with vanishing first Chern class. For instance, if the sum of line bundle contains two line bundles which are dual to one another, say $L_2 =L_1^{^*}$, 
this implies that $c_1\left(L_1 \oplus L_2 \right) = 0$, and further $c_1\left(L_3\oplus L_4 \oplus L_5\right) =0$. In this situation, the structure group of $V$ would be $S\left( U(1)^2\right)\times S\left( U(1)^3\right)\cong U(1)^3$, a group whose commutant in $E_8$ is the phenomenologically unattractive group $SU(6)\times U(1)^3$. In order to avoid such situations in which the structure group is a subgroup of $S\left(U(1)^5\right)$, we must require that the first Chern class of any partial sum of line bundles does not vanish:
\begin{equation}\label{no_subgroup_first}
c_1\left(\bigoplus_{a\in S} L_a \right) \neq 0\ \ \ \text{for any }S\text{ strict subset of }\{1,\ldots,5\}
\end{equation}

As explained in \cite{Anderson:2013xka}, accidental isomorphisms of the type $V\cong V^{^*}$ can render the structure group to be non-unitary. For a rank $5$ bundle, if $V\cong V^{^*}$, the set of transition functions which defines the structure group is contained either in $SO(5)$ or in $Sp(4)$. For a sum of five line bundles (and, in fact, for any sum of line bundles containing an odd number of terms), $V\cong V^{^*}$ can happen only if at least one of the line bundles is trivial. Thus, in order to avoid this situation, one needs to impose that $V$ contains no trivial line bundles.

\subsection{The GUT spectrum} \label{spec}

If the conditions of Section \ref{gut_group} are satisfied, a sum of five line bundles breaks the $E_8$ symmetry to $SU(5)\times S\left(U(1)^5\right)$. In this case, the computation of the spectrum of the heterotic line bundle model can be carried out as follows. 

The low energy limit of the $E_8\times E_8$ heterotic string is $\cN=1$, 10-dimensional supergravity coupled to $E_8\times E_8$ super Yang-Mills theory. The theory contains: a Yang-Mills supermultiplet, consisting of a vector potential $A$ and a gaugino $\chi$; a supergravity multiplet consisting of the vielbein $e$, a Neveu-Schwarz two-form $B$, a dilaton $\phi$, a gravitino $\psi$ and a dilatino $\lambda$. The fermionic fields in the 4-dimensional compactification of the theory on a Calabi-Yau manifold $X$ arise from the decomposition of the gaugino $\chi$, transforming in the $\mathbf{248}$ adjoint representation of $E_8$. Under the symmetry breaking $E_8\rightarrow H\times G$, the adjoint representation of $E_8$ branches as
\begin{equation} 
\mathbf{248}_{E_8} \longrightarrow \left(1, \text{Ad}(G) \right) \oplus \bigoplus_i \left(R_i, r_i \right)
\end{equation}

As before, $G$ denotes the remnant symmetry group in the GUT theory, where the gauge fields transform in the adjoint representation of $G$ and the matter fields transform in the representations $r_i$ of $G$. In order to be observable at low energy, the matter fields must be massless modes of the Dirac operator on $X$ - otherwise, their mass would be comparable to the compactification scale and thus, experimentally inaccessible. It can be shown that the number of massless modes corresponding to a certain representation $r_i$ equals the rank of a certain bundle-valued cohomology group on $X$: 
\begin{equation}
n_{r_i}  = \text{dim } H^1(X, V_{R_i}) = h^1(X, V_{R_i}) 
\end{equation}
where $V_{R_i}$ are vector bundles associated to a principal $H$-bundle and carrying the representation $R_i$ of $H$. These bundles can be obtained by taking the dual or appropriate tensor products of the vector bundle carrying the fundamental representation of $H$. 

To be more concrete, let us start with the well-known branching rule for $E_8\rightarrow SU(5) \times SU(5)$: 
\begin{equation}\label{branching_rule}
\mathbf{248} \longrightarrow (\mathbf{1}, \mathbf{24}) \oplus (\mathbf{24},\mathbf{1}) \oplus (\mathbf{10}, \overline{\mathbf{5}}) \oplus(\mathbf{5}, \mathbf{10}) \oplus (\overline {\mathbf{10}} , \mathbf{5}) \oplus (\overline{\mathbf{5}}, \overline {\mathbf{10}})
\end{equation}

The fundamental representation $\mathbf 5$ of $H=SU(5)$ is carried by the vector bundle $V$. From the above branching rule, it follows that the number of ${\mathbf{10}}$ multiplets in the $SU(5)$ GUT is equal to~$h^1(X,V)$. Similarly, the number of ${\mathbf{\overline{10}}}$ multiplets is given by $h^1(X,V^{^*}) = h^2(X,V)$. A vector bundle carrying the $\mathbf{10}$ representation is $\wedge^2 V$, thus the number of $\mathbf{\overline{5}}$ multiplets is equal to $h^1(X,\wedge^2 V)$. Similarly, the number of $\mathbf{5}$'s is equal to $h^1(X,\wedge^2 V^{^*}) = h^2(X,\wedge^2 V)$. Finally, the number of singlets is determined by the first cohomology group of a vector bundle carrying the adjoint representation of $SU(5)$. This vector bundle is $End(V)\cong V\otimes V^{^*}$. It follows that the number of singlet fields equals $h^1(X,V\otimes V^{^*})$. The singlets correspond to bundle moduli and their VEVs parametrize the bundle moduli space. 

\vspace{10pt}
Now let us turn to the breaking $E_8\rightarrow S\left( U(1)^5\right)\times \left (S\left( U(1)^5\right) \times SU(5) \right)$. As before, the low-energy theory contains the $SU(5)$ multiplets $\mathbf{10}, \mathbf{\overline{5}}$, their conjugates $\mathbf{\overline{10}}, \mathbf{5}$ and bundle moduli singlets $\mathbf{1}$. Now, however, these multiplets receive $S\left( U(1)^5\right)$ charges, which can be represented by vectors $\mathbf{q} = \left( q_1,\ldots, q_5\right)$. The group $S\left( U(1)^5\right)$ consists of elements $\left( e^{i\theta_1},\ldots, e^{i\theta_5}\right)$, such that the sum of the phases $\theta_1+\ldots+\theta_5=0$. Due to this determinant condition, two $S\left( U(1)^5\right)$ representations, labelled by $\mathbf{q}$ and $\mathbf{q'}$ have to be identified, if $\mathbf{q}-\mathbf{q'} \in \IZ\mathbf{n}$, where $\mathbf{n} = \left(1,1,1,1,1\right)$. 


How does the above branching rule \ref{branching_rule} change? Let $\{\mathbf{e}_a\}_{a=1,\ldots,5}$ denote the standard basis in five dimensions. Then the fundamental representation of (the internal) $SU(5)$ splits as a direct sum of five 1-dimensional $U(1)$ representations with charges $\mathbf{e}_a$. These are the charges of the corresponding $\mathbf{10}$-multiplets of the external $SU(5)$ and remain in the broken phase. Thus, compared with the spectrum of standard $SU(5)$ GUT discussed above, the multiplet content of line bundle models splits into sub-sectors, labelled by different $S\left( U(1)^5\right)$ charges. The number of $\mathbf{10}$-multiplets in the $a$-subsector, $n\left(\mathbf{10}_{\mathbf{e}_a}\right)$ is equal to $h^1(X,L_a)$.  

\begin{table}[!h]
\vspace{12pt}
\begin{center}
\begin{tabular}{|l|l|l|l|l|l|l|l|}
\hline
\varstr{14pt}{9pt} repr. & cohomology & total number & required for MSSM \\ \hline\hline
\varstr{14pt}{9pt} ${\bf 1}_{{\bf e}_a - {\bf e}_b}$ & $H^1(X, L_a \otimes L_b^{^*})$  &  $\sum_{a,b} h^1(X, L_a \otimes L_b^{^*}) = h^1(X, V \otimes V^{^*})$ & \;\;\;\;\; - \\ \hline
\varstr{14pt}{9pt} ${\bf 5}_{-{\bf e}_a -{\bf e}_b}$ & $H^1(X, L_a^{^*} \otimes L_b^{^*})$  & $\sum_{a<b} h^1(X, L_a^{^*} \otimes L_b^{^*}) =h^1(X, \wedge^2 V^{^*}) $ & \;\;\;\;\;$n_h$\\ \hline
\varstr{14pt}{9pt} ${\bf \overline{5}}_{{\bf e}_a+{\bf e}_b}$ & $H^1(X, L_a \otimes L_b)$  & $\sum_{a<b} h^1(X, L_a \otimes L_b) =h^1(X, \wedge^2 V) $ & \;\;\;\;\;$3 |\Gamma| + n_h$\\ \hline
\varstr{14pt}{9pt} ${\bf 10}_{{\bf e}_a}$ &$H^1(X, L_a)$ & $\sum_a h^1 (X,L_a) = h^1 (X,V)$& \;\;\;\;\;$3 | \Gamma|$\\ \hline
\varstr{14pt}{9pt} ${\bf  \overline{10}}_{-{\bf e}_a}$ & $H^1(X, L_a^{^*})$ & $\sum_a h^1(X,L_a^{^*}) = h^1(X,V^{^*})$&\;\;\;\;\; 0
\\ \hline 
 \end{tabular}
 \vskip 0.4cm
\parbox{15cm}{\caption{\it\small The spectrum of $SU(5)$ GUT models derived from the heterotic line bundle construction. In the final column, $|\Gamma|$~stands for the order of the fundamental group of $X$ and  $n_h$ represents the number of $\mathbf{5}-\overline{\mathbf{5}}$ Higgs fields.}\label{spectrum}}
 \end{center}
 \vspace{-8pt}
 \end{table}
 
 Different representations of the external $SU(5)$ have different $U(1)$ charges. Table (\ref{spectrum}) shows these charges for each of the different $SU(5)$ multiplets, as well as the cohomology group whose dimension determines the number of such fields present in the four dimensional theory.  One can notice from the table that, for example, the ${\bf 10}$'s of $SU(5)$ have charge one under precisely one of the five $U(1)$ symmetries, while the ${\bf \overline{5}}$ multiplets have charge one with respect to two of the Abelian factors. Apart from such rules, the precise pattern of charges across the spectrum (including bundle moduli) is model dependent. This is of major phenomenological importance: invariance under the (global remnant of the) $S\left( U(1)^5\right)$ symmetry constrains the allowed operators in the low-energy theory. Indeed, one can easily envisage situations in which the pattern of charges is such that, e.g.~proton decay operators are forbidden. 
 
The final column of the table shows the number of each $SU(5)$ representation that we require in the GUT model such that after quotienting the Calabi-Yau manifold and adding suitable Wilson lines, we obtain the spectrum of the MSSM. As before, $\Gamma$ denotes a freely acting finite group by which we quotient the Calabi-Yau manifold and $|\Gamma|$ is its order.

\subsection{The Green-Schwarz mechanism and the extra $U(1)$ symmetries}

In the four dimensional theory, the extra $U(1)$s have non-zero triangle contributions to the mixed  anomaly between an $S\left(U(1)^5 \right)$ gauge boson and two $G_{\text{GUT}}$ bosons, as well as to the cubic anomaly between three $S\left(U(1)^5 \right)$ gauge bosons. The anomaly is only apparent and these contributions are cancelled by a four-dimensional version of the Green-Schwarz mechanism - which is what one would expect since the underlying string theory is not anomalous. 

The anomaly generated by such triangle graphs is cancelled by a diagram in which an $S\left(U(1)^5 \right)$ gauge boson mixes with axion-like fields which in turn couple to the gauge bosons involved in the compactification. These axionic fields participating in the anomaly cancellation get eaten up into massive $U(1)$ gauge bosons and disappear from the low-energy spectrum. 

The mass matrix for the $S\left(U(1)^5\right)$ gauge bosons, given in \cite{Anderson:2011ns,Anderson:2012yf}, is:
\begin{equation}
 M_{ab} = G_{ij} \,c_1^i (L_a)\, c_1^j(L_b) 
 \end{equation}
where $G_{ij} = -\partial_i\partial_j \left( \mathrm{ln}\, \kappa\right)$ is the K\"ahler moduli space metric and $\kappa = d_{ijk} t^i t^j t^k$ is proportional to the Calabi-Yau volume. Since the matrix $G_{ij}$ is positive-definite (and thus non-degenerate), the number of massless $U(1)$ gauge bosons is given by 
\begin{equation} 
4 - \mathrm{rank}\, C
\end{equation} 
where $C$ is a matrix whose columns are the first Chern classes of the five line bundles.  As explained below, poly-stability of $V$ implies that at most $h^{1,1}(X)$ vectors corresponding to the first Chern classes of the individual line bundles are linearly independent. Thus the number of massless $U(1)$s is greater than $4 - h^{1,1}(X)$. As such, $h^{1,1}(X)=5$ is the smallest Hodge number for which all the~$U(1)$ gauge bosons can receive masses from the Green-Schwarz mechanism. 

In the cases in which the Green-Schwarz mechanism leaves massless $U(1)$ gauge bosons in the spectrum, one can invoke a second mechanism. By moving away from the locus where the bundle is split a sum of line bundles, the structure group becomes non-Abelian, thus removing the extra $U(1)$ symmetries from the low energy gauge group. 

\section{The Manifolds}
\label{Sec3}
Historically the first class of Calabi-Yau three-folds explicitly constructed \cite{Candelas:1987kf}, complete intersections in products of projective spaces (CICYs) have often served as the starting point in heterotic model building \cite{Greene:1986jb, Greene:1986bm, Braun:2009qy, Braun:2005ux, Braun:2005bw, Braun:2005nv, Bouchard:2005ag, Anderson:2007nc, Anderson:2008uw, Anderson:2009mh, Braun:2011ni}. The systematic computer-based scan for standard models presented here continues this tradition.

\subsection{CICY Manifolds}
\vspace{-10pt}
It is known that all CICY manifolds can be classified by finitely many configuration matrices, whose entries represent multi-degrees of homogeneous polynomials in products of projective spaces  \cite{Green:1987rw}. Each configuration represents a family of Calabi-Yau manifolds, with smooth generic members. By specialising the coefficients of the defining homogeneous polynomials, one can construct Calabi-Yau manifolds $X$ that are symmetric under the action of some finite group $\Gamma$. If this action is fixed-point-free, the quotient manifold $X/\Gamma$ is smooth. This technique of constructing quotient Calabi-Yau three-folds with non-trivial fundamental group was used at length in the previous chapter. The list of CICY threefolds, classified in \cite{Candelas:1987kf, Candelas:1987du}, contains $7868$ configurations whose Euler number ranges between $-200$ and $0$. The Hodge numbers $h^{1,1}$ and $h^{1,2}$ associated with these manifolds where computed in \cite{Green:1987cr}, leading to $265$ different pairs. Due to Wall \cite{Wall:1966}, it is known that the homotopy type of Calabi-Yau three-folds is classified by the Hodge numbers, the triple intersection numbers $d_{ijk} = \int_X J_i\wedge J_j \wedge J_k$ and the vector $c_2\cdot J_i = \int_X c_2 \wedge J_i$. This classification leads to at least $2590$ configurations being topologically distinct. 

\vspace{3pt}
The choice of CICY three-folds made in the present model building project  is based on two crucial features of these manifolds. In the first place, there exists a systematic classification of all linearly realised freely acting discrete symmetries on the CICY manifolds in the database \cite{Candelas:2008wb, Braun:2010vc}. More accurately, Braun's classification \cite{Braun:2010vc} provides a list of all such symmetries which descend from a linearly acting symmetry on the ambient space. However, a given Calabi-Yau manifold can frequently be embedded in many different products of projective spaces and  the question of whether a certain action is realised in a linear or in a non-linear manner depends on the particular embedding of the Calabi-Yau manifold. The symmetry classification of \cite{Braun:2010vc} is carried out for a limited selection of possible ambient spaces for each Calabi-Yau. 

As discussed in the previous section, a knowledge of such symmetries is an essential ingredient in breaking the GUT group to the Standard Model gauge group. We note that in constructing the GUT models presented here, it is in fact only knowledge of the possible orders of the available groups, and thus the possible values of $|\Gamma|$, which is required.

\vspace{3pt}
The second feature of the CICY's which makes them particularly suitable for the current work is related to their relative simplicity. The embedding of Calabi-Yaus in such simple ambient spaces means that computations of the cohomology of line bundles over these manifolds can be effectively automatised on a computer \cite{cicypackage}. This is particularly true when the line bundles on the Calabi-Yau threefold are restrictions of line bundles on the ambient space. This happens precisely when the CICYs are presented in a `favourable' embedding.

\subsection{Favourable configurations}\label{fav_config}
\vspace{-10pt}
Favourable embeddings can be described in many equivalent ways. For example, on manifolds such as those considered here, isomorphism classes of line bundles are completely classified by their first Chern class. This is an element of the second cohomology group $H^2(X,\IZ)$ of the base space $X$. Favourable CICYs can be defined to be those whose second cohomology descends entirely from the second cohomology of the embedding space. In such a case, all of the line bundles on $X$ are restrictions of line bundles over the ambient product of projective spaces. For this property to hold for a given description of a CICY, certain requirements have to be satisfied, as discussed below.

Let $X$ be a CICY embedded in the product of projective spaces ${\cal A}=\prod_{r=1}^m\mathbb{P}^{n_r}$, as the common zero locus of certain polynomials. These polynomials are sections of the normal bundle ${\cal N}$ on ${\cal A}$. Let us denote the restriction of ${\cal N}$ to $X$ by $N={\cal N}|_X$ and also introduce the bundle $S=\bigoplus_{r=1}^m{\cal O}_X({\bf e}_r)^{\oplus (n_r+1)}$, where ${\bf e}_r$ are the standard unit vectors in $m$ dimensions. 

The tangent bundle $TX$ can be obtained from the two short exact sequences
\begin{equation}
 0\rightarrow TX\rightarrow T{\cal A}|_X\rightarrow N\rightarrow 0\;,\quad
 0\rightarrow {\cal O}_X^{\oplus m}\rightarrow S\rightarrow T{\cal A}|_X\rightarrow 0\; .
\end{equation}
 Noting that $H^{1,1}(X)\cong H^2(X,TX)$ and $H^3(X,TX)\cong H^{0,1}(X)=0$ the two associated long exact sequences lead to the following relations for the second cohomology of X
 \begin{eqnarray}
  H^{1,1}(X)\!\!&\cong&\!\! {\rm Coker}\left(H^1(X,S)\rightarrow H^1(X,N)\right)\oplus{\rm Ker}\left(H^2(X,T{\cal A}|_X)\rightarrow H^2(X,N)\right)\ \ \ \ \ \\
  H^2(X,T{\cal A}|_X)\!\!&\cong&\!\! H^2(X,S)\oplus {\rm Ker}\left(\mathbb{C}^m\rightarrow H^3(X,S)\right)\; .
\end{eqnarray}  
The part of $H^{1,1}(X)$ which descends from the second ambient space cohomology corresponds to the~$\mathbb{C}^m$ term in the second equation. Hence, the precise conditions for $X$ to be favourable are
\begin{equation}
  {\rm Coker}\left(H^1(X,S)\rightarrow H^1(X,N)\right)=0\;,\quad  H^2(X,S)=0\; . \label{favour}
\end{equation} 
In particular, this means that a CICY with $h^{1,1}(X)>m$ or $h^1(X,S)<h^1(X,N)$ or $h^2(X,S)>0$ is not favourable. A sufficient, however slightly too strong, condition for $X$ to be favourable is
\begin{equation}
 h^1(X,N)=h^2(X,S)=0\; ,\label{favourprac}
\end{equation} 
where the first of these conditions guarantees that the Coker in \eqref{favour} vanishes. Eq.~\eqref{favourprac} can be checked relatively easily since it only involves cohomologies of line bundles on $X$. I will adopt this weaker criterion as a definition for favourability.

\vspace{20pt}
\section{The Bundles}\label{bundle_sec}
Recall that our bundles $V\rightarrow X$ are taken to be sums of five line bundles 
\begin{equation}\label{Vdef}
\displaystyle V=\bigoplus_{a=1}^{5} L_a\; .
\end{equation}
A single line bundle is specified by its first Chern class and, hence, by a set of $h^{1,1}(X)$  integers. Given this, a sum of $5$ line bundles is specified by a matrix of integers with $h^{1,1}(X)$ rows and $5$ columns. In our systematic investigation, we have scanned over $\sim\!\!10^{40}$ such matrices and selected approximately $35,000$ bundles which lead to phenomenologically consistent $SU(5)$ GUTs. In the following subsections we list the criteria that these $35,000$ models satisfy.

\vspace{-5pt}
\subsection{Topological constraints}
\vspace{-8pt}
I will require that,
\vspace{-10pt}
\begin{eqnarray} \label{C1}
c_1(V) = 0 \;,  
\end{eqnarray}
as well imposing Eq.~\eqref{no_subgroup_first}, such that the structure group of $V$ is $S(U(1)^5)$ which leads to a GUT group $G = SU(5)\times S\left(U(1)^5\right)$. Apart from this group theoretical advantage, imposing~$c_1(V)=0$ guarantees the existence of a spin structure on $V$.

In addition, the integrability condition on the Bianchi Identity for the Neveu-Schwarz two form leads to the following constraint on the vector bundle $V$.
\vspace{-5pt}
\beq\label{anom_canc}
{\rm ch}_2(TX)-{\rm ch}_2(V) -{\rm ch}_2(\tilde V)= [C]
\eeq
\vspace{-27pt}

Here $[C]\in H^4(X)$ is the Poincar\'e dual to the effective holomorphic curve class  wrapped by a five-brane and~$\tilde V$ is the hidden-sector bundle (which we will take to be trivial). The simplest way to guarantee that this condition can be satisfied is to require that
$c_2(TX)-c_2(V)\in $ Mori cone of~$X$, where I have used that $c_1(V)=0$. In this case, an effective curve class which saturates the condition~\eqref{anom_canc} for a trivial hidden bundle $\widetilde{V}$ exists (although, typically, solutions with non-trivial $\widetilde{V}$ can be found as well). For favourable CICYs, we have a basis $\{J_r\}$ of (1,1)-forms on $X$, obtained from hyperplane classes of the embedding projective spaces, such that the K\"ahler forms $J=t^rJ_r$ correspond to positive values, $t^r>0$, of the K\"ahler parameters $t^r$. The above Mori cone condition can then be written as
\begin{eqnarray} \label{C2}
 \int_X\left( c_2(TX)-c_2(V)\right)\wedge\,J_r \geq0,\text{ for all }r\in\{1,\ldots,h^{1,1}(X)\}\; .
\end{eqnarray} 

\subsection{Constraints from stability}
\vspace{-10pt}
Demanding an $\mathcal N=1$ supersymmetric vacuum in four dimensions leads to the requirement that the gauge connection on $V$ satisfies the hermitian Yang-Mills equations at zero slope. By the Donaldon-Uhlenbeck-Yau theorem this is possible if and only if $V$ is holomorphic, has vanishing slope and is polystable. 

The slope of a vector bundle $V$ defined as
$$ \mu(V)  = \frac{1}{\mathrm{rk}(V)} \int_X c_1(V)\wedge J\wedge J =  \frac{1}{\mathrm{rk}(V)} \sum_{r,s,t =1}^{h^{1,1}(X)} d_{rst}\, c_1^r(V) t^s t^t\; ,$$
where  $\displaystyle d_{rst}=\int_X J_r\wedge J_s\wedge J_t$ are the triple intersections on $X$.  \\[-8pt]

For the case of interest, $V$ is a direct sum of line bundles and $c_1(V)=0$. The vanishing slope condition~$\mu(V)=0$ is therefore automatically satisfied. In addition, these sums of line bundles are automatically holomorphic. On the other hand, poly-stability reduces to the requirement that,
\begin{eqnarray} \label{C3}
\exists \, t^r \;\;\; \textnormal{such that} \;\;\; \mu(L_a)|_{t^r}=0  \;\; \forall a
\end{eqnarray} 
somewhere in the interior of the K\"ahler cone ($t^r>0 \;\; \forall r$). 
\vspace{10pt}

Finally, we note that for slope(poly)-stable bundles on a Calabi-Yau threefold there is a positivity condition on the second Chern class, given by the so-called Bogomolov bound \cite{huybrechts2010geometry}. For $SU(n)$ bundles this takes the simple form
\beq\label{bogomolov}
\int_{X} c_2(V) \wedge J \geq 0
\eeq
and $J$ is any K\"ahler form for which $V$ is poly-stable.

\subsection{Constraints from the GUT spectrum}
The $SU(5)\times S\left(U(1)^5\right)$ GUT spectrum has already been discussed in section \ref{spec}. In order to secure a chiral asymmetry of 3 after taking the quotient of $X$ by $\Gamma$, we must require that $h^1(X,V)-h^2(X,V)=h^1(X,\wedge^2V)-h^2(X,\wedge^2V)=3|\Gamma|$. Since for a poly-stable bundle the zeroth and the top cohomologies vanish, the chiral asymmetry conditions can be formulated in terms of the indices
\begin{eqnarray} \label{C4}
\text{ind}(V) = \text{ind}(\wedge^2V) = -3|\Gamma| \; .
\end{eqnarray}

In fact, for an $SU(5)$ bundle, $\text{ind}(V) = \text{ind}(\wedge^2V)$, so one needs to check only one chiral asymmetry. Furthermore, in order to exclude anti-families, we require the absence of $\overline{\mathbf{10}}$ multiplets, and hence that
\begin{eqnarray} \label{C5}
h^2(X,V)=0
\end{eqnarray}

Finally, in order to safeguard the presence of at least one Higgs doublet, it is necessary to demand the existence of at least one $\mathbf{5}-\mathbf{\overline{5}}$ pair which is expressed by the requirement
\begin{eqnarray} \label{C6}
h^2(X,\wedge^2 V) >0 \; .
\end{eqnarray}

\subsection{Equivariance and the doublet-triplet splitting problem}
\vspace{-8pt}
In addition to the above constraints, we demand that, for each $a<b$
\begin{eqnarray} \label{C7}
\text{ind}(L_a\otimes L_b) \leq 0
\end{eqnarray}

In the case where each of the line bundles composing $V$ are individually equivariant, this condition is necessary for it to be possible to project out all Higgs triplets upon the addition of a Wilson line. If the line bundles composing $V$ are individually equivariant, then $V$ descends to a simple sum of line bundles on the quotient space $\hat{X}$. In such a case there is a relation between the index of the line bundle products on $X$ and $\hat{X}$.
\vspace{-12pt}
\beq
\text{ind}(\widetilde L_a\otimes \widetilde L_b) = \frac{1}{|\Gamma|}\, \text{ind}(L_a\otimes L_b)
\eeq 

All indices involved must, of course, be integers and in the case where the line bundles are individually equivariant the size of $\text{ind}(L_a\otimes L_b)$ will be such as to ensure that this is true. Given this relationship, if~$\text{ind}(L_a\otimes L_b)>0$ then so is $\text{ind}(\widetilde L_a\otimes \widetilde L_b)$. This ensures that in such cases there is at least one complete set of ${\bf 5}$ degrees of freedom of $SU(5)$ in the four dimensional effective theory - leading to the presence of Higgs triplets. This result is unaffected by the presence of Wilson lines as the undesirable particle content is protected by an index which such gauge configurations do not affect.

For line bundle sums with non-trivial equivariant blocks the situation is more complicated since divisibility of the index only applies to each equivariant block rather than to individual line bundles. In the simplest such case, an equivariant block is formed by two or more same line bundles which are permuted by the equivariant structure. More complicated equivariant blocks can consist of different line bundles which are mapped into each other, typically subject to an additional permutation of their integer entries. At any rate, bundle isomorphisms between the relevant line must exist in this case so that they must have the same index. In conclusion, line bundles within equivariant block must have the same index and, hence, if this index is positive so is the index of the equivariant block. Then, a generalization of the above index argument to the entire block leads to the same conclusion, namely the inevitable presence of Higgs triplets. Hence, the condition~\eqref{C7} should be imposed in all cases.

\section{The Scanning Algorithm}
\label{sec:alg}
If a sum of five line bundles passes the criteria (\ref{C1} - \ref{C7}), it leads to a consistent four dimensional GUT theory which, with appropriate Wilson line breaking, will lead to heterotic standard models. As mentioned above, a sum of five line bundles is specified by $5\cdot h^{1,1}(X)$ integers. For the manifolds in the CICY database which are favourable and admit known linear free actions of discrete groups, the values of the Hodge number $h^{1,1}(X)$ are restricted~\footnote{Manifolds with $h^{1,1}(X)=1$ cannot lead to consistent line bundle models since the slope zero condition~\eqref{C3} cannot be satisfied.} to the range $2\leq h^{1,1}(X)\leq 6$. Thus, we are interested in investigating bundles described by matrices of between $10$ and $30$ integers, and deciding when they obey the criteria we have described.

One could envisage a scan over all line bundle sums with entries between, say, $-10$ and $10$, for the manifolds with $h^{1,1}(X)=6$. This would require us to check $\sim\!\!10^{30}$ matrices representing sums of line bundles. For comparison, a year has $\sim\!3\cdot 10^{7}$ seconds. It is rather clear that such an attempt would be impossible if one desired to explicitly construct each such line bundle matrix and then check the necessary criteria. A better approach is based on the observation that the criteria (\ref{C1} - \ref{C7}) impose certain conditions on individual line bundles, as well as on partial sums of line bundles. These restrictions are of four kinds: stability related, index related, conditions stemming from the integrability of the heterotic Bianchi identity and restrictions on cohomology. 

\vspace{8pt}
The constraint from stability imposes that each collection of up to five line bundles $\{L_a\}$ can only describe a heterotic vacuum if there exists a point in the interior of the K\"ahler cone, $t^r>0$, such that simultaneously for all $a$, 
\beq 
\sum_{r,s,t =1}^{h^{1,1}(X)} d_{rst}\, c_1^{\,r}(L_a)\, t^{\,s} t^{\,t} = 0 .
\eeq
This is condition \eqref{C3} of the previous section.

Deciding whether a quadratic equation in several variables has positive solutions can be a computationally intensive question. Establishing the existence of common positive solutions for a collection of such equations is an even more formidable problem. On the other hand, given that the K\"ahler cone for our manifolds is given by $t^r>0$ for all $r=1,\ldots ,h^{1,1}(X)$, a fairly strong, and obviously necessary, condition for the existence of positive solutions to the slope-zero equation for each line bundle $L_a$ is that the matrix $(M_a)_{st}= d_{rst}\,c_1^r(L_a)$ has both positive and negative entries. For any subset $\{L_{a_1},\ldots,L_{a_n}\}\subset \{L_1,\ldots ,L_5\}$ of our five line bundles to have common positive solutions, it is necessary that any linear combination of the matrices $\{M_{a_1},\ldots,M_{a_n}\}$ has both positive and negative entries. In practice, we consider linear combinations with integer coefficients between $-5$ and $5$. This turns out to be a remarkably effective way of eliminating line bundle sums that are not poly-stable. 

For the line bundle sums that pass this necessary criterion we explicitly find common solutions to the slope-zero equations in the interior of the K\"ahler cone. In a great majority of the cases we are able to find exact solutions, while in the remaining cases (most of which appear for the $h^{1,1}=6$ manifolds), we have to resort to numerical methods.

\vspace{8pt}
The index-related criteria impose, for any subset of the five line bundles, that the sum of their indices is negative and greater or equal than $-3|\Gamma|$. This criterion simply follows from equation \eqref{C4} - we do not want more than three generations of standard model particles. We must also check the index based criteria given by equation \eqref{C7}. Indices of line bundles can be computed very rapidly in terms of their defining integers using the following standard formula:
\beq
\begin{aligned}
\text{ind}(L) & = \frac{1}{12} \left( 2\,c_1(L)^3 + c_1(L)\,c_2(TX)\right)\\
& = \sum_{r,s,t =1}^{h^{1,1}(X)} d_{rst}  \left( \frac{1}{6} \,c_1^{\,r}(L)\,c_1^{\,s}(L)\,c_1^{\,t}(L) + \frac{1}{12} c_1^{\,r}(L)\,c_2^{\,s\,t}(TX) \right)
\end{aligned}
\eeq

\vspace{8pt}
The condition \eqref{C2} stemming from the integrability of the heterotic Bianchi Identity constrains the full sum of five line bundles. This can be rewritten in terms of the integers describing the line bundles as:
\vspace{-8pt}
\beq
\int_X c_2(TX)\wedge J_r \geq \int_X c_2(V)\wedge J_r = \frac{1}{2}\,d_{rst}\sum_{a=1}^5 c_1^{\,s}(L_a)\, c_1^{\,t}(L_a)
\eeq

The restrictions on the cohomology of the line bundle sums (\ref{C5} - \ref{C6}) require a larger amount of computational resources to implement than do the simple checks already described. Therefore, at first, we only take into account the remaining constraints (\ref{C1} - \ref{C4}) and (\ref{C7}). This stage of the scan leads to GUT models with the correct chiral asymmetry, but does not exclude the possibility of having $\overline{\mathbf{10}}$ multiplets or no Higgs doublets at all. In the second stage of the scan, we attempt to eliminate the models containing anti-families. In $94\%$ of the cases we are able to compute the required cohomology and thus decide upon the fate of the corresponding models. Finally, we eliminate the models that have no Higgs doublets, with a rate of decidability of $88\%$. The computation of the cohomology of line bundles over CICYs is reviewed, for example, in Ref.~\cite{Anderson:2013qca}.

\vspace{1pt}
Below, I will schematically present the algorithm used in this automated search. The input parameters are the Calabi-Yau data (configuration matrix, intersection numbers, $c_2(TX)$, a list of row permutations that leave the configuration matrix unchanged); the order of a freely acting discrete group $\Gamma$ and the maximal value for a line bundle entry, $k_{\text{max}}$. The list of permutations present in the Calabi-Yau data is used in order to eliminate redundant line bundle sums, that is, line bundle sums that can be related to one another by a trivial re-labeling of the ambient space projective factors. The algorithm outputs a list of {\tt \itshape Models} represented as matrices of integers whose columns stand for the first Chern classes of the $5$ line bundles.

\vspace{1pt}

{\tt
\begin{itemize}
\item[1.]assemble {\itshape List\_1} containing line bundles satisfying:
\begin{itemize}
\item[$i)$] $-3|\Gamma| \leq \text{ind}\,(L)\leq0$ and 
\item[$ii)$]$\mu(L)=0$, somewhere in the interior of the K\"ahler cone\\[-14pt]
 \end{itemize}

\item[2.] obtain {\itshape List\_1r}$\,\subset\,${\itshape List\_1} by removing all redundant line bundles;

\item[3.]for each $L_{i_1}\in \text{\itshape List\_1r}$ assemble $\text{\itshape List\_2}\,(L_{i_1}) \subset \text{\itshape List\_1}$ containing line bundles such that, for every $L_{i_2}\in \text{\itshape List\_2}\,(L_{i_1})$ the following relations hold: 
\begin{itemize}
\item[$i)$] $-3|\Gamma| \leq \text{ind}\,(L_{i_1})+\text{ind}\,(L_{i_2})$;
\item[$ii)$] $-3|\Gamma| \leq \text{ind}\left(\wedge^2\left(L_{i_1}\oplus L_{i_2}\right)\right) = \text{ind}\,(L_{i_1}\otimes L_{i_2})\leq 0$;
\item[$iii)$]$\mu(L_{i_1})=\mu(L_{i_2})=0$, somewhere in the interior of the K\"ahler cone\\[-14pt]
 \end{itemize}

\item[4.]given $L_{i_1}\in \text{\itshape List\_1r}$, for each $L_{i_2}\in \text{\itshape List\_2}\,(L_{i_1})$ assemble $\text{\itshape List\_3}\,(L_{i_1},L_{i_2}) \subset \text{\itshape List\_2}\,(L_{i_1})$,
such that any  $L_{i_3}\in \text{\itshape List\_3}\,(L_{i_1},L_{i_2})$ satisfies:
\begin{itemize}
\item[$i)$] $-2\,k_{\text{max}} \leq c_1^{\,r}(L_{i_1})+c_1^{\,r}(L_{i_2})+c_1^{\,r}(L_{i_3})\leq2\,k_{\text{max}}$, for all $ r\in\{1,\ldots,h^{1,1}(X)\}$
\item[$ii)$] $-3|\Gamma| \leq \text{ind}\,(L_{i_1})+\text{ind}\,(L_{i_2})+\text{ind}\,(L_{i_3})$;
\item[$iii)$] $-3|\Gamma| \leq \text{ind}(L_{i_1}\otimes L_{i_3})\leq 0$; $-3|\Gamma| \leq \text{ind}(L_{i_2}\otimes L_{i_3})\leq 0$;
\item[$iv)$] $-3|\Gamma| \leq \text{ind}\left(\wedge^2\left(L_{i_1}\oplus L_{i_2}\oplus L_{i_3}\right)\right) = \text{ind}\,(L_{i_1}\otimes L_{i_2})+\text{ind}\,(L_{i_1}\otimes L_{i_3})+\text{ind}(L_{i_2}\otimes L_{i_3})\leq 0$;
\item[$v)$]$\mu(L_{i_1})=\mu(L_{i_2})=\mu(L_{i_3})=0$, somewhere in the interior of the K\"ahler cone\\[-14pt]
 \end{itemize}
 
 \item[5.]given $L_{i_1},L_{i_2}$ and $L_{i_3}$ as above, select from $\text{\itshape List\_3}\,(L_{i_1},L_{i_2})$ those line bundles $L_{i_4}$, such that the line bundle $L_{i_5}$ defined by $c_1(L_{i_1}\oplus L_{i_2}\oplus L_{i_3}\oplus L_{i_4}\oplus L_{i_5})=0$ satisfies: 
\begin{itemize}
\item[$i)$] $-k_{\text{max}} \leq c_1^{\,r}(L_{i_5})\leq k_{\text{max}}$
\item[$ii)$] $-3|\Gamma| \leq \text{ind}\,(L_{i_5})\leq 0$;
\item[$iii)$]$\mu(L_{i_5})=0$ somewhere in the interior of the K\"ahler cone\\[-14pt]
 \end{itemize}
 
 \item[6.] given $L_{i_1},L_{i_2},L_{i_3},L_{i_4}$ and $L_{i_5}$ as above, check: 
 \begin{itemize}
\item[$i)$] $-3|\Gamma| = \text{ind}\,(L_{i_1})+\text{ind}\,(L_{i_2})+\text{ind}\,(L_{i_3})+\text{ind}\,(L_{i_4})+\text{ind}\,(L_{i_5})$;
\item[$ii)$] $-3|\Gamma| = \text{ind}\,\left(\wedge^2 \left( L_{i_1}\oplus L_{i_2}\oplus L_{i_3}\oplus L_{i_4}\oplus L_{i_5} \right) \right)$;
\item[$iii)$] $\text{ind}(L_{i_a}\otimes L_{i_b})\leq 0$ for all pairs $a<b$ that have not been checked so far;
\item[$iv)$]$\mu(L_{i_1})\!=\!\mu(L_{i_2})\!=\!\mu(L_{i_3})\!=\!\mu(L_{i_4})\!=\!\mu(L_{i_5})\!\!\!\!\
=\!0$,  in the interior of the K\"ahler cone\\[-14pt]
\item[$v)$] $\displaystyle 2\,d_{rst}\,c_2^{\,s\,t}(TX) \geq d_{rst}\,\sum_{a=1}^5 c_1^{\,s}(L_{i_a}) \,c_1^{\,t}(L_{i_a})$
 \end{itemize}
 
 if these requirements are satisfied, append $L_{i_1}\oplus L_{i_2}\oplus L_{i_3}\oplus L_{i_4}\oplus L_{i_5}$ to {\itshape Models}
 
 \item[7.] remove all redundant line bundle sums from {\itshape Models}.

 \item[8.] for the remaining {\itshape Models}, check stability by explicitly finding points in the K\"ahler cone where all line bundles have slope zero. 
  
 \item[9.] eliminate models with $\overline{\mathbf{10}}$ multiplets
      
 \item[10.] eliminate models with no $\mathbf{5}-\mathbf{\overline{5}}$ pairs
 
 \end{itemize}
}

\vspace{0.5cm}

\section{An Example}
Using the above algorithm, one can obtain a great multitude of line bundle models (see the following section). The line bundle database can be accessed at \cite{database}. At the moment this thesis is being written, the database is simply a listing of viable line bundle sums for each manifold and symmetry group. For illustration, I have picked the following example, based on the Cicy with number 7447, defined by the configuration matrix and line bundle sum
\beq
X~=~~
\cicy{\IP^1 \\   \IP^1\\ \IP^1\\ \IP^1\\ \IP^1}
{ ~\bf{1}& \bf{1} \!\!\!\!\\
  ~\bf{1} & \bf{1}\!\!\!\! & \\
  ~\bf{1} & \bf{1}\!\!\!\! & \\
  ~\bf{1} & \bf{1} \!\!\!\!& \\
  ~\bf{1} & \bf{1}\!\!\!\!}_{-80}^{5,45}\
\hskip0.35in
V~=~~
\cicy{ \\ \\ \\ \\ }
{ -1 & -2 & ~~1 & ~~1 & ~~1\\
~~0 & -2 & -1 & ~~1 & ~~2\\
~~0 & ~~2 & -1 & ~~1 & -2\\
 ~~0 & ~~2 & ~~0 & ~~0 & -2\\
 ~~1 & ~~0 &~~ 0 & -2 & ~~1\\}\; .
\eeq

According to Ref.~\cite{Braun:2010vc}, the manifold $X$ can be smoothly quotiented by a group of order $4$. The columns of the second matrix correspond to the first Chern classes of the five line bundles composing $V$. The dimension $h^\bullet(X,V) = \left(h^0(X,V), h^1(X,V), h^2(X,V), h^3(X,V) \right)$ of the bundle cohomologies for $V$ are explicitly given by
\vspace{-8pt}
\begin{align} 
h^\bullet(X,V)\ \, & =\  (0,12,0,0) \\
h^\bullet (X,\wedge^2V)&=\ (0, 15, 3, 0)
\end{align}
\vspace{-27pt}

The model has a chiral asymmetry of $12$, which, after quotienting, is reduced to $3$. It contains a number of $\bf{5}-\overline{\bf{5}}$ pairs, which after introducing Wilson lines lead to one (or possibly more than one) pair of Higgs doublets. 

\vspace{10pt}
The above example is interesting as it satisfies the anomaly cancellation condition without the addition of any 5-branes. In this case, 
\vspace{-8pt}
\beq
c_2(TX).J_i = c_2(V).J_i = \left(24, 24, 24, 24, 24\right) 
\eeq
\vspace{-27pt}

As the ranks of $V$ and $TX$ are the same, and their second Chern classes match, one could study the interesting problem of deforming $V$ to $TX$, which would bring us back to the standard embedding. The database contains 348 such models which saturate the inequality (\ref{C2}).

In the next section, I will discuss the results of running the above algorithm in addition to the comprehensive nature of the list of models obtained.

\section{Results and Finiteness}\label{sec:results}

The number of models over a certain manifold admitting discrete symmetries of a fixed order is an increasing and saturating function of the maximal line bundle entry in modulus. This can be observed for all the pairs~$\left( X,|\Gamma|\right)$ that we have considered, as shown in the tables below \ref{tabel:finiteness1}-\ref{tabel:finiteness4}. However, we believe that these results reflect a more general phenomenon.  In practice, we have applied the above algorithm to all pairs $(X,|\Gamma|)$ of favourable CICYs with the orders $|\Gamma|$ of a freely-acting symmetries and all line bundle sums~\eqref{Vdef} with $|c_1^r(L_a)|\leq k_{\rm m}$, for a fixed upper bound $k_{\rm m}$. In each case, the number of viable models has then been determined for increasing values of $k_{\rm m}$ until saturation occurred. As a practical criterion for the onset of saturation we have required the number of models to remain unchanged for three consecutive values of $k_{\rm m}$. The results, before imposing the absence of $\overline{\bf 10}$ multiplets, Eq.~\eqref{C5}, and the presence of Higgs multiplets, Eq.~\eqref{C6}, can be found in the subsequent tables. The Calabi-Yau manifolds, $X$, are specified by a number, given in the first column of the tables below, which represents their position in the standard list of CICYs compiled in Refs.~\cite{Candelas:1987kf,Green:1987cr} and explicitly accessible here~\cite{cicylist2}. As it is evident from the tables, all viable models consist of line bundles satisfying
\begin{equation}
 |c_1^r(L_a)|\leq 10\; .
\end{equation} 
As one would expect, their number increases dramatically with $h^{1,1}(X)$, the number of K\"ahler parameters. For $h^{1,1}(X)=1,2,3,4,5,6$ we find $0,0,6,552,21731,41036$ models, respectively, for a total of $63325$ models, the number already quoted in the introduction. When the two further constraints~\eqref{C5} and \eqref{C6} are imposed this number reduces to $44343$ and $34989$, as already indicated in Table~\ref{basicstat}. The number of models at each stage, for all pairs $(X,|\Gamma|)$ is tabulated in Section~\ref{sec:distribution}. The complete list of these models can be accessed here~\cite{database}.

We note that the average number of viable models per pair $(X,|\Gamma|)$ as a function of $h^{1,1}(X)$ is approximately given by $0.3,20,530,4560$ for $h^{1,1}=3,4,5,6$, respectively. Very roughly, this corresponding to an increase of one order of magnitude per additional K\"ahler parameter.
At this point it is tempting to speculate about the total number of standard models, that is, models with the MSSM spectrum, in string theory. Known Calabi-Yau three-folds have Hodge numbers in the range of up to $h^{1,1}(X)\leq 500$. If the increase by an order of magnitude observed at small $h^{1,1}(X)$ continues to such large values the number of string standard models is enormous. However, a line bundle sum is determined by $5\cdot h^{1,1}(X)$ integers and it seems likely that the three-family constraint becomes more difficult to satisfy for a large number of these integers. We would, therefore, expect the increase to slow down at larger $h^{1,1}(X)$. Currently, we do not see any way of checking this by explicit scanning since $h^{1,1}(X)=6$ marks out the reach of present computational power.

{\setstretch{1.7}
\vspace{20pt}
\begin{footnotesize}
\begin{longtable}{| c ||  *{7}{c|}}
\captionsetup{width=14cm}
\caption{\label{tabel:finiteness1}Number of models as a function of $k_{\text{max}}$ on CICYs with $h^{1,1}(X)=3$:}\\
\hline
$\ \ X,\,|\Gamma|\ \ $ & $k_{\text{max}}=1$ & $k_{\text{m}}=2$ & $k_{\text{m}}=3$ & $k_{\text{m}}=4$ & $k_{\text{m}}=5$ & $k_{\text{m}}=6$ & $k_{\text{m}}=7$ \\
\hline
\endfirsthead
\multicolumn{7}{c}%
{\tablename\ \thetable\ -- \textit{Continued from previous page}} \\
\hline
\hline
\endhead
\hline \multicolumn{2}{r}{\textit{Continued on next page}} \\
\endfoot
\hline
\endlastfoot
7484, 4 & 0& 0& 0& 1& 1& 1 & \\ \hline
7669, 3 & 0& 0& 2& 2& 2 & & \\ \hline
7669, 9 & 0& 0& 1& 1& 1 & & \\ \hline
7735, 8 & 0& 0& 0& 0& 1& 1& 1 \\ \hline
7745, 8 & 0& 0& 0& 0& 1& 1& 1 \\
\end{longtable}
\end{footnotesize}

\vspace{20pt}
\begin{footnotesize}
\begin{longtable}{| c ||*{9}{c|}}
\captionsetup{width=14cm}
\caption{\label{tabel:finiteness2}Number of models as a function of $k_{\text{max}}$ on CICYs with $h^{1,1}(X)=4$:}\\
\hline
$\ \ X,\,|\Gamma|\ \ $ & $k_{\text{max}}=1$ & $k_{\text{m}}=2$ & $k_{\text{m}}=3$ & $k_{\text{m}}=4$ & $k_{\text{m}}=5$ & $k_{\text{m}}=6$ & $k_{\text{m}}=7$ & $k_{\text{m}}=8$ & $k_{\text{m}}=9$\\
\hline
\endfirsthead
\multicolumn{7}{c}%
{\tablename\ \thetable\ -- \textit{Continued from previous page}} \\
\hline
$\ \ X,\,|\Gamma|\ \ $ & $k_{\text{max}}=1$ & $k_{\text{m}}=2$ & $k_{\text{m}}=3$ & $k_{\text{m}}=4$ & $k_{\text{m}}=5$ & $k_{\text{m}}=6$ & $k_{\text{m}}=7$ & $k_{\text{m}}=8$ & $k_{\text{m}}=9$\\
\hline
\endhead
\hline \multicolumn{2}{r}{\textit{Continued on next page}} \\
\endfoot
\hline
\endlastfoot
6784, 2&0 & 0 & 2 & 10 & 12 & 12 & 12 & &\\ \hline
6784, 4&0 & 6 & 38 & 50 & 62 & 70 & 70 & 70 & \\ \hline
6828, 2&0 & 0 & 1 & 5 & 6 & 6 & 6 & & \\ \hline
6828, 4&0 & 3 & 19 & 25 & 31 & 35 & 35 & 35 &\\ \hline
6831, 2&0 & 1 & 2 & 2 & 2 & & & &\\ \hline
7204, 2&0 & 2 & 14 & 22 & 22 & 22 & & &\\ \hline
7218, 2&0 & 1 & 7 & 11 & 11 & 11 & & &\\ \hline
7241, 2&0 & 1 & 7 & 11 & 11 & 11 & & &\\ \hline
7245, 2&0 & 1 & 4 &  4 & 4 & & & &\\ \hline
7247, 3&0 & 19 & 57 & 59 & 59 & 59 & & &\\ \hline
7270, 2&0 & 2 & 14 &  22 & 22 & 22 & & &\\ \hline
7403, 2&0 & 3 & 6 & 6 & 6 & & & &\\ \hline
7435, 2&0 & 0 & 0 & 2 & 2 & 2 & & &\\ \hline
7435, 4&0 & 0 & 5 & 8 & 9 & 10 & 10 & 10 &\\ \hline
7462, 2&0 & 0 & 0 & 6 & 6 & 6 & & &\\ \hline
7462, 4&0 & 0 & 15 & 24 & 27 & 30 & 30 & 30 &\\ \hline
7468, 2&0 & 5 & 7 & 7 & 7 & & & &\\ \hline
7491, 2&0 & 0 & 0 & 2 & 2 & 2 & & &\\ \hline
7491, 4&0 & 0 & 5 & 8 & 9 & 10 & 10 & 10 &\\ \hline
7522, 2&0 & 0 & 0 & 6 & 6 & 6 & & &\\ \hline
7522, 4&0 & 0 & 15 & 24 & 27 & 30 & 30 & 30 &\\ \hline
7719, 2&0 & 4 & 14 & 26 & 26 & 26 & & &\\ \hline
7736, 2&0 & 2 & 7 & 13 & 13 & 13 & & &\\ \hline
7742, 2&0 & 2 & 7 & 13 & 13 & 13 & & &\\ \hline
7862, 2&0 & 5 & 7 & 10 & 10 & 10 & & &\\ \hline
7862, 4&0 & 9 & 46 & 54 & 58 & 58 & 58 & &\\ \hline
7862, 8&0 & 3 & 40 & 53 & 58 & 62 & 64 & 64 & 64\\ \hline
7862, 16&0 & 0 & 0 & 1 & 4 & 5 & 5 & 5 &\\
\end{longtable}
\end{footnotesize}

\vspace{20pt}
\begin{footnotesize}
\begin{longtable}{| c ||*{9}{c|}}
\captionsetup{width=14cm}
\caption{\label{tabel:finiteness3}Number of models as a function of $k_{\text{max}}$ on CICYs with $h^{1,1}(X)=5$:}\\
\hline
$\ \ X,\,|\Gamma|\ \ $ & $k_{\text{max}}=1$ & $k_{\text{m}}=2$ & $k_{\text{m}}=3$ & $k_{\text{m}}=4$ & $k_{\text{m}}=5$ & $k_{\text{m}}=6$ & $k_{\text{m}}=7$ & $k_{\text{m}}=8$ & $k_{\text{m}}=9$\\
\hline
\endfirsthead
\multicolumn{7}{c}%
{\tablename\ \thetable\ -- \textit{Continued from previous page}} \\
\hline
$\ \ X,\,|\Gamma|\ \ $ & $k_{\text{max}}=1$ & $k_{\text{m}}=2$ & $k_{\text{m}}=3$ & $k_{\text{m}}=4$ & $k_{\text{m}}=5$ & $k_{\text{m}}=6$ & $k_{\text{m}}=7$ & $k_{\text{m}}=8$ & $k_{\text{m}}=9$\\
\hline
\endhead
\hline \multicolumn{2}{r}{\textit{Continued on next page}} \\
\endfoot
\hline
\endlastfoot
5256, 2&0 & 575 & 727 & 775 & 779 & 779 & 779 & &\\ \hline
5256, 4&0 & 672 & 1857 & 2085 & 2173 & 2180 & 2180 & 2180 &\\ \hline
5301, 2&0 & 144 & 182 & 194 & 195 & 195 & 195 & &\\ \hline
5301, 4&0 & 169 & 466 & 523 & 545 & 547 & 547 & 547 &\\ \hline
5452, 2&0 & 574 & 726 & 774 & 778 & 778 & 778 & &\\ \hline
5452, 4&0 & 672 & 1854 & 2083 & 2171 & 2177 & 2177 & 2177 &\\ \hline
6024, 3&0 & 303 & 510 & 513 & 513 & 513 & & &\\ \hline
6204, 2&0 & 62 & 116 & 122 & 125 & 125 & 125 & &\\ \hline
6225, 2&0 & 147 & 221 & 231 & 232 & 232 & 232 & &\\ \hline
6715, 2&0 & 96 & 148 & 184 & 184 & 184 & & &\\ \hline
6715, 4&0 & 165 & 690 & 812 & 844 & 848 & 848 & 848 &\\ \hline
6724, 2&0 & 19 & 34 & 36 & 39 & 39 & 39 & &\\ \hline
6732, 2&0 & 434 & 778 & 880 & 880 & 880 & & &\\ \hline
6770, 2&0 & 216 & 307 & 329 & 331 & 331 & 331 & &\\ \hline
6777, 2&0 & 434 & 778 & 880 & 880 & 880 & & &\\ \hline
6788, 2&0 & 96 & 148 & 184 & 184 & 184 & & &\\ \hline
6788, 4&0 & 165 & 690 & 812 & 844 & 848 & 848 & 848 &\\ \hline
6802, 2&0 & 432 & 775 & 877 & 877 & 877 & & &\\ \hline
6804, 2&0 & 59 & 154 & 169 & 173 & 173 & 173 & &\\ \hline
6834, 2&0 & 218 & 390 & 441 & 441 & 441 & & &\\ \hline
6836, 2&0 & 24 & 37 & 46 & 46 & 46 & & &\\ \hline
6836, 4&0 & 43 & 175 & 206 & 214 & 215 & 215 & 215 &\\ \hline
6836, 8&0 & 6 & 94 & 120 & 131 & 133 & 137 & 137 & 137\\ \hline
6836, 16&0 & 0 & 0 & 0 & 2 & 3 & 3 & 3 &\\ \hline
6890, 2&0 & 860 & 1546 & 1750 & 1750 & 1750 & & &\\ \hline
6896, 2&0 & 218 & 390 & 441 & 441 & 441 & & &\\ \hline
6927, 2&0 & 144 & 222 & 276 & 276 & 276 & & &\\ \hline
6927, 4&0 & 244 & 1030 & 1212 & 1260 & 1266 & 1266 & 1266 &\\ \hline
6927, 8&0 & 34 & 554 & 706 & 770 & 782 & 806 & 806 & 806\\ \hline
6947, 2&0 & 24 & 37 & 46 & 46 & 46 & & &\\ \hline
6947, 4&0 & 43 & 175 & 206 & 214 & 215 & 215 & 215 &\\ \hline
6947, 8&0 & 6 & 94 & 120 & 131 & 133 & 137 & 137 & 137\\ \hline
6947, 16&0 & 0 & 0 & 0 & 2 & 3 & 3 & 3 &\\ \hline
7279, 2&0 & 128 & 204 & 212 & 218 & 218 & 218 & &\\ \hline
7447, 2&0 & 56 & 87 & 93 & 93 & 93 & & &\\ \hline
7447, 4&0 & 214 & 377 & 419 & 428 & 430 & 432 & 432 & 432\\ \hline
7447, 10&0 & 6 & 58 & 72 & 81 & 82 & 83 & 83 & 83\\ \hline
7487, 2&0 & 277 & 430 & 459 & 459 & 459 & & &\\ \hline
7487, 4&0 & 1052 & 1851 & 2058 & 2101 & 2111 & 2121 & 2121 & 2121\\
\end{longtable}
\end{footnotesize}

\vspace{20pt}
\begin{scriptsize}
\begin{longtable}{| c ||*{10}{c|}}
\captionsetup{width=14cm}
\caption{\label{tabel:finiteness4}Number of models as a function of $k_{\text{max}}$ on CICYs with $h^{1,1}(X)=6$:}\\
\hline
$\ X,\,|\Gamma|\ $ & $k_{\text{max}}=1$ & $k_{\text{m}}=2$ & $k_{\text{m}}=3$ & $k_{\text{m}}=4$ & $k_{\text{m}}=5$ & $k_{\text{m}}=6$ & $k_{\text{m}}=7$ & $k_{\text{m}}=8$ & $k_{\text{m}}=9$ & \myalign{m{1.4cm}|}{$k_{\text{m}}=10,$ $\ \ \ \ 11,\,12,\,13$}\\
\hline
\endfirsthead
\multicolumn{7}{c}%
{\tablename\ \thetable\ -- \textit{Continued from previous page}} \\
\hline
$\ \ X,\,|\Gamma|\ \ $ & $k_{\text{max}}=1$ & $k_{\text{m}}=2$ & $k_{\text{m}}=3$ & $k_{\text{m}}=4$ & $k_{\text{m}}=5$ & $k_{\text{m}}=6$ & $k_{\text{m}}=7$ & $k_{\text{m}}=8$ & $k_{\text{m}}=9$ & \myalign{m{1.3cm}|}{$k_{\text{m}}\!=\!10,$ $\ \ \ \ 11,\,12$}\\
\hline
\endhead
\hline \multicolumn{2}{r}{\textit{Continued on next page}} \\
\endfoot
\hline
\endlastfoot
\varstr{10pt}{5pt} 3413, 3&0& 2278& 2897& 2906& 2906& 2906 & & & &\\ \hline  
\varstr{10pt}{5pt}4190, 2&11& 766&1175& 1243& 1246& 1247& 1249& 1249& 1249 &\\ \hline  
\varstr{10pt}{5pt}5273, 2&29& 4895& 7149& 7738& 7799& 7810& 7810& 7810 & &\\ \hline  
\varstr{10pt}{5pt}5302, 2&0& 4314& 5978& 6360& 6369& 6369& 6369 & & &\\ \hline  
\varstr{10pt}{5pt}5302, 4&0& 11705& 16988& 17687& 17793& 17838& 17868& 17868& 17868 &\\ \hline  
\varstr{10pt}{5pt}5425, 2&0& 2381& 3083& 3305& 3337& 3337& 3337 & & &\\ \hline  
\varstr{10pt}{5pt}5958, 2&0& 148& 224& 240& 253& 253& 253 & & &\\ \hline  
\varstr{10pt}{5pt}6655, 5&0& 92& 178& 189& 194& 194& 198& 201& 202& 203\\ \hline  
\varstr{10pt}{5pt}6738, 2&1& 2733& 4116& 4346& 4386& 4393& 4399& 4399& 4399 &\\
\end{longtable}
\end{scriptsize}

}

\section[Distribution of Models]{The Distribution of Models According to $\left(X,|\Gamma|\right)$}\label{sec:distribution}

\vspace{20pt}
\begin{footnotesize}
\begin{longtable}{| c || c | c | c | c |}
\captionsetup{width=14cm}
\caption{Number of models on CICYs with $h^{1,1}(X)=3$:}\\
\hline
\myalign{| c||}{$\ \ X,\ |\Gamma|\ \ $} &
\myalign{m{2.2cm}|}{$\ $GUT models} &
\myalign{m{3.5cm}|}{ $\ \ \ $ no $ \overline{\mathbf{10}}$ multiplets$\ \ \ $ }&
\myalign{m{3.5cm}|}{$\ \ \ \ \ \ \ $ no $ \overline{\mathbf{10}}\,$s  and $\ \ \ \ \ \ \ $ at least one $\mathbf{5}-\overline{\mathbf{5}}$ pair}&
\myalign{m{3.5cm}|}{$\ \ \ \ \ \ $ no $ \overline{\mathbf{10}}\,$s  and $\ \ \ \ \ \ $ equivariance check for individual line bundles}\\
\hline
\endfirsthead
\multicolumn{5}{c}%
{\tablename\ \thetable\ -- \textit{Continued from previous page}} \\
\hline
\myalign{| c||}{$\ \ X,\ |\Gamma|\ \ $} &
\myalign{m{2.2cm}|}{$\ $GUT models} &
\myalign{m{3.5cm}|}{ $\ \ \ $ no $ \overline{\mathbf{10}}$ multiplets$\ \ \ $ }&
\myalign{m{3.5cm}|}{$\ \ \ \ \ \ \ $ no $ \overline{\mathbf{10}}\,$s  and $\ \ \ \ \ \ \ $ at least one $\mathbf{5}-\overline{\mathbf{5}}$ pair}&
\myalign{m{3.5cm}|}{$\ \ \ \ \ \ $ no $ \overline{\mathbf{10}}\,$s  and $\ \ \ \ \ \ $ equivariance check for individual line bundles}\\
\hline
\endhead
\hline \multicolumn{2}{r}{\textit{Continued on next page}} \\
\endfoot
\hline
\endlastfoot
7484, 4  & 1 & 1 & 1 & 1\\ \hline
7669, 3 & 2 & 2 & 0 (2) & 2 \\ \hline
7669, 9 & 1 & 1 & 0 (1) & 1 \\ \hline
7735, 8 & 1 & 1& 1 &  0 \\ \hline
7745, 8 & 1 & 1 & 1 & 0 
\end{longtable}
\end{footnotesize}

\vspace{10pt}
\begin{footnotesize}
\begin{longtable}{| c || c | c | c | c |}
\captionsetup{width=14cm}
\caption{Number of models on CICYs with $h^{1,1}(X)=4$:}\\
\hline
\myalign{| c||}{$\ \ X,\ |\Gamma|\ \ $} &
\myalign{m{2.2cm}|}{$\ $GUT models} &
\myalign{m{3.5cm}|}{ $\ \ \ $ no $ \overline{\mathbf{10}}$ multiplets$\ \ \ $ }&
\myalign{m{3.5cm}|}{$\ \ \ \ \ \ \ $ no $ \overline{\mathbf{10}}\,$s  and $\ \ \ \ \ \ \ $ at least one $\mathbf{5}-\overline{\mathbf{5}}$ pair}&
\myalign{m{3.5cm}|}{$\ \ \ \ \ \ $ no $ \overline{\mathbf{10}}\,$s  and $\ \ \ \ \ \ $ equivariance check for individual line bundles}\\
\hline
\endfirsthead
\multicolumn{5}{c}%
{\tablename\ \thetable\ -- \textit{Continued from previous page}} \\[8pt]
\hline
\myalign{| c||}{$\ \ X,\ |\Gamma|\ \ $} &
\myalign{m{2.2cm}|}{$\ $GUT models} &
\myalign{m{3.5cm}|}{ $\ \ \ $ no $ \overline{\mathbf{10}}$ multiplets$\ \ \ $ }&
\myalign{m{3.5cm}|}{$\ \ \ \ \ \ \ $ no $ \overline{\mathbf{10}}\,$s  and $\ \ \ \ \ \ \ $ at least one $\mathbf{5}-\overline{\mathbf{5}}$ pair}&
\myalign{m{3.5cm}|}{$\ \ \ \ \ \ $ no $ \overline{\mathbf{10}}\,$s  and $\ \ \ \ \ \ $ equivariance check for individual line bundles}\\
\hline
\endhead
\hline \multicolumn{2}{r}{\textit{Continued on next page}} \\
\endfoot
\hline
\endlastfoot
6784, 2 & 12 & 10  & 10  & 10 \\ \hline
 6784, 4 & 70 & 59  & 59  & 55 \\ \hline
 6828, 2 & 6 & 6  & 6  & 6 \\ \hline
 6828, 4 & 35 & 33  & 33  & 31 \\ \hline
 6831, 2 & 2 & 2  & 2  & 2 \\ \hline 
 7204, 2 & 22 & 14  & 14  & 14 \\ \hline
 7218, 2 & 11 & 11  & 11  & 11 \\ \hline
 7241, 2 & 11 & 9  & 9  & 9 \\ \hline
7245, 2 & 4 & 4  & 4  & 4 \\ \hline
 7247, 3 & 59 & 42 (14) & 22 (4) & 38 \\ \hline
 7270, 2 & 22 & 18  & 18  & 18 \\ \hline
 7403, 2 & 6 & 4 (2) & 0 (3) & 2 \\ \hline
 7435, 2 & 2 & 2  & 2  & 2 \\ \hline 
 7435, 4 & 10 & 9  & 9  & 7 \\ \hline
 7462, 2 & 6 & 6  & 6  & 6 \\ \hline
 7462, 4 & 30 & 16  & 16  & 14 \\ \hline
 7468, 2 & 7 & 6  & 5  & 6 \\ \hline
 7491, 2 & 2 & 2  & 2  & 2 \\ \hline 
7491, 4 & 10 & 4  & 4  & 4 \\ \hline
 7522, 2 & 6 & 6  & 6  & 6 \\ \hline
 7522, 4 & 30 & 21  & 21  & 17 \\ \hline
 7719, 2 & 26 & 24  & 24  & 24 \\ \hline
 7736, 2 & 13 & 12  & 12  & 12 \\ \hline
 7742, 2 & 13 & 12  & 12  & 12 \\ \hline
 7862, 2 & 10 & 10  & 8  & 10 \\ \hline
 7862, 4 & 58 & 53  & 46  & 44 \\ \hline 
7862, 8 & 64 & 52  & 36  & 10 \\ \hline
 7862, 16 & 5 & 5  & 4  & 0 \\ \hline
\end{longtable}
\end{footnotesize}

\vspace{10pt}
\begin{footnotesize}
\begin{longtable}{| c || c | c | c | c |}
\captionsetup{width=14cm}
\caption{Number of models on CICYs with $h^{1,1}(X)=5$:}\\
\hline
\myalign{| c||}{$\ \ X,\ |\Gamma|\ \ $} &
\myalign{m{2.2cm}|}{$\ $GUT models} &
\myalign{m{3.5cm}|}{ $\ \ \ $ no $ \overline{\mathbf{10}}$ multiplets$\ \ \ $ }&
\myalign{m{3.5cm}|}{$\ \ \ \ \ \ \ $ no $ \overline{\mathbf{10}}\,$s  and $\ \ \ \ \ \ \ $ at least one $\mathbf{5}-\overline{\mathbf{5}}$ pair}&
\myalign{m{3.5cm}|}{$\ \ \ \ \ \ $ no $ \overline{\mathbf{10}}\,$s  and $\ \ \ \ \ \ $ equivariance check for individual line bundles}\\
\hline
\endfirsthead
\multicolumn{5}{c}%
{\tablename\ \thetable\ -- \textit{Continued from previous page}} \\[8pt]
\hline
\myalign{| c||}{$\ \ X,\ |\Gamma|\ \ $} &
\myalign{m{2.2cm}|}{$\ $GUT models} &
\myalign{m{3.5cm}|}{ $\ \ \ $ no $ \overline{\mathbf{10}}$ multiplets$\ \ \ $ }&
\myalign{m{3.5cm}|}{$\ \ \ \ \ \ \ $ no $ \overline{\mathbf{10}}\,$s  and $\ \ \ \ \ \ \ $ at least one $\mathbf{5}-\overline{\mathbf{5}}$ pair}&
\myalign{m{3.5cm}|}{$\ \ \ \ \ \ $ no $ \overline{\mathbf{10}}\,$s  and $\ \ \ \ \ \ $ equivariance check for individual line bundles}\\
\hline
\endhead
\hline \multicolumn{2}{r}{\textit{Continued on next page}} \\
\endfoot
\hline
\endlastfoot
5256, 2 & 763 & 625 (12) & 480 (65) & 625 \\ \hline
 5256, 4 & 2128 & 1812 (23) & 1485 (167) & 1444 \\ \hline
 5301, 2 & 191 & 178 (3) & 87 (40) & 178 \\ \hline
 5301, 4 & 534 & 504 (6) & 323 (82) & 406 \\ \hline
 5452, 2 & 762 & 547 (11) & 497 (25) & 547 \\ \hline
 5452, 4 & 2122 & 1624 (17) & 1518 (71) & 1278 \\ \hline
 6024, 3 & 509 & 244 (69) & 215 (29) & 237 \\ \hline
 6204, 2 & 119 & 96 (14) & 76 (17) & 93 \\ \hline
 6225, 2 & 229 & 137 (21) & 118 (17) & 133 \\ \hline
 6715, 2 & 184 & 170 (0) & 138 (4) & 170 \\ \hline
 6715, 4 & 847 & 711 (4) & 539 (76) & 457 \\ \hline
 6724, 2 & 39 & 32 (7) & 20 (10) & 21 \\ \hline
 6732, 2 & 880 & 667 (6) & 532 (60) & 667 \\ \hline
 6770, 2 & 330 & 271 (0) & 197 (39) & 271 \\ \hline
 6777, 2 & 880 & 587 (6) & 549 (32) & 587 \\ \hline
 6788, 2 & 184 & 155 (0) & 147 (4) & 155 \\ \hline
 6788, 4 & 848 & 621 (4) & 579 (28) & 397 \\ \hline
 6802, 2 & 877 & 786 (6) & 524 (128) & 786 \\ \hline
 6804, 2 & 141 & 108 (4) & 99 (5) & 101 \\ \hline
 6834, 2 & 441 & 371 (3) & 283 (47) & 371 \\ \hline
 6836, 2 & 46 & 37 (0) & 36 (1) & 37 \\ \hline
 6836, 4 & 214 & 151 (1) & 147 (4) & 97 \\ \hline
 6836, 8 & 136 & 109 (0) & 97 (9) & 14 \\ \hline
 6836, 16 & 3 & 3 (0) & 2 (1) & 0 \\ \hline
 6890, 2 & 1750 & 1245 (12) & 1091 (83) & 1245 \\ \hline
 6896, 2 & 441 & 421 (3) & 232 (88) & 421 \\ \hline
 6927, 2 & 276 & 243 (0) & 218 (6) & 243 \\ \hline
 6927, 4 & 1264 & 983 (6) & 856 (67) & 628 \\ \hline
 6927, 8 & 798 & 659 (5) & 510 (79) & 81 \\ \hline
 6947, 2 & 46 & 45 (0) & 30 (1) & 45 \\ \hline
 6947, 4 & 214 & 196 (1) & 105 (35) & 127 \\ \hline
 6947, 8 & 136 & 125 (0) & 44 (19) & 21 \\ \hline
 6947, 16 & 3 & 3 (0) & 2 (1) & 0 \\ \hline
 7279, 2 & 218 & 109 (49) & 96 (10) & 108 \\ \hline
 7447, 2 & 93 & 89 (0) & 45 (15) & 89 \\ \hline
 7447, 4 & 430 & 396 (2) & 182 (77) & 306 \\ \hline
 7447, 5 & 0 & 0 (0) & 0 (0) & 0 \\ \hline
 7447, 10 & 81 & 76 (0) & 12 (19) & 0 \\ \hline
 7447, 20 & 0 & 0 (0) & 0 (0) & 0 \\ \hline
 7487, 2 & 459 & 319 (0) & 261 (28) & 319 \\ \hline
 7487, 4 & 2115 & 1505 (8) & 1257 (94) & 1136 \\
\end{longtable}
\end{footnotesize}

\vspace{10pt}
\begin{footnotesize}
\begin{longtable}{| c || c | c | c | c |}
\captionsetup{width=14cm}
\caption{Number of models on CICYs with $h^{1,1}(X)=6$:}\\
\hline
\myalign{| c||}{$\ \ X,\ |\Gamma|\ \ $} &
\myalign{m{2.2cm}|}{$\ $GUT models} &
\myalign{m{3.5cm}|}{ $\ \ \ $ no $ \overline{\mathbf{10}}$ multiplets$\ \ \ $ }&
\myalign{m{3.5cm}|}{$\ \ \ \ \ \ \ $ no $ \overline{\mathbf{10}}\,$s  and $\ \ \ \ \ \ \ $ at least one $\mathbf{5}-\overline{\mathbf{5}}$ pair}&
\myalign{m{3.5cm}|}{$\ \ \ \ \ \ $ no $ \overline{\mathbf{10}}\,$s  and $\ \ \ \ \ \ $ equivariance check for individual line bundles}\\
\hline
\endfirsthead
\multicolumn{5}{c}%
{\tablename\ \thetable\ -- \textit{Continued from previous page}} \\[8pt]
\hline
\myalign{| c||}{$\ \ X,\ |\Gamma|\ \ $} &
\myalign{m{2.2cm}|}{$\ $GUT models} &
\myalign{m{3.5cm}|}{ $\ \ \ $ no $ \overline{\mathbf{10}}$ multiplets$\ \ \ $ }&
\myalign{m{3.5cm}|}{$\ \ \ \ \ \ \ $ no $ \overline{\mathbf{10}}\,$s  and $\ \ \ \ \ \ \ $ at least one $\mathbf{5}-\overline{\mathbf{5}}$ pair}&
\myalign{m{3.5cm}|}{$\ \ \ \ \ \ $ no $ \overline{\mathbf{10}}\,$s  and $\ \ \ \ \ \ $ equivariance check for individual line bundles}\\
\hline
\endhead
\hline \multicolumn{2}{r}{\textit{Continued on next page}} \\
\endfoot
\hline
\endlastfoot
 3413, 3 & 1737 & 709 (516) & 599 (98) & 698 \\ \hline
 4190, 2 & 1145 & 540 (195) & 473 (57) & 429 \\ \hline
 5273, 2 & 6753 & 4154 (934) & 3292 (701) & 3757 \\ \hline
 5302, 2 & 6294 & 4130 (246) & 3291 (456) & 4130 \\ \hline
 5302, 4 & 17329 & 13242 (82) & 10174 (1678) & 9235 \\ \hline
 5425, 2 & 3128 & 1946 (533) & 1358 (409) & 1802 \\ \hline
 5958, 2 & 246 & 215 (23) & 103 (66) & 179 \\ \hline
 6655, 5 & 161 & 143 (15) & 67 (64) & 1 \\ \hline
 6738, 2 & 4243 & 1846 (743) & 1599 (169) & 1763 \\
\end{longtable}
\end{footnotesize}

\chapter{Line Bundle Models on the Tetraquadric Hypersurface}\label{TQ1}

\section{Introduction}
Smooth Calabi-Yau compactifications of the heterotic string and M-theory represent one of the classic and most trodden avenues from string theory to the low energy physics \cite{Candelas:1985en}. Despite the great interest received over the years, however, this approach has led to a relatively  small number of models that can adequately address the most basic phenomenologically relevant questions, such as exhibiting the Standard Model particle content without any exotics \cite{Braun:2005ux, Braun:2005bw, Bouchard:2005ag, Anderson:2009mh, Braun:2011ni}. The paucity of such models, owing to the considerable mathematical complexity involved in the analysis of the compactification geometry, has made impossible a genuine undertake of finer phenomenological questions. 

This unpromising situation has recently changed.  In a series of publications \cite{Anderson:2011ns, Anderson:2012yf, Anderson:2013xka, He:2013ofa}, a salient class of $E_8\times E_8$ heterotic compactifications has been advanced. These correspond to rank four or five Abelian vector bundles over smooth Calabi-Yau three-folds with non-trivial fundamental group. This construction has been presented in the previous chapter. Constructing the polystable, holomorphic vector bundle as a direct sum of line bundles comes with a number of far-reaching advantages. On one hand, the split nature of the bundle facilitates an algorithmic implementation of the various consistency and physical constraints. On the other hand, it leads to a rich phenomenology, such that one can readily envisage situations in which physical questions beyond the particle spectrum, such as proton stability or the structure of the holomorphic Yukawa couplings can be simultaneously addressed. 

More concretely, in \cite{Anderson:2011ns, Anderson:2012yf} over 200 $SU(5)$ GUT models were constructed on discrete quotients of complete intersection Calabi-Yau manifolds. All these had precisely three generations of GUT families, no anti-families, at least one $5 - \overline{5}$ pair of Higgs fields and no other charged matter. Completing the vector bundle with Wilson lines in order to break the GUT gauge group, augmented the number of models exhibiting the exact spectrum of MSSM by an order of magnitude. This preliminary search for standard-like model was perfected to an exhaustive scan presented in \cite{Anderson:2013xka}, and also in Chapter~\ref{LineBundles}. This scan led to some $35,000$ $SU(5)$  models over the same class of Calabi-Yau manifolds. Finally, in \cite{He:2013ofa}, over $100$ $SU(5)$ models and about $29,000$ $SO(10)$ GUT models were constructed over $14$~Calabi-Yau hypersurfaces embedded in toric varieties. These large numbers are indicative of the huge potentiality of the line bundle construction. 

In the present and the following chapter I will elaborate on three points. Firstly, it was noticed in Chapter~\ref{LineBundles} that the number of viable models reaches a certain saturation limit after repeatedly increasing the range of integers defining the line bundles. I will revisit this question here and address it from two different perspectives for the particular case in which the Calabi-Yau manifold is a hypersurface embedded in a product of four $\IC\IP^1$ spaces, henceforth referred to as the tetraquadric hypersurface. The second objective is to present in detail a semi-realistic model based on the tetraquadric manifold. Finally, as discussed below for the case of $SU(5)$ models, the particle spectrum of the ensuing low-energy theory contains a number of singlet fields whose vacuum expectation values parametrize the bundle moduli space. Turning on VEVs for some of the singlet fields corresponds to moving away from the locus where the holomorphic bundle splits into a sum of line bundles. If the line bundle construction is to be taken seriously, it is compelling to understand the moduli space of non-Abelian bundles in which these handy models live. In the next chapter, I will explore the moduli space around the particular model mentioned above by applying the monad construction of vector bundles. 

These seemingly dissimilar questions run along a common thread and illustrate several inquests into the moduli space of heterotic compactifications built on a specific manifold, the tetraquadric hypersurface.  Coming from afar, we identify a substantial, though finite, number of points in this moduli space representing Abelian bundles which lead to phenomenologically appealing low-energy theories. In the second step, Section~\ref{sec:model}, we focus on one of these points, so that in the last part, Section~\ref{sec:monads}, we can zoom out again, this time in order to explore the space of non-Abelian deformations around the chosen point. 

The discussion runs on two, even three, levels. On one hand, we have the high energy theory, described in terms of the compactification data: a Calabi-Yau three-fold supplied with a holomorphic, poly-stable bundle. On the other hand, this geometrical set-up leads to a four-dimensional, low-energy supersymmetric GUT, whose gauge group can further be broken to that of the Standard Model. Frequently, we switch back and forth between the high-energy theory level and the GUT level. The Standard Model appears only briefly in Section~\ref{sec:model}, simply to illustrate the virtues of the line bundle model chosen as a case study. The comparison between the distinguished Abelian model and its non-Abelian deformations is carried out both at the high energy level and at the GUT level, with a meaningful interplay between the two pictures.

\section{The Line Bundle Construction}
The line bundle construction has been discussed in the previous chapter, as well as in Refs.~\cite{Anderson:2011ns, Anderson:2012yf}. Below, I will summarise the GUT spectrum emerging in this set-up.  The GUT spectrum will be a central point in the comparison between Abelian and non-Abelian compactifications in Chapter~\ref{ModuliSpace}. 
After reviewing the $SU(5)$ GUT spectrum, in Section~\ref{sec:conditions}, I will present a number of general aspects about Abelian compactifications on the tetraquadric manifold.

\subsection{A brief review of $SU(5)$ line bundle models}
\vspace{-4pt}
As before, I will discuss $E_8\times E_8$ heterotic compactifications leading to $SU(5)$ GUT models. For this purpose, in the line bundle approach, we choose the poly-stable, holomorphic bundle in the visible sector to be a direct sum of five line bundles with vanishing first Chern class:
\begin{equation}
V = \bigoplus_{a=1}^5 L_a\, , \ \ \ \ c_1(V) =0
\end{equation}

This choice of the internal gauge field generically leads to an $SU(5)$ GUT group, completed with several global $U(1)$ symmetries.\footnote{For a complete discussion on the GUT group emerging in line bundle constructions, see Ref.~\cite{Anderson:2013xka}.} Invariance under these global $U(1)$ transformations leads to significant constraints on the allowed operators in the 4 dimensional effective supergravity. 

The consistency requirements, such as the anomaly cancellation and the conditions imposed on the vector bundle by supersymmetry can be checked in a straightforward manner. Supersymmetry requires that the vector bundle is holomorphic (automatic, in the present case) and slope poly-stable. In general a laborious task, checking poly-stability for a sum of line bundles reduces to the question of deciding whether a set of quadratic equations (corresponding to the vanishing slope for each line bundle) have common solutions in the K\"ahler cone. 
%
The particle content of the subsequent GUT model is computed in terms of certain bundle-valued cohomology groups on the Calabi-Yau manifold~$X$, as detailed in Table~\ref{spectrum2}. The low-energy theory contains the $SU(5)$ multiplets $\mathbf{10}, \mathbf{\overline{5}}$, their conjugates $\mathbf{\overline{10}}, \mathbf{5}$ and bundle moduli singlets $\mathbf{1}$. These multiplets receive $S\left( U(1)^5\right)$ charges, represented by vectors $\mathbf{q} = \left( q_1,\ldots, q_5\right)$. Table ~\ref{spectrum2} shows these charges for each of the different $SU(5)$ multiplets, as well as the cohomology group whose dimension determines the number of multiplets present in the four dimensional theory. 

\begin{table}[!h]
\vspace{12pt}
\begin{center}
\begin{tabular}{|l|l|l|l|l|l|l|l|}
\hline
\varstr{14pt}{9pt} repr. & cohomology & total number & required for MSSM \\ \hline\hline
\varstr{14pt}{9pt} ${\bf 1}_{{\bf e}_a - {\bf e}_b}$ & $H^1(X, L_a \otimes L_b^{^*})$  &  $\sum_{a,b} h^1(X, L_a \otimes L_b^{^*}) = h^1(X, V \otimes V^{^*})$ & \;\;\;\;\; - \\ \hline
\varstr{14pt}{9pt} ${\bf 5}_{-{\bf e}_a -{\bf e}_b}$ & $H^1(X, L_a^{^*} \otimes L_b^{^*})$  & $\sum_{a<b} h^1(X, L_a^{^*} \otimes L_b^{^*}) =h^1(X, \wedge^2 V^{^*}) $ & \;\;\;\;\;$n_h$\\ \hline
\varstr{14pt}{9pt} ${\bf \overline{5}}_{{\bf e}_a+{\bf e}_b}$ & $H^1(X, L_a \otimes L_b)$  & $\sum_{a<b} h^1(X, L_a \otimes L_b) =h^1(X, \wedge^2 V) $ & \;\;\;\;\;$3 |\Gamma| + n_h$\\ \hline
\varstr{14pt}{9pt} ${\bf 10}_{{\bf e}_a}$ &$H^1(X, L_a)$ & $\sum_a h^1 (X,L_a) = h^1 (X,V)$& \;\;\;\;\;$3 | \Gamma|$\\ \hline
\varstr{14pt}{9pt} ${\bf  \overline{10}}_{-{\bf e}_a}$ & $H^1(X, L_a^{^*})$ & $\sum_a h^1(X,L_a^{^*}) = h^1(X,V^{^*})$&\;\;\;\;\; 0
\\ \hline 
 \end{tabular}
 \vskip 0.4cm
\parbox{15cm}{\caption{\it\small The spectrum of $SU(5)$ GUT models derived from the heterotic line bundle construction. In the final column, $|\Gamma|$~stands for the order of the fundamental group of $X$ and  $n_h$ represents the number of $\mathbf{5}-\overline{\mathbf{5}}$ Higgs fields.}\label{spectrum2}}
 \end{center}
 \vspace{-8pt}
 \end{table}

\subsection{Highlights in the moduli space: line bundle models on the tetraquadric}\label{sec:conditions}
The singlet fields are associated with harmonic representatives of the bundle-valued cohomology group $H^1(X,V\otimes V^{^*} ) =H^1(X,\text{End}(V))$. But the 1-cocycle cohomology of $X$ with coefficients in $\text{End}(V)$ corresponds to deformations of $V$, making the singlet fields into bundle moduli. The locus where $V$ splits as a sum of line bundles corresponds to vanishing vacuum values for all the singlet fields. Turning on non-trivial VEVs for some of the singlets corresponds to moving away from the split locus, thus reaching into the moduli space of non-Abelian bundles. We will explore such deformations of split bundles in Section~\ref{sec:monads}. Before that, however, let us discuss the class of split bundles on the tetraquadric hypersurface that lead to $SU(5)$ GUT models with three families of particles, no anti-families and at least one pair of $\mathbf{5}-\overline{\mathbf{5}}$ Higgs fields.

\vspace{8pt}
So far, we have mentioned the tetraquadric manifold in a rather loose manner. We make this more precise here. We start with the class of Calabi-Yau hypersurfaces embedded in a product of four $\IC\IP^1$ spaces, defined as the zero locus of some homogeneous polynomial that is quadratic in the homogeneous coordinates of each $\IC\IP^1$ space, hence the name `tetraquadric'. Manifolds in this class have Euler number $\eta = - 128$ and Hodge numbers $h^{1,1}(X)=4$ and $h^{2,1}(X)=68$. This information is summarised by the following configuration matrix:
\begin{equation}
X~=~~
\cicy{\IC\IP^1 \\   \IC\IP^1\\ \IC\IP^1\\ \IC\IP^1}
{ ~\bf{1} \!\!\!\!\\
  ~\bf{1}\!\!\!\! & \\
  ~\bf{1}\!\!\!\! & \\
  ~\bf{1}\!\!\!\!}_{-128}^{4,68}\
\end{equation}

At certain loci in the complex structure moduli space, the tetraquadric hypersurface admits free actions of finite groups of orders $2, 4, 8$ and $16$. These groups are: $ \IZ_2;\  \IZ_2\times \IZ_2,\ \IZ_4;\  \IZ_2\times \IZ_4,\ \IZ_8,\ \IH;$ $\IZ_4\times \IZ_4,\ \IZ_4 \rtimes \IZ_4,\ \IZ_8\times \IZ_2,\ \IZ_8\rtimes \IZ_2$ and $\IH\times \IZ_2$. Being at one or another of these special loci corresponds to different choices of coefficients for the monomials composing the defining polynomial, as discussed in \cite{Candelas:2008wb, Candelas:2010ve}. In other words, saying that the tetraquadric manifold $X$ admits free quotients by a finite group $\Gamma$ implies a partial fixing of the complex structure of $X$. 

The K\"ahler form on $X$ gets also partially fixed in the compactification, as follows. It can be checked that the second cohomology of $X$ descends entirely from that of the ambient space and hence, the restrictions of the four standard K\"ahler forms on the embedding $\IP^1$ spaces, to the Calabi-Yau hypersurface define a basis for the $(1,1)$-cohomology of $X$. We denote these $(1,1)$-forms, as well as their restrictions, by $\{J_i, 1\leq i\leq4 \}$. A K\"ahler form on $X$ is defined by a set of four real parameters $t^i$, as $J=t^iJ_i$. The supergravity approximation of the heterotic string is valid for $t^i\gg 1$, thus the region of interest is the positive K\"ahler cone. The partial fixing of the K\"ahler form is due to the requirement that the holomorphic bundle $V$ is poly-stable. More concretely, in the case of rank 5 bundles with vanishing first Chern class realised as direct sums of line bundles, poly-stability requires the simultaneous vanishing of the slopes 
\beq
\mu(L_a) = d_{ijk}\, c_1^i(L_a)\, t^j\, t^k
\eeq
for all line bundles $L_a$ in $V =  \oplus_a L_a$. Here $d_{ijk}$ are the triple intersection numbers of $X$. The first Chern classes of the $5$ line bundles composing $V$ are not linearly independent due to the vanishing condition $c_1(V)=0$. If $4$ of these first Chern classes are linearly independent, the locus where $V$ is poly-stable is at most a set of points; if only 3 of them are independent, the poly-stability locus is at most $1$-dimensional and so on. At first, this might seem too restrictive. However, as we elaborate in Section~\ref{sec:monads}, deforming the K\"ahler form away from the locus where the split bundle $V$ is poly-stable can still lead to consistent and viable compactifications if $V$ is simultaneously deformed to some non-Abelian bundle. This is a crucial observation and represents the main motive of this study. 


Let us now present an overview on the class of line bundle compactifications on the tetraquadric manifold.
In Ref.~\cite{Anderson:2013xka} a powerful algorithm for generating large numbers of $SU(5)$ GUT models was outlined. Following this algorithm, a total of $94$ viable models were found on the tetraquadric manifold. The construction  required the knowledge of a simply connected Calabi-Yau manifold $X$ (here $X$ represents the tetraquadric hypersurface) admitting smooth quotients by a finite group $\Gamma$ and equipped with a rank 5 sum of line bundles $V$ satisfying the following requirements:  
\begin{itemize}
\item[$i)$] poly-stability: there exists a point in the positive K\"ahler cone such that $\mu(L_a) = 0$ simultaneously for all line bundles $L_a$ in the sum $V = \oplus_a L_a$
\item[$ii)$] condition for spinors: $c_1(V) = 0$
\item[$iii)$] anomaly cancellation condition with five-branes and trivial hidden sector: $c_2(V).J_i \leq c_2(TX).J_i$, for all $1\leq i \leq h^{1,1}(X)$
\item[$iv)$] three chiral families: $\text{ind}(V) = \text{ind}(\wedge^2 V) = -3 |\Gamma|$ and absence of anti-families: $h^1(X,V^{^*})=0$
\item[$v)$] the presence of $\mathbf{5}-\overline{\mathbf{5}}$ pairs: $h^1(X,\wedge^2 V^{^*})>0$ and a condition necessary in order to project out all Higgs triplets after breaking the GUT group: $\text{ind}\left(L_a\otimes L_b \right)\leq 0$ for all $a<b$
\end{itemize}

All the 94 $SU(5)$ models found in the automated search \cite{Anderson:2013xka} have the correct field content in order to induce low-energy models with the exact matter spectrum of the supersymmetric standard model plus a number of fields -- bundle moduli -- uncharged under the Standard Model gauge group. The details of the automated scan are presented in Table (\ref{tqmodels}). The complete dataset of viable line bundle sums can be accessed here \cite{lbdatabase}.

\vspace{4pt}
\begin{longtable}{| c || c | c | c | c | c |}
\hline
\myalign{| c||}{\varstr{21pt}{16pt}$\ \ \ \  |\Gamma| $ $\ \ \ \ $} &
\myalign{m{3.2cm}|}{$\ $ GUT models} &
\myalign{m{3.5cm}|}{ $\ \ \ \ $ no $ \overline{\mathbf{10}}$ multiplets$\ \ \ $ }&
\myalign{m{3.5cm}|}{$\ \ \ \ \ \ \ $ no $ \overline{\mathbf{10}}\,$s  and $\ \ \ \ \ \ \ $ at least one $\mathbf{5}-\overline{\mathbf{5}}$ pair} & 
$\ \ \ k_{\text{max}}\ \ \ $
\\ \hline\hline
\varstr{14pt}{8pt} 2 & 10 & 10 & 8 & 4 \\
 \hline
\varstr{14pt}{8pt} 4 & 58 & 53 & 46 & 5 \\
 \hline
\varstr{14pt}{8pt} 8 & 64 & 52 & 36 & 7 \\
 \hline
\varstr{14pt}{8pt} 16 & 5 & 5 & 4 & 6 \\
 \hline
 \captionsetup{width=16cm}
 \caption{\itshape Statistics on the number of models on the tetraquadric manifold. $k_{\text{max}}$ represents the first maximal value for the line bundle entries (in modulus) beyond which no further models could be found in the automated search.}\label{tqmodels}
\end{longtable}

It has been noticed in \cite{Anderson:2013xka} that the number of viable models reaches a certain saturation limit after repeatedly increasing the range of integers, indicated in Table \ref{tqmodels} by $k_{\text{max}}$, which define the first Chern class of the line bundles. We would like to understand this behaviour in more detail. Before that, however, let us discuss a number of topological aspects in relation to the tetraquadric hypersurface, as well as line bundle sums constructed over this manifold.

\section{The Tetraquadric Hypersurface}\label{app:KahlerCone}
Let $X$ denote a generic (smooth) hypersurface embedded in a product of four $\IP^{1}$ spaces, $\cA=\prod_{a=1}^4 \IP^1$ defined as the zero locus of a polynomial of multi-degree (2,2,2,2) in the homogeneous coordinates of the four projective spaces. 
Let $\{J_i, 1\leq i\leq4\}$ denote the standard K\"ahler forms on the four embedding $\IP^{1}$ spaces. The restrictions $\left.J_i\right|_X$ span $H^2(X, TX)$. In the following, we will use the same notation $J_i$ when referring to the restrictions of the K\"ahler forms to $X$. A K\"ahler form on $X$ is defined by a set of four real parameters $t^i$, as $J=t^iJ_i$. The supergravity approximation of the heterotic string is valid for $t^i\gg 1$, thus the region of interest is the positive K\"ahler cone. For further reference, denote this cone by
\beq
C_{{\bf t}} =\left\{ {\bf t} \in \IR^4\ \left|\ t^i\geq 0,\, 1\leq i\leq 4\right. \right\} 
\eeq

The second Chern class of $X$ is given by $c_2(X).J_i = (24,24,24,24)$. The triple intersection numbers have the following simple expression: 
\beq
d_{ijk} = \int_X J_i\wedge J_j\wedge J_k = \begin{cases} 2 & \mbox{ if } i\neq j, j\neq k \\ 0 &\mbox{ otherwise } \end{cases}
\eeq

This leads to the following simple expressions for the volume $\kappa = d_{ijk}\, t^i\,t^j\,t^k$ and its first derivatives $\kappa_i = d_{ijk}\,t^j\,t^k$:
\begin{align}
\kappa & = 12\, \left( t_1\,t_2\,t_3 + t_1\,t_2\,t_4 + t_1\,t_3\,t_4 + t_2\,t_3\,t_4\right)\\
\kappa_{i_1} & = 4\, \left( t_{i_2}\,t_{i_3} + t_{i_2}\,t_{i_4} +t_{i_3}\,t_{i_4} \right) 
\end{align}
for $\left(i_1, i_2, i_3, i_4 \right)$ any permutation of $(1,2,3,4)$. The derivatives $\kappa_i$ turn out to be useful for the purpose of discussing the slope of vector bundles, defined as $\mu(V) = d_{ijk}\, c_1^i(V)\, t^j\, t^k = c_1^i(C)\,\kappa_i$. 

\vspace{12pt}
Poly-stability imposes certain constraints on the slope, which restrict the allowed K\"ahler cone. Since it is easier to discuss these constraints in the $\kappa_i$ variables, let us define the dual K\"ahler cone as the image of $C_{\bf t}$ under the map $f({\bf t}) = \left( \kappa_1, \kappa_2, \kappa_3, \kappa_4\right) /4$. Denote the new coordinates by ${\bf s}\in \IR$ and define the vectors ${\bf n}_i = {\bf n}-{\bf e}_i$, where ${\bf n} = (1,1,1,1)/2$ and ${\bf e}_i$ are the standard basis vectors in $\IR^4$. The dual K\"ahler cone, $f(C_{{\bf t}})$ is contained as a dense subset in the cone: 
\beq 
C_{{\bf s}} = \left\{ {\bf s}\in\IR^4 \ \left|\  {\bf n}_i\cdot {\bf s} \geq 0,\ {\bf e}_i\cdot {\bf s} \geq 0,\ 1\leq i\leq 4\right.\right\}
\eeq

The inclusion $f(C_{{\bf t}}) \subset C_{{\bf s}}$ is straightforward: since $t^i\geq 0$ for all $1\leq i\leq 4$, ${\bf e}_i\cdot f({\bf t}) = \kappa_i/4 \geq 0$ and ${\bf n}_i\cdot f({\bf t}) = \sum_{j\neq i} t_i\,t_j \geq 0$. In order to prove that the inclusion is dense, we need to show that almost every point ${\bf s}\in C_{{\bf s}}$ has at least one pre-image in $C_{\bf t}$. For this purpose, it is useful to define a new set of coordinates $x_i = {\bf n}_i\cdot{\bf s}$. Then ${\bf s}\in C_{\bf s}$ is equivalent with $x_i\geq 0$ and ${\bf n}_i\cdot {\bf x}\geq 0$, since ${\bf e}_i\cdot {\bf s}  =\left( {\bf n}-{\bf n}_i\right) \cdot {\bf s} = \left( \frac{1}{2}\sum{\bf n}_j-{\bf n}_i\right) \cdot {\bf s} = \frac{1}{2}\sum x_j - x_i = {\bf n}_i \cdot {\bf x}$. Without loss of generality, we will assume in the following that $x_1\leq x_2\leq x_3 \leq x_4$. 

If ${\bf t}$ is a pre-image of ${\bf x}$, then 
\beq 
{\bf n}_i\cdot f({\bf t})  = t_i \left(\sum_{j=1}^4 t_j - t_i \right) \stackrel{!}{=} x_i
\eeq

This can happen if we write 
\beq
t_i = \tau - \epsilon_i \sqrt{\tau^2 - x_i}\ \ \ \ \ \ \tau = \frac{1}{2} \sum_{i=1}^4 \epsilon_i \sqrt{\tau^2-x_i} 
\eeq
where $\epsilon_i$ are signs and we choose $\epsilon_1 = \epsilon_2 = \epsilon_3 =1$. Deciding whether ${\bf x}$ has a pre-image ${\bf}$ amounts to deciding whether the function $g: [\sqrt{x_4}, \infty) \rightarrow \IR$, defined by
\beq 
g(\tau)  = \tau - \frac{1}{2} \sum_{i=1}^4 \epsilon_i \sqrt{\tau^2-x_i}
\eeq
vanishes somewhere. Since $g$ is continuous, the existence of a vanishing point is guaranteed if $g\left(\sqrt{x_4}\right)$ and $g(\tau)_{\tau\rightarrow\infty}$ have different signs. We can use the freedom of choosing $\epsilon_4$ in order to arrange for this. Note, however, that $g\left(\sqrt{x_4}\right)$ does not depend on this choice.

\vspace{12pt}
If $g\left( \sqrt{x_4}\right)=0$, we already have a vanishing point.  This happens if $x_1=x_2=0$ and $x_3=x_4$. Otherwise, if $g\left( \sqrt{x_4}\right)>0$, we can choose $\epsilon_4=1$, such that $g(\tau)_{\tau\rightarrow\infty}<0$. If $g\left( \sqrt{x_4}\right)<0$ we choose $\epsilon_4 = -1$. However, this final case needs a more extended discussion. In this case, for large~$\tau$,
\beq 
g(\tau) \simeq \frac{1}{4\,\tau}\left(x_1 + x_2 + x_3 - x_4\right) + \frac{1}{16\,\tau^3}\left(x_1^2 + x_2^2 + x_3^2 - x_4^2\right)+\ldots
\eeq
If $x_1 + x_2 + x_3 - x_4>0$, i.e.~${\bf n}_4\cdot {\bf x}>0$, then $g(\tau)_{\tau\rightarrow\infty} >0$, which guarantees the existence of a vanishing point in the interval $[\sqrt{x_4}, \infty)$. If ${\bf n}_4\cdot {\bf x}=0$ and $x_2 > 0$ (the second condition excludes the case $x_1=x_2=0$ discussed above), $g(\tau)_{\tau\rightarrow\infty} < 0$, thus the above criterion cannot be used in order to decide whether $g$ has any vanishing points. In fact, in these cases it is easy to check that the equations $f({\bf t}) = {\bf s}$ are inconsistent. Thus the corresponding points, which lie on the boundary of $C_{\bf s}$ do not belong to the dual K\"ahler cone.

\section{Topological Identities for Line Bundle on the Tetraquadric}\label{app:lbtopology}

A line bundle $L$ is determined by its first Chern class, $c_1(L) = k^i J_i$. This is also reflected by the notation $L = \cO_X({\bf k})$. The Chern character  for a single line bundle is given by
\beq
\text{ch}(L) = e^{\,c_1(L)}  = 1 + c_1(L) + \frac{1}{2} c_1(L)^2+ \frac{1}{3!} c_1(L)^3 
\eeq

This implies, on the tetraquadric, 
\begin{align}
\text{ch}_2(L).J_i &= \int_X \frac{1}{2} c_1(L)^2 \wedge J_i = \frac{1}{2}\,d_{ijk}\,k^j\,k^k = 2\sum_{\stackrel{j<k}{j,k\neq i}} k^j\,k^k\\
\int_X \text{ch}_3(L) & = \frac{1}{3!}\, d_{ijk}\,k^i\,k^j\,k^k = 2\sum_{i<j<k} k^i\,k^j\,k^k
\end{align}

For a sum of line bundles, $\displaystyle V =\bigoplus_{a=1}^r \cO_X({\bf k}_a)$, the Chern characters are defined additively:
\begin{align}
\text{ch}_1(V) &= \left(\sum_{a=1}^r k^i_a \right) J_i\\
\label{eq:c2}\text{ch}_2(V).J_i &= \frac{1}{2}\,d_{ijk}\sum_{a=1}^r\,k_a^i\,k_a^j=  2\sum_a\sum_{\stackrel{j<k}{j,k\neq i}} k_a^j\,k_a^k\\
\int_X \text{ch}_3(V) & = \frac{1}{3!}\, d_{ijk}\sum_{a=1}^r\,k_a^i\,k_a^j\,k_a^k = 2\sum_{a=1}^r\sum_{i<j<k} k^i\,k^j\,k^k
\end{align}

The index of $V$ on a Calabi-Yau manifold $X$, for which $\text{Td}(TX) = 1 +c_2(TX)/12$ can be computed using the index theorem:
\beq
\begin{aligned} 
\text{ind} (V) & = \int_X \text{ch}(V)\, \text{Td}(TX) = \int_X \text{ch}_3(V) + \frac{1}{12}\, \text{ch}_1(V)\,c_2(TX)\\
& = \frac{1}{6} \sum_{a=1}^r \left( d_{ijk}\,k_a^i\,k_a^j\,k_a^k +\frac{1}{2}\, k_a^i \left(c_2(TX).J_i \right) \right) \\
& = 2\sum_{a=1}^r\sum_{i<j<k} k^i\,k^j\,k^k + 2\sum_{a=1}^r \sum_{i} k_a^i
\end{aligned}
\eeq

The slope of a line bundle is defined as
\beq 
\mu(L) = \int_X c_1(L) \wedge J\wedge J = d_{ijk}\,k^i\,t^j\,t^k = k^i\kappa_i
\eeq
and will play a central role in deciding stability of vector bundles.

\section{Finiteness of the Class of Line Bundle Models}\label{sec:finiteness}
The automated scan presented in Chapter~\ref{LineBundles} searched for viable line bundle models constructed over complete intersection Calabi-Yau manifolds in products of projective spaces. These compactifications lead to $SU(5)$ GUT models, which have all the necessary features in order to descend to standard-like models after quotienting the Calabi-Yau manifold by the free action of a finite group $\Gamma$ and introducing discrete Wilson lines. 
An important result of this computer scan was the observation that given a CICY manifold admitting discrete symmetries of a fixed order, the number of consistent and physically viable models is an increasing and saturating function of the maximal line bundle entry in modulus. A similar observation was made in \cite{He:2013ofa}, in relation to $SU(5)$ and $SO(10)$ GUT models constructed over non-simply connected Calabi-Yau hypersurfaces embedded in toric varieties. In Figure~\ref{SaturationPlots} we illustrate this observation for the tetraquadric manifold and symmetry orders $2$, $4$, $8$ and $16$.

All the models constructed in \cite{Anderson:2013xka} satisfy the following bound on the entries of the first Chern class of each line bundle: 
\begin{equation}\label{eq:boundc1}
 |c_1^r(L_a)|\leq 10\; 
\end{equation} 

This empiric bound comes as a result of imposing the various requirements on the line bundle sums enumerated in Section~\ref{sec:conditions}. Analysing the manner in which these constraints conspire in order to give the bound \ref{eq:boundc1} is untidy. We carry out an inquiry of this kind for the tetraquadric manifold in Section~\ref{app:finiteness}. In the next section we work out a bound on the line bundle integers for rank 2 line bundle sums. In Section~\ref{sec:physbound} we propose a transparent argument to derive a bound for line bundle sums of arbitrary rank; however, in order to make the argument feasible, we have to restrict the K\"ahler cone by imposing two constraints from Physics: $t^i>1$, linked to the validity of the supergravity approximation, and finite Calabi-Yau volume.

\vspace{-20pt}

\begin{flushright}
$${\includegraphics[width=3.1in]{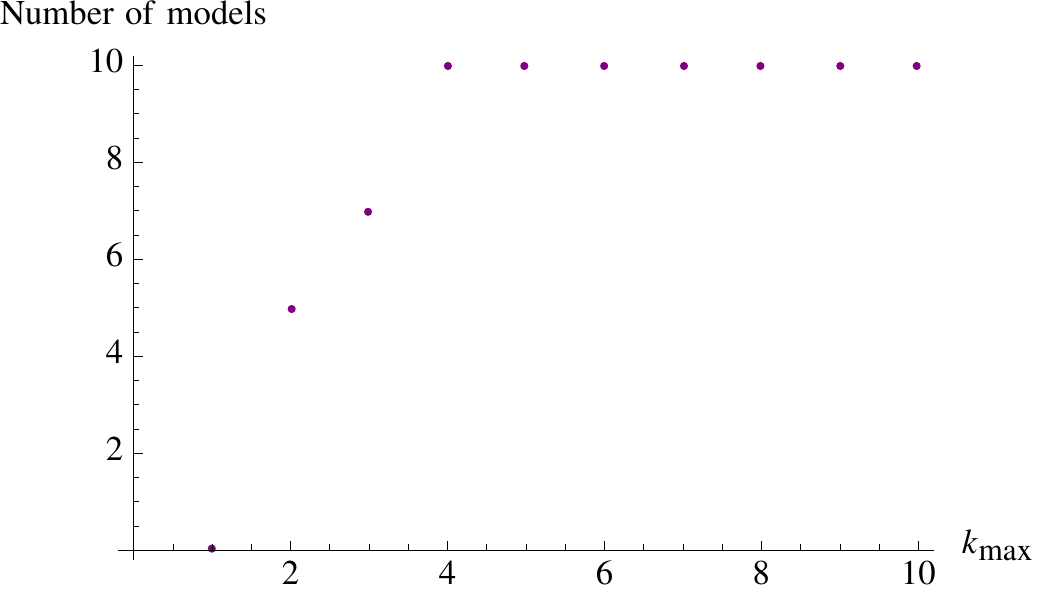}}
\hskip 2pt \lower 0pt\hbox{\includegraphics[width=3.1in]{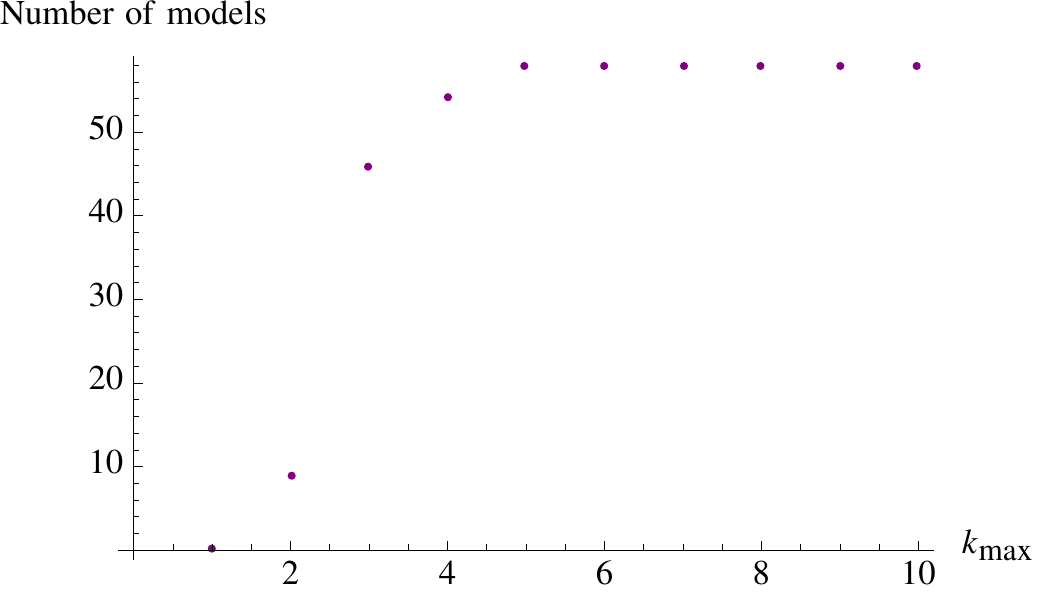}}
$$
\end{flushright}
\vspace{-21pt}
\begin{flushright}
$${\includegraphics[width=3.1in]{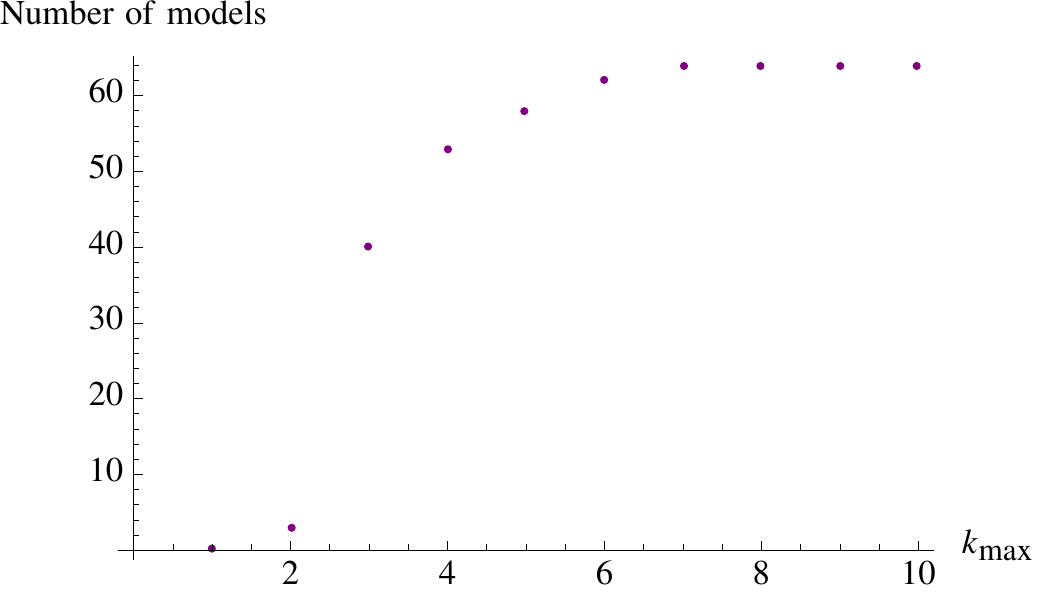}}
\hskip 2pt \lower 0pt\hbox{\includegraphics[width=3.1in]{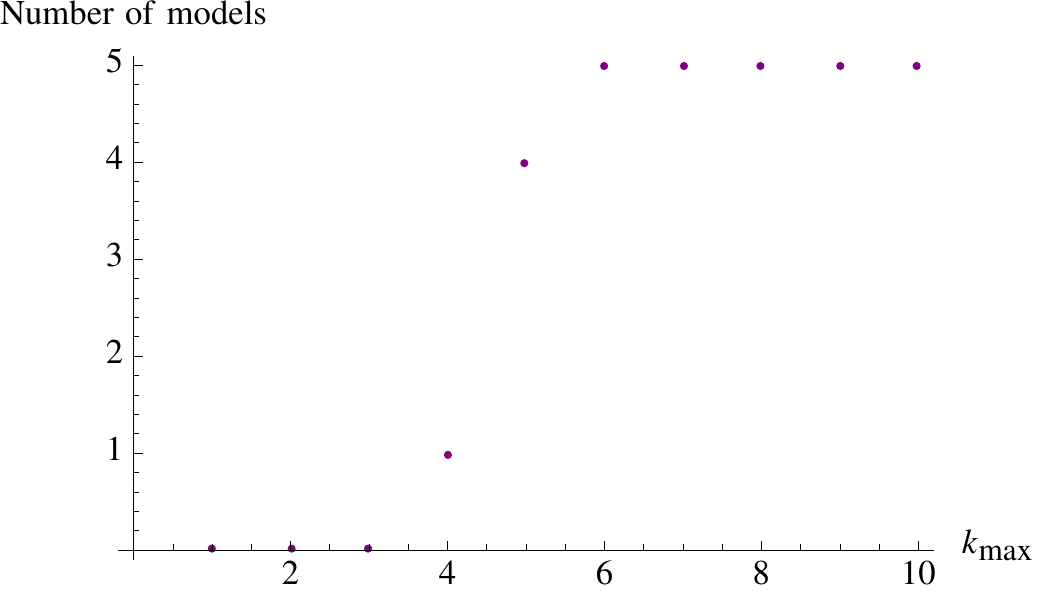}}
$$
\end{flushright}
\vspace{-21pt}
\begin{figure}[h!]
\vspace{12pt}
\begin{center}
 \captionsetup{width=16cm}
\caption{\itshape The plots show the number of line bundle models (before imposing the absence of $\overline{\mathbf{10}}$ multiplets and the existence of $\mathbf{5}-\overline{\mathbf{5}}$ pairs) on the tetraquadric manifold as a function of the maximal line bundle entry in modulus. The four plots correspond to $|\Gamma| = 2, 4, 8$ and $16$.}\label{SaturationPlots}
\end{center}
\vspace{-20pt}
\end{figure}

\subsection{A workable example: rank two line bundle sums}\label{sec:tqbound}
Rank two line bundle sums with vanishing first Chern class have the form $V = L\oplus L^{^*}$. We would like to show that the number of such bundles on the tetraquadric manifold $X$ satisfying $\mu(L)=0$ and $c_2(V).J_i \leq c_2(TX).J_i = 24$,  for all $i$ in the range $1\leq i\leq 4$, is finite. In particular, using the notation $L = \cO_X(\mathbf{k})$, we are able to show that the inequalities $-7<k^i<7$ must hold for all the components $k^i$ of $\mathbf{k}$.
We start by writing the slope of a line bundle $L =\cO(\mathbf{k})$ as 
\beq
\mu(L) = \int_X c_1(L)\wedge J \wedge J = d_{ijk}\,k^i\,t^j\,t^k = \kappa_i k^i 
\eeq
where the `dual' K\"ahler parameters are defined as $\kappa_i = d_{ijk}\,t^j\,t^k$. The equation $\mu(L)=0$ defines a hyperplane in the $\kappa_i$ variables, thus (poly)-stability is most easily discussed in these variables. However, we first need to find out what the positive K\"ahler cone $t^i>0$ becomes in these new variables. For the tetraquadric manifold, the triple intersection numbers have the following simple expression: 
\beq
d_{ijk} = \int_X J_i\wedge J_j\wedge J_k = \begin{cases} 2 & \mbox{ if } i\neq j, j\neq k \\ 0 &\mbox{ otherwise } \end{cases}
\eeq
and hence,
$\kappa_{i_1}  = 4\, \left( t_{i_2}\,t_{i_3} + t_{i_2}\,t_{i_4} +t_{i_3}\,t_{i_4} \right) $,
for $\left(i_1, i_2, i_3, i_4 \right)$ any permutation of $(1,2,3,4)$. 
Let us define the dual K\"ahler cone as the image of $C_{\bf t}$ under the map $f({\bf t}) = \left( \kappa_1, \kappa_2, \kappa_3, \kappa_4\right) /4$. Denote the new coordinates by ${\bf s}\in \IR$ and define the vectors ${\bf n}_i = {\bf n}-{\bf e}_i$, where ${\bf n} = (1,1,1,1)/2$ and ${\bf e}_i$ are the standard basis vectors in $\IR^4$. In Section~\ref{app:KahlerCone} we have shown that the dual K\"ahler cone, $f(C_{{\bf t}})$ is contained as a dense subset in the cone: 
\vspace{-4pt}
\beq 
C_{{\bf s}} = \left\{ {\bf s}\in\IR^4 \ \left|\  {\bf n}_i\cdot {\bf s} \geq 0,\ {\bf e}_i\cdot {\bf s} \geq 0,\ 1\leq i\leq 4\right.\right\}
\vspace{-4pt}
\eeq

I would like to answer the question which line bundles allow for a vanishing slope somewhere in the positive K\"ahler cone of the tetraquadric manifold. To this end, let us introduce the cone $\check{C}_{\mathbf k}$, dual to $C_{\mathbf s}$ in the standard way by
$\check{C}_{\mathbf k} = \{\, {\mathbf k} \in \IR^4 \left | {\mathbf k\cdot \mathbf s} \geq 0, \forall \mathbf s\in C_{\mathbf s} \right. \}$.
%
Introducing the vectors ${\mathbf e}_{ij} = {\mathbf e}_i + {\mathbf e}_j$, it is straightforward to show that
\vspace{-4pt}
\beq
\check{C}_{\mathbf k} = \{\, {\mathbf k} \in \IR^4 \left | {\mathbf k\cdot \mathbf e}_{ij} \geq 0, \forall i<j \right. \}
\vspace{-4pt}
\eeq

Now, it can be shown  that for a given ${\mathbf k} \in \IZ^4$, the equation ${\mathbf k\cdot \mathbf s}  = 0$ has a solutions with ${\mathbf s}$ in the interior $\mathring{C}_{\mathbf s}$, if and only if $\mathbf k \notin \check{C}_{\mathbf k} \cup \left( -\check{C}_{\mathbf k}\right)$.
This means that a line bundle $L = \cO_X(\mathbf k)$ allows for a zero-slope solution somewhere in the interior of the K\"ahler cone if and only if $\mathbf k$ has two components $k_i, k_j$ with $k_i + k_j< 0$ and two components $k_l,
k_m$ with $k_l + k_m<0$. Thus, in order to satisfy the slope zero condition, we can have two possibilities, up to permuting components:
\vspace{-4pt}
\beq\label{ineq1}
\begin{aligned}
&k_1 + k_2 > 0 \text{ and } k_1 + k_3 < 0 \text{ or } \\
&k_1 + k_2 > 0 \text{ and } k_3 + k_4 < 0 
\end{aligned}
\vspace{-4pt}
\eeq

Further, from the second Chern class condition we have that $c_2(V).J_i \leq 24$. But for $V = L\oplus L^{^*}$, we have $c_2(V) = -\text{ch}_2(V) = - 2\,\text{ch}_2(L)$ and, according to (\ref{eq:c2}), $\text{ch}_2(L).J_i = 2\left( k_j\,k_l + k_m(k_i + k_j)\right)$, with $i\neq j, l, m$. Thus, we obtain an inequality
\vspace{-9pt}
\beq\label{ineq2}
k_i k_j + k_m(k_i + k_j)\geq -6
\eeq
for each triplet of different $i, j, m$. It can be shown, for example using Mathematica, that the system of integer
inequalities given by (\ref{ineq1}) and (\ref{ineq2}) has a finite number of solutions $\mathbf k = (k_1,k_2,k_3,k_4)$ which all satisfy $-7< k_i <7$.

\subsection{A bound rooted in Physics}\label{sec:physbound}

A more elegant way to translate this finiteness question into a mathematical statement is to consider, on a given Calabi-Yau manifold $X$, the line bundle sums $V=\oplus_a L_a$ with a fixed total Chern character $\text{ch}(V)$, such that the slopes of all constituent line bundles simultaneously vanish somewhere in the interior of the positive K\"ahler cone. We would like to show that the number these bundles is finite. 

To this end, consider the usual K\"ahler moduli space metric \cite{Candelas:1990pi}:
\begin{equation}
G_{ij} = \frac{1}{2\,\text{Vol}(X)} \int_X J_i\wedge \star J_j = -3 \left( \frac{\kappa_{ik}}{\kappa} - \frac{2\kappa_i\kappa_j}{3\kappa^2}\right) 
\end{equation}
where $\text{Vol}(X)$ is the Calabi-Yau volume with respect to the K\"ahler metric, $\kappa = d_{ijk}\, t^i\,t^j\,t^k$, $\kappa_i = d_{ijk}\,t^j\,t^k$ and $\kappa_{ij}=d_{ijk}\,t^k$. The proportionality between $\kappa$ and the Calabi-Yau volume is $\text{Vol}(X)=\kappa/6$. The line bundles $L_a$ are defined by their first Chern class, $L_a = \cO_X(\vec{k}_a)$. Let $J$ denote the K\"ahler form on $X$. With these notations, the slope zero-conditions can be written as \beq
\mu(L_a) = \int_X c_1(L_a)\wedge J \wedge J = d_{ijk}\,k_a^i\,t^j\,t^k = \kappa_i k^i_a = 0
\eeq

Then consider the sum 
\begin{equation}\label{eq:bound1}
\sum_a \vec{k}_a^T G \vec{k}_a = -\frac{3}{\kappa} d_{ijk} \sum_a k_a^i k_a^j t^k = -\frac{6}{\kappa} t^i \text{ch}_{2i}(V) \leq \frac{6}{\kappa} |\vec{t}| |\text{ch}_{2i}(V) |
\end{equation}

Thus 
\begin{equation}\label{eq:bound2}
\sum_a \vec{k}_a^T\, \frac{\kappa\, G}{6\,|\vec t |}\, \vec{k}_a  \leq |\text{ch}_{2i}(V) |
\end{equation}

For a fixed K\"ahler class $\vec{t}$ in the interior of the K\"ahler cone, $\kappa$ (which is proportional to the Calabi-Yau volume) is bounded from below and $G_{ij}$ is positive definite with eigenvalues bounded from below. Hence, for a fixed $\text{ch}_2(V)$, the above inequality constraints the available integer vectors $k_a$ to a finite set. Denoting by $\lambda_{\text{min}}$ the lower bound on the eigenvalues of $\widetilde G  = \kappa\, G / 6 |\vec t|$, we obtain the following bound:
\beq
\sum_a \vec{k}_a^T\, \vec{k}_a  \leq\frac{ |\text{ch}_{2i}(V) |}{\lambda_{\text{min}}}
\eeq

 This statement applies to all Calabi-Yau three-folds. However, it has a limitation which is relevant for the physics application we are discussing. In physics, we are not interested in fixing the K\"ahler class. Rather, we are looking for line bundle sums $V$ which satisfy the slope zero conditions anywhere in the interior of the K\"ahler cone. In this case, the K\"ahler class may approach the boundary of the K\"ahler cone in which case the eigenvalues of $\widetilde G$ are no longer bounded from below. 

In this case, the above argument has to be supplemented by the physical requirements that the Calabi-Yau volume is finite and that the supergravity limit $t^i\gg1$ holds. Typically, the Calabi-Yau volume is of order 1, and we can require $\text{Vol}(X)\lesssim 10$. This bound, together with the requirement $t^i\geq1$, defines a finite region in the K\"ahler cone in which we can study the eigenvalues of $\widetilde G$. 

This approach can be carried out for any Calabi-Yau manifold, leading to a bound on the entries of viable line bundle models. In the present paper we implement this method for the tetraquadric manifold. In this case, the entries of $\widetilde G $ have the simple expression: 
\beq
\widetilde G_{ij} = \frac{1}{\sqrt{t_1^2+t_2^2+t_3^2+t_4^2}}\, \left({\displaystyle \sum_{a<b<c} t_a\,t_b\,t_c}\right)^{-1}\, {\displaystyle \sum_{\stackrel{a<b}{a,b\neq i,j}} t_a^2\, t_b^2}
\eeq
and, since $c_1(V)=0$, we have $\text{ch}_{2}(V)=-c_2(V)$, which is bounded by $c_2(TX)$ due to,  the anomaly cancellation condition: $c_2(V).J_i \leq c_2(TX).J_i$. For the tetraquadric manifold, 
\beq
c_2(TX).J_i = (24,24,24,24)
\eeq
leading to $ |\text{ch}_{2i}(V) | \leq 48$. We approach numerically the problem of finding the minimal eigenvalue of $\widetilde G$ in the region defined by $t^i\geq 1$ and $\text{Vol}(X)\lesssim 10$, thus deriving for different values of $\text{Vol}(X)\lesssim 10$ the bond
\beq\label{eq:bound}
\sum_a \vec{k}_a^T\, \vec{k}_a  \leq\frac{ 48}{\lambda_{\text{min}}}
\eeq

\begin{figure}[h!]
\vspace{2pt}
\begin{center}
\includegraphics[width=4in]{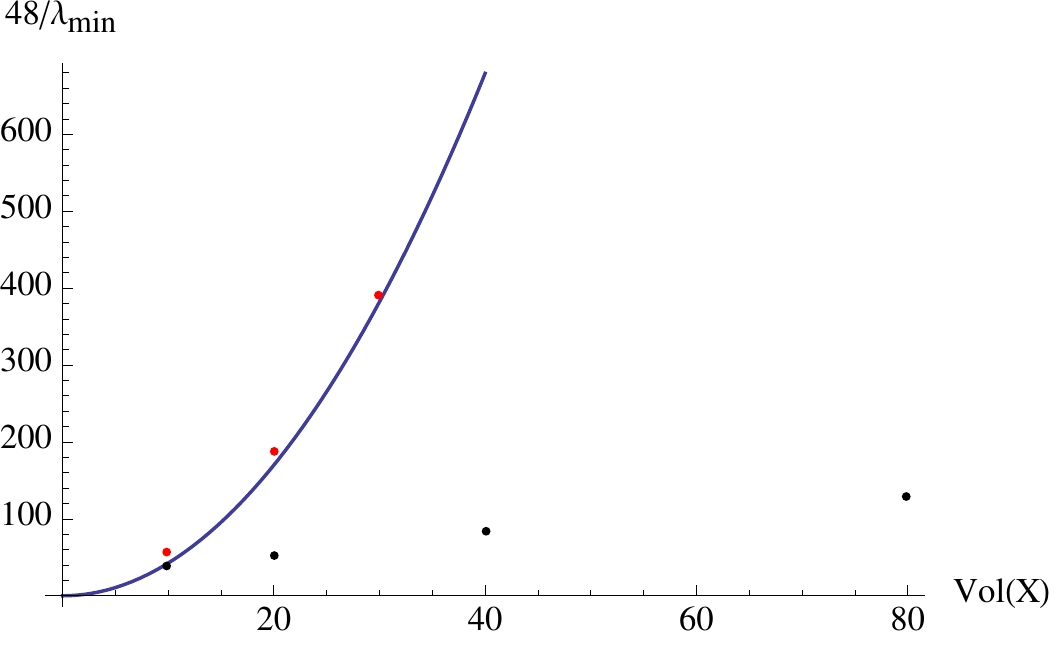}
\captionsetup{width=16cm}
\caption{\itshape The plot shows the dependence of the bound $48/\lambda_{\text{min}}$ (vertical axis) on the Calabi-Yau volume (horizontal axis). The red points represent the values obtained by numerical methods. The blue curve represents the best fit with a parabola passing through the origin. The black points represent the maximal value of $\sum_a \vec{k}_a^T\, \vec{k}_a$ among the rank 5 line bundle models constructed in \cite{Anderson:2013xka} over the tetraquadric manifold for $|\Gamma| = 2,4,8\text{ and }16$. For, e.g.,~$|\Gamma| = 2$ we have set $\text{Vol}(X) \simeq 10$, where $X$ represents the tetraquadric manifold which, in this case, serves as a double cover for the quotient manifold on which the physical models live.}\label{BoundVsVolume2}
\end{center}
\end{figure}

\vspace{10pt}
In Figure~\ref{BoundVsVolume2} we plot the bound $48/\lambda_{\text{min}}$ for $\text{Vol}(X)\lesssim 50$. We also include the maximal value of $\sum_a \vec{k}_a^T\, \vec{k}_a$ among the viable models constructed in \cite{Anderson:2013xka} over the tetraquadric manifold for each value of the group order $|\Gamma| = 2,4,8\text{ and }16$. Since the physical models live on the quotient manifold $X/ \Gamma$, the volume bound has to be imposed for $X/ \Gamma$ and not for $X$. Estimating $\text{Vol}(X/ \Gamma) \simeq 5$, we get $\text{Vol}(X) \simeq 5 |\Gamma|$. The plot shown in Figure~\ref{BoundVsVolume2}, illustrates the fact that the theoretical bound~(\ref{eq:bound}) is stronger than the actual bound found in the computer search. 
This discrepancy is not surprising. The bound (\ref{eq:bound}) applies to line bundle sums of arbitrary rank, while the bounds found from the results of the computer scan take into account only rank five bundles. The experience derived from the work \cite{He:2013ofa} shows that rank four bundles can have larger line bundle entries, while satisfying the same topological constraints. 

\vspace{12pt}
In Section~\ref{app:finiteness} I will present an argument which shows the way in which the different constraints imposed on a line bundle sum conspire in order to render the number of viable models finite. For that end, however, we need a precise knowledge of line bundle cohomology on the tetraquadric manifold.

\section{Line Bundle Cohomology on the Tetraquadric}\label{app:tqcoh}
Let $X$ denote a hypersurface embedded in a product of four $\IP^{1}$ spaces, $\cA=\prod_{a=1}^4 \IP^1$ defined by a polynomial of multi-degree (2,2,2,2) in the coordinates of the four projective spaces. If the polynomial is sufficiently generic, $X$ is smooth and Calabi-Yau and $h^{1,1}(X)=4$. Moreover, this embedding is favourable (in the sense of \cite{Anderson:2013xka}), such that the second cohomology of $X$ descends entirely from that of the embedding space. As such, all possible line bundles on $X$ can be obtained as restrictions of line bundles on $\cA$ to $X$. Let $\vec{k}$ denote the first Chern class of a line bundle. For the ease of expression, it will be assumed that $k_1\leq k_2 \leq k_3\leq k_4$ in the following discussion.

\vspace{20pt}
The ranks of the ambient space cohomology groups can be obtained using Bott's formula:
\beq\label{bott}
h^q(\IP^n,\cO(k)) =
\left\{\ba{lll}
~~{k+n \choose n} & q = 0 & k\geq 0,\\[12pt]
{-k-1 \choose -k-n-1} & q = n & k\leq-n-1,\\[9pt]
~~~~~0 & \mbox{otherwise} & 
\ea\right.
\eeq
and  K\"unneth formula, which gives the cohomology of line bundles over direct products of
projective spaces $\cA = \IP^{n_1} \times \ldots \times \IP^{n_m}$:
\beq\label{kunneth1}
H^q(\cA, \cO\big(k_1, \ldots, k_m)\big) =
\bigoplus_{q_1+\ldots+q_m = q} H^{q_1}\big(\IP^{n_1},\cO(k_1)\big) \times
\ldots \times H^{q_m}\big(\IP^{n_m},\cO(k_m)\big) \ ,
\eeq

\vspace{10pt}
Thus, for a product of four projective spaces $\cA=\prod_{a=1}^4 \IP^1$, we obtain: 
\begin{align}
\mathrm{dim}\, H^0\big{(}\cA, \calO (\vec{k}) \big{)} & = \prod_{a=1}^4 (k_a +1)^+\\
\mathrm{dim}\, H^1\big{(}\cA, \calO (\vec{k}) \big{)} & =  (-k_1 -1)^+ \prod_{a=2}^4 (k_a +1)^+ \\[4pt]
\mathrm{dim}\, H^2\big{(}\cA, \calO (\vec{k}) \big{)} & =  (-k_1 -1)^+ (-k_2 -1)^+ (k_3 +1)^+(k_4 +1)^+\\[4pt]
\mathrm{dim}\, H^3\big{(}\cA, \calO (\vec{k}) \big{)} & = (k_4 +1)^+\prod_{a=1}^3(-k_a -1)^+ \\
\mathrm{dim}\, H^4\big{(}\cA, \calO (\vec{k}) \big{)} & = \prod_{a=1}^4 (-k_a -1)^+
\end{align}
where the notation $f^+$ indicates the positive part of a function $f^+(x)= \mathrm{max}\,(f(x),0)$.
 
\vspace{21pt}
Line bundle cohomology on $X$ can be computed using the Koszul resolution sequence:
\beq
0\ \longrightarrow\ \cO_{\cA}(\vec{k}) \otimes \cN_X^{^{*}} \longrightarrow\ \cO_{\cA}(\vec{k})\ \longrightarrow\ \cO_{X}(\vec{k})\ \longrightarrow\ 0
\eeq
where $\cN_X^{^{*}}$ is the dual to the normal bundle $\cN_X = \cO_{\cA}(2,2,2,2)$. The associated long exact sequence in cohomology reads: 
\begin{equation}
\begin{array}{ccccccccc}
0& \longrightarrow & H^0(\cA,\cO(\vec{k}) \otimes \cN_X^{^{*}})& \longrightarrow& H^0(\cA,\cO(\vec{k})) & \longrightarrow&H^0(X,\cO(\vec{k}))&\longrightarrow&\\[4pt]
& \longrightarrow & H^1(\cA,\cO(\vec{k}) \otimes \cN_X^{^{*}})& \longrightarrow& H^1(\cA,\cO(\vec{k})) & \longrightarrow&H^1(X,\cO(\vec{k}))&\longrightarrow&\\[4pt]
& \longrightarrow & H^2(\cA,\cO(\vec{k}) \otimes \cN_X^{^{*}})& \longrightarrow& H^2(\cA,\cO(\vec{k})) & \longrightarrow&H^2(X,\cO(\vec{k}))&\longrightarrow&\\[4pt]
& \longrightarrow & H^3(\cA,\cO(\vec{k}) \otimes \cN_X^{^{*}})& \longrightarrow& H^3(\cA,\cO(\vec{k})) & \longrightarrow&H^3(X,\cO(\vec{k}))&\longrightarrow&0\\
\end{array}
\end{equation}

Thus we can express the cohomology groups on $X$ as:
\beq
\begin{aligned}
H^{q}\big(X,\cO(\vec{k})\big)\  =\ \ & \text{Coker}\left( H^q(\cA,\cO(\vec{k}) \otimes \cN_X^{^{*}}) \longrightarrow H^q(\cA,\cO(\vec{k}))  \right) \ \oplus\\
\oplus\ \ & \text{Ker}\left( H^{q+1}(\cA,\cO(\vec{k})\otimes \cN_X^{^{*}}) \longrightarrow H^{q+1}(\cA,\cO(\vec{k}))  \right)
\end{aligned}
\eeq

\vspace{-4pt}
In general, the ranks of these maps are maximal, in which case 
\begin{equation}
\begin{aligned}
\mathrm{dim}\, H^0\big{(}X, \calO (\vec{k}) \big{)}  & = \Big( \prod_{a=1}^4 (k_a +1)^+  - \prod_{a=1}^4 (k_a -1)^+\Big)^ + \\
& + \Big((-k_1 +1)^+ \prod_{a=2}^4 (k_a -1)^+ -(-k_1 -1)^+ \prod_{a=2}^4 (k_a +1)^+ \Big)^ + \\[7pt]
\mathrm{dim}\, H^1 \big{(}X, \calO (\vec{k}) \big{)}  & = \Big((-k_1 -1)^+ \prod_{a=2}^4  (k_a +1)^+- (-k_1 +1)^+ \prod_{a=2}^4 (k_a -1)^+  \Big)^ + \\
 & + \Big( \prod_{a=1}^2(-k_a+1)^+\prod_{a=3}^4(k_a-1)^+ -  \prod_{a=1}^2(-k_a-1)^+\prod_{a=3}^4(k_a+1)^+ \Big)^+ \\[7pt]
\mathrm{dim}\, H^2\big{(}X, \calO (\vec{k}) \big{)}  & =  \Big( \prod_{a=1}^2(-k_a-1)^+\prod_{a=3}^4(k_a+1)^+ -  \prod_{a=1}^2(-k_a+1)^+\prod_{a=3}^4(k_a-1)^+ \Big)^+ \\
& + \Big( (k_4 -1)^+\prod_{a=1}^3(-k_a + 1)^+ - (k_4 +1)^+\prod_{a=1}^3(-k_a -1)^+  \Big)^+ \\[7pt]
\mathrm{dim}\, H^3\big{(}X, \calO (\vec{k}) \big{)} &  = \Big( (k_4 +1)^+\prod_{a=1}^3(-k_a -1)^+ - (k_4 -1)^+\prod_{a=1}^3(-k_a + 1)^+ \Big)^+ \\
& - \Big( \prod_{a=1}^4 (-k_a +1)^+- \prod_{a=1}^4 (-k_a -1)^+ \Big)^+
\end{aligned}
\end{equation}

\vspace{-4pt}
However, even in the cases when these ranks are non-maximal we are able to write down closed-form expressions for the ranks of the cohomology groups on $X$. This is due to the fact that the ranks of the maps involved exhibit a surprising regularity.  These formulae are given below and hold for any non-negative integer $p$. We use the notation $C_n^k$ for the binomial coefficient indexed by $n$ and $k$. We distinguish the following cases:
\vspace{2pt}
\begin{itemize}
\item[$i)$] $k_1\leq -(4+2p)$, $k_2= k_3 = -(4+2p)$ and $k_4=2+p$:
\end{itemize}
\begin{equation}
\begin{aligned}
\mathrm{dim}\, H^0\big{(} X, \calO (\vec{k}) \big{)}  &= \mathrm{dim}\, H^1\big{(}X, \calO (\vec{k}) \big{)}  = 0 \\[8pt]
\mathrm{dim}\, H^2\big{(}X, \calO (\vec{k}) \big{)}  & = 48\,C^{3}_{p+3} + (p+1)\,(-k_1-(4+2p)-1)\\
\mathrm{dim}\, H^3\big{(}X, \calO (\vec{k}) \big{)}  & = (k_4 +1)^+\prod_{a=1}^3(-k_a -1)^+ \\
-\, \Big((k_4 -1)^+ &\prod_{a=1}^3(-k_a + 1)^+ - \big(48\,C^{3}_{p+3} + (p+1)\,(-k_1-(4+2p)-1) \big)  \Big)
\end{aligned}
\end{equation}

\vspace{4pt}
\begin{itemize}
\item[$ii)$] $k_1, k_2< -(4+2p)$, $k_3\leq-(4+2p)$ and $k_4=2+p$ or \\[4pt] $k_1,k_2< -(4+2p)-2$, $k_3 = -(4+2p)+1$ and $k_4=2+p$:
\end{itemize}
\begin{equation}
\begin{aligned}
\mathrm{dim}\, H^0\big{(}X, \calO (\vec{k}) \big{)}  &= \mathrm{dim}\, H^1\big{(}X, \calO (\vec{k}) \big{)}  = 0 \\[8pt]
\mathrm{dim}\, H^2\big{(}X, \calO (\vec{k}) \big{)}  &= 48\,C^{3}_{p+3}\\
\mathrm{dim}\, H^3\big{(}X, \calO (\vec{k}) \big{)}  & = (k_4 +1)^+\prod_{a=1}^3(-k_a -1)^+ - \Big((k_4 -1)^+\prod_{a=1}^3(-k_a + 1)^+ -48\,C^{3}_{p+3}  \Big)
\end{aligned}
\end{equation}

\vspace{8pt}
The last two cases represent the dual version of first two. Although very similar, we include these formulae here for the sake of completeness:
\begin{itemize}
\item[$iii)$] $k_1=-2-p$, $k_2=k_3=4+2p$, $k_4\geq 4+2p$:
\end{itemize}
\begin{equation}
\begin{aligned}
\mathrm{dim}\, H^0\big{(}X, \calO (\vec{k}) \big{)}  & = (-k_1 +1)^+ \prod_{a=2}^4 (k_a -1)^+\\
 - \Big((-k_1  - &1)^+ \prod_{a=2}^4 (k_a +1)^+ - \big(48\,C^{3}_{p+3} + (p+1)\,(k_4-(4+2p)-1) \big)  \Big)\\[8pt]
\mathrm{dim}\, H^1\big{(}X, \calO (\vec{k}) \big{)}  & = 48\,C^{3}_{p+3} + (p+1)\,(k_4-(4+2p)-1)\\[8pt]
\mathrm{dim}\, H^2\big{(}X, \calO (\vec{k}) \big{)}  &= \mathrm{dim}\, H^3\big{(} \calO_X (\vec{k}) \big{)} = 0
\end{aligned}
\end{equation}

\vspace{8pt}
\begin{itemize}
\item[$iv)$] $k_1=-2-p$, $k_2\geq 4+2p$, $k_3, k_4> 4+2p$ or \\[4pt] $k_1=-2-p$, $k_2=4+2p-1$, $k_3, k_4> 4+2p+2$:
\end{itemize}
\begin{equation}
\begin{aligned}
\mathrm{dim}\, H^0\big{(} \calO_X (\vec{k}) \big{)}  &= (-k_1 +1)^+ \prod_{a=2}^4 (k_a -1)^+ - \Big((-k_1 -1)^+ \prod_{a=2}^4 (k_a +1)^+ - 48\,C^{3}_{p+3}  \Big)\\
\mathrm{dim}\, H^1\big{(} \calO_X (\vec{k}) \big{)}  &= 48\,C^{3}_{p+3}\\[8pt]
\mathrm{dim}\, H^2\big{(} \calO_X (\vec{k}) \big{)}  &=  \mathrm{dim}\, H^3\big{(} \calO_X (\vec{k}) \big{)} = 0
\end{aligned}
\end{equation}

\vspace{12pt}
\section{Finiteness: a Different Perspective}\label{app:finiteness}

In Section~\ref{sec:finiteness} we have tried to understand the observation, made in Chapter~\ref{LineBundles}, that the number of consistent and physically viable line bundle models constructed over a certain manifold $X$ admitting discrete symmetries $\Gamma$ of a fixed order is an increasing and saturating function of the maximal line bundle entry in modulus. This phenomenon could be observed for all the pairs $(X,|\Gamma|)$ studied in Chapter~\ref{LineBundles}. In this section we will argue that the combined effect of cohomology constraints, poly-stability and the bound on the second Chern class imposed by the anomaly cancellation condition limits the range of allowed line bundle integers.

 \subsection{Bounds from cohomology}
 
In order to ensure the correct chiral asymmetry and the absence of $\overline{\bf{10}}$-multiplets form the $SU(5)$ GUT spectrum, we must require the following pattern for the bundle cohomology: 
\beq
 h^\bullet(X,V)\ \,  =\  (0,3\,|\Gamma|,0,0)  
\eeq

\vspace{0pt}
In Section~\ref{app:tqcoh} we have seen explicit formulae for line bundle cohomology on the tetraquadric hypersurface. By using these, and imposing the above cohomology pattern we can immediately exclude the following line bundles (where I assume $k_1\leq k_2\leq k_3\leq k_4$):
\begin{itemize}
\item[-] the trivial line bundle $\calO_X (0,0,0,0)$;
\item[-] all semipositive line bundles $\calO_X (k_1,k_2,k_3,k_4)$ with $k_i\geq 0$;
\item[-] $\calO_X (-p,0,0,k_4)$ with $p\geq 1$ and $k_4\geq 2$;
\item[-] $\calO_X (-1,k_2,k_3,k_4)$ with $k_2,k_3,k_4\geq 2$;
\item[-] all special cohomology cases presented in Section \ref{app:tqcoh}
\item[-] $\calO_X (k_1,k_2, k_3,k_4)$ with $k_1,k_2<0$ and $k_3, k_4=0$
\item[-] $\calO_X (k_1,k_2,k_3,k_4)$ with $k_1,k_2,k_3<0$ and $k_4\geq0$
\end{itemize}

Thus we are left with the following classes of line bundles:
\begin{itemize}
\item[-] $\calO_X (-p,0,0,1)$ with $p\geq 1$
\item[-] $\calO_X (-p,0,k_3,k_4)$ with $k_3,k_4>0$ 
\item[-] $\calO_X (-p,k_2,k_3,k_4)$ with $k_2<4+2p-1$ 
\item[-] $\calO_X (-p,k_2,k_3,k_4)$ with $k_2=4+2p-1$ and $k_3<4+2p+2$ 
\item[-] $\calO_X (k_1,k_2,k_3,k_4)$ with $k_1,k_2<0$ and $k_3,k_4>0$ 
\end{itemize} 

\vspace{10pt}
Moreover, for any line bundle $L$ in a viable line bundle sum, we must require $h^1\big{(} X,L \big{)} \leq 3| \Gamma |$, where $\Gamma$ is a freely acting group on the tetraquadric hypersurface. According to \cite{Braun:2010vc}, the available group orders are $| \Gamma | \in \{2,4,8,16\}$. So  $h^1\big{(} X,L\big{)} \leq 48$. This imposes further bounds on the allowed line bundle entries. It follows that the allowed line bundles (assuming $k_1\leq k_2\leq k_3\leq k_4$) fall into several categories: 
\begin{itemize}
\item[$i)$] one negative and one positive entry:
	\begin{itemize}
	\item[-] $\calO_X ([-25,-1],0,0,1)$. Cohomology: $h^\bullet \big{(} \calO_X (-p,0,0,1) \big	{)} = \big(0,2(p-1),0,0\big)$
	\end{itemize}
	
\item[$ii)$] one negative and two positive entries:
	\begin{itemize}
	\item[-] $\calO_X (-1,0,1,[1,\infty))$. Cohomology: $h^\bullet \big{(} \calO_X (-1,0,1,q) \big{)} = \big(0,0,0,0\big)$
	\item[-] $\calO_X (-k_1,0,k_3,k_4)$, $n_i\in \mathbb Z_+$. Cohomology: $h^\bullet \big{(} \calO_X (-k_1,0,k_3,k_4) \big{)} =$\\ $ \big(0,(-k_1 -1)^+ (k_3 +1)^+(k_4 +1)^+ + (-k_1 +1)^+ (k_3 -1)^+(k_4 -1)^+ ,0,0\big)$. \\Bound on the entries: $k_1, k_3, k_4 \leq 25$
	\end{itemize}
	
\item[$iii)$] one negative and three positive entries:
	\begin{itemize}
	\item[-] $\calO_X ([-7,-1],1,1,1)$. Cohomology: $h^\bullet \big{(} \calO_X (-p,1,1,1) \big{)} = \big(0,8(p-1),0,0\big)$
	\item[-] $\calO_X (-1,1,[1,\infty),[1,\infty))$. Cohomology: $h^\bullet \big{(} \calO_X (-1,1,p,q) \big{)} = \big(0,0,0,0\big)$
	\item[-] $\calO_X (-p,k_2,k_3,k_4)$, $p>1,k_i\in\mathbb Z_+$.  Bound on the entries: for $p=2$, $k_i\leq 11$; for $p=3$, $k_i\leq 5$; for $p=4$, $k_i\leq 3$; for $p=5$, $k_i\leq 2$; for $p=6, 7$, $k_i = 1$; 
	\end{itemize}

\item[$iv)$] two negative entries and one positive entry:
	\begin{itemize}
	\item[-] $\calO_X ((-\infty,-1],-1,0,1)$. Cohomology: $h^\bullet \big{(} \calO_X (-p,-1,0,1) \big{)} = \big(0,0,0,0\big)$
	\end{itemize}
	
\item[$v)$] two negative and two positive entries:
\vspace{-8pt}
	\begin{itemize}
	\item[-] $\calO_X (-p,-q,p,q)$, $p,q\in \mathbb Z_+$. Cohomology: $h^\bullet \big{(} \calO_X (-p,-q,p,q) \big{)} = \big(0,0,0,0\big)$
	\item[-] $\calO_X ((-\infty,-1],(-\infty,-1],[1,\infty),[1,\infty))$. Cohomology: $h^\bullet \big{(} \calO_X (-p,-q,q,p) \big{)} = \big(0,0,0,0\big)$
	\item[-] $\calO_X (-k_1,-k_2,k_3,k_4)$, $k_i\in\mathbb Z_+$. Cohomology: $h^\bullet \big{(} \calO_X (-k_1,-k_2,k_3,k_4) \big{)} =$\\ $ \big(0,(-k_1 +1)^+ (-k_2 +1)^+ (k_3 -1)^+(k_4 -1)^+ -(-k_1 -1)^+ (-k_2 -1)^+ (k_3 +1)^+(k_4 +1)^+,0,0\big) $ \newline
	Bound on the entries (for cases not covered above): $n_i\leq 13$
	\end{itemize}
	
\item[$vi)$] three negative entries and one positive entry:
\vspace{-8pt}
	\begin{itemize}
	\item[-] $\calO_X ((-\infty,-1],(-\infty,-1],-1,1)$. Cohomology: $h^\bullet \big{(} \calO_X (-p,-q,-1,1) \big{)} = \big(0,0,0,0\big)$
	\end{itemize}
\end{itemize}
 
\vspace{2pt} 
\subsection{Bound from stability}\label{sec:boundstab}
Poly-stability for a sum of five line bundles $V = \bigoplus_a L_a$ reduces to the question of finding simultaneous solutions for the equations $\mu(L_a) = 0$, i.e.~finding a vector $(\kappa_1,\kappa_2,\kappa_3,\kappa_4)$ in the dual K\"ahler cone, such that for all $1\leq a \leq 5$,
\vspace{-8pt}
\beq
0=\mu(L_a) =\kappa_1 k_a^1 +\kappa_2 k_a^2 +\kappa_3 k_a^3 +\kappa_4 k_a^4 
\vspace{-8pt}
\eeq

However, due to the condition $c_1(V)=0$, only four of these equations are independent. By successively multiplying the above equation by $k_a^1,\ldots, k_a^4$, we obtain:
\begin{equation}
\begin{aligned}
0 & = \kappa_1 \left( k_a^1 \right)^2 +\kappa_2 k_a^2 k_a^1+\kappa_3 k_a^3 k_a^1+\kappa_4 k_a^4k_a^1 \\
0 & = \kappa_1 k_a^1k_a^2 +\kappa_2  \left( k_a^2 \right)^2 +\kappa_3 k_a^3k_a^2 +\kappa_4 k_a^4 k_a^2 \\
0 & = \kappa_1 k_a^1k_a^3 +\kappa_2 k_a^2 k_a^3+\kappa_3  \left( k_a^3 \right)^2 +\kappa_4 k_a^4k_a^3 \\
0 & = \kappa_1 k_a^1 k_a^4+\kappa_2 k_a^2k_a^4 +\kappa_3 k_a^3 k_a^4+\kappa_4  \left( k_a^4 \right)^2 
\end{aligned}
\end{equation}

Add these up and sum over all $a$: 
\vspace{-8pt}
\beq
\begin{aligned} 
0& = \kappa_1\left( \sum_{a=1}^5 \left( k_a^1 \right)^2 + \frac{-c_2^1+c_2^2+c_2^3+c_2^4}{2} \right)   + \kappa_2\left( \sum_{a=1}^5 \left( k_a^2 \right)^2 + \frac{+c_2^1-c_2^2+c_2^3+c_2^4}{2} \right)  \\
& + \kappa_3\left( \sum_{a=1}^5 \left( k_a^3 \right)^2 + \frac{+c_2^1+c_2^2-c_2^3+c_2^4}{2} \right)   + \kappa_4\left( \sum_{a=1}^5 \left( k_a^4 \right)^2 + \frac{+c_2^1+c_2^2+c_2^3-c_2^4}{2} \right)
\end{aligned}
\eeq
where
\vspace{-8pt}
\beq 
c_2 ^1 :=   c_2 (V) . J_1 = \sum_{a=1}^5  k_a^2 k_a^3 +k_a^2 k_a^4 + k_a^3 k_a^4 \ \ \ \ \text{ and so on.}
\vspace{-8pt}
\eeq

Since $\kappa_i$ are all positive, and $c_2^i$ are bounded from above by the anomaly cancellation condition (see Section~\ref{sec:conditions}), it follows that $ \sum_a \left( k_a^1 \right)^2$, $ \sum_a \left( k_a^2 \right)^2$, $ \sum_a \left( k_a^3 \right)^2$ and $ \sum_a \left( k_a^4 \right)^2$ cannot be large at the same time. That is, at least one of the rows in the matrix representing the sum of line bundles has to contain only small numbers. 

More significantly, in order to satisfy the above equation with large line bundle entries, some (but not all) of the $\kappa_i$ parameters have to become small. This corresponds to having some (but again, not all) of the K\"ahler parameters $t^i$ arbitrarily small. However, this falls out of the supergravity approximation of the heterotic string theory, valid only when all the K\"ahler parameters satisfy $t^i\gg 1$. 

\subsection{Combined bounds}
The physical constraints on the cohomologies of $V$ limit the range of line bundle integers in almost all cases. However, these constraints leave three types of line bundles with large integers in modulus, having trivial cohomology. Up to permutations, these are:
\vspace{-7pt}
\begin{itemize}
\item[-] line bundles with one large integer: $\cO_X(-1,0,1, \pm p)$\vspace{-4pt}
\item[-] line bundles with two large integers: $\cO_X(-1,1,\pm q, \pm p)$\vspace{-4pt}
\item[-] line bundles with four large integers: $\cO_X(-p,-q,q,p)$
\end{itemize}

The third type of line bundles are excluded by stability and the $c_2(V)$ constraint discussed above, in Section \ref{sec:boundstab}. Moreover, line bundles of the form $\cO_X(-1,1, p, q)$ where both $p$ and $q$ are either positive or negative do not satisfy the slope zero condition in the positive K\"ahler cone. Thus the only allowed type of line bundles with large entries are of two types: 
\vspace{-8pt}
\begin{itemize}
\item[-] two large integers: $\cO_X(p,-q,-1, 1)$, $p, q\geq 0$ or \vspace{-4pt}
\item[-] $\cO_X(-p,-q,q,p)$ where only $p\geq 0$ can be arbitrarily large
\end{itemize}

\vspace{0pt}
We would like to argue that such line bundles, with arbitrarily large entries, cannot enter a viable line bundle model. Before doing that, however, we give an example of an infinite family of line bundle sums satisfying all the topological constraints required from a viable model:  
\begin{equation}
V~=~~
\cicy{ \\ \\ \\ \\ }
{ - p & p-2 & ~~0 & ~~1 & ~~1~ \\
~~0 & ~~0 & -1 & ~~2 & -1~ \\
-1 & -1 & -1 & ~~1 &~~2 ~\\
 ~~1 & ~~1 &~~ 1 & -2 & -1 ~\\}\; 
\end{equation}

The bundles in this class satisfy, independent of the value of $p\in\IZ$:
\vspace{-4pt}
\begin{align}
c_1(V) &= 0\\  
c_2(V).J_i &= \left(-20, -14, -16, 14\right) \\
\text{ind}(V) &= \text{ind}(\wedge^2 V) = 12
\end{align}
\vspace{-12pt}

However, the slopes of the above line bundles cannot be simultaneously put to zero in the interior of the K\"ahler cone, except when $p=1$. 
\vspace{8pt}

The following family contains poly-stable line bundle sums for any $p\in 3\IZ$:
\begin{equation}
V~=~~
\cicy{ \\ \\ \\ \\ }
{ - p & p/3 & p/3 & p/3 & ~~0~ \\
 ~~0 & -1 & -1 & -1 & ~~3~ \\
-1 & ~~0 & ~~0 & ~~0 &~~1 ~\\
 ~~1 & ~~1 &~~ 1 & -2 & -4 ~\\}\; 
\end{equation}
and has 
\vspace{-20pt}
\begin{align}
c_1(V) &= 0\\  
\text{ind}(V) &= \text{ind}(\wedge^2 V) = 24
\end{align}

However, the second and the third entries in $c_2(V).J_i$ depend linearly on $p$.
The two examples above illustrate the conflict between, on one hand, trying to satisfy the slope zero condition and on the other hand trying to keep $c_2(V).J_i$ bounded. If one starts with a line bundle of the type $\cO_X(-p,-q,q,p)$, one is forced by the slope zero condition to continue with line bundles of the same kind. However, this always leaves at least two entries in $c_2(V).J_i$ large. Alternatively, if one starts with a line bundle $\cO_X(-p,q,-1,1)$, this will produce at least one large entry in $c_2(V).J_i$. Trying to make up for that renders the slope zero condition impossible to satisfy.

\chapter{Exploring the Moduli Space of Line Bundle Compactifications}\label{ModuliSpace}
 \vspace{-14pt}
In this chapter, I present a line bundle model that is convenient enough in order to discuss its deformations to non-Abelian bundles. In Section~\ref{sec:model}, I present the full GUT spectrum associated with the chosen line bundle model and the allowed operators. In the second step, Section~\ref{sec:monads}, I discuss the space of non-Abelian deformations around the chosen line bundle sum using the monad construction. 
 \vspace{-12pt}

\section{A Line Bundle Model on the Tetraquadric Manifold}\label{sec:model}
 \vspace{-8pt}
Let us consider the following line bundle sum:
 \vspace{-12pt}
\begin{equation}\label{eq:example}
U~=~~
\cicy{ \\ \\ \\ \\ }
{ -1 & -1 & ~~0 & ~~1 & ~~1~ \\
~~0 & -3 & ~~1 & ~~1 & ~~1~ \\
~~0 & ~~2 & -1 & -1 & ~~0 ~\\
 ~~1 & ~~2 &~~ 0 & -1 & -2 ~\\}\; 
\end{equation}
where the columns of the matrix represent the first Chern classes of the line bundles. 

The dimensions of the relevant cohomology groups (computed, e.g.~using the formulae of Section~\ref{app:tqcoh}) are given by
 \vspace{-8pt}
\beq
\begin{aligned} 
h^\bullet(X,U)\ \, & =\  (0,12,0,0) \\
h^\bullet (X,\wedge^2U)&=\ (0, 15, 3, 0)\\
h^\bullet (X,U\otimes U^{^*})&=\ (5, 60, 60, 5)
\end{aligned}
\eeq
Thus, according to Table~\ref{spectrum2}, the model leads to an $SU(5)$ GUT with $15$ multiplets transforming in the conjugate representation $\mathbf{\overline{5}}$, $3$~multiplets transforming in the fundamental representation $\mathbf{5}$ and $12$ multiplets transforming in the $\mathbf{10}$~representation. The GUT model contains also a number of $60$ singlets, representing bundle moduli. By quotienting the tetraquadric manifold by a group of order $4$, this leads to a model with chiral asymmetry equal to the number of families in the Standard Model. 
 \vspace{-8pt}

\subsection{The particle spectrum at the Abelian locus}
\vspace{-8pt}
The ensuing discussion will require the precise data describing how these multiplets split into sub-sectors, according to their $S\left(U(1)^5\right)$ charges. We accomplish this by computing the cohomologies of each individual line bundle, as well as those for pairs of line bundles. 

\vspace{12pt}
\begin{table}[!h]
\hspace{.25cm}
\parbox{.45\linewidth}{
\centering
\begin{tabular}{| c | c | c |}
\hline
\varstr{14pt}{9pt} repr. & cohomology & multiplets \\ \hline\hline
\varstr{14pt}{9pt} $~{\bf 1}_{{\bf e}_2 - {\bf e}_1}~$ & $~H^1(X, L_2 \otimes L_1^{^*})~$  &  $~~12\, S_{2,1}~~$ \\ \hline
\varstr{14pt}{9pt} ${\bf 1}_{{\bf e}_2 - {\bf e}_3}$ & $H^1(X, L_2 \otimes L_3^{^*})$  &  $20\, S_{2,3}$ \\ \hline
\varstr{14pt}{9pt} ${\bf 1}_{{\bf e}_2 - {\bf e}_4}$ & $H^1(X, L_2 \otimes L_4^{^*})$  &  $12\, S_{2,4}$ \\ \hline
\varstr{14pt}{9pt} ${\bf 1}_{{\bf e}_5 - {\bf e}_1}$ & $H^1(X, L_5 \otimes L_1^{^*})$  &  $12\, S_{5,1}$ \\ \hline
\varstr{14pt}{9pt} ${\bf 1}_{{\bf e}_5 - {\bf e}_3}$ & $H^1(X, L_5 \otimes L_3^{^*})$  &  $4\, S_{5,3}$ \\ \hline
 \end{tabular}
}
\hspace{.5cm}
\parbox{.45\linewidth}{
\centering
\begin{tabular}{| c | c | c |}
\hline
\varstr{14pt}{9pt} $~{\bf \overline{5}}_{{\bf e}_2+{\bf e}_4}~$ &~ $H^1(X, L_2 \otimes L_4)~$  & $~~4\, \overline{\mathbf 5}_{2,4}~~$ \\ \hline
\varstr{14pt}{9pt} ${\bf \overline{5}}_{{\bf e}_2+{\bf e}_5}$ & $H^1(X, L_2 \otimes L_5)$  & $3\, \overline{\mathbf 5}_{2,5}$ \\ \hline
\varstr{14pt}{9pt} ${\bf \overline{5}}_{{\bf e}_4+{\bf e}_5}$ & $H^1(X, L_4 \otimes L_5)$  & $8\, \overline{\mathbf 5}_{4,5}$ \\ \hline\hline
\varstr{14pt}{9pt} ${\bf 5}_{-{\bf e}_2 -{\bf e}_5}$ & $H^1(X, L_2^{^*} \otimes L_5^{^*})$  & $3\, \mathbf{5}_{2,5}$ \\ \hline
\varstr{14pt}{9pt} ${\bf 10}_{{\bf e}_2}$ &$H^1(X, L_2)$ & $8\,\mathbf{10}_2$\\ \hline
\varstr{14pt}{9pt} ${\bf 10}_{{\bf e}_5}$ &$H^1(X, L_5)$ & $4\,\mathbf{10}_5$\\ \hline
 \end{tabular}
}\vspace{10pt}
\caption{The GUT spectrum, including the $S\left(U(1)^5 \right)$ charges. }
 \vspace{-8pt}
\end{table}

There are several phenomenologically appealing features of this particular model. Combined with a $\IZ_2\times \IZ_2$ symmetry, the above line bundle sum leads to a model with one Higgs pair, a rank-2 up Yukawa coupling matrix at the abelian locus, a vanishing $\mu$-term everywhere, vanishing dimension 4 and 5 proton-decay operators and a spectrum of singlets which indicates the existence of non-abelian deformations. In Section~\ref{sec:monads}, I will use this model as a case study for the problem of embedding a line bundle sum into the moduli space of non-abelian bundles by relating line bundle sums to monad bundles. Let us first discuss the allowed operators in the low-energy theory derived from this particular line bundle sum. 
 \vspace{-8pt}

\subsection{The allowed operators at the abelian locus} \label{sec:operators}
\vspace{-8pt}
The $SU(5)$ GUT group can be broken to the gauge group of the Standard Model by introducing discrete Wilson lines. To achieve this, one needs to quotient the original Calabi-Yau threefold by the free action of some finite group. In order to obtain a consistent model on the quotient manifold, the holomorphic vector bundle must allow for an equivariant structure with respect to the action considered on the base manifold. The detailed construction on the quotient manifold will not be discussed here. The interested reader might want to consult Refs.~\cite{Anderson:2011ns, Anderson:2012yf}. 

Regardless of the quotienting details, a number of statements can be made in relation to the particle spectrum in the resulting standard-like models.  After quotienting the model described by the line bundle sum (\ref{eq:example}) by a finite group of order $4$, the number of $\mathbf{10}$ multiplets reduces from $12$ to $3$: $\mathbf{10}_{2}, \mathbf{10}_{2}$ and $\mathbf{10}_{5}$. The low-energy theory will also contain three multiplets $\mathbf{\overline{5}}_{2,4}, \mathbf{\overline{5}}_{4,5}$ and $\mathbf{\overline{5}}_{4,5}$ and whatever remains from the three vector-like pairs of $\mathbf{5}_{2,5}$ and $\mathbf{\overline{5}}_{2,5}$ multiplets after Wilson line breaking. As discussed in \cite{Anderson:2012yf}, the Higgs triplets can be projected out and at least one pair of Higgs doublets survives. 
It follows that we can discuss the general structure of the operators allowed by the global $U(1)$ symmetries before explicitly breaking the GUT group to the Standard Model gauge group. The Standard Model multiplets keep the same $S\left( U(1)^5\right)$ charges as the $SU(5)$ multiplets from which they descend. For this reason, and for the sake of simplicity, I will denote the Standard Model families by $SU(5)$ representations
The generic form for the superpotential can be organised into several terms \cite{Anderson:2012yf}, as follows:
\vspace{-12pt}
\begin{equation}
 W=W_{\rm Y}+W_{\rm R}+W_5+W_{\rm sing}+W_{\rm np}\;  \label{eq:W}
 \vspace{-16pt}
\end{equation}

$W_{\rm np}$ stands for all the non-perturbative contributions. $W_{\rm Y}$ contains the standard Yukawa couplings and the $\mu$-term. $W_{\rm R}$ consists of the R-parity violating terms and $W_5$ consists of the dimension five operators in standard model fields. $W_{\rm sing}$ collects all the pure singlet field terms. Schematically, these perturbative parts can be written as
\vspace{-12pt}
\begin{align}
 W_{\rm Y}&=\mu\, H \bar{H}+Y^{(d)}_{pq}\,H\,\bar{\bf 5}^p{\bf 10}^q+Y^{(u)}_{pq}\,\bar{H}\,{\bf 10}^p{\bf 10}^q\label{WYuk}\\
W_{\rm R}&=\rho_p\, \bar{H}L^p+\lambda_{pqr}\,\bar{\bf 5}^q\bar{\bf 5}^q{\bf 10}^r\label{WR}\\
W_5&=\lambda_{pqrs}'\,\bar{\bf 5}^p{\bf 10}^q{\bf 10}^r{\bf 10}^s\label{Wprot5}\\
W_{\rm sing}&=\tau_{\alpha\beta\gamma}\,S^\alpha S^\beta S^\gamma\; .
 \end{align} 
\phantom{}
\vspace{-38pt}

The above couplings are functions of the complex structure moduli and the bundle moduli~$S^\alpha$. The locus, in bundle moduli space, where the vector bundle splits as a direct sum of line bundles corresponds to $S^\alpha =0$. Moving away from this Abelian locus corresponds to non-trivial vacuum values of these fields. According to the discussion of Section~\ref{spec}, the operators allowed by the $S\left( U(1)^5\right)$ symmetry must contain multiplets whose charges sum up to a multiple of $(1,1,1,1,1)$. 

For example, dimension four operators that induce a fast proton decay have the form $\bar{\bf 5}_{a,b}\,\bar{\bf 5}_{c,d}\,{\bf 10}_{e}$. Such operators have $S\left( U(1)^5\right)$ charge $\mathbf{e}_a+\mathbf{e}_b+\mathbf{e}_c+\mathbf{e}_d+\mathbf{e}_e$ and in order to be allowed, $a, b, c, d$ and $e$ must be different. For our model, this cannot happen, thus $\lambda_{pqr}$ must be zero at the Abelian locus, corresponding to $S^\alpha =0$. Away from the Abelian locus, such operators might re-appear by singlet VEV insertions of the type $S^{\alpha_1}\ldots S^{\alpha_n}\,\bar{\bf 5}\,\bar{\bf 5}\,{\bf 10}$. By considering the available $S\left( U(1)^5\right)$ charges for the singlet fields, it is easy to see that all such operators are forbidden, thus $\lambda_{pqr}$ is zero everywhere. Following a similar argument, it turns out that dimension five operators, as well as all pure singlet field terms are also forbidden everywhere. 
Let us now discuss the remaining terms: the Higgs $\mu$-term, the Yukawa couplings and R-parity violating terms of the form $\bar{H}\,L$ which generate too large neutrino masses. 
The up-Higgs field $\bar{H}_{a,b}$ descends from a $\mathbf{5}_{a,b}$ multiplet and thus has $S\left( U(1)^5\right)$ charge $Q(\bar{H}_{a,b})= -\mathbf{e}_a - \mathbf{e}_b$. Similarly, the down-Higgs $H_{a,b}$ descends from a $\overline{\mathbf{5}}_{a,b}$ multiplet with charge $Q(H_{a,b})= \mathbf{e}_a+ \mathbf{e}_b$. The lepton field $L_{a,b}$ descends from a $\overline{\mathbf{5}}_{a,b}$ multiplet and, accordingly, has charge  $Q(L_{a,b})= \mathbf{e}_a+ \mathbf{e}_b$. 

The Higgs doublets come from the same sector of $S\left( U(1)^5\right)$ charges. Thus the combination $H\,\bar H$ is neutral. Moreover, since for our model there are no completely uncharged singlet fields, it follows that the only $\mu$-term, allowed by the $S\left( U(1)^5\right)$ symmetry is $H\,\bar H$. However, the presence of such a term is in contradiction with the cohomology computations, which indicate the existence of massless Higgs doublets at the Abelian locus. We conclude that the term $H\,\bar H$ is absent from the superpotential. 

The structure of the Yukawa couplings is also independent of singlet VEV insertions. The perturbative down-type Yukawa matrix vanishes identically everywhere. The up-Yukawa matrix has two eigenvalues of order one and a zero eigenvalue:
\beq
Y^u~=~~
\left( 
\begin{array}{c c c}
 0 & 0 & 1\\
 0 & 0 & 1\\
 1 & 1 & 0\\
\end{array}
\right)\
\hskip0.35in
Y^d~=~~
\left( 
\begin{array}{c c c}
 0 & 0 & 0\\
 0 & 0 & 0\\
 0 & 0 & 0\\
\end{array}
\right)\; .
\eeq
In an ideal situation, one would have all perturbatively generated Yukawa couplings identically zero, except the top Yukawa coupling, generated at order one. However, even in the present situation, one can hope that the hierarchy between top and charm quarks can be restored by non-perturbative effects. 

Finally, an R-parity violating term with a singlet VEV insertion of the form 
 \vspace{-12pt}
\beq\label{HLterm}
S_{2,4}\,\bar{H}\,L
 \vspace{-12pt}
\eeq 
is allowed by the $S\left( U(1)^5\right)$ symmetry. This term will generate a mass for the up-Higgs field in a certain direction away from the Abelian locus. Since the singlet field $S_{2,4}$ comes from the cohomology $H^1(X,L_2\otimes L_4)$, we expect that the term (\ref{HLterm}) is generated for deformation of the bundle that have both $L_2$ and $L_4$ incorporated into non-Abelian pieces. This observation is consistent with the particle spectrum derived from the monad bundles discussed below.

\section{Monad Bundles}\label{sec:monads}
Monad bundles provide a relatively straightforward way to explore the moduli space of non-abelian bundles around a split locus. Monad bundles are defined via a short exact sequence:
\begin{equation} \label{eq:monad}
0\ \longrightarrow\ V \longrightarrow\ B\ \stackrel{f}{\longrightarrow}\ C\ \longrightarrow\ 0  
\end{equation}
where $\displaystyle B = \bigoplus_{i=1}^{\text{rk}(B)} b_i$ and $\displaystyle C = \bigoplus_{i=1}^{\text{rk}(C)} c_i$ are sums of line bundles, $f\in \text{Hom}(B,C)\cong H^0(C\otimes B^{^*})$ and $V\cong \text{Ker}(f)$. The map $f$ can be represented by a matrix whose entries are sections of $H^0(c_i\otimes b_j^{^*})$. Heterotic compactifications involving monad bundles have been recently studied in  \cite{Anderson:2007nc, Anderson:2008uw, Anderson:2009mh,He:2009wi, He:2011rs}. In this section we would like to construct the monad sequence such that, for a certain choice of the defining map $f$, $V$ splits into a sum of line bundles. For example, $U$ can be the line bundle sum given in definition~(\ref{eq:example}). 

\subsection{The monad construction}\label{sec:monadconstruction}
An obvious way to construct the monad sequence would be to take $B = U\oplus \widetilde B$ where $\widetilde B$ is a sum of line bundles of equal rank with $C$ satisfying $\text{ch}(\widetilde{B}) =\text{ch}(C)$ and build the monad map $f$ in a block diagonal structure $(g,h)$, where $g$ corresponds to $U$-part of $B$ and $h$ corresponds to $\widetilde B$. In this case, at the locus (in bundle moduli space) where $g=0$, $U\subset \text{Ker}(f)$. The problem with this construction is that $\text{Ker}(f)$ is not a vector bundle when $f=(0,g)$: the matrix $h$ degenerates on a co-dimension one locus in the embedding space, corresponding to $\text{det} (h)=0$, which, generically, intersects the Calabi-Yau hypersurface. 

\vspace{10pt}
One way to avoid that the degeneracy locus of the monad map intersects the Calabi-Yau manifold is the following. Start with a monad realisation of the structure sheaf $\cO_X$: 
\begin{equation}\label{eq:structure_sheaf}
0\ \longrightarrow\ \cO_X \longrightarrow\ \widetilde{B}_a\ \stackrel{f_a}{\longrightarrow}\ \widetilde{C}_a\ \longrightarrow\ 0  
\end{equation}
with $\widetilde{B}_a$ and $\widetilde{C}_a$ sums of line bundles satisfying 
\begin{equation}
\text{rk} (\widetilde{B}_a) = \text{rk}(\widetilde{C}_a) +1 \ \ \ \text{and}\ \ \ c_1(\widetilde{B}_a) = c_1(\widetilde{C}_a)
\end{equation}
Apart from the trivial realisation $\widetilde{B}_a = \cO_X$ and $\widetilde{C}_a = 0$, one can also consider
\begin{equation}
\widetilde{B}_a = \cO_X(0,0,0,p_a)\oplus \cO_X(0,0,0,q_a) \ \ \ \text{and} \ \ \ \widetilde{C}_a = \cO_X(0,0,0,p_a+q_a)
\end{equation}
where $p_a$ and $q_a$ are positive integers. For this choice, the map $f_a = (f_{a_1},f_{a_2})$ contains two polynomials of multi-degrees $(0,0,0,p_a)$ and $(0,0,0,q_a)$. Generically, this map has rank 1. In order to reduce it to rank 0, both polynomials have to vanish. However, this cannot happen in~$\IC\IP^1$. Now twist the exact sequence (\ref{eq:structure_sheaf}) by $L_a$:
\vspace{-4pt}
\begin{equation}
0\ \longrightarrow\ L_a \longrightarrow\ L_a\otimes \widetilde{B}_a\ \stackrel{f_a}{\longrightarrow}\ L_a\otimes \widetilde{C}_a\ \longrightarrow\ 0  
\vspace{-4pt}
\end{equation}
and sum over all $a$ to obtain the short exact sequence
\begin{equation}\label{eq:structure_sheaf}
0\ \longrightarrow\ V \longrightarrow\ B\ \stackrel{f}{\longrightarrow}\ C\ \longrightarrow\ 0  
\end{equation}
where 
\vspace{-8pt}
\begin{equation}
B = \bigoplus_a L_a\otimes \widetilde{B}_a \ \ \ \text{and}\ \ \  C = \bigoplus_a L_a\otimes \widetilde{C}_a 
\end{equation}

At the locus where the monad map is block-diagonal, 
\vspace{-4pt}
\begin{equation}
f=\text{diag}\left(f_1, \ldots f_{\text{rk}(U)} \right)
\vspace{-4pt}
\end{equation}
the bundle $V$ splits into the sum of line bundles $U$.

\subsection{A concrete example}
In the following, we will consider the moduli space of monad bundles defined by the following two sums of line bundles:
\beq\label{eq:bandc}
B~=~~
\cicy{ \\ \\ \\ \\ }
{ - 1 & -1 & -1 & ~~0 &~~1& ~~1& ~~1~ \\
 ~~0 & -1 & -1 & ~~1 &~~1& ~~1& ~~1~ \\
~~0 & ~~2 & ~~2 & -1 &-1 &~~0 &~~0~\\
 ~~1 & ~~2 &~~ 2 & ~~0 &-1& ~~0 &~~0~\\}\
\hskip0.35in
C~=~~
\cicy{ \\ \\ \\ \\ }
{ - 1 & ~~1~ \\
 ~~1 & ~~1~ \\
~~2 & ~~0 ~\\
 ~~2 & ~~2~\\}\; .
\eeq

At the split locus, the bundle $V$ is the sum of line bundles presented in Eq.~(\ref{eq:example}):
\begin{equation}\label{eq:splitlocus}
U~=~~
\cicy{ \\ \\ \\ \\ }
{ -1 & -1 & ~~0 & ~~1 & ~~1~ \\
~~0 & -3 & ~~1 & ~~1 & ~~1~ \\
~~0 & ~~2 & -1 & -1 & ~~0 ~\\
 ~~1 & ~~2 &~~ 0 & -1 & -2 ~\\}\; 
\end{equation}

As discussed in the previous section, this particular line bundle sum leads to a viable model when the tetraquadric manifold is quotiented by a free action of $\IZ_2\times\IZ_2$ acting on the coordinates of each $\IP^1$ factor with generators 
\begin{equation}
\left( 
\begin{array}{cc}
1 & ~~0\\
0& -1 
\end{array}
\right)\ 
\hskip0.35in
\left( 
\begin{array}{cc}
0 & ~~1\\
1 &~~0 
\end{array}
\right)\; .
\end{equation}

A generating set for the line bundles which admit an equivariant structure with respect to this symmetry is given by $\{\cO_X,\cO_X(2,0,0,0),\cO_X(1,1,0,0)\text{ and permutations thereof}\}$, such that all equivariant line bundles can be obtained by conjugation and tensor products of the above line bundles. By an equivariant structure of a bundle $V$ with respect to an action of a group $G$ on the base manifold $X$, we mean an action of $G$ on $V$ which commutes with the projection $V\longrightarrow X$. 

Since $B$ and $C$ contain only equivariant line bundles, the monad bundle $V$ is itself equivariant under $\IZ_2\times\IZ_2$, leading to a well defined bundle on the quotient manifold $X/\left(\IZ_2\times\IZ_2\right)$.

\section{Stability of Monad Bundles}

The question that we would like to address now is whether monad bundles $V$ defined as in Section~\ref{sec:monadconstruction} can provide a family of consistent and physically viable non-abelian models. A first requirement for preserving $\cN=1$ supersymmetry is that $V$ is poly-stable and has slope 0. The slope of a bundle $V$ is defined as
\begin{equation}
 \mu(V) = \frac{1}{\text{rk}(V)} \int_X c_1(V)\wedge J\wedge J  = d_{ijk}\, c_1^i(V)\, t^j\, t^k = c_1^i(V)\,\kappa_i
\end{equation}
where  $\kappa_i = d_{ijk}\,t^j\,t^k$.
A bundle $V$ is called slope-stable if $\mu(F) < \mu(V)$ for all (coherent) sub-sheafs $F \subset V$ with $0 < \text{rk}(F) < \text{rk}(V)$;  it is called poly-stable if it is a direct sum of stable bundles, all with the same slope. For a poly-stable bundle $V$, its dual $V^{^*}$ and 
 $V \otimes L$ for any line bundle $L$ are also poly-stable. A stable bundle $V$ with $\text{rk}(V)>1$ and vanishing slope must satisfy $H^0(X, V) = H^3(X, V) = 0$. This has to be so, since $H^0(X, V) = H^0(X,V\otimes \cO^{^*}) \cong \text{Hom}_X\,(\cO_X,V)$ and $\mu(\cO_X) =0$. Thus if $H^0(X, V)\neq 0$, a constant map taking sections to sections makes $\cO_X$ be a proper sub-sheaf of $V$ and since $\mu(\cO_X) = \mu(V)$, it de-stabilises $V$. A similar argument for $V^{^*}$ leads to the condition $H^3(X, V) = 0$. Furthermore, for a stable bundle $V$ with vanishing slope, it follows that $H^0(X,\wedge^k V) = H^3(X,\wedge^k V) = 0$.

\subsection{Checking Stability}
Slope-stability, as defined above, is a property dependent on the K\"ahler parameters $t^i$. In order to produce a supersymmetric vector bundle $V$, we have to impose that $V$ is poly-stable and has slope 0 somewhere in the K\"ahler cone, defined by $t^i\geq 0$. By construction, the monad bundle has vanishing first Chern class, thus $\mu(V) =0$ is automatic. 

One can find the region in the K\"ahler cone where $V$ is stable by first identifying all line
bundles $L$ on $X$ which inject into at least one of the wedge powers $\wedge^k V$ , for $k = 1, \ldots, \text{rk}(V )-1$. Then we have to restrict the K\"ahler cone to the locus at which $\mu(L)<0$ for any injecting line bundle $L$. This can be most conveniently discussed is in the dual, $\kappa_i$ variables. In Appendix~\ref{app:KahlerCone} we show that the dual K\"ahler cone is a dense subset of 
\vspace{-4pt}
\beq
 C_s=\{\mathbf{s}\in \mathbb R ^4 \ | \ \mathbf{s.e}_i\geq 0 \text{ and } \mathbf{s.n}_i\geq 0 \}\; .
 \eeq

\vspace{-8pt}
The fact that the cone $C_{{\bf s}}$ is defined through hyperplane inequalities simplifies the discussion of stability. For any line bundle $L = \cO_X(-{\bf k})$, we add to the set of normal vectors $\{ \mathbf{e}_i, \mathbf{n}_i\}$, the vector~${\bf k}$. Hence the cone where $V$ is (poly)-stable will be defined by 
\beq
\begin{aligned} 
C_s=\{\mathbf{s}\in \mathbb R ^4 \ | & \ \mathbf{s.e}_i\geq 0 \text{ and } \mathbf{s.n}_i\geq 0 \text{ and } \mathbf{s}.\mathbf{k} \geq 0\\
& \text{ for any } L^{\,\!^*} = \mathcal{O}(\mathbf{k}),\, L \text{ injects in } \wedge^k V,\ k = 1, \ldots , \text{rk}(V )-1 \}  
\end{aligned}
\eeq

Note that we have written ${\bf s}.{\bf k}\geq 0$, instead of ${\bf s}.{\bf k} > 0$. Requiring the strict inequality guarantees that $V$ is stable. On the other hand, at the locus ${\bf s}.{\bf k}= 0$, the bundle $V$ can still be poly-stable, as a direct sum $V = L\oplus V'$. In fact, this will turn out to be the case for our example discussed below. The structure group of such bundles is $S\left(U(4)\times U(1) \right)\subset SU(5)$. 



\subsection{Criteria for Stability of Monad Bundles}\label{sec:stab_criteria}
If a line bundle $L$ is a proper sub-bundle of $\wedge^k V$, then 
\vspace{-4pt}
\begin{equation}
\text{Hom}_X\,\big( L,\wedge^k V\big)\cong H^0 \big( X, \wedge ^k V\otimes L^{^{\!*}} \big) \neq 0
\vspace{-8pt}
\end{equation}

This is a necessary but not sufficient condition for a sub-line bundle. $\text{Hom}_X\big(L,\wedge^k(V )\big) = 0$ guarantees that there are no maps between $L$ and $V$ . However, if $\text{Hom}_X\big(L,\wedge^k(V )\big) \neq 0$ this does not necessarily imply the existence of an injection from $L$ into $V$. Below, we derive several criteria for deciding whether $\text{Hom}_X\big(L,\wedge^k(V )\big)$ is trivial or not. We will use the fact that, for $SU(n)$ bundles, there exists the isomorphism
\vspace{-4pt}
\beq
\wedge^{n-k}V \cong \wedge^k V^{^*}
\vspace{-4pt}
\eeq

\vspace{2pt}
1. The case $k=1$, $L\hookrightarrow V$. By twisting the monad sequence (\ref{eq:monad}) with the line bundle $L^{^{\!*}}$, we obtain a short exact sequence which yields the following long exact sequence in cohomology:
\beq
0\ \longrightarrow\ \text{Hom}_X\big(L,V\big) \longrightarrow\ H^0\big(B\otimes L^{^{\!*}}\big)\ \longrightarrow\ H^0\big(C\otimes L^{^{\!*}}\big)\ \longrightarrow\ \ldots 
\vspace{-8pt}
\eeq
Thus 
\vspace{-8pt}
\beq\label{eq:hom}
 \text{Hom}_X\big(L,V\big) \cong \text{Ker}\left( H^0\big(B\otimes L^{^{\!*}}\big)\ \longrightarrow\ H^0\big(C\otimes L^{^{\!*}}\big)\right)
\eeq

If $h^0\big(B\otimes L^{^{\!*}}\big) = 0$, then $L$ does not inject into $V$. If 
\vspace{-4pt}
\beq\label{eq:cohcondiiton1}
h^0\big(B\otimes L^{^{\!*}}\big)> h^0\big(C\otimes L^{^{\!*}}\big)
\vspace{-4pt}
\eeq 
then $\text{Hom}_X\big(L,V\big)$ is non-trivial. In this case, we will treat $L$ as a de-stabilising line bundle and restrict the K\"ahler cone accordingly.  In the cases in which these simple cohomology checks are not conclusive, we will need to explicitly find the kernel of the map (\ref{eq:hom}).

\vspace{12pt}
2. The case $k=2$, $L\hookrightarrow \wedge^2 V$. Start with the second exterior power sequence:
\vspace{-4pt}
\beq\label{eq:sequence2}
0\ \longrightarrow\ \wedge^2V \longrightarrow\ \wedge^2B\ \longrightarrow\ B\otimes C\ \longrightarrow\ S^2C\ \longrightarrow\ 0
\vspace{-4pt}
\eeq
then twist this with $L^{^{\!*}}$ and split the resulting sequence into two short exact sequences. The associated long exact sequences in cohomology start as:
\begin{align}
&0\ \longrightarrow\ \text{Hom}_X\big(L,\wedge^2V\big) \longrightarrow\ H^0\big(\wedge^2B\otimes L^{^{\!*}}\big)\ \longrightarrow\ H^0\big(Q)\ \longrightarrow\ \ldots \\
&0\ \longrightarrow\ H^0\big(Q)\ \longrightarrow\ H^0\big(B\otimes C \otimes L^{^{\!*}}\big) \longrightarrow\ H^0\big(S^2C\otimes L^{^{\!*}}\big)\ \longrightarrow\ \ldots
\end{align}
Thus 
\vspace{-20pt}
\begin{align}\label{eq:hom2}
 \text{Hom}_X\big(L,\wedge^2V\big)& \cong \text{Ker}\left( H^0\big(\wedge^2B\otimes L^{^{\!*}}\big)\ \longrightarrow\ H^0\big(Q\big)\right)\\
 H^0(Q)&\cong \text{Ker}\left( H^0\big(B\otimes C\otimes L^{^{\!*}}\big)\ \longrightarrow\ H^0\big(S^2C\otimes L^{^{\!*}}\big)\right)
\end{align}

As before, there are two simple cohomology checks that can be performed in this case. If $h^0\big(\wedge^2B\otimes L^{^{\!*}}\big)=0$, then $L$ does not inject into $\wedge^2 V$. However, $\text{Hom}_X\big(L,\wedge^2V\big)$ is non-trivial, if 
\vspace{-8pt}
\beq\label{eq:cohcondiiton2}
h^0\big(\wedge^2B\otimes L^{^{\!*}}\big)> h^0\big(B\otimes C\otimes L^{^{\!*}}\big)
\eeq

\vspace{12pt}
3. The case $k=3$, $L\hookrightarrow \wedge^3 V\cong \wedge^2 V^{^*}$. There are two relevant sequences in this case. First, start with the dual sequence for $\wedge^2V$ twisted up with $L^{^{\!*}}$:
\vspace{-4pt}
\beq\label{eq:sequence3}
0\ \longrightarrow\ S^2C^{^*}\otimes L^{^{\!*}} \longrightarrow\ B^{^*}\otimes C^{^*} \otimes L^{^{\!*}}\ \longrightarrow\ \wedge^2 B^{^*} \otimes L^{^{\!*}}\ \longrightarrow\ \wedge^2V\otimes L^{^{\!*}}\ \longrightarrow\ 0
\vspace{-4pt}
\eeq
and split this into two short exact sequences with associated long exact sequences in cohomology:
\vspace{-4pt}
\begin{align*}
&0\ \longrightarrow\ H^0\big( S^2C^{^*}\otimes L^{^{\!*}} \big) \longrightarrow\ H^0\big(B^{^*}\otimes C^{^*} \otimes L^{^{\!*}}\big)\ \longrightarrow\ H^0\big(Q)\ \longrightarrow\ H^1\big( S^2C^{^*}\otimes L^{^{\!*}} \big) \longrightarrow\ \ldots \\
&0\ \longrightarrow\ H^0\big(Q)\ \longrightarrow\ H^0\big(\wedge^2 B^{^*} \otimes L^{^{\!*}}\big) \longrightarrow\ \text{Hom}_X\big(L,\wedge^2V^{^*}\big)\ \longrightarrow\ H^1\big(Q)\ \longrightarrow\ \ldots
\vspace{-4pt}
\end{align*}

\vspace{-8pt}
From here we can infer that $\text{Hom}_X\big(L,\wedge^3V\big) = 0$ if $h^0\big(\wedge^2 B^{^*} \otimes L^{^{\!*}}\big) = h^1\big( S^2C^{^*}\otimes L^{^{\!*}} \big)=h^2\big( S^2C^{^*}\otimes L^{^{\!*}} \big) = 0$. 

\vspace{8pt}
3'. Alternatively we can start with the third exterior power sequence:
\vspace{-4pt}
\beq\label{eq:sequence4}
0\ \longrightarrow\ \wedge^3V \longrightarrow\ \wedge^3B\ \longrightarrow\ \wedge^2B\otimes C\ \longrightarrow\ B\otimes S^2C\ \longrightarrow\ S^3C\ \longrightarrow\ 0
\vspace{-4pt}
\eeq
then split this sequence into three short exact sequences, whose long exact sequences in cohomology lead to the following indentifications:
\vspace{-4pt}
\begin{align}
 \text{Hom}_X\big(L,\wedge^3V\big)& \cong \text{Ker}\left( H^0\big(\wedge^3B\otimes L^{^{\!*}}\big)\ \longrightarrow\ H^0\big(Q_1\big)\right)\\
 H^0(Q_1)&\cong \text{Ker}\left( H^0\big(\wedge^2B\otimes C\otimes L^{^{\!*}}\big)\ \longrightarrow\ H^0\big(Q_2\big)\right)\\
  H^0(Q_2)&\cong \text{Ker}\left( H^0\big(B\otimes S^2C\otimes L^{^{\!*}}\big)\ \longrightarrow\ H^0\big(S^3C\otimes L^{^{\!*}}\big)\right)
  \vspace{-4pt}
\end{align}

\vspace{-8pt}
Thus, if $h^0\big(\!\wedge^3B\otimes L^{^{\!*}}\big)=0$, then $L$ does not inject into $\wedge^3V$. On the other hand, if 
\vspace{-4pt}
\beq\label{eq:cohcondiiton3}
h^0\big(\wedge^3B\otimes L^{^{\!*}}\big)> h^0\big(\wedge^2B\otimes C\otimes L^{^{\!*}}\big)
\vspace{-4pt}
\eeq 
then $\text{Hom}_X\big(L,\wedge^3V\big)$ is non-trivial. 

\vspace{12pt}
4. The case $k=4$, $L\hookrightarrow \wedge^4 V\cong V^{^*}$. In this case, start with the dual monad sequence twisted up with $L^{^{\!*}}$: 
\vspace{-4pt}
\begin{equation} \label{eq:dualmonad}
0\ \longrightarrow\ C^{^*}\otimes L^{^{\!*}} \longrightarrow\ B^{^*}\otimes L^{^{\!*}}\ \longrightarrow\ V^{^*}\otimes L^{^{\!*}}\ \longrightarrow\ 0  
\vspace{-4pt}
\end{equation}
and obtain the long exact sequence in cohomology:
\vspace{-4pt}
\beq
0\ \longrightarrow\ H^0\big( C^{^*}\otimes L^{^{\!*}} \big) \longrightarrow\ H^0\big(B^{^*} \otimes L^{^{\!*}}\big)\ \longrightarrow\ \text{Hom}_X\big(L,V^{^*})\ \longrightarrow\ H^1\big( C^{^*}\otimes L^{^{\!*}} \big) \longrightarrow\ \ldots
\vspace{-4pt}
\eeq
This implies the following identification:
\vspace{-8pt}
\beq
\text{Hom}_X\big(L,V^{^*}) \cong \text{Coker}\left(H^0\big( C^{^*}\otimes L^{^{\!*}} \big) \longrightarrow H^0\big(B^{^*} \otimes L^{^{\!*}}\big)  \right) \oplus \text{Ker}\left(H^1\big( C^{^*}\otimes L^{^{\!*}} \big) \longrightarrow H^1\big(B^{^*} \otimes L^{^{\!*}}\big)  \right)
\vspace{4pt}
\eeq

If $h^0\big(B^{^*} \otimes L^{^{\!*}}\big)=h^1\big( C^{^*}\otimes L^{^{\!*}} \big) = 0$, then $L$ does not inject into $V^{^*}$. However, $\text{Hom}_X\big(L,V^{^*})$ is non-trivial if 
\vspace{-12pt}
\begin{align}\label{eq:cohcondiiton4}
h^0\big( C^{^*}\otimes L^{^{\!*}} \big)&<h^0\big( B^{^*}\otimes L^{^{\!*}} \big) \text{  or }\\ 
\label{eq:cohcondiiton5}
h^1\big( C^{^*}\otimes L^{^{\!*}} \big) &>h^1\big( B^{^*}\otimes L^{^{\!*}} \big)
\vspace{-4pt}
\end{align}

\subsection{Results on the stability of $V$}\label{sec:stability1}
First, we can find those line bundles that inject into $V$ or a higher exterior power of it, based on the cohomology criteria (\ref{eq:cohcondiiton1}), (\ref{eq:cohcondiiton2}), (\ref{eq:cohcondiiton3}), (\ref{eq:cohcondiiton4}), (\ref{eq:cohcondiiton5}). Among these line bundles, some destabilise certain regions of the dual K\"ahler cone, leaving as a region of stability the following hyperplane
\beq\label{eq:hyperplane}
\begin{aligned}
  C_{\text{\bfseries s}}=\{\mathbf{s}\in \mathbb R ^4 \ | &\ \mathbf{s.}(1,1,-1,-1)= 0 \text{ and } \mathbf{s.}(1,0,0,0)\geq 0  \text{ and }\\ 
 &\ \mathbf{s.}(0,0,1,0)\geq 0  \text{ and } \mathbf{s.}(-1,1,0,0)\geq 0  \text{ and } \mathbf{s.}(1,1,-2,0)\geq 0 \}\; .
 \end{aligned}
\eeq 

This result was obtained by scanning over all line bundles with entries between $-3$ and $3$. Increasing this range of integers further did not change the region of stability defined above. 
As noted above, since the stability region is restricted to a hyper-plane, the monad bundle~$V$ is poly-stable only as a direct sum $ L_4 \oplus V'$ with structure group $S\left(U(4)\times U(1) \right)\subset SU(5)$. Here $L_4 =\cO_X(1,1,-1,-1)$.

\vspace{8pt}
In the second step, we need to worry about those line bundles for which the cohomology criteria of Section~\ref{sec:stab_criteria} are inconclusive. This discussion requires an understanding of the maps between different cohomology groups induced by the monad map and we postpone it for Section~\ref{sec:stability2}.

\section{The Particle Spectrum of Non-Abelian Models}\label{sec:monadspectrum}
\vspace{-3pt}
We now turn to the question of analysing the spectrum of the low energy theory obtained from compactifying the heterotic string on the tetraquadric Calabi-Yau threefold with a gauge background specified by a monad bundle of the type discussed above. For this, we need to understand the cohomology of $V$ and $\wedge^2 V$. 

The monad sequence
\vspace{-8pt}
\begin{equation} 
0\ \longrightarrow\ V \longrightarrow\ B\ \stackrel{f}{\longrightarrow}\ C\ \longrightarrow\ 0  
\vspace{-8pt}
\end{equation}
leads to the following long exact sequence in cohomology:
\vspace{-8pt}
\begin{equation}
\begin{array}{ccccccccc}
0& \longrightarrow & H^0(X,V)& \longrightarrow& H^0(X,B) & \longrightarrow&H^0(X,C)&\longrightarrow&\\
& \longrightarrow & H^1(X,V)& \longrightarrow& H^1(X,B) & \longrightarrow&H^1(X,C)&\longrightarrow&\\
& \longrightarrow & H^2(X,V)& \longrightarrow& H^2(X,B) & \longrightarrow&H^2(X,C)&\longrightarrow&\\
& \longrightarrow & H^3(X,V)& \longrightarrow& H^3(X,B) & \longrightarrow&H^3(X,C)&\longrightarrow&0\\
\end{array}
\vspace{-8pt}
\end{equation}

The dimensions of the cohomology groups entering in this long exact sequence are:
\vspace{-8pt}
\begin{equation}
\begin{array}{ccccccccc}
0& \longrightarrow & h^0(X,V)& \longrightarrow& \ \ \ 8 \ \ \  & \longrightarrow& \ \ \ 12 \ \ \  &\longrightarrow&\\
& \longrightarrow & h^1(X,V)& \longrightarrow& \ \ \ 8 \ \ \  & \longrightarrow& \ \ \ 0 \ \ \  &\longrightarrow&\\
& \longrightarrow & h^2(X,V)& \longrightarrow& \ \ \ 0 \ \ \ & \longrightarrow& \ \ \ 0 \ \ \  &\longrightarrow&\\
& \longrightarrow & h^3(X,V)& \longrightarrow& \ \ \ 0 \ \ \  & \longrightarrow& \ \ \ 0 \ \ \  &\longrightarrow&0\\
\end{array}
\vspace{-4pt}
\end{equation}

Thus 
\vspace{-4pt}
\begin{equation}
\begin{aligned}
H^0(X,V) & \cong \text{Ker}\left( H^0(X,B) \longrightarrow H^0(X,C)  \right)\\ 
H^1(X,V) & \cong \text{Coker}\left( H^0(X,B) \longrightarrow H^0(X,C)  \right) \oplus H^1(X,B)
\end{aligned}
\eeq

Using an explicit representation for the cohomology groups $H^0(X,B)$ and $H^0(X,C)$, as well as for the map between these groups induced by the monad map, it follows that 
\begin{equation}
\begin{aligned}
\text{dim}\, \text{Ker} & \left( H^0(X,B)  \longrightarrow H^0(X,C)  \right) = 0\\
\text{dim} \, \text{Coker} & \left( H^0(X,B)  \longrightarrow H^0(X,C)  \right) = 4
 \end{aligned}
\end{equation}

It follows that
\beq
 h^\bullet(X,V)\ \,  =\  (0,12,0,0)  
\eeq

\vspace{12pt}
Earlier on we have decided that the monad bundle $V$ can be stable only as a sum $V=L_4\oplus V'$, where $L_4=\cO_X(1,1,-1,-1)$. This corresponds to a particular choice of the monad map. As such the above cohomology calculations, performed with a generic monad map, must change. The non-split bundle $V'$ is defined by a monad sequence
\begin{equation}\label{eq:monadseqnew}
0\ \longrightarrow\ V' \longrightarrow\ B'\ \longrightarrow\ C\ \longrightarrow\ 0  
\end{equation}
\vspace{-12pt}
where
\beq\label{eq:bandc2}
B'~=~~
\cicy{ \\ \\ \\ \\ }
{ - 1 & -1 & -1 & ~~0 & ~~1& ~~1~ \\
 ~~0 & -1 & -1 & ~~1& ~~1& ~~1~ \\
~~0 & ~~2 & ~~2 & -1 &~~0 &~~0~\\
 ~~1 & ~~2 &~~ 2 & ~~0 & ~~0 &~~0~\\}\
\hskip0.35in
C~=~~
\cicy{ \\ \\ \\ \\ }
{ - 1 & ~~1~ \\
 ~~1 & ~~1~ \\
~~2 & ~~0 ~\\
 ~~2 & ~~2~\\}\; .
\eeq

Since $L_4=\cO_X(1,1,-1,-1)$ has trivial cohomology, it follows that $V$  and $V'$ have the same cohomology. A calculation similar to the above one shows that:
\beq
 h^\bullet(X,V)\ \,  = h^\bullet(X,V')\ \,  =\  (0,12,0,0)  
\eeq

\vspace{12pt}
We also need to compute the cohomology of $\wedge^2V = \left(L_4\otimes V'\right) \, \oplus\, \wedge^2 V'$. The cohomology of $L\otimes V'$ can be easily obtained by twisting the monad sequence (\ref{eq:monadseqnew}) with $L_4$:
\begin{equation}
0\ \longrightarrow\ L_4\otimes V' \longrightarrow\ L_4\otimes B'\ \longrightarrow\ L_4\otimes C\ \longrightarrow\ 0  
\end{equation}

It follows that
\beq
 h^\bullet(X,L_4\otimes V')\ \,  =\  (0,12,0,0)  
\eeq
so this part of $\wedge^2V$ takes care of the chiral asymmetry. 

\vspace{8pt}
In order to compute the cohomology of $\wedge^2 V'$, we need to use the second exterior power of the monad sequence
\beq
0\ \longrightarrow\ \wedge^2V' \longrightarrow\ \wedge^2B'\ \longrightarrow\ B'\otimes C\ \longrightarrow\ S^2C\ \longrightarrow\ 0
\eeq
which can be split in two short exact sequences, whose associated long exact sequences read:
\beq\label{eq:2sequences}
\begin{array}{ccccccccc|ccccccccc}
0 \ \ &   \!\!\!\!\longrightarrow\!\!\!\!\!\!\! & \wedge^2 V' &  \!\!\!\!\!\!\!\longrightarrow\!\!\! &\wedge^2 B' & \!\!\longrightarrow\!\!\!\!\!\!\! & Q & \!\!\!\!\!\!\longrightarrow\!\!\!\! &0\ \ \ \ \  &\ \ \ \ \  0\ \ &  \!\!\!\!\longrightarrow\!\!\!\!\!\!\! & Q & \!\!\!\!\!\!\longrightarrow\!\!\!\!  & B'\otimes C & \!\!\!\!\longrightarrow\!\!\!\! & S^2 C & \!\!\!\!\longrightarrow\!\!\!\!  &0  \\   \\[-12pt]
& & h^0(\wedge^2V')&&53&&h^0(Q)&& & & & h^0(Q)&&150&&96&&\\
& & h^1(\wedge^2V') &&85&&h^1(Q)&& & & & h^1(Q)&&134&&48&&\\
& & h^2(\wedge^2V')&&0&&h^2(Q)&& & & & h^2(Q)&&0&&0&&\\
& & h^3(\wedge^2V')&&0&&h^3(Q)&& & & & h^3(Q)&&0&&0&&\\
\end{array}
\eeq

The second short exact sequence implies that 
\begin{equation}
\begin{aligned}
H^0(X,Q) & \cong \text{Ker}\left( H^0(X,B'\otimes C) \longrightarrow H^0(X,S^2C)  \right)\\ 
H^1(X,Q) & \cong \text{Coker}  \left( H^0(X,B\otimes C) \longrightarrow H^0(X,S^2C)  \right) \\
 & \oplus \text{Ker}\left( H^1(X,B'\otimes C) \longrightarrow H^1(X,S^2C)  \right) \\ 
H^2(X,Q) & \cong \text{Coker}\left( H^1(X,B'\otimes C) \longrightarrow H^1(X,S^2C)  \right) \\
H^3(X,Q) & \cong 0
\end{aligned}
\eeq

The computation of these cohomology groups follows several stages. In the first step, we need to find the map between the line bundle sums $B'\otimes C$ and $S^2C$ induced by the monad map (\ref{eq:monadseqnew}). In the second step, we construct the induced map between various cohomology groups of the same order and compute their ranks. A comprehensive exposition on the computation of cohomology groups and of ranks of maps between them goes beyond the scope of the present paper. The interested reader can find in Ref.~\cite{Anderson:2013qca} an outline of the basic techniques for computing line bundle cohomology on complete intersection Calabi-Yau manifolds in products of projective spaces. 

We find that the map $H^0(X,B'\otimes C) \longrightarrow H^0(X,S^2C)$ has rank $94$,  while the map between $H^1(X,B'\otimes C) \longrightarrow H^1(X,S^2C)$ has maximal rank $48$. This leads to 
\beq
 h^\bullet(X,Q)\ \,  =\  (56,88,0,0)  
\eeq

The final step consists in determining the cohomology of $\wedge^2V'$. The first long exact sequence in cohomology in (\ref{eq:2sequences}) implies:
\begin{equation}
\begin{aligned}
H^0(X,\wedge^2V') & \cong \text{Ker}\left( H^0(X,\wedge^2B') \longrightarrow H^0(X,Q)  \right)\\ 
H^1(X,\wedge^2V') & \cong \text{Coker}  \left( H^0(X,\wedge^2B') \longrightarrow H^0(X,Q)  \right) \\
 & \oplus \text{Ker}\left( H^1(X,\wedge^2B') \longrightarrow H^1(X,Q)  \right) \\ 
H^2(X,\wedge^2V') & \cong \text{Coker}\left( H^1(X,\wedge^2B') \longrightarrow H^1(X,Q)  \right) \\
H^3(X,\wedge^2V') & \cong 0
\end{aligned}
\eeq

Computing these maps requires several layers of extra complication. To start with, the map $H^0(X,\wedge^2B') \longrightarrow H^0(X,Q)$ is induced by the bundle map $\wedge^2B' \longrightarrow Q$ which itself has to be determined from the monad map (\ref{eq:monadseqnew}). However, since 
\beq
H^0(X,Q)  \cong \text{Ker}\left( H^0(X,B'\otimes C) \longrightarrow H^0(X,S^2C)  \right)
\eeq
it follows that $H^0(X,Q)$ is a subspace of $H^0(X,B'\otimes C)$ and thus the map $H^0(X,\wedge^2B') \longrightarrow H^0(X,Q)$ is equivalent with the map $H^0(X,\wedge^2B') \longrightarrow H^0(X,B'\otimes C)$, with the single difference that for the latter, the target space is larger. Computing the rank of $H^0(X,\wedge^2B') \longrightarrow H^0(X,Q)$, we obtain 53. This leaves us with the following tableaux of dimensions:
\beq
\begin{array}{ccccccccc}
0 \ \ &   \longrightarrow & \wedge^2 V' &  \longrightarrow  &\wedge^2 B' & \longrightarrow& Q & \longrightarrow &0\ \ \ \ \\   \\[-12pt]
& & 0 &&53&& 56 &&  \\
& & 3+K &&85&& 88 & & \\
& & C &&0&&0&&  \\
& & 0 &&0&&0&&  \\
\end{array}
\eeq
where $K = \text{dim}\, \text{Ker}\left( H^1(X,\wedge^2B') \longrightarrow H^1(X,Q)  \right)$ and $C = \text{dim}\, \text{Coker}\left( H^1(X,\wedge^2B') \longrightarrow H^1(X,Q)  \right)$. From exactness, it follows that $C = 3 + K$. Computing the rank of this map comes with an extra level of complication, since $H^1(X,Q)$ is a direct sum of a subspace of $H^0(X,S^2C)$ and a subspace of $H^1(X, B'\otimes C)$. As such, the map $H^1(X,\wedge^2B') \longrightarrow H^1(X,Q)$ requires the knowledge of a co-boundary map.  We are currently developing the techniques needed in order to deal with such cases. 

\vspace{8pt}
Fortunately, for the present case, the information acquired so far is enough to make an important statement. We have obtained that
\beq
 h^\bullet(X,\wedge^2 V) \ \, = \  h^\bullet(X, L\otimes V') \ +\  h^\bullet(X,\wedge^2 V') \ \,  =\  (0,15+K,3+K,0)  
\eeq
At the split locus $V=U$, as given in Eq.~(\ref{eq:splitlocus}), and $h^\bullet(X,\wedge^2 U)\ \,  =\  (0,15,3,0)$. This means that moving away from the split locus within the hyperplane defined in Eq.~(\ref{eq:hyperplane}), the number of vector-like $\mathbf{5}-\overline{\mathbf{5}}$ pairs does not decrease. From a low-energy point of view, moving away from the split locus corresponds to higgsing some of the bundle moduli. In this process, it can happen that some of the massless states become massive. In any case we do not expect extra massless states to appear in the spectrum, in particular, the number of $\mathbf{5}-\overline{\mathbf{5}}$ pairs cannot increase. Thus we conclude that $K=0$. 

This result guarantees the presence of massless Higgs fields in the low-energy theory and is in agreement with the observation made at the end of Section \ref{sec:operators}. Indeed, stability forced us to considers monad bundles of the form $V = L_4\oplus V'$. Since $L_4$ splits, we are still at a locus where the VEV of the singlet field $S_{2,4}$, present in the allowed mass term $S_{2,4}\,\bar{H}\,L$, is zero. 

\vspace{8pt}
Had we not restricted our computation to the hyperplane given in Eq.~(\ref{eq:hyperplane}), the information about $\wedge^2V$ would read:
\vspace{-4pt}
\beq\label{eq:2sequences}
\begin{array}{ccccccccc|ccccccccc}
0 \ \ &   \!\!\!\!\longrightarrow\!\!\!\!\!\!\! & \wedge^2 V &  \!\!\!\!\!\!\!\longrightarrow\!\!\! &\wedge^2 B & \!\!\longrightarrow\!\!\!\!\!\!\! & Q & \!\!\!\!\!\!\longrightarrow\!\!\!\! &0\ \ \ \ \  &\ \ \ \ \  0\ \ &  \!\!\!\!\longrightarrow\!\!\!\!\!\!\! & Q & \!\!\!\!\!\!\longrightarrow\!\!\!\!  & B \otimes C & \!\!\!\!\longrightarrow\!\!\!\! & S^2 C & \!\!\!\!\longrightarrow\!\!\!\!  &0  \\   \\[-12pt]
& & 0 && 61 && 68 && & & & 68 &&162 && 96&&\\
& & 7 + K  &&93&&88&& & & & 88 &&134&&48&&\\
& & C &&0&& 0 && & & & 0 &&0&&0&&\\
& & 0 &&0&& 0 && & & & 0 &&0&&0&&\\
\end{array}
\vspace{-4pt}
\eeq
From exactness, $K = 5 + C$. We thus obtain $h^\bullet(X,\wedge^2 V) = (0,12+C,C,0)$, which is inconclusive in regard to the survival of massless $\mathbf{5}-\overline{\mathbf{5}}$ pairs away from the split locus.

\section{More on the Stability of $V$}\label{sec:stability2}
In Section~\ref{sec:stability1} we have argued that the monad bundles $V$ defined by Eq.~(\ref{eq:bandc}) can only be poly-stable as a split $L\oplus V'$ where $L=\cO_X (1,1,-1,-1)$ and $V'$ is a monad bundle defined by Eq.~(\ref{eq:bandc2}). Relying on the cohomology criteria (\ref{eq:cohcondiiton1}), (\ref{eq:cohcondiiton2}), (\ref{eq:cohcondiiton3}), (\ref{eq:cohcondiiton4}), (\ref{eq:cohcondiiton5}), we have argued that the region of stability, presented in the dual K\"ahler parameters, is contained in the three-dimensional cone (\ref{eq:hyperplane}):
\beq
\begin{aligned}
 C_{\text{\bfseries s}}=\{\mathbf{s}\in \mathbb R ^4 \ | &\ \mathbf{s.}(1,1,-1,-1)= 0 \text{ and } \mathbf{s.}(1,0,0,0)\geq 0  \text{ and }\\ 
 &\ \mathbf{s.}(0,0,1,0)\geq 0  \text{ and } \mathbf{s.}(-1,1,0,0)\geq 0  \text{ and } \mathbf{s.}(1,1,-2,0)\geq 0 \}\; .
 \end{aligned}
\eeq 

In this section we would like to argue that there are no further de-stabilising line bundles. The present situation is slightly changed: we have to deal with the $SU(4)$--bundle $V'$ and check whether there exist any line bundles injecting in $V'$, $\wedge^2 V'$ or $\wedge^3 V'\simeq V^{^*}$ that destabilise the cone $ C_{\text{\bfseries s}}$. 

As discussed in Section~\ref{sec:stab_criteria}, if a line bundle $L$ injects into $\wedge^k V'$, then $H^0\left(X, \wedge^k V' \otimes L^{^*} \right)$ is non-trivial. For the line bundles that potentially inject into $V'$ we need to compute 
\beq
 H^0\left(X, V' \otimes L^{^*} \right) \cong \text{Ker}\left( H^0\big(B'\otimes L^{^{\!*}}\big)\ \longrightarrow\ H^0\big(C\otimes L^{^{\!*}}\big)\right)
\eeq
where $B'$ and $C$ are defined in Eq.~(\ref{eq:bandc2}). We have computed this kernel for all the line bundles with entries between $-3$ and $3$, obtaining the following set of injecting line bundles:

\begin{table}[!h]
\vspace{12pt}
\begin{center}
\begin{tabular}{ c c c c c }
\varstr{14pt}{9pt} (-3, -3, -3, 2), &(-3, -3, -2, 2), &(-3, -3, -1, 2), &(-3, -3, 0, 1), &(-3, -3, 0, 2) \\ 
\varstr{14pt}{9pt}(-3, -2, -3, 1), &(-3, -2, -2, 1), &(-3, -1, -3, 1), &(-3, 1, -3, -1), &(-3, 1, -2, -1), \\
\varstr{14pt}{9pt}(-3, 1, -1, -3), &(-3, 1, -1, -2), &(-3, 1, -1, -1), &(-3, 1, 0, -3), &(-3, 1, 0, -2), \\
\varstr{14pt}{9pt}(-3, 2, -3, -3), &(-2, -3, -3, 2), &(-2, -3, -2, 2), &(-2, -3, -1, 2), &(-2, -3, 0, 1), \\
\varstr{14pt}{9pt}(-2, -3, 0, 2), &(-2, -2, -3, 1), &(-2, -1, -3, 1), &(-2, 1, -3, -1), &(-2, 1, -2, -1), \\
\varstr{14pt}{9pt}(-2, 1, -1, -3), &(-2, 1, -1, -2), &(-2, 1, -1, -1), &(-2, 1, 0, -3), &(-2, 1, 0, -2), \\
\varstr{14pt}{9pt}(-1, -3, -3, 1), &(-1, -3, -3, 2), &(-1, -3, -2, 1), &(-1, -3, -2, 2), &(-1, -3, -1, 1), \\
\varstr{14pt}{9pt}(-1, -3, -1, 2), &(-1, -3, 0, 1), &(-1, -3, 0, 2), &(-1, 1, -3, -1), &(-1, 1, -2, -1), \\
\varstr{14pt}{9pt}(-1, 1, -1, -3), &(-1, 1, -1, -2), &(-1, 1, -1, -1), &(-1, 1, 0, -3), &(-1, 1, 0, -2)
\end{tabular}
 \end{center}
 \vspace{-8pt}
 \end{table}

However, none of these line bundles de-stabilises the cone $C_{\text{\bfseries s}}$ and we believe that by increasing the range of integers the situation would remain unchanged. 

\vspace{12pt}
For the line bundles that potentially inject into $\wedge^2V'$ we need to compute 
\beq
 H^0\left(X, \wedge^2V' \otimes L^{^*} \right) \cong \text{Ker}\left( H^0\big(\wedge^2B'\otimes L^{^{\!*}}\big)\! \longrightarrow \text{Ker}\big( H^0\big(B'\otimes C\otimes L^{^{\!*}}\big)\! \longrightarrow\! H^0\big(S^2C\otimes L^{^{\!*}}\big)\big)\!\right)
\eeq
This computation is similar to that performed in Section~\ref{sec:monadspectrum} in order to decide the existence of vector-like $\mathbf{5}-\overline{\mathbf{5}}$ pairs. Although computationally challenging, we have computed the cohomology $H^0\left(X, \wedge^2V' \otimes L^{^*} \right)$ for all the line bundles with entries between $-1$ and $1$, finding no injecting line bundles.

\vspace{21pt}
Finally, for the line bundles that potentially inject into $\wedge^3V'\cong V^{^*}$ we need to compute the cohomology group $ H^0\left(X, V'^{^*} \otimes L^{^*} \right)$, which according to the discussion in Section~\ref{sec:stab_criteria} can be written as a direct sum:
\beq
 \text{Coker}\left(H^0\big( C^{^*}\otimes L^{^{\!*}} \big) \longrightarrow H^0\big(B^{^*} \otimes L^{^{\!*}}\big)  \right) \oplus \text{Ker}\left(H^1\big( C^{^*}\otimes L^{^{\!*}} \big) \longrightarrow H^1\big(B^{^*} \otimes L^{^{\!*}}\big)  \right)
\vspace{4pt}
\eeq

We have performed this computation for all the line bundles with entries between $-3$ and $3$, obtaining the following set of injecting line bundles:
\begin{table}[!h]
\begin{center}
\begin{tabular}{ c c c c c }
\varstr{14pt}{9pt} (-3, -3, 0, 1),\, &(-3, -2, -3, 1),\, &(-3, -2, -2, 1),\, &(-3, -2, -1, 1),\, &(-3, -2, 0, 1)
\end{tabular}
 \end{center}
 \vspace{-12pt}
 \end{table}

As before, none of these line bundles de-stabilises the cone $C_{\text{\bfseries s}}$.

{\setstretch{1.33}
\chapter{Conclusion}\label{Conclusion}

The scope of this thesis was two-fold. In the first half, I have discussed several mathematical aspects of Calabi-Yau manifolds, followed, in the second half, by the exposition of an effective approach to string phenomenology based on smooth Calabi-Yau compactifications of the heterotic string. In the following, I will summarise the main results of the previous chapters and discuss a few directions which, I hope, will be the subject of future work. 

In Chapter \ref{ToricCY}, we looked at the class of compact Calabi-Yau threefolds realised as hypersurfaces in toric varieties. Such manifolds correspond to (triangulations of) four-dimensional reflexive polytopes; these have been completely classified by Kreuzer and Skarke. Among the 4-polytopes in the Kreuzer-Skarke list, there are many which contain three-dimensional reflexive sub-polytopes, both as a slice and as a projection. Such 4-polytopes correspond to $K3$ fibrations over $\IC\IP^1$, for which the three-dimensional sub-polytope describes the $K3$ fiber. The $K3$-slice divides the 4-polytope into two halves, a top and a bottom. By finding all possible tops over a given $K3$ slice, one can construct a large number of $K3$ fibrations. In Chapter \ref{ToricCY}, I have presented this construction for three different $K3$-polytopes, which were, in turn, elliptic fibrations over~$\IC\IP^1$. I~have also explained that, under certain assumptions on the $K3$-polytope, the operation of mixing and matching tops enjoys a remarkable additive property for the Hodge numbers. This property underlies much of the fractal-like structure present in the Hodge plot associated with the 4-polytopes in the Kreuzer-Skarke list.

An interesting question which was not addressed in our discussion is related to the geometrical interpretation of the polytope surgery described in Chapter \ref{ToricCY} and briefly summarised above. One could start by asking what is the toric interpretation of a top. Certainly, given a top and its dual (see the discussion in Section \ref{ComposingTops}), we can construct the fan (and the corresponding toric variety) associated with the top, assign a polynomial with the dual bottom, and construct a hypersurface as the zero locus of this polynomial.  However, since in this case the fan does not fill the whole space, the toric variety is non-compact. Accordingly, the hypersurface is non-compact; however, it retains the fibration structure, this time over $\IC$ (see \cite{Bouchard:2006ah}). For the sake of clarity, say we discuss $K3$-fibrations. If the top comes represents half of a reflexive polytope, it will encode the geometry of the Calabi-Yau manifold away from the preimage of the point $\infty$ of the $\IC\IP^1$ base. Thus the process of exchanging the top of a reflexive polytope with a different one should correspond to a manifold surgery.

A different avenue which descends from the work presented in Chapter \ref{ToricCY} is that of constructing large classes of elliptically fibered Calabi-Yau four-folds, relevant for phenomenological applications of F-theory. This work is currently in progress.

\vspace{20pt}

In Chapter \ref{Z3Quotients}, we looked at Calabi-Yau threefolds that admit free actions of finite groups. In particular, we discussed the class of complete intersection Calabi-Yau (CICY) threefolds embedded in products of projective spaces that admit smooth quotients by $\IZ_3$. These manifolds, as well as their quotients, form a web which is connected by conifold transitions. 

Discrete symmetries of Calabi-Yau threefolds are essential for string phenomenology, in particular for compactifications of the heterotic string. Such symmetries appear in the model building programme discussed in Chapter \ref{LineBundles}.  For phenomenological applications, it would be very important to have a complete classification of freely acting discrete symmetries on the larger class of toric Calabi-Yau three-folds. 

\vspace{20pt}
In the second half of this thesis, I have discussed the construction of heterotic models on favourable complete intersection Calabi-Yau (CICY) manifolds with freely-acting discrete symmetries and holomorphic vector bundles realised as direct sums of line bundles. In Chapter~\ref{LineBundles} I have outlined the construction of line bundle models, emphasising its versatility in regard to automatisation and its phenomenological advantages. The split nature of the vector bundle provides a geometrical set-up that is relatively simple to deal with from a computational point of view and allows for an algorithmic construction of heterotic models. Moreover, the GUT models resulted from heterotic compactifications with rank five line bundle sums contain, apart from an $SU(5)$ gauge group, several global $U(1)$ symmetries which can eliminate undesired operators, such as dimension 4 and 5 proton decay operators. 

In the final part of Chapter~\ref{LineBundles}, I have presented the results of a comprehensive scan based on the line bundle algorithm. The scan considered $68$ CICY manifolds with $1<h^{1,1}(X)<7$. These manifolds are part of the standard list of CICY three-folds~~\cite{Candelas:1987kf,Green:1987cr} available at~\cite{database} and are the only favourable ones that admit linearly realised free actions of finite groups. For the purpose of the scan only the order of such finite groups mattered. By considering rank five line bundle sums over these manifolds, the scan ran over about $10^{40}$ configurations. For each manifold and finite group order the scan ran until no more consistent and physically viable GUT models could be found, leading to a database of $63,325$ such models. This provides us with a computational proof of the finiteness of the class of models realised within this framework. This class consists of models with poly-stable line bundle sums whose chiral asymmetries have the correct values to produce a standard model upon taking the quotient by the freely-acting symmetry and including the Wilson line. Requiring, in addition, the absence of $\overline{\bf 10}$ multiplets and the presence of at least one ${\bf 5}$--$\overline{\bf 5}$ pair to account for Higgs doublets, the number of viable models is reduced to about 35,000. In turn, these GUT models will lead to a significantly larger number of standard models, possibly by an order of magnitude, once the GUT group is broken by the inclusion of Wilson lines. The line bundle database represents the largest set of string GUT models to date. As outlined in the beginning of Chapter~\ref{LineBundles}, it is crucial to be in possession of large datasets of models exhibiting the right symmetries and particle spectrum, before one starts looking at the more detailed features such as proton decay, the $\mu$-problem and the structure of Yukawa-couplings. 

Of course, the task ahead involves constructing the standard models associated to these GUT models. A number of technical obstacles have to be overcome in order to complete this task, notably devising and implementing a complete algorithm for computing (equivariant) line bundle cohomology on CICYs. This work is currently in progress. 

\vspace{20pt}
In Chapter~\ref{TQ1}, I discussed the finiteness of the class of consistent and physically viable line bundle models on the tetraquadric hypersurface. Two arguments were presented. The first line of argument, presented in Section~\ref{sec:physbound} and applicable to line bundle sums of arbitrary rank relied on two physical assumptions which restrict the K\"ahler cone. These assumptions are connected to the validity of the supergravity approximation and the physical bound on the Calabi-Yau volume. The second argument made use of stability and topological constraints only. It is hard to see how this second argument can generalise beyond the tetraquadric manifold. However, the question related to the finiteness of the class of bundles that are poly-stable somewhere in the interior of the positive K\"ahler cone  for a fixed total Chern class remains as an important and interesting mathematical problem. To our knowledge, this question is currently open.  

\vspace{20pt}
The final chapter contained a study of the moduli space of non-Abelian bundles around the locus where the bundle splits as a direct sum of line bundles. The monad construction, used in order to investigate the moduli space of non-Abelian bundles, provided a full  description of poly-stable $S\left(U(4)\times U(1) \right)$--bundles leading to GUT models with the correct field content in order to induce standard-like models. These deformations represent a class of consistent models that has co-dimension one in K\"ahler moduli space. In this chapter, the effort concentrated around proving stability for monad bundles and deriving the particle spectrum, in particular deciding whether the Higgs fields remain perturbatively massless after deforming the line bundle sum to arbitrary, $S\left(U(4)\times U(1) \right)$--monad bundles. 
Extension sequences might provide a description of the full moduli space of $SU(5)$--bundles, which I hope, will be the subject of future work.

\subsection*{Addendum (five years later)}
The ideas presented in this thesis have been developed in various directions over the following years. The nested fibration structures described in Chapter \ref{ToricCY} and found in abundance in the Kreuzer-Skarke list played an important role in the study of G-theory \cite{Candelas:2014jma, Candelas:2014kma}. Generalising F-theory, the framework known as G-theory can describe non-geometric compactifications of type IIB string theory with flux potentials. 
The work presented in Chapter \ref{Z3Quotients} was continued in Refs.~\cite{Candelas:2015amz, Constantin:2016xlj, Candelas:2016fdy} which completed the study of free quotients of CICY threefolds. 

The class of heterotic line bundle models discussed in Chapters \ref{LineBundles}, \ref{TQ1} and \ref{ModuliSpace} provided a fruitful ground for the study of a number of topics in string phenomenology topics: 
understanding the relevance of Abelian fluxes in heterotic compactifications in order to address well-known problems of supersymmetric GUT models, such as fast proton decay, the $\mu$-problem and obtaining a hierarchy of Yukawa couplings \cite{Buchbinder:2014qda, Buchbinder:2014sya}; 
understanding the role of enhanced symmetry loci in the moduli space for certain classes of effective field theories derived from the heterotic string \cite{Buchbinder:2013dna, Buchbinder:2014sya}; 
understanding the role played by intermediate grand unified theories and Wilson lines in string derived standard models, in particular understanding that direct breaking from $E_8$ to $G_{\rm SM}$ cannot lead to a correct physical spectrum \cite{Anderson:2014hia}; 
realising phenomenologically viable QCD axions in string models \cite{Buchbinder:2014qca}; 
realising Yukawa unification for a single family of particles, in contrast with traditional GUTs where Yukawa unification is enforced for all families \cite{Buchbinder:2016jqr};
understanding why there are three families of particles and how the answer to this question might be related to the smallness of the gauge unification coupling \cite{Constantin:2015bea}; 
devising an analytic method to calculate the matter field Kähler metric, and hence physical Yukawa couplings, in heterotic compactifications on smooth Calabi-Yau three-folds with Abelian internal gauge fields which relies on the localisation of matter field normalisation integrals in the presence of large internal gauge flux \cite{Blesneag:2018ygh}. The existence of explicit line bundle valued cohomology formulae on Calabi-Yau manifolds illustrated in Chapter~\ref{TQ1} for the tetra-quadric three-fold was found in \cite{Constantin:2018hvl} to be a more general phenomenon. The origin of these formulae remains an open question. 

}

\bibliographystyle{utcaps}

\providecommand{\href}[2]{#2}\begingroup\raggedright\endgroup

\end{document}